\documentclass[pdftex,twocolumn,epjc3]{svjour3}          
\pdfoutput=1

\smartqed  

\RequirePackage{graphicx}
\RequirePackage{mathptmx}      
\RequirePackage{flushend}
\RequirePackage[numbers,sort&compress]{natbib}
\RequirePackage[colorlinks,citecolor=blue,urlcolor=blue,linkcolor=blue]{hyperref}
\RequirePackage{atlasphysics}
\RequirePackage{array}
\RequirePackage{multirow}
\RequirePackage{makeidx}
\RequirePackage{amsmath}
\RequirePackage{amssymb}
\RequirePackage[refpage]{nomencl}
\RequirePackage{subfigure}

\makenomenclature 

\journalname{CERN-PH-EP-2011-078~~~~~~~~~~~~~Submitted to Eur. Phys. J. C~~~~~~~~~~~~~~~~~~~~}

\begin{document}
\title{Performance of the ATLAS Trigger System in 2010\thanksref{t1}}

\author{The ATLAS Collaboration}

\thankstext[$\star$]{t1}{e-mail: atlas.publications@cern.ch}

\institute{CERN, 1211 Geneva 23, Switzerland}

\date{Received: date / Accepted: date}

\maketitle

\begin{abstract}
Proton-proton collisions at \sTev{7}\ and heavy ion collisions at \sHi\ were produced by the LHC and recorded using the ATLAS experiment's trigger system in 2010.   The LHC is designed with a maximum \linebreak bunch crossing rate of 40~MHz and the ATLAS trigger system is designed to record approximately 200 of these per second.   The trigger system selects events by rapidly identifying signatures of muon, electron, photon, tau lepton, jet, and $B$ meson candidates, as well as using global event signatures, such as missing transverse energy. An overview of the ATLAS trigger system, the evolution of the system during 2010 and the performance 
of the trigger system components and selections based on the 2010 collision data are shown. A brief outline of plans for the trigger system in 2011 is presented.
\end{abstract}

\section{Introduction}\label{sec:introduction}
\def \figurepath{.}

\begin{sloppypar}
ATLAS~\cite{DetectorPaper} is one of two general-purpose experiments 
recording LHC~\cite{LHCPaper} collisions  
to study the Standard Model (SM) and search for physics beyond the SM. 
The LHC is designed to operate at a centre of mass energy of \sTev{14} in proton-proton (\pp) collision mode with an instantaneous luminosity $\Lum = \Lumi{34}$ and at \sHi\ in heavy-ion (\pbpb) collision mode with $\Lum = \Lumi{31}$. 
The LHC  started single-beam 
operation in 2008 and achieved first collisions in 2009.  During a prolonged 
period of \pp\ collision operation in 2010 at \sTev{7}, ATLAS collected 45~\ipb of data
with luminosities ranging from 
\Lumi{27} to $2\cdot\Lumi{32}$. The \pp\ running was
followed by a short period of heavy ion running 
at \sHi\, in which ATLAS collected 9.2~\imub of \pbpb\ collisions.
\end{sloppypar}

Focusing mainly on the \pp\ running, the performance of the ATLAS trigger system during 2010 LHC operation is 
presented in this paper.  The ATLAS trigger system is designed to record events at approximately 200~Hz from the LHC's 40~MHz bunch crossing rate.    The system has three levels; the first level ({\it L1}) is a hardware-based system using information from the calorimeter and muon sub-\linebreak detectors, the second ({\it L2}) and third (\emph{Event Filter}, {\it EF}) levels   are software-based systems using information from all sub-detectors.  Together, L2 and EF are called the 
\emph{High Level Trigger} ({\it HLT}). 
\nomenclature{\bf HLT}{High Level Trigger comprising L2 and the EF}
\nomenclature{\bf L2}{Second level trigger}
\nomenclature{\bf EF}{Event Filter, the third level trigger}

For each bunch crossing, the 
trigger system verifies if at least one of hundreds of conditions 
(\emph{triggers}) is satisfied.   The triggers  
are based on identifying combinations of candidate physics objects (\nomenclature{\bf signature}{a candidate physics object identified by the trigger system}\emph{signatures}) such as 
electrons, photons, muons, jets, jets with $b$-flavour tagging ($b$-jets) 
or specific $B$-physics decay modes.  In addition, there are triggers for inelastic \pp\ collisions
(\emph{minbias}) and triggers based on global event properties such as 
missing transverse energy (\MET) and summed transverse energy (\SumET).
 
In Section~\ref{sec:overview}, following a  brief introduction to the ATLAS detector, 
an overview of the ATLAS trigger system is given and the terminology  used in the remainder of the paper is explained.
Section~\ref{sec:commissioning} presents a description of the trigger system commissioning with cosmic 
rays, single-beams, and collisions.   Section~\ref{sec:level1} provides a brief description of the L1
trigger system.   Section~\ref{sec:reconstruction} introduces 
the reconstruction algorithms used in the HLT to process information from 
the calorimeters, muon spectrometer, and inner detector tracking detectors. 
The performance of the trigger signatures, including rates and efficiencies, is described in Section~\ref{sec:signatures}.  
Section~\ref{sec:global} describes the overall performance of the trigger system.  
The plans for the trigger system operation in 2011 are described in Section~\ref{sec:outlook}.

\section{Overview}\label{sec:overview}
\def \figurepath{.}
The ATLAS detector~\cite{DetectorPaper} shown in Fig.~\ref{fig:atlas}, has a cylindrical geometry\footnote{The ATLAS coordinate system has its origin at the nominal interaction point at the centre of the detector and the $z$-axis coincident with the beam pipe, such that pseudorapidity $\eta \equiv -\ln ( \tan \tfrac{\theta}{2} )$. The positive x-axis is defined as pointing from the interaction point towards the centre of the LHC ring and the positive y-axis is defined as pointing upwards. The azimuthal degree of freedom is denoted $\phi$.} which covers almost the entire solid angle around the nominal interaction point.  Owing to its cylindrical geometry, detector components are described as being part of the 
\emph{barrel} if they are in the central region of pseudo-rapidity or part of the \emph{end-caps} if they are in the forward regions. The ATLAS detector is composed of the following sub-detectors: 
\begin{figure*}[!ht]
    \centering
    \includegraphics[width=0.9\textwidth]{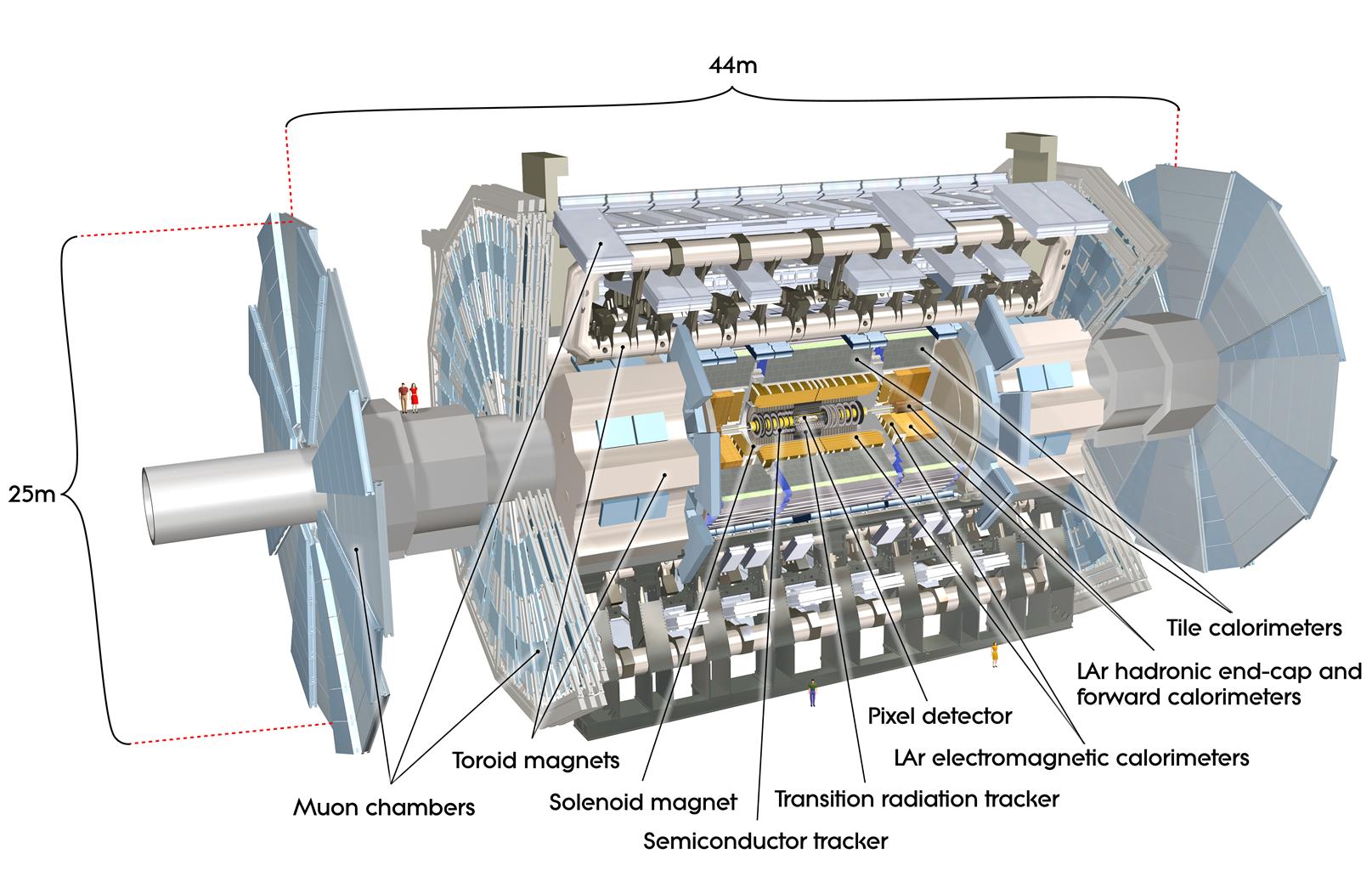} 
    \caption{The ATLAS detector} 
    \label{fig:atlas} 
\end{figure*}
\begin{description}
  \setlength{\itemsep}{1pt}
  \setlength{\parskip}{0pt}
  \setlength{\parsep}{0pt}
\item \emph{Inner Detector:} \nomenclature{\bf ID}{Inner Detector, comprising the pixel detector, SCT and TRT} The 
\emph{Inner Detector} tracker ({\it ID}) consists of a silicon
\emph{pixel} detector nearest the beam-pipe, 
surrounded by a \nomenclature{\bf SCT}{SemiConductor Tracker} \emph{SemiConductor Tracker} ({\it SCT}) and a 
\nomenclature{\bf TRT}{Transition Radiation Tracker} \emph{Transition Radiation Tracker} ({\it TRT}).  Both 
the Pixel and SCT cover the region $|\eta|<2.5$, while the TRT covers $|\eta|<2$.  The ID is contained in a 2 Tesla solenoidal magnetic field. 
Although not used in the L1 trigger system, tracking information is a key ingredient of the HLT.
\begin{sloppypar}
\item \emph{Calorimeter:} The calorimeters cover the region $|\eta|<4.9$ and consist of \emph{electromagnetic} 
\nomenclature{\bf EM calorimeter}{Electromagnetic calorimeter} ({\it EM}) and \emph{hadronic} 
\nomenclature{\bf HCAL}{Hadronic Calorimeter} ({\it HCAL}) calorimeters.
The EM, \emph{Hadronic End-Cap} 
\nomenclature{\bf HEC}{Hadronic End-Cap calorimeter} 
({\it HEC}) and \emph{Forward Calorimeters} 
\nomenclature{\bf FCal}{Forward Calorimeter} ({\it FCal}) use a Liquid Argon and absorber technology 
\nomenclature{\bf LAr calorimeter}{Liquid Argon calorimeter} ({\it LAr}). The central hadronic calorimeter is
based on steel absorber interleaved with plastic scintillator 
\nomenclature{\bf Tile calorimeter}{central hadronic calorimeter}
({\it Tile}). A 
\nomenclature{\bf presampler}{a thin calorimeter located in front of the EM calorimeter} 
\emph{presampler} is installed in front of the
EM calorimeter for $|\eta|<1.8$.   There are two separate readout paths: one with coarse granularity (\emph{trigger towers}) used by L1, and one with fine granularity used by the HLT and offline reconstruction.
\end{sloppypar}
\item \emph{Muon Spectrometer:}  The \emph{Muon Spectrometer} 
\nomenclature{\bf MS}{Muon Spectrometer, comprising the RPC, TGC, MDT and CSC}
(\emph{MS}) detectors are 
mounted in and around air core toroids that generate an average field of 0.5~T in the barrel and 1~T in the end-cap regions.
Precision tracking information is provided by 
\nomenclature{\bf MDT}{Monitored Drift Tubes}
\emph{Monitored Drift Tubes} ({\it MDT}) over the
region $|\eta| < 2.7$ ($|\eta| < 2.0$ for the innermost layer) and by \emph{Cathode Strip Chambers} 
\nomenclature{\bf CSC}{Cathode Strip Chambers}
({\it CSC}) in the region $2 < |\eta| < 2.7$. Information is provided 
to the L1 trigger system by the \emph{Resistive Plate Chambers} 
\nomenclature{\bf RPC}{Resistive Plate Chambers} 
({\it RPC}) in the barrel ($|\eta|< 1.05$)
and the \emph{Thin Gap Chambers} 
\nomenclature{\bf TGC}{Thin Gap Chambers} 
({\it TGC}) in the end-caps ($1.05 <|\eta|< 2.4$).

\item \emph{Specialized detectors:}  
\nomenclature{\bf BPTX}{electrostatic beam pick-up devices}
 Electrostatic beam pick-up devices ({\it BPTX}) are located  at  $z=\pm$175~m. 
The 
\nomenclature{\bf BCM}{Beam Conditions Monitor}
\emph{Beam Conditions Monitor} ({\it BCM}) consists of two stations containing diamond sensors located at $z=\pm$1.84~m, corresponding to $|\eta| \simeq 4.2$.
There are two forward detectors, the 
\nomenclature{\bf LUCID}{forward detector for luminosity measurement}
{\it LUCID} Cerenkov 
counter covering $5.4<|\eta|<5.9$ and the 
\nomenclature{\bf ZDC}{Zero Degree Calorimeter}
 \emph{Zero Degree Calorimeter} ({\it ZDC}) covering $|\eta|>8.3$. The 
\nomenclature{\bf MBTS}{Minimum Bias Trigger Scintillators}
 \emph{Minimum Bias Trigger Scintillators} ({\it MBTS}), consisting of two scintillator wheels with 32 
counters mounted in front of the calorimeter end-caps, cover $2.1<|\eta|<3.8$.  

\end{description}

When operating at the design luminosity of \Lumi{34} the 
LHC will have a 
40~MHz bunch crossing rate, with an average of 25 interactions per bunch 
crossing. The 
purpose of the 
\nomenclature{\bf trigger system}{the hardware and software implementing the trigger selection}
trigger system is to reduce this input rate to an output rate of about 
200~Hz for 
recording and offline processing. This limit, corresponding to an average data rate of
$\sim$300~MB/s, is determined by the computing resources for
offline storage and processing of the data. It is possible to record data at 
significantly higher rates for short periods of time.  
For example, during 2010 running there were physics benefits from running 
the trigger system with output rates of up to $\sim$600~Hz.
During runs with instantaneous luminosity 
$\sim\Lumi{32}$, the average event size was $\sim$1.3~MB.

\begin{figure}[!h]
    \centering
    \includegraphics[height=7.5cm]{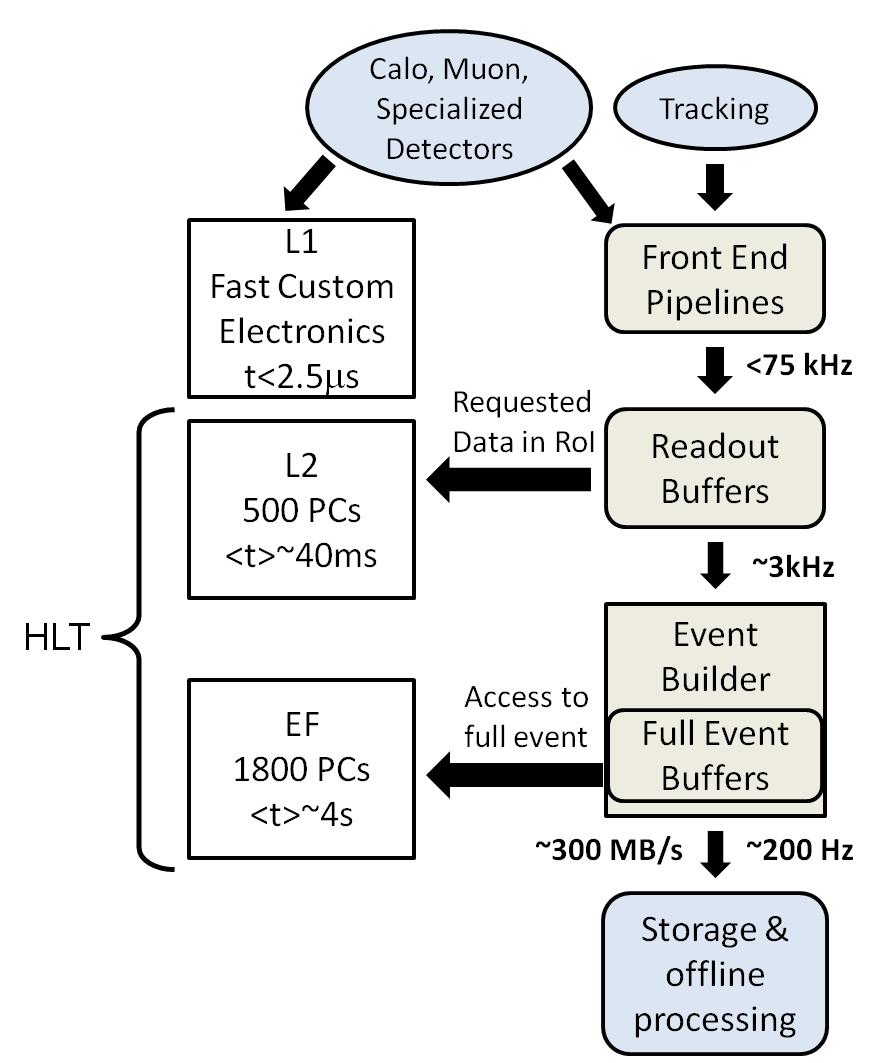} 
    \caption{Schematic of the ATLAS trigger system} 
    \label{fig:TriggerSchematic} 
\end{figure}

A schematic diagram of the ATLAS trigger system is shown in 
Fig.~\ref{fig:TriggerSchematic}. Detector signals 
are stored in front-end pipelines pending a decision from the 
\nomenclature{\bf L1}{first level trigger}
L1 trigger system. In order to achieve a latency of less than 2.5~$\mu$s, the L1 trigger system is 
implemented in fast custom electronics. The L1 trigger system is designed to reduce 
the rate to a maximum of 75~kHz. In 2010 running, the maximum L1 rate did not exceed
30~kHz. 
\nomenclature{\bf RoI}{Region of Interest, geometrical regions of the detector identified by L1}
In addition to performing the first selection step, the 
L1 triggers identify 
\emph{Regions of Interest (RoIs)} within the detector to be 
investigated by the HLT. 

The 
HLT consists of farms of commodity processors connected by  
fast dedicated 
networks (Gigabit and 10 Gigabit Ethernet). During 2010 running, the 
HLT processing farm consisted of about 800 nodes  
configurable as either L2 or EF 
and 300 dedicated EF nodes. Each node consisted of eight processor cores, the majority with a 2.4~GHz clock speed. The system is designed to expand to about 500 L2 nodes and 1800 EF nodes for running at LHC design luminosity. 
When an event is accepted by the L1 trigger
\nomenclature{\bf L1 accept}{trigger decision at L1 to accept the event for further investigation at the HLT} (referred to as an \emph{L1 accept}), data from each detector are transferred to the 
detector-specific Readout Buffers (ROB)
\nomenclature{\bf ROB}{Readout Buffer}
, which store the event in fragments pending the L2 decision. One or more ROBs are grouped
into Readout Systems (ROS) \nomenclature{\bf ROS}{Readout Systems} which are connected to the HLT networks. 
The L2 selection is based on 
fast custom algorithms processing partial event data 
within the RoIs identified by L1. The L2 processors request data from the ROS 
corresponding to detector elements inside each RoI, reducing the amount of data to be 
transferred and 
processed in L2 to 2--6\% of the total data volume.
The L2 triggers reduce the rate to $\sim$3~kHz with 
an average processing time of $\sim$40~ms/event.   Any event with an L2 processing time exceeding 5~s is recorded as a \emph{timeout} event.
During runs with instantaneous luminosity $\sim\Lumi{32}$, the average processing time of L2 was $\sim$50~ms/event (Section~\ref{sec:global}).  

The Event Builder assembles all event fragments from the ROBs for events accepted
by L2, providing full event information to the EF. The EF is mostly based on offline 
algorithms invoked from custom interfaces for running in the trigger system.  The EF is
designed to reduce the rate to ~$\sim$200~Hz with an average processing time of $\sim$4~s/event.  Any event with an EF processing time exceeding 180~s is recorded as a \emph{timeout} event.
During runs with instantaneous luminosity $\sim\Lumi{32}$, the average processing time of EF was $\sim$0.4~s/event (Section~\ref{sec:global}).

\begin{figure}[!h]
  \centering
    \includegraphics[height=7.5cm]{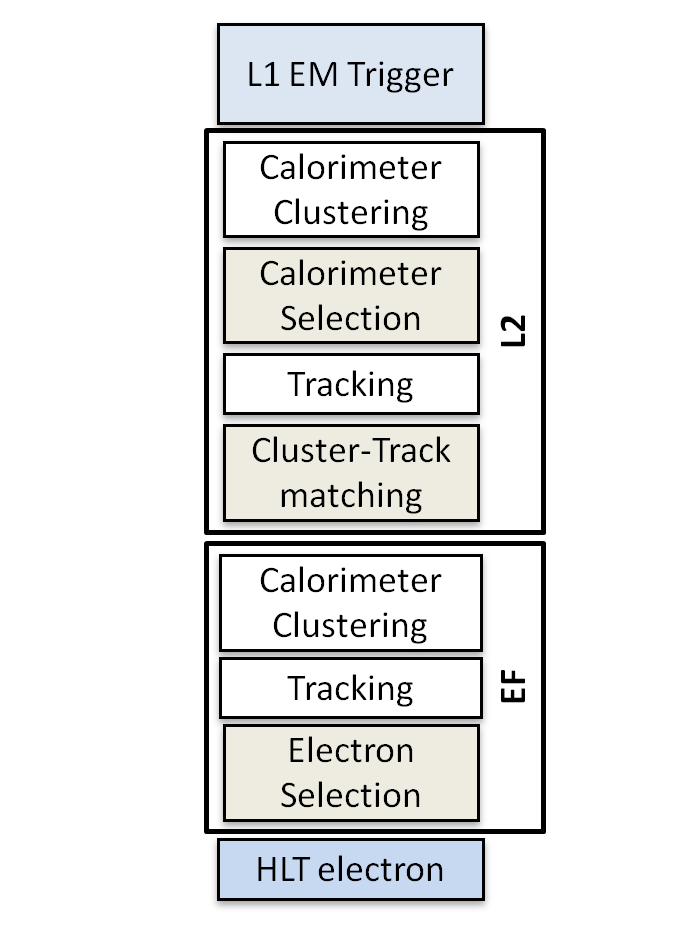} 
    \caption{Electron trigger chain} 
    \label{fig:ElectronChain} 
\end{figure}

\begin{table*}[!ht]
\begin{center}
  \caption{The key trigger objects, the shortened names used to represent 
them in the trigger menu at L1 and the HLT, and the L1 thresholds used for each trigger signature in the menu at \Lum=\Lumi{32}. Thresholds are applied to \et\ for calorimeter triggers and \pT\ for muon triggers}
    \begin{tabular}{lllrrrrrrrr}
      \hline 
           & \multicolumn{2}{l}{Representation} & \multicolumn{8}{c}{}     \\
      
      Trigger Signature            & L1    & HLT  & \multicolumn{8}{l}{L1 Thresholds [\GeV]} \\
      \hline \hline
      electron                  & EM    & e    & 2  & 3  & 5  & 10 & 10i & 14 & 14i & 85  \\
      photon                    & EM    & g    & 2  & 3  & 5  & 10 & 10i & 14 & 14i & 85  \\
      muon                      & MU    & mu   & 0  & 6  & 10 & 15 & 20 & \multicolumn{3}{r}{}\\
      jet                       & J     & j    & 5  & 10 & 15 & 30 & 55  & 75 & 95  & 115 \\
      forward jet               & FJ    & fj   & 10 & 30 & 55 & 95 & \multicolumn{4}{r}{} \\
      tau                       & TAU   & tau  & 5  & 6  & 6i & 11 & 11i & 20 & 30  & 50  \\
      \MET\ & XE    & xe   & 10 & 15 & 20 & 25 & 30  & 35 & 40  & 50  \\
      \SumET\              & TE    & te   & 20 & 50 & 100 & 180 &\multicolumn{4}{r}{}\\
      total jet energy          & JE    & je   & 60 & 100& 140 & 200 &\multicolumn{4}{r}{}\\
      $b$ jet\footnotemark[1] &   ---  & b   &  & &  &  &\multicolumn{4}{r}{}\\
      MBTS                      & MBTS  & mbts & \multicolumn{8}{r}{} \\
      BCM                       & BCM   & --- & \multicolumn{8}{r}{} \\
      ZDC                       & ZDC   & --- & \multicolumn{8}{r}{} \\
      LUCID                     & LUCID & --- & \multicolumn{8}{r}{} \\
      Beam Pickup (BPTX)        & BPTX  & --- & \multicolumn{8}{r}{} \\
      \hline 
    \end{tabular} 
  \label{tab:ExampleMenu}
  \end{center}
\end{table*}
\footnotetext[1]{The HLT b-jet trigger requires a jet trigger at L1, see Section~\ref{sec:bjet}.}

Data for events selected by the trigger system
are written to inclusive data \emph{streams} based on the trigger type.  
There are four primary physics 
\nomenclature{\bf stream}{a subset of recorded data based on trigger conditions}
streams, \emph{Egamma}, \emph{Muons}, 
\emph{JetTauEtmiss}, \emph{MinBias}, plus several additional calibration streams.  Overlaps and rates for these streams are shown in Section~\ref{sec:global}.
About 10\% of events are written 
to an \emph{express} stream where prompt offline reconstruction 
provides calibration and Data Quality 
\nomenclature{\bf DQ}{Data Quality} (DQ) 
information prior to the reconstruction 
of the physics streams. 
In addition to writing complete events to a stream, it is also 
possible to write partial information from 
one or more sub-detectors into a stream. 
Such events, used for detector calibration, are
written to the calibration streams.

The trigger system is configured via a trigger 
\nomenclature{\bf menu}{defines the configuration of the trigger system} 
\emph{menu} which defines trigger 
\nomenclature{\bf chain}{a sequence of reconstruction and selection steps} 
\emph{chains} that start from a L1 trigger and specify a sequence of reconstruction and 
selection steps for the specific trigger signatures required in the trigger chain.  A trigger chain is often referred to simply as a trigger.
\nomenclature{\bf trigger}{a condition that, when satisfied, causes the event to be recorded}
Figure~\ref{fig:ElectronChain} shows an 
illustration of a trigger chain to select electrons.   Each chain is composed of 
\nomenclature{\bf FEX}{Feature Extraction algorithm that reconstructs features such as calorimeter clusters} \emph{Feature Extraction (FEX)} algorithms which create the objects (like calorimeter clusters) and 
\nomenclature{\bf HYPO}{Hypothesis algorithm that applies selection criteria} 
\emph{Hypothesis (HYPO)} algorithms that apply selection criteria to the objects (e.g. transverse momentum greater than 20~\GeV).  
\nomenclature{\bf caching}{features extracted in one chain can be re-used in another chain}
\emph{Caching} in the trigger system allows features extracted from one chain to be re-used in another chain, reducing both the data access and processing time of the trigger system.

Approximately 500 triggers are defined in the current trigger menus.
Table~\ref{tab:ExampleMenu} shows the key physics objects identified by the  
trigger system and gives the shortened representation used in the trigger menus. Also shown are the 
L1 thresholds applied to \emph{transverse energy} (\et ) for calorimeter triggers
and \emph{transverse momentum} (\pT ) for muon triggers. 
The menu is composed of a number of different classes of trigger:
\begin{description}
  \setlength{\itemsep}{1pt}
  \setlength{\parskip}{0pt}
  \setlength{\parsep}{0pt}
\item \emph{Single object triggers:} used for final states with at least one characteristic object.  For example, a 
single muon trigger with a nominal $6\GeV$ threshold is referred to in the trigger menu as mu6. 

\item \emph{Multiple object triggers:} used for final states with two or more characteristic objects of the same type.
  For example, di-muon triggers for selecting \Jmm\ decays.  Triggers requiring a multiplicity of two or more 
are indicated in the trigger menu by prepending the multiplicity to the trigger name, as in, 2mu6.

\item \emph{Combined triggers:} used for final states with two or more characteristic objects of different types. For
example, a $13\GeV$ muon plus $20\GeV$ missing transverse energy (\MET) trigger for selecting \Wmunu\ decays would be denoted mu13\_xe20. 
\nomenclature{\bf combined trigger}{trigger composed of two or more signatures of different types, e.g. muon plus electron}

\item \emph{Topological triggers:} used for final states that require selections based
on information from two or more RoIs. For example the \Jmm\ trigger
combines tracks from two muon RoIs. 
\nomenclature{\bf topological trigger}{trigger that combines information from two or more RoIs} 
\end{description}

\noindent When referring to a particular level of a trigger, the level (L1, L2 or EF) appears as a prefix, so L1\_MU6 refers to the L1 trigger item with a $6\GeV$ threshold and L2\_mu6 refers to the L2 trigger item with a $6\GeV$ threshold. 
A name without a level prefix refers to the whole trigger chain.

Trigger rates can be controlled by changing thresholds or applying different sets of selection cuts.  
The selectivity of a set of cuts applied to a given trigger object in the menu
is represented by the terms \emph{loose}, \emph{medium}, and \emph{tight}.  This selection criteria is suffixed to the trigger name, for example e10\_medium.
Additional requirements, such as \emph{isolation}, can also be imposed to reduce the
rate of some triggers.   Isolation is a measure of the amount of energy or number of particles near a signature.  For example, the amount of transverse energy (\et) deposited in the calorimeter within 
$\dR \equiv \sqrt{(\Delta\eta)^2 +(\Delta\phi)^2}<0.2$ of a muon is a measure of the muon isolation.  
Isolation is indicated in the trigger menu by an i appended to the trigger name (capital I for L1), 
for example L1\_EM20I or e20i\_tight.  Isolation was not used in any primary  triggers in 2010 (see below).

\nomenclature{\bf prescale}{parameter, N, applied to a trigger such that only 1 in N events passing the trigger is accepted}
Prescale factors can be applied to each L1 trigger 
and each HLT chain, such that only 1 in N events passing the trigger causes an event to be accepted 
at that trigger level. Prescales can also be set so as to disable specific chains.
Prescales control the rate and composition 
of the express stream. A series of L1 and HLT prescale sets, covering a range of 
luminosities, are defined to accompany each menu. These prescales are auto-generated
based on a set of rules that take into account the priority for
each trigger within the following categories:

\begin{description}
  \setlength{\itemsep}{1pt}
  \setlength{\parskip}{0pt}
  \setlength{\parsep}{0pt}
\item \emph{Primary triggers:} principal physics triggers, which should not be prescaled.
\begin{sloppypar}
\item \emph{Supporting triggers:} triggers important to support the primary triggers, e.g. orthogonal 
triggers for efficiency measurements or lower \et\ threshold, prescaled versions of primary triggers.
\end{sloppypar}
\item \emph{Monitoring and Calibration triggers:} to collect data to ensure the correct operation of the trigger and detector, including detector calibrations.
\end{description}
\nomenclature{\bf primary trigger}{principal physics trigger, which should not be prescaled}
\nomenclature{\bf supporting trigger}{trigger important to support the primary triggers}
\nomenclature{\bf monitoring trigger}{collects data to ensure the correct operation of the trigger system}

Prescale changes are applied as luminosity drops during an LHC fill,
in order to maximize the bandwidth for physics, while ensuring a constant
rate for monitoring and calibration triggers.  Prescale changes can be applied at any point during a run at the beginning of a new \emph{luminosity block} ({\it LB}).  A 
\nomenclature{\bf LB}{Luminosity Block, a fundamental unit of time for the luminosity measurement that corresponded to approximately 120 seconds in 2010} luminosity block is the fundamental unit of time for the luminosity measurement and  was approximately 120 seconds
in 2010 data-taking. 

\begin{sloppypar}
Further flexibility is provided by defining 
\nomenclature{\bf bunch group}{a specific set of requirements on the LHC bunches colliding in ATLAS} 
\emph{bunch groups}, which
allow triggers
to include specific requirements on the LHC bunches colliding in ATLAS.
These requirements include paired (colliding) bunches for physics triggers and empty bunches 
for cosmic ray, random noise and pedestal triggers. More complex schemes are
possible, such as requiring unpaired bunches separated by at least 75~ns from any 
bunch in the other beam.
\end{sloppypar}

\begin{table}[!ht]
  \begin{center}
  \caption{Data-taking periods in 2010 running}
    \begin{tabular}{cccc}
      \hline
      Period & Dates  & $\int$ \Lum [\ipb] & Max. \Lum [\cms]\\
\hline \hline
 \emph{ proton-proton }     & & &\\
	A &30/3 - 22/4  & 0.4~$\times$~10$^{-3}$ & 2.5~$\times$~10$^{27}$ \\
	B &23/4 - 17/5  & 9.0~$\times$~10$^{-3}$ & 6.8~$\times$~10$^{28}$ \\
	C &18/5 - 23/6  & 9.5~$\times$~10$^{-3}$ & 2.4~$\times$~10$^{29}$ \\
	D &24/6 - 28/7  & 0.3 & 1.6~$\times$~10$^{30}$\\
	E &29/7 - 18/8  & 1.4 & 3.9~$\times$~10$^{30}$\\
	F &19/8 - 21/9  & 2.0 & 1.0~$\times$~10$^{31}$\\
	G &22/9 - 07/10  & 9.1 & 7.1~$\times$~10$^{31}$\\
	H &08/10 - 23/10  & 9.3 & 1.5~$\times$~10$^{32}$\\
	I &24/10 - 29/10  & 23.0 & 2.1~$\times$~10$^{32}$\\
\emph { heavy ion } &  & & \\
        J &8/11 - 6/12  & 9.2~$\times$~10$^{-6}$ & 3.0~$\times$~10$^{25}$\\
      \hline 
    \end{tabular}
  \label{tab:DataPeriods}
  \end{center}
\end{table}

\subsection{Datasets used for Performance Measurements}

\begin{figure}[!ht]
\centering
  \includegraphics[width=0.5\textwidth]{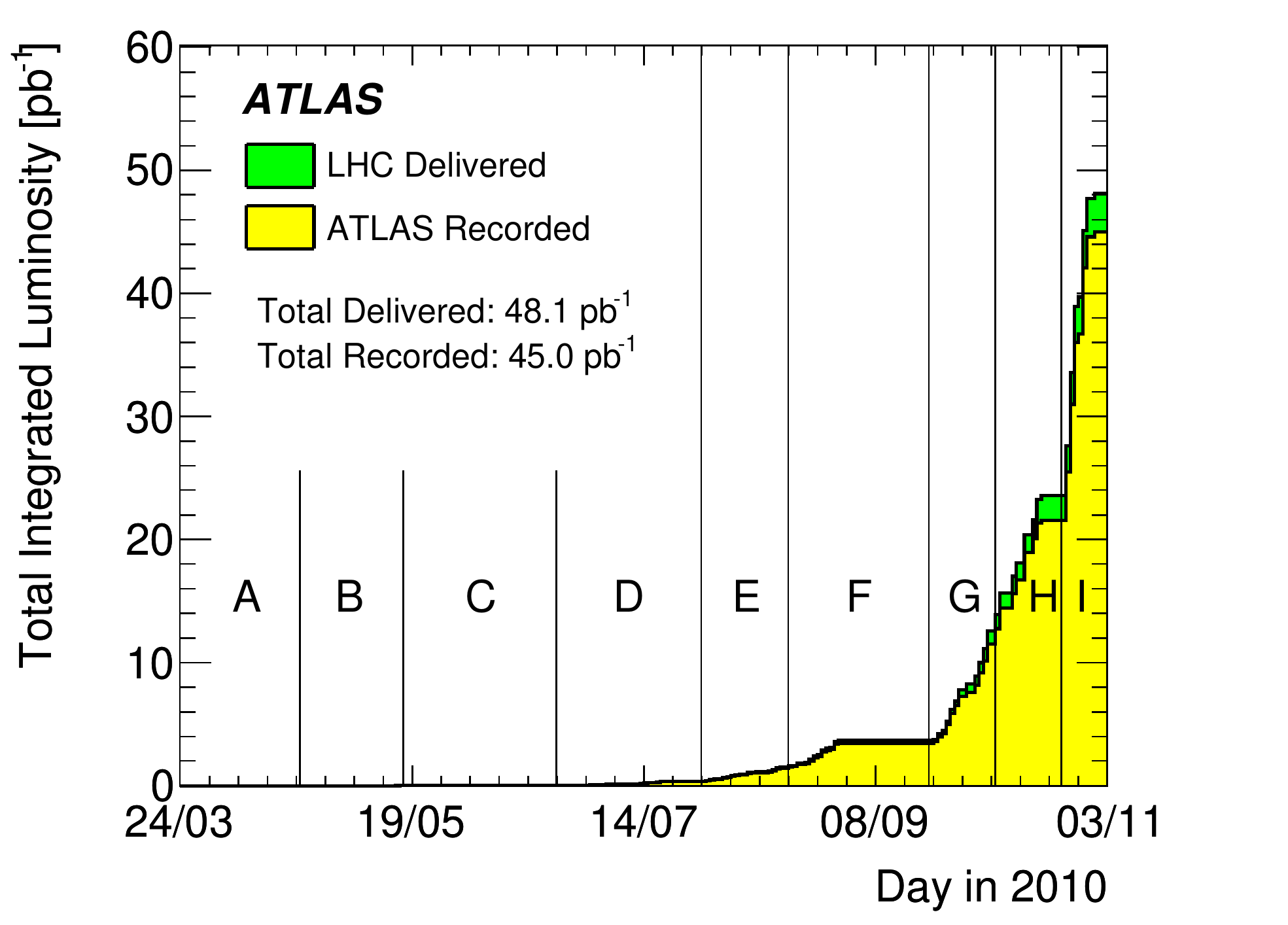}
  \caption{Profile with respect to time of the cumulative luminosity recorded by ATLAS during stable beams in \sTev{7} \pp\ collisions (from the online luminosity measurement).  The letters A-I refer to the data taking periods. }
  \label{fig:DQsumLumi}
\end{figure}

During 2010 the LHC delivered  a total integrated luminosity of 
48.1~\ipb\ to ATLAS during stable beams in \sTev{7} \pp\ collisions, of which
45~\ipb\ was recorded. Unless otherwise stated, 
the analyses presented in this publication are based on the full 2010
dataset.
To ensure the quality of data, events are required to pass 
data quality (DQ) conditions that include stable beams and good status for the relevant 
detectors and triggers. 
The cumulative luminosities delivered by the LHC and recorded by ATLAS
are shown as
a function of time in Fig.~\ref{fig:DQsumLumi}. 

In order to compare trigger performance between data and MC simulation, a number of MC samples
were generated. The MC samples used were produced
using the \linebreak PYTHIA~\cite{pythia64} event generator with a parameter set~\cite{ATLAS-MC09-Tune} tuned to 
describe the underlying event and minimum bias data
from Tevatron measurements at 0.63 TeV and 1.8 TeV. The generated events were processed through a
GEANT4~\cite{geant4} based simulation of the ATLAS detector~\cite{AtlasSim}. 

In some cases, where explicitly mentioned, performance results are shown for a 
subset of the data corresponding to
a  specific period of time. The 2010 run was split into 
data-taking 
\nomenclature{\bf period}{a new data-taking period is defined when there is 
a significant change in the detector conditions or instantaneous luminosity}
\emph{periods}; a new period being defined when there was 
a significant change in the detector conditions or instantaneous luminosity.  The data-taking periods are 
summarized in Table~\ref{tab:DataPeriods}. 
The rise in luminosity during the year was accompanied by an 
increase in the
number of proton bunches injected into the LHC ring. From the end of September (Period G onwards)
the protons were injected in 
\nomenclature{\bf bunch trains}{a number of proton bunches separated, in 2010, by 150~ns}
\emph{bunch trains} each consisting of
a number of proton bunches separated by 150~ns.

\section{Commissioning}\label{sec:commissioning}
\def \figurepath{.}
\begin{sloppypar}
In this section, the steps followed to commission the trigger are outlined and
the trigger menus employed during the commissioning phase are described. 
The physics trigger menu, deployed in July 2010,
is also presented and the evolution of the menu during the subsequent 2010 data-taking
period is described.
\end{sloppypar}

\subsection{Early Commissioning}
The commissioning of the ATLAS trigger system started before the first LHC beam using
cosmic ray events and, to commission L1, test pulses injected into the detector
front-end electronics.  
To exercise the data acquisition system and HLT, simulated collision data were
inserted into the ROS and processed through the whole online 
chain. This procedure provided the first full-scale
test of the HLT selection software running on the online system. 

\begin{sloppypar}
The L1 trigger system was exercised for the first time with beam during single beam
commissioning runs 
in 2008. Some of these runs included so-called 
\nomenclature{\bf splash events}{events for which the proton beam was intentionally brought into collision  
with the collimators upstream from the experiment in order to generate very large particle 
multiplicities that could be used for detector commissioning}
\emph{splash events}
for which the proton beam was intentionally brought into collision with the
collimators upstream from the experiment in order to generate very large particle 
multiplicities that could be used for
detector commissioning. During this short period of single-beam data-taking, 
the HLT algorithms were tested offline. 
\end{sloppypar}

Following the single beam data-taking in 2008, there
was a period of cosmic ray data-taking, during which the HLT algorithms
ran online. In addition to testing the selection algorithms used for collision 
data-taking, triggers  
specifically developed for cosmic ray data-taking were included.
The latter were used to select and record a very large sample of cosmic ray events, 
which were invaluable for 
the commissioning and alignment of the detector sub-systems such as the inner detector 
and the muon spectrometer~\cite{CosmicMuons}. 

\subsection{Commissioning with colliding beams}

\begin{sloppypar}
Specialized commissioning trigger menus were developed for the early collision 
running in 2009 and 2010. 
These menus consisted mainly of L1-based triggers since the initial low interaction rate, of 
the order of a few Hz, allowed all events passing L1 to be recorded.
Initially, the L1 MBTS trigger (Section~\ref{sec:minbias}) was unprescaled
and acted as the primary physics trigger, recording all interactions. Once the luminosity
exceeded $\sim2\cdot\Lumi{27}$, the L1 MBTS trigger was prescaled and the lowest threshold
muon and calorimeter triggers became the primary physics triggers. 
With further luminosity increase, these triggers were
also prescaled and higher threshold triggers, which were included in the
commissioning menus in readiness, became the primary physics triggers. 
A coincidence with filled bunch crossing was required for the physics triggers.
In addition, the menus contained non-collision triggers which required a 
coincidence with an empty or unpaired bunch crossing. 
For most of the lowest threshold physics triggers,
a corresponding non-collision trigger was included in the menus to be used for 
background studies.
The menus also contained a large number of supporting triggers needed 
for commissioning the L1 trigger system.
\end{sloppypar}

In the commissioning menus, event streaming was based on the L1 trigger categories.
Three main inclusive physics streams were recorded: {\it L1Calo} for calorimeter-based triggers, 
{\it L1Muon} for triggers coming from the muon system 
and {\it L1MinBias} for events triggered by minimum bias detectors such as MBTS, LUCID and ZDC. 
In addition to these L1-based physics streams, the express stream was also recorded.  Its content 
evolved significantly during the first weeks
of data-taking. In the early data-taking, it comprised a random 10-20\% of all triggered events in order to
exercise the offline express stream processing system. Subsequently, the content was changed to enhance
the proportion of electron, muon, and jet triggers.
Finally, a small set of triggers of each trigger type was sent to the express stream.  
For each individual trigger, the fraction contributing to the express stream
was adjustable by means of dedicated prescale values. The use of the express stream 
for data quality assessment and for calibration prior to offline 
reconstruction of the physics streams was commissioned during this period. 

\begin{table*}[!ht]
  \begin{center}
  \caption{
    Main calibration streams and their average event size per event.  The average event size in the physics streams is 1.3 MB}
    \begin{tabular}{llc}
     \hline 
      Stream      & Purpose & Event size [kB/event]\\ 
      \hline \hline
      LArCells & LAr detector calibration & 90\\
      beamspot & Online beamspot determination & 54 \\
      & based on Pixel and SCT detectors & \\
      IDTracks & ID alignment & 20\\
      PixelNoise, SCTNoise & Noise of the silicon detectors & 38\\
      Tile & Tile calorimeter calibration & 221\\
      Muon & Muon alignment & 0.5\\
      CostMonitoring & HLT system performance information & 233 \\ 
                     & including algorithm CPU usage & \\
      \hline 
    \end{tabular}
  \label{tab:CalibStreams}
  \end{center}
\end{table*}

\subsubsection{HLT commissioning}
The HLT commissioning proceeded in several steps.  During the very first collision data-taking at 
\sGev{900} in 2009, no HLT algorithms were run online.  Instead they were exercised offline 
on collision events recorded in the express stream. 
Results were carefully checked to confirm that the trigger algorithms were functioning 
correctly and the algorithm execution 
times were evaluated to verify that timeouts would not occur during online running. 

\begin{table*}[!htb]
  \begin{center}
  \caption[Main lowest threshold L1 triggers in the commissioning menus.]{
    Preliminary bandwidth allocations defined as guidelines to the various trigger groups, at three luminosity points,
for an  EF trigger rate of $\sim$200~Hz}
    \begin{tabular}{lcccc}
     \hline
           & Luminosity [\cms] & $10^{30}$  & $10^{31}$ & $10^{32}$  \\ 
      Trigger Signature && Rate [Hz] & Rate [Hz] & Rate [Hz] \\ 
      \hline \hline
      Minimum bias && 20 & 10 & 10 \\
      Electron/Photon && 30 & 45 & 50  \\
      Muon && 30 & 30 & 50  \\
      Tau && 20 & 20 & 15  \\
      Jet and forward jet && 25 & 25 & 20  \\
      $b$-jet && 10 & 15 & 10  \\
      $B$-physics && 15 & 15 & 10  \\
      \MET\ and \SumET\ && 15 & 15 & 10  \\
      Calibration triggers && 30 & 13 & 13 \\ 
      \hline 
    \end{tabular}
  \label{tab:Bandwidth}
  \end{center}
\end{table*}

After a few days of running offline, and having verified that the algorithms behaved as expected, the 
HLT algorithms 
were deployed online in 
\nomenclature{\bf monitoring mode}{HLT algorithms executing online but not influencing the trigger decision}  
{\em monitoring mode}.  In this mode, 
the HLT algorithms ran
online, producing trigger objects (e.g. calorimeter clusters and tracks) and a trigger decision 
at the HLT; however events 
were selected based solely on their L1 decision. Operating first in monitoring mode allowed each 
trigger to be validated before the trigger was put into {\em active rejection mode}. 
Recording the HLT objects and decision in each event allowed the efficiency of each 
trigger chain to be measured with 
respect to offline reconstruction.  In addition 
a {\em rejection factor}, defined as input rate over output rate, could be evaluated for  
each trigger chain at L2 and EF. 
Running the HLT algorithms online also allowed the online trigger monitoring system to be exercised 
and commissioned under real circumstances.  

\begin{sloppypar}
Triggers can be set in monitoring or active rejection mode individually. 
This important feature allowed individual triggers to be put into active rejection mode 
as luminosity increased and trigger rates exceeded allocated maximum values. 
The first HLT trigger to be enabled for active rejection was a minimum bias trigger chain (mbSpTrk)
based on a random bunch crossing trigger at
L1 and an ID-based selection on track multiplicity at the HLT (Section~\ref{sec:minbias}). This trigger was
already in active rejection mode in 2009.
\end{sloppypar}

\begin{figure}[!h]
  \centering
  \includegraphics[width=0.45\textwidth]{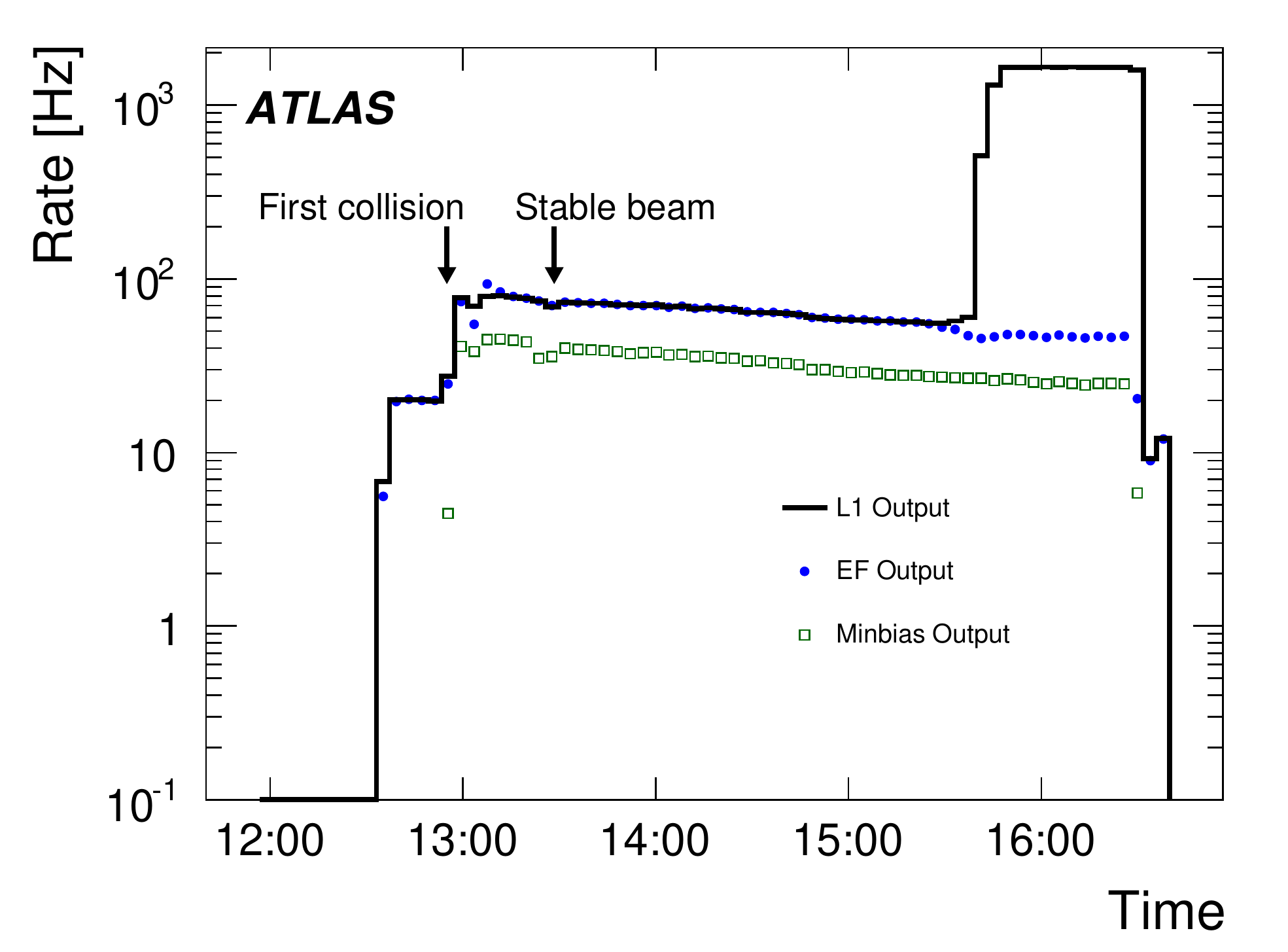}
  \caption{L1 and EF trigger rates during the first \sTev{7}\ \pp\ collision run}
  \label{fig:FirstCollisions}
\end{figure}

Figure~\ref{fig:FirstCollisions} illustrates the enabling of active HLT rejection during
the first  \sTev{7}\ collision run, in March 2010. 
Since the HLT algorithms were disabled 
at the start of the run, the L1 and EF trigger rates were initially the same.
The HLT algorithms were turned on, following rapid validation from offline processing, 
approximately two hours after the start of collisions, at about 15:00. 
All trigger chains were in monitoring mode apart 
from the mbSpTrk chain, which was in active rejection mode. However the random L1 trigger
that forms the input to the mbSpTrk chain was disabled for the first part of the run
and so the L1 and EF trigger rates remained the same until around 15:30 when
this random L1 trigger was enabled. At this time there was  
a significant increase in the L1 rate, but the
EF trigger rate stayed approximately constant due to the rejection by the 
mbSpTrk chain.

\begin{table*}[!htb]
  \begin{center}
  \caption[Lowest threshold triggers in commissioning menus.]{
    Examples of \pt\ thresholds and selections for the lowest unprescaled triggers in the physics menu at three luminosity values}
    \begin{tabular}{lcccc}
     \hline 
      &Luminosity [$cm^{-2}s^{-1}$]: & $3\times 10^{30}$& $2 \times 10^{31}$ &$2 \times 10^{32}$\\
      Category      && \multicolumn{3}{c}{\pt\ threshold [\GeV], selection}\\ 
      \hline \hline
      Single muon & &4, none & 10, none & 13,tight \\
      Di-muon & &4, none & 6, none & 6,loose \\
      Single electron & &10, medium &15, medium & 15, medium  \\
      Di-electron  & &3, loose & 5, medium& 10, loose  \\
      Single photon  & &15, loose & 30, loose& 40, loose  \\
      Di-photon  & &5, loose& 15, loose& 15, loose  \\
      Single tau  & &20, loose& 50, loose& 84, loose  \\
      Single jet  & &30, none& 75, loose& 95, loose  \\
      \MET & &25, tight& 30, loose& 40,loose\\
      $B$-physics & & mu4\_DiMu & mu4\_DiMu & 2mu4\_DiMu\\
      \hline
    \end{tabular}
  \label{tab:LowestUnprescaled}
  \end{center}
\end{table*}

During the first months of 2010 data-taking, the LHC peak luminosity increased from 
\Lumi{27}\ to \linebreak \Lumi{29}.
This luminosity was sufficiently low to allow the HLT to continue to run in monitoring mode and trigger 
rates were controlled by applying prescale factors at L1. 
Once the peak luminosity delivered by the LHC reached 1.2$\times$\Lumi{29}, it was necessary
to enable HLT rejection for the highest rate L1 triggers.
As luminosity progressively increased, more triggers were put into active rejection mode.

In addition to physics and commissioning triggers, a set of  HLT-based calibration
chains were also activated to produce dedicated data streams for detector calibration and monitoring. 
Table~\ref{tab:CalibStreams} lists the main calibration streams. These contain partial event
data, in most cases data fragments from one sub-detector, in contrast to the physics streams which 
contain 
information from the whole detector.

\subsection{Physics Trigger Menu}

The end of July 2010 marked a change in emphasis from commissioning to physics.  
A physics trigger menu was 
deployed for the first time, designed for luminosities from \linebreak
\Lumi{30} to \Lumi{32}. The physics trigger menu continued to evolve during 2010 to adapt to 
the LHC conditions. 
In its final form, it consisted of more than 470 triggers, the majority of which were primary 
and supporting physics triggers.

\begin{figure}[!htb]
  \centering
  \subfigure[]{
  \includegraphics[width=0.45\textwidth]{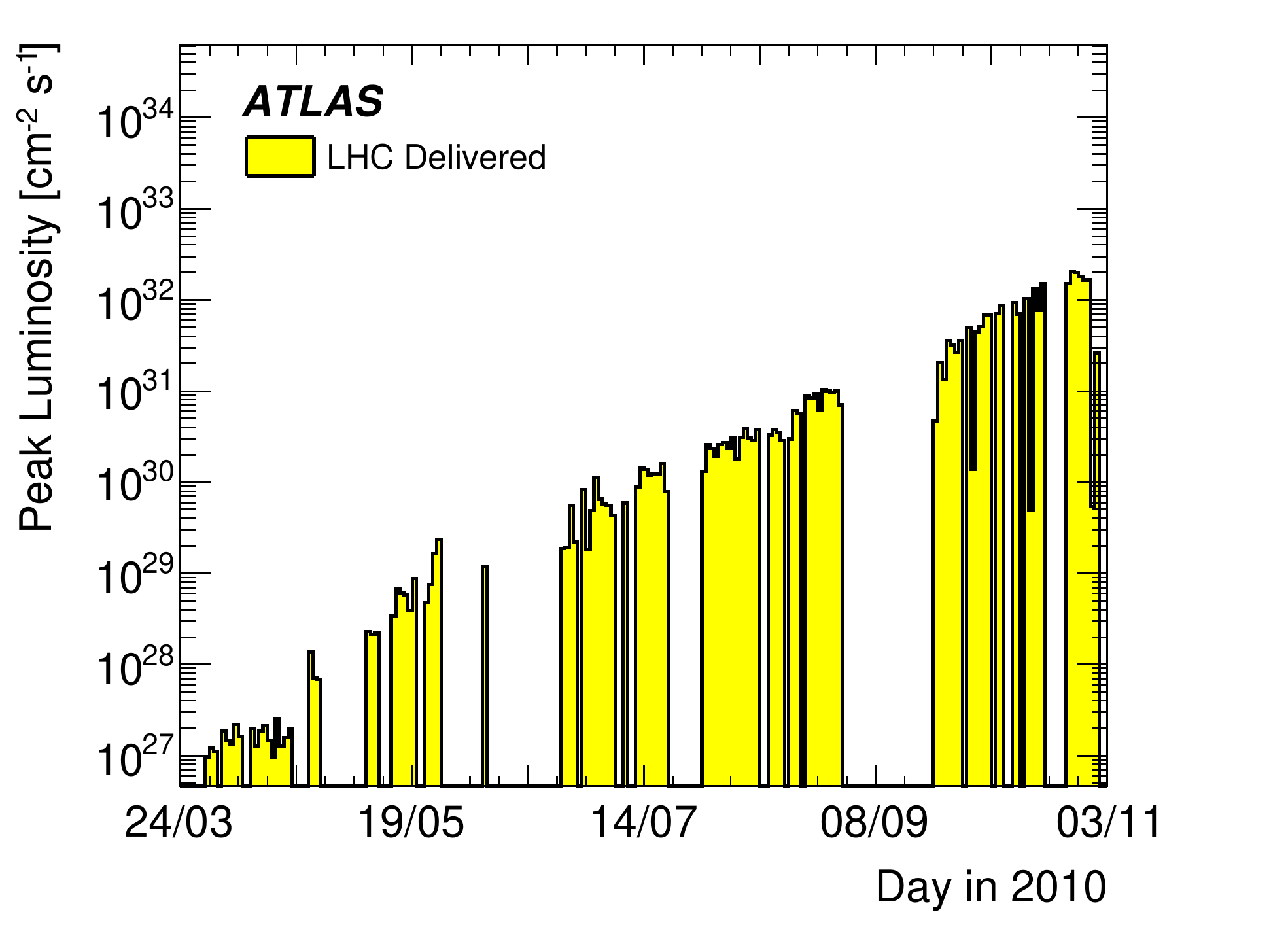}
  \label{fig:DQinfoLumi}
  }
  \subfigure[]{
  \includegraphics[width=0.45\textwidth]{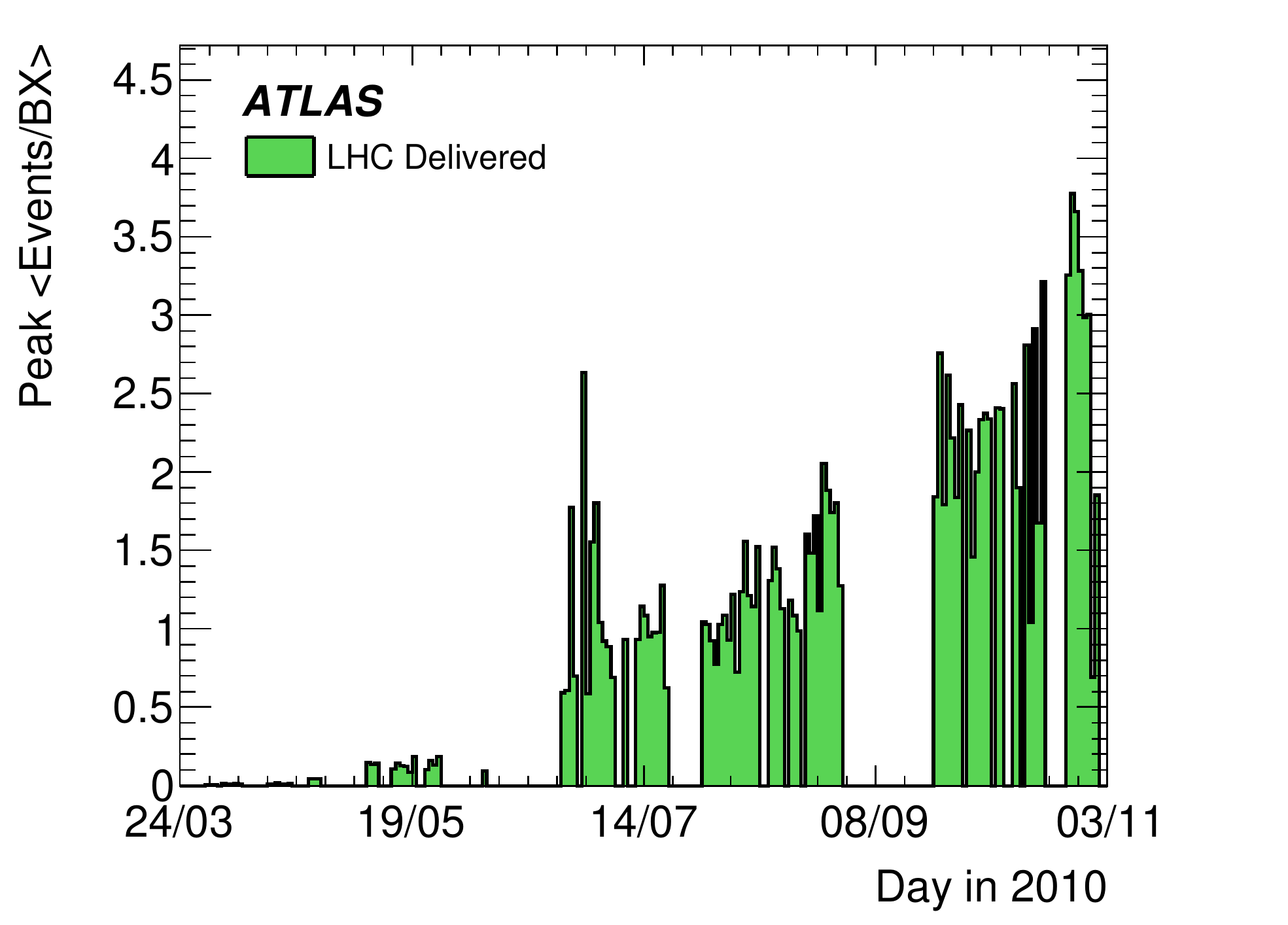}
  \label{fig:DQinfoNBx}
  }
  \caption{Profiles with respect to time of \subref{fig:DQinfoLumi} the maximum instantaneous luminosity per day and \subref{fig:DQinfoNBx} the peak mean number of interactions per bunch crossing (assuming a total inelastic cross section of 71.5 mb) recorded by ATLAS during stable beams in \sTev{7} \pp\ collisions.  Both plots use the online luminosity measurement}
  \label{fig:DQinfo}
\end{figure}
\begin{figure}[!htb]
  \centering
  \includegraphics[width=0.45\textwidth]{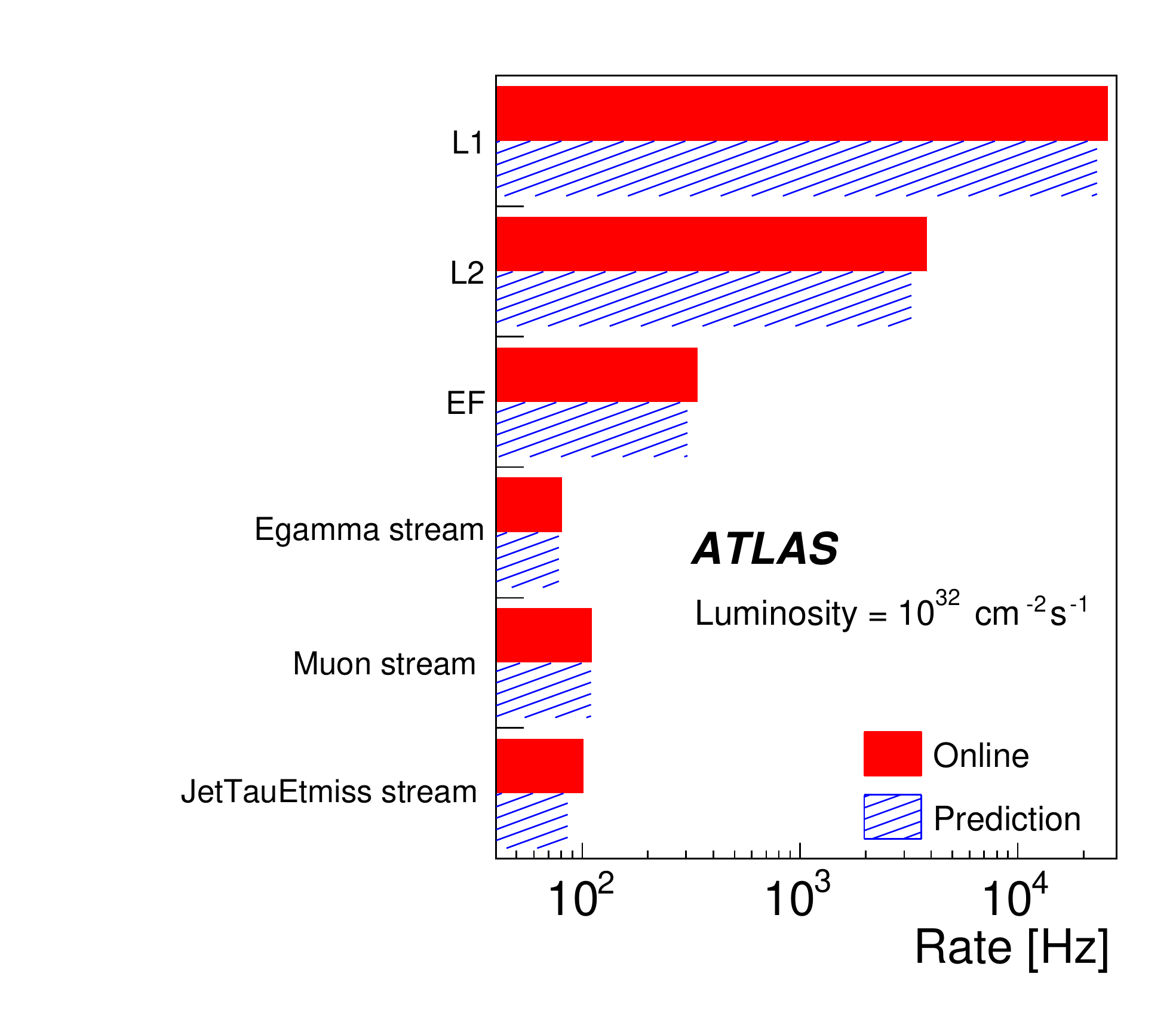}
  \caption{Comparison of online rates (solid)  
    with offline rate predictions (hashed) at luminosity \Lumi{32} for L1, L2, EF 
    and main physics streams}
  \label{fig:RatePredictions}
\end{figure}

\begin{sloppypar}
In the physics menu, L1 commissioning items were removed, allowing for the addition of higher threshold
physics triggers in preparation for increased luminosity. At the same time, combined triggers 
based on a logical ``and'' between two 
L1 items were introduced into the menu.
Streaming based on the HLT decision was 
introduced and the corresponding L1-based streaming was disabled. 
In addition to calibration and express streams, data were recorded in the physics streams 
presented in Section~\ref{sec:overview}.
At the same time, preliminary bandwidth allocations were defined as guidelines for all trigger 
groups, as listed in Table~\ref{tab:Bandwidth}. 
\end{sloppypar}

\begin{sloppypar}
The maximum instantaneous luminosity per day is shown in Fig.~\ref{fig:DQinfoLumi}.
As luminosity increased and the trigger rates approached the limits imposed by offline 
processing, primary and supporting triggers continued to evolve 
by progressively tightening the HLT selection cuts and by prescaling the lower \et\ threshold triggers.
Table~\ref{tab:LowestUnprescaled} shows the 
lowest unprescaled threshold of various trigger signatures for three luminosity values.
\end{sloppypar}

\begin{sloppypar}
In order to prepare for higher luminosities, 
tools to optimize prescale factors became very important. 
For example, the rate prediction tool uses \emph{enhanced bias} data (data recorded with a very 
loose L1 trigger selection and no HLT selection) as input.  
Initially, these data were collected in dedicated enhanced bias runs using the lowest trigger 
thresholds, which were
unprescaled at L1, and no HLT selection.  Subsequently, enhanced bias triggers were added
to the physics menu to collect the data sample during normal physics data-taking.
\end{sloppypar}

Figure~\ref{fig:RatePredictions} shows a comparison between online rates at \Lumi{32}
and predictions based on extrapolation from enhanced bias data collected at lower luminosity. 
In general online rates agreed with predictions within 10\%. 
The biggest discrepancy was seen in rates from the {\it JetTauEtmiss} stream, as a result of the 
non-linear
scaling of \MET\ and \SumET\ trigger rates with luminosity, as shown 
later in Fig.~\ref{fig:l1caloratesscaling}. This non-linearity is due to 
\emph{in-time pile-up},
defined as the effect of multiple \pp\ interactions in a bunch crossing.  
The maximum mean number of interactions per bunch 
crossing, which reached 3.5 in 2010, is shown as a function of day in Fig.~\ref{fig:DQinfoNBx}.
In-time pile-up had the most significant effects on the \MET, \SumET (Section~\ref{sec:met}), and minimum bias (Section~\ref{sec:minbias}) signatures.   \emph{Out-of-time pile-up} is defined as the effect of an earlier bunch crossing on the detector signals for the current bunch crossing.  Out-of-time pile-up did not have a significant effect in the 2010 \pp\ data-taking because the bunch spacing was 150~ns or larger.


\section{Level 1}\label{sec:level1}

The Level 1 (L1) trigger decision is formed by the Central Trigger Processor~(CTP)
based on information from the calorimeter trigger towers and dedicated 
triggering layers in the muon system.  An overview of the CTP, L1 calorimeter, and L1 muon systems  
and their performance follows.   The CTP also takes input from
the MBTS, LUCID and ZDC systems, described in 
Section~\ref{sec:minbias}.

\begin{figure*} [!htb]
  \centering
  \includegraphics[width=0.7\textwidth]{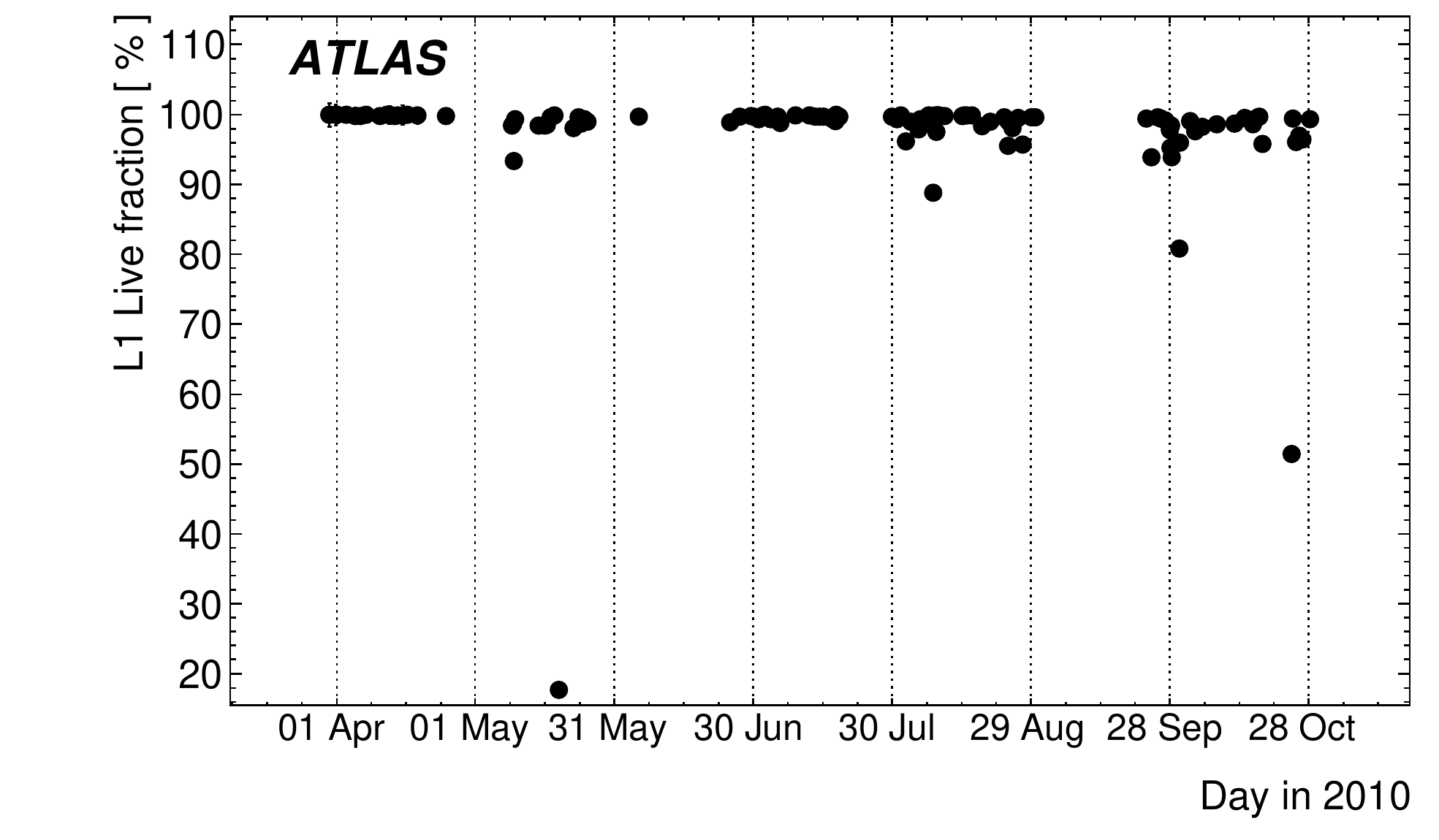}
  \caption{L1 live fraction per run throughout the full data-taking period of
    2010, as used for luminosity estimates for correcting trigger
    dead-time effects. The live fraction is derived from the trigger
    L1\_MBTS\_2}
  \label{fig:l1deadtime}
\end{figure*}

\subsection{Central Trigger Processor}\label{sec:L1ctp}
\def \figurepath{.}\nomenclature{\bf CTP}{Central Trigger Processor}
The CTP~\cite{DetectorPaper,Ask:2008zz} forms the L1 trigger decision
by applying the multiplicity requirements and prescale factors specified in the
trigger menu to the inputs from the L1 trigger systems.
The CTP also provides random triggers and can apply 
specific LHC bunch crossing requirements.  
The L1 trigger decision is distributed, together with 
timing and control signals, to all ATLAS sub-detector readout systems.

The timing signals are defined
with respect to the LHC bunch crossings.  A \nomenclature{\bf bunch crossing}{a 25~ns time-window 
centred on the instant at which a proton bunch may traverse the
ATLAS interaction point}
\emph{bunch crossing} is defined as a 25~ns time-window 
centred on the instant at which a proton bunch may traverse the
ATLAS interaction point.    Not all bunch crossings contain protons; those 
that do are called \emph{filled bunches}.
In 2010, the minimum spacing 
between filled bunches was 150~ns.
In the nominal LHC configuration, 
there are a maximum of 3564 bunch crossings per LHC revolution.  
Each bunch crossing is given a bunch 
crossing identifier (\emph{BCID}) from 0 to 3563.
A \emph{bunch group} consists of a 
numbered list of BCIDs during which the CTP 
generates an internal trigger signal. 
The bunch groups are used to apply
specific requirements to triggers 
such as \emph{paired} (colliding) bunches for physics triggers, single (one-beam) bunches 
for background triggers, and empty bunches for cosmic ray, noise 
and pedestal triggers.  

\subsubsection{Dead-time}
Following an L1 accept the CTP introduces dead-time, by vetoing subsequent triggers, 
to protect front-end readout \linebreak buffers from overflowing. 
This \emph{preventive dead-time}
mechanism limits the minimum time between two consecutive L1 accepts 
(\emph{simple dead-time}), and restricts the
number of L1 accepts allowed in a given period (\emph{complex dead-time}).
In 2010 running, the simple dead-time was set to 125~ns and the complex
dead-time to 8 triggers in $80~\mu$s. 
This preventative dead-time is in addition to 
\emph{busy dead-time} which can be introduced by ATLAS sub-detectors to
temporarily throttle the trigger rate.

The CTP monitors the total
L1 trigger rate and the rates of individual L1 triggers. These rates are
monitored before 
and after prescales and after dead-time related vetoes have been applied.
One use of this information is to  
provide a measure of the L1 dead-time, which
needs to be accounted for when determining
the luminosity. The L1 dead-time correction is determined from the
\emph{live fraction}, defined as the ratio of trigger rates after CTP 
vetoes to the corresponding trigger rates before vetoes.
Figure~\ref{fig:l1deadtime} shows the live fraction
based on the L1\_MBTS\_2 trigger (Section~\ref{sec:minbias}),
the primary trigger used for these corrections in 2010.
The
bulk of the data were recorded with live fractions in excess of
98\%. As a result of the relatively low L1 trigger rates and
a bunch spacing  that was relatively
large ($\geq 150$~ns) compared to the nominal LHC spacing (25~ns), the preventive dead-time was 
typically below $10^{-4}$ and no bunch-to-bunch variations in dead-time existed.
 
\begin{figure*} [ht!]
  \centering
  \includegraphics[width=\textwidth]{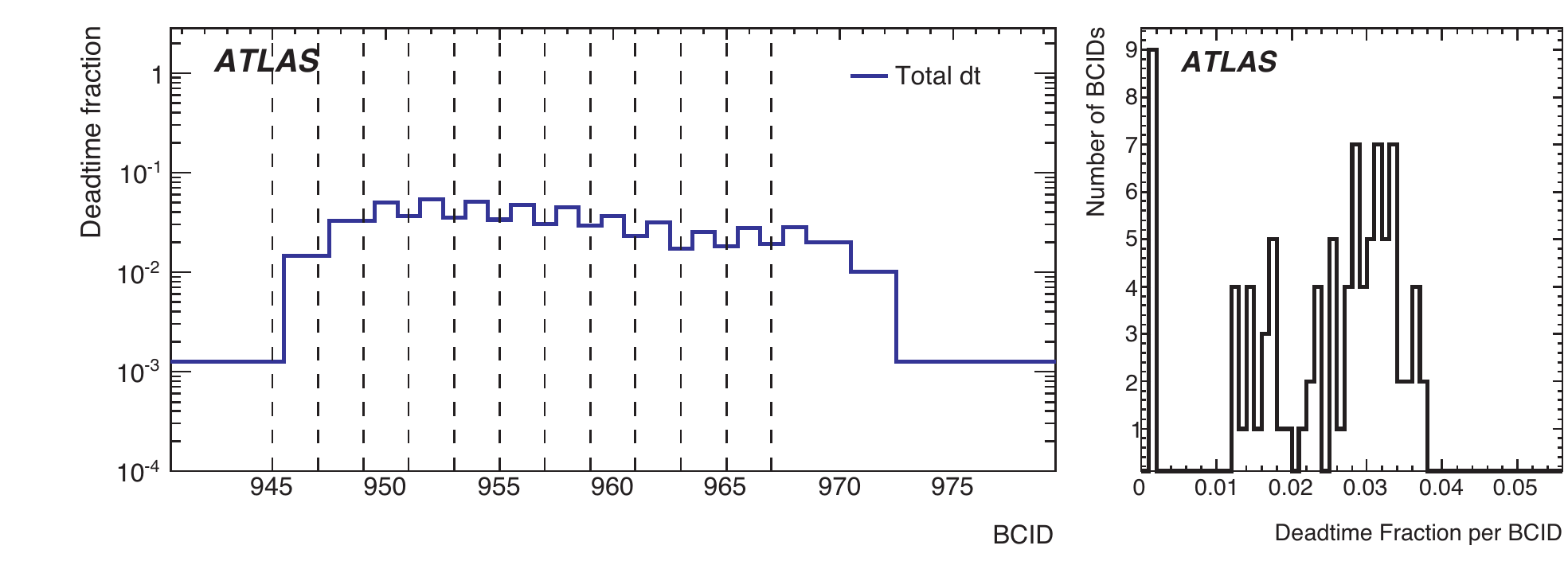}
  \caption{L1 dead-time fractions per bunch crossing 
for a LHC test fill with a 50~ns bunch spacing. 
The dead-time fractions as a function of
  BCID~(left), taking a single bunch train of 12 bunches  as an example and 
a histogram of the individual dead-time fractions for each 
paired bunch crossing~(right). The paired bunch crossings (odd-numbered BCID from 945 to 967) 
are indicated by vertical dashed lines on the left hand plot} 

  \label{fig:l1perbunchdeadtime}
\end{figure*}

Towards the end of the 2010 data-taking a test was performed with a fill of bunch
trains with 50~ns spacing, the running mode expected for the bulk of 2011 data-taking. 
The dead-time measured during this test is shown as a function of BCID in Fig.~\ref{fig:l1perbunchdeadtime}, taking a single bunch train as an example.
The first bunch of the train (BCID 945) is only
subject to sub-detector dead-time of $\sim$0.1\%, while the following bunches in the train 
(BCIDs 947 to 967) are subject to up to 4\% dead-time as a result of the preventative
dead-time generated by the CTP. The variation in dead-time between bunch crossings will 
be taken into account when calculating the dead-time corrections to luminosity in 
2011 running.

\subsubsection{Rates and Timing}

\begin{figure*}[!ht]
  \centering
  \includegraphics[width=0.7\textwidth]{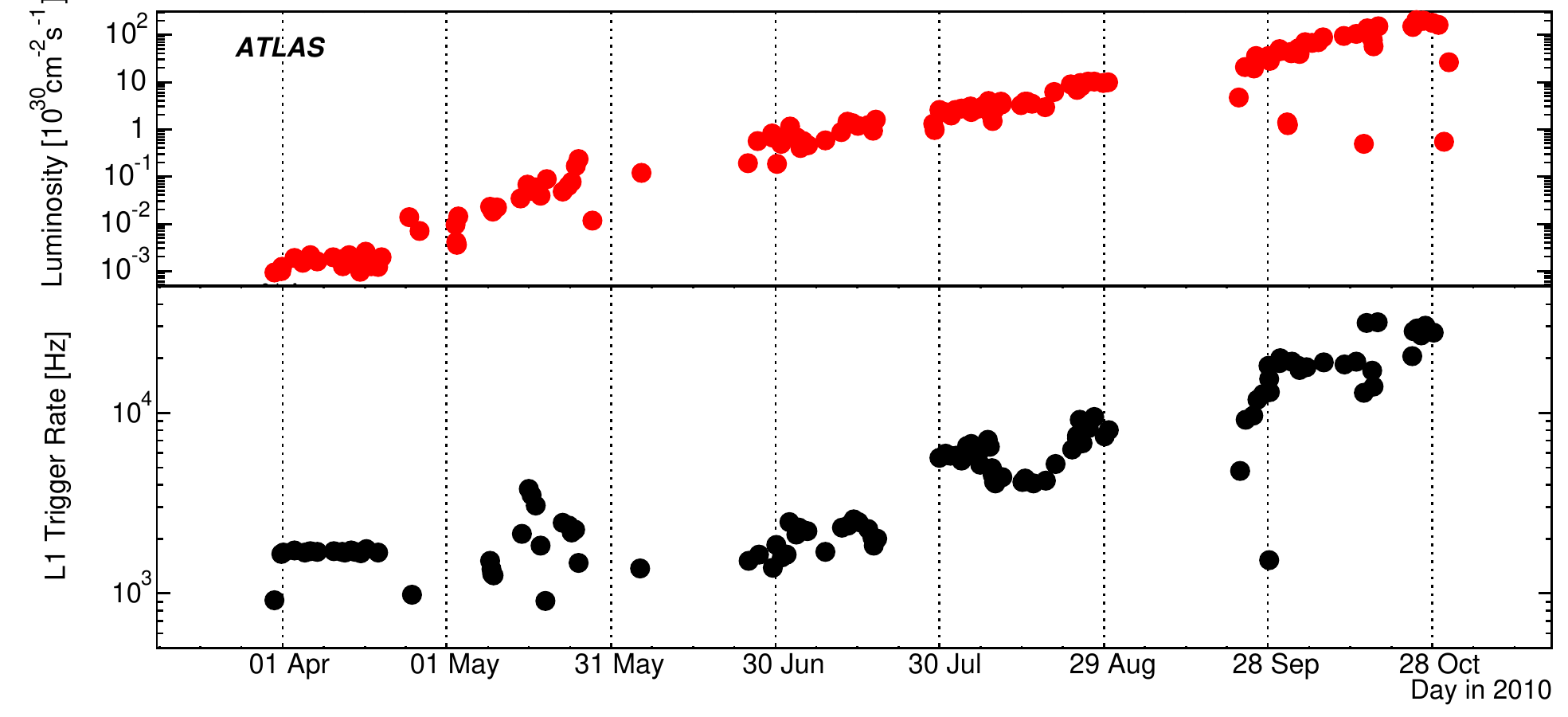}
  \caption{Evolution of the L1 trigger rate throughout 
   2010~(lower panel), compared to the instantaneous
    luminosity evolution~(upper panel)}
  \label{fig:l1rate}
\end{figure*}

Figure~\ref{fig:l1rate} shows the
trigger rate for the whole data-taking period of 2010, compared to the
luminosity evolution of the LHC. The individual rate
points are the average L1 trigger rates in ATLAS runs with stable beams,
and the luminosity points correspond to peak
values for the run. The increasing selectivity of the trigger during the course of
2010 is illustrated by the fact that the L1 trigger rate increased by one order of 
magnitude; whereas, the peak instantaneous luminosity increased by
five orders of magnitude. The L1 trigger system was operated at a maximum trigger
rate of just above 30~kHz, leaving more than a
factor of two margin to the design rate of 75~kHz. 

\begin{figure}[!ht]
  \centering
  \includegraphics[width=0.45\textwidth]{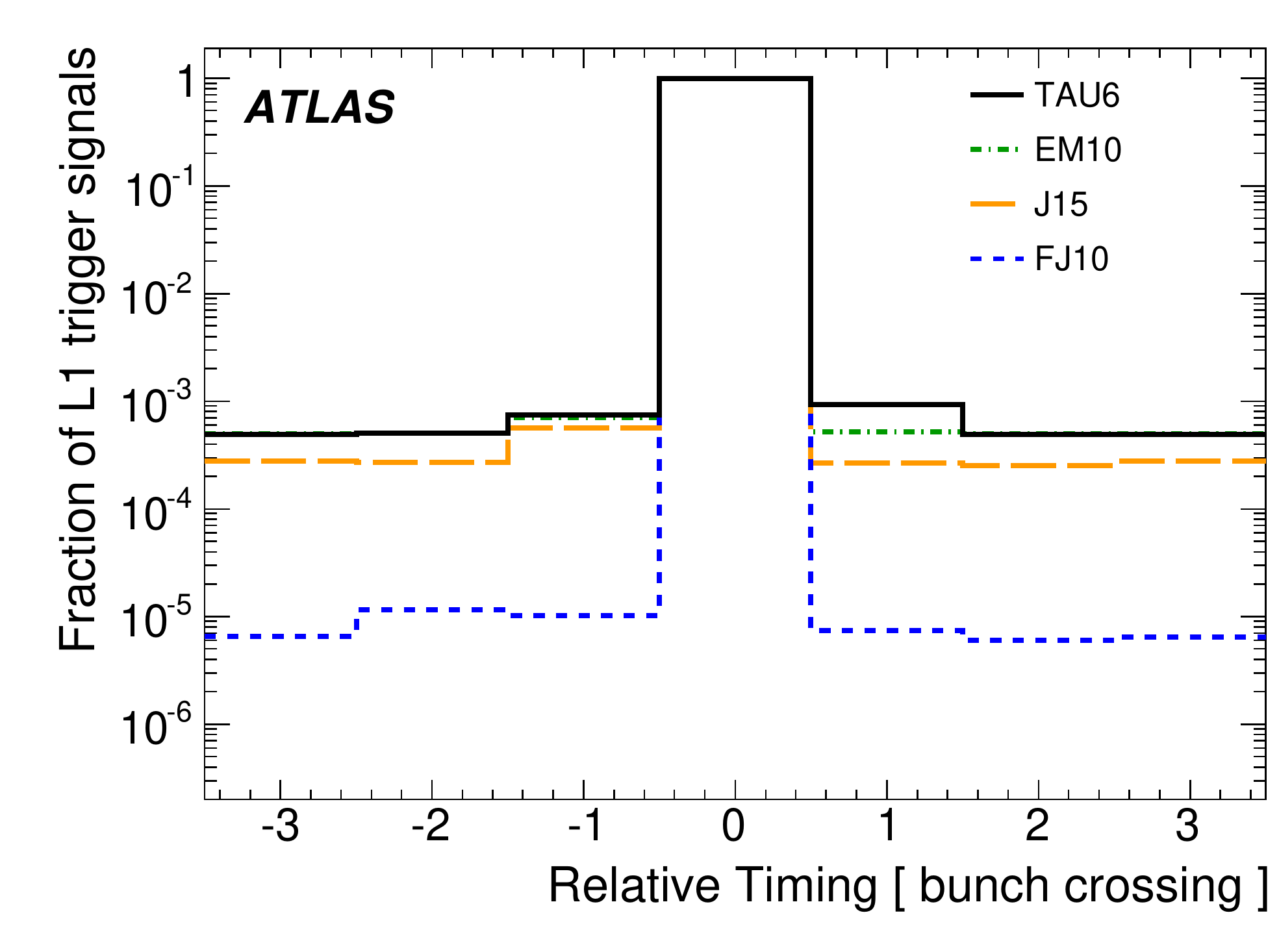}
  \caption{L1 trigger signals in units of bunch crossings for a number of
    triggers. The trigger signal time is plotted relative
    to the position of filled paired bunches for a particular data-taking 
    period towards the end of the 2010 \pp\ run}
  \label{fig:l1timing}
\end{figure}

The excellent level of synchronization of L1 trigger signals in time is
shown in Fig.~\ref{fig:l1timing} for a selection of  L1
triggers. The plot represents a snapshot taken at the end of October 2010.
Proton-proton collisions in nominal filled paired
bunch crossings are defined to occur in the central bin at 0.  As a result of
mistiming caused by alignment of the calorimeter pulses that are longer than a single 
bunch crossing, trigger signals may appear in bunch crossings preceding or
succeeding the central one. In all cases mistiming effects are below $10^{-3}$.
The timing alignment procedures for the L1 calorimeter and L1 muon triggers
are described in Section~\ref{sec:L1calo} and Section~\ref{sec:L1muon} respectively.


\subsection{L1 Calorimeter Trigger}\label{sec:L1calo}
\def \figurepath{.}
The L1 calorimeter trigger \cite{L1CaloPaper} is based on 
inputs from the electromagnetic and hadronic calorimeters
covering the region $|\eta|<4.9$.  It provides triggers for localized objects
(e.g. electron/photon, tau and jet) and global transverse energy triggers.
The pipelined processing and logic is performed in a series 
of custom built hardware modules with a latency of less than
1~$\mu$s.  The architecture, calibration and performance 
of this hardware trigger are described in the following sub-sections.

\subsubsection{L1 Calorimeter Trigger Architecture}

The L1 calorimeter trigger decision is based on dedicated
analogue trigger signals provided by the \linebreak[3] ATLAS calorimeters 
independently from the signals read out and used at the
HLT and offline. 
\nomenclature{\bf trigger tower}{projective region of the calorimeter over which signals are summed
to provide input to L1}
Rather than using the full granularity of the calorimeter, the L1 decision is based 
on the information from analogue sums of calorimeter elements within
projective regions, called \emph{trigger towers}. The trigger towers
have a size of approximately $\Delta\eta \times \Delta\phi = 0.1 \times 0.1$ 
in the central part of the calorimeter, $|\eta|<2.5$, and are larger and less regular 
in the more forward region. 
Electromagnetic and hadronic calorimeters have separate trigger towers.

The 7168 analogue inputs must first be digitized and then associated to a particular 
LHC bunch crossing.
Much of the tuning of the timing
and transverse energy calibration was performed during the 2010 data-taking 
period since the final adjustments could only be determined with colliding
beam events.
Once digital transverse energies per LHC bunch crossing are formed, two separate
processor systems, working in parallel, run the trigger algorithms.  
One system, the \emph{cluster processor} uses the full L1 trigger granularity information in
the central region to look for small localized clusters typical of
electron, photon or tau particles. The other, the \emph{jet and energy-sum 
processor},  uses $2\times2$ sums of trigger towers, called jet elements, to identify
jet candidates and form global transverse energy sums: missing 
transverse energy, total transverse energy and jet-sum transverse energy.
The magnitude of the objects and sums are compared to programmable 
thresholds to form the trigger decision.  The thresholds used in 2010 are shown in Table~\ref{tab:ExampleMenu}
in Section~\ref{sec:overview}.

\begin{figure} [ht!]
  \centering
  \includegraphics[width=0.35\textwidth]{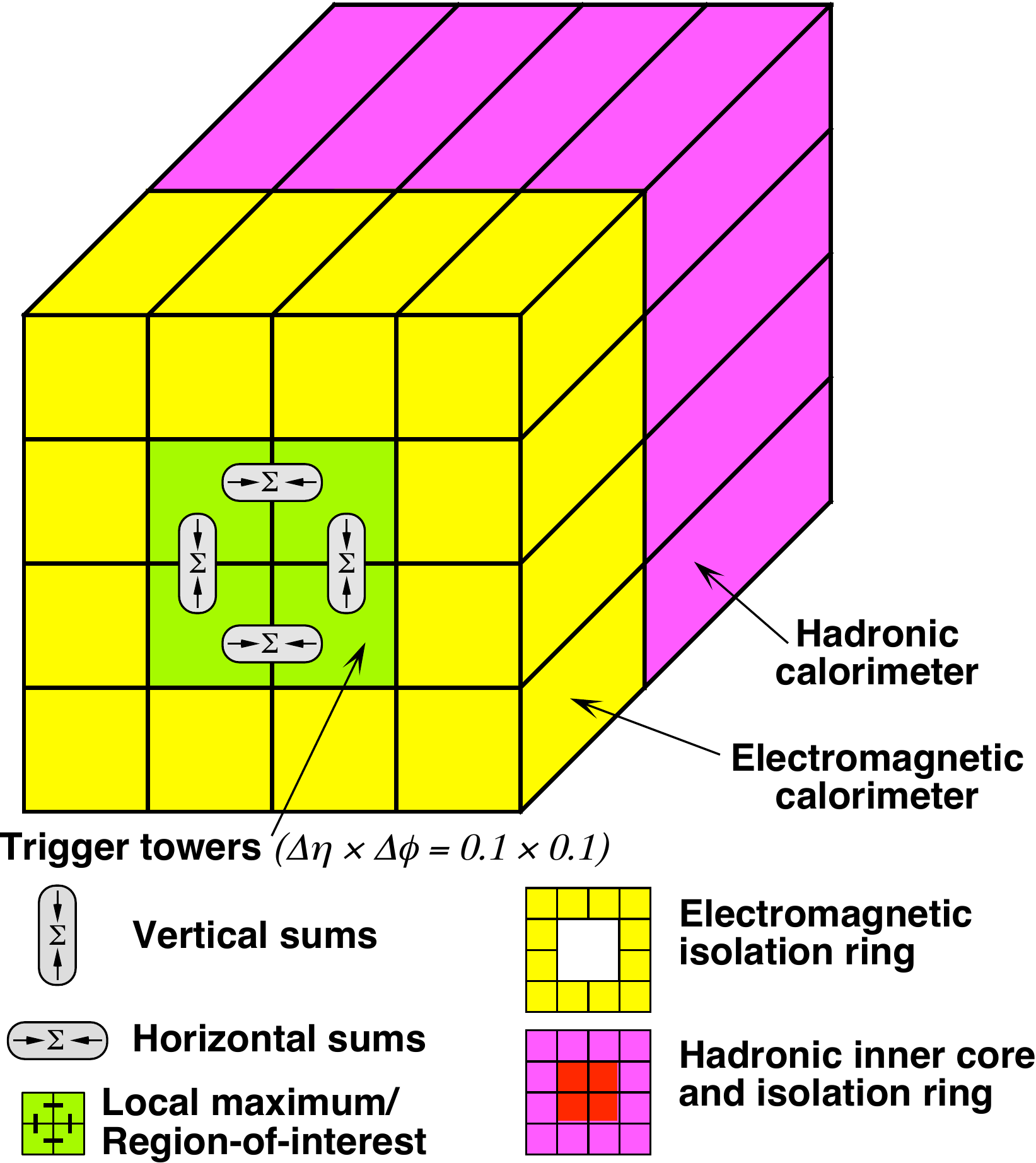}
  \caption{Building blocks of the electron/photon and tau algorithms with the sums to be compared to programmable thresholds}
  \label{fig:l1caloalgorithms}
\end{figure}

The details of the algorithms can be found elsewhere~\cite{L1CaloPaper} and only
the basic elements are described here. Figure~\ref{fig:l1caloalgorithms} illustrates 
the electron/photon and tau triggers as an example.
The electron/photon trigger 
algorithm identifies an Region of Interest as a $2\times2$ trigger tower cluster
in the electromagnetic calorimeter for which the transverse energy  sum from at least one of the 
four possible pairs of nearest neighbour towers ($1\times2$ or $2\times1$) 
exceeds a pre-defined threshold. Isolation-veto thresholds can be
set for the 12-tower surrounding ring in the electromagnetic calorimeter, as well as 
for hadronic tower sums in a central 
$2\times2$ core behind the cluster and the 12-tower hadronic ring around it. 
Isolation requirements were not applied in 2010 running. 
Jet RoIs are defined as $4\times4$, $6\times6$ or $8\times8$ trigger tower windows for which
the summed electromagnetic and hadronic transverse energy exceeds pre-defined thresholds
and which surround a
$2\times2$ trigger tower 
core that is a local maximum.
The location of this local maximum also defines the coordinates of the jet RoI. 

The real-time output to the CTP consists of more than 100 bits per bunch crossing, 
comprising the coordinates and threshold bits for each of the RoIs 
and the counts of the number of objects (saturating at seven) that satisfy each
of the electron/photon, tau and jet criteria.

\subsubsection{L1 Calorimeter Trigger Commissioning and Rates}
After commissioning with cosmic ray and collision data, 
including event-by-event 
checking of L1 trigger results against offline emulation of the L1 trigger logic, the 
calorimeter trigger processor ran stably and without any algorithmic errors. 
Bit-error rates in digital links were less than 1 in $10^{20}$. 
Eight out of 7168 trigger 
towers were non-operational in 2010 due to failures in inaccessible analogue 
electronics on the detector. Problems with detector high and low voltage led to 
an additional $\sim$1\% of trigger towers with low or no response. After 
calibration adjustments, L1 calorimeter trigger conditions remained essentially 
unchanged for 99\% of the 2010 proton-proton integrated luminosity.

\begin{figure}[ht!]
  \centering
  \includegraphics[width=0.45\textwidth]{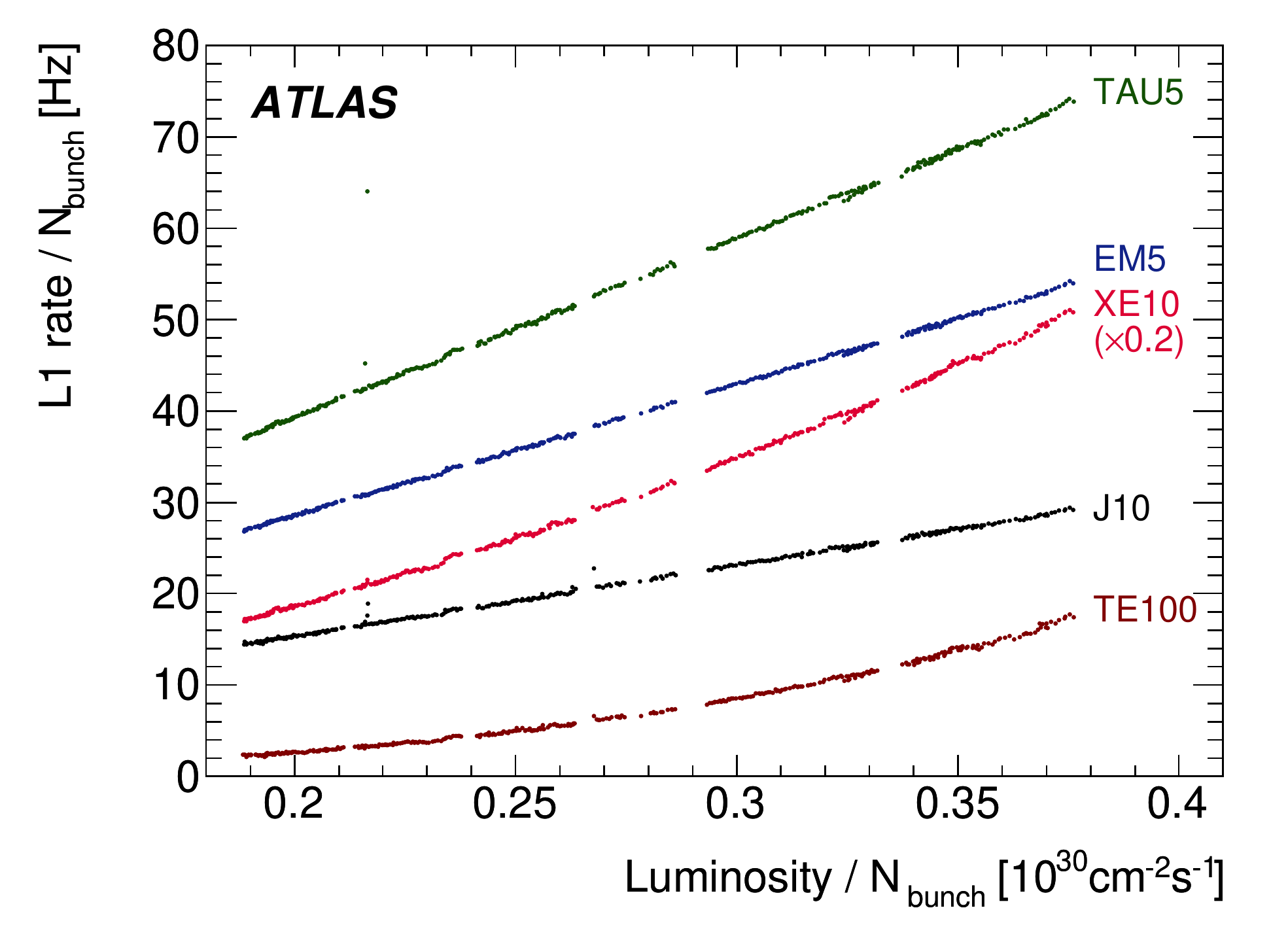}
  \caption{L1 trigger rate scaling for some low threshold trigger items
           as a function of luminosity per bunch crossing. The rate for XE10 has been scaled by 0.2}
  \label{fig:l1caloratesscaling}
\end{figure}

The scaling of the L1
trigger rates with luminosity is shown in 
Fig.~\ref{fig:l1caloratesscaling} for some of the low-threshold 
calorimeter trigger items.  
The localised objects, such as electrons and jet candidates, 
show an excellent linear scaling relationship with luminosity over a 
wide range of luminosities and time. 
Global quantities such as the missing transverse energy and 
total transverse energy triggers also scale in a smooth way, but are not linear as they are
strongly affected by in-time pile-up which was present in the later running periods.

\subsubsection{L1 Calorimeter Trigger Calibration}

In order to assign the calorimeter tower signals to the correct bunch crossing, a task
performed by the bunch crossing identification logic, the signals 
must be synchronized to the LHC clock phase with nanosecond precision.
The timing synchronization was first established with calorimeter pulser systems and
cosmic ray data and then refined using the first beam delivered to the
detector in the splash events (Section~\ref{sec:commissioning}). 
During the earliest data-taking in 2010 the correct bunch crossing was determined
for events with transverse energy above about
5~\GeV. Timing was incrementally improved, and for
the majority of the 2010 data the timing of most towers was better than
${\pm}2$~ns, providing close to ideal performance.  

\begin{figure}[!ht]
\centering
  \includegraphics[width=0.45\textwidth]{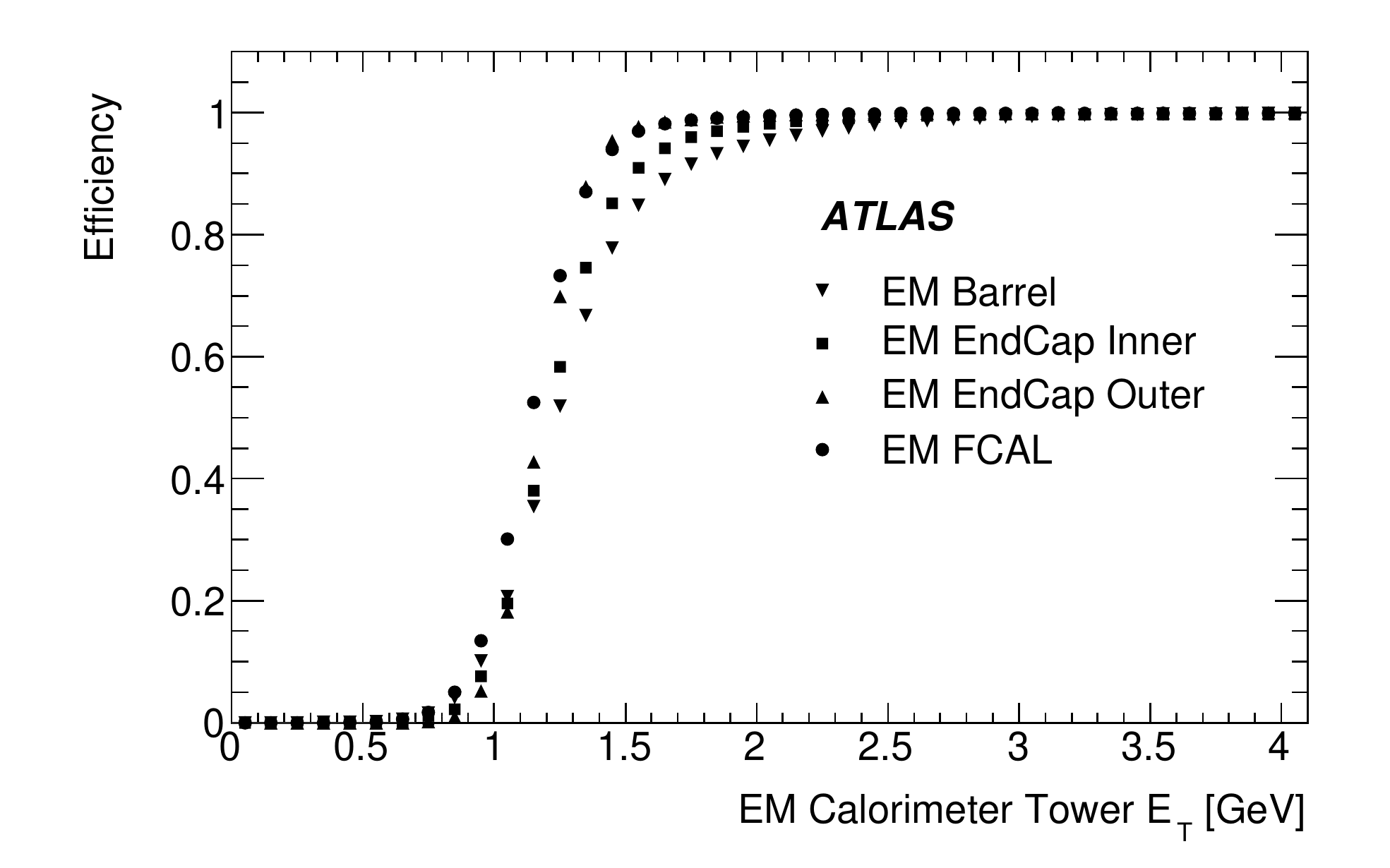}
  \caption{Efficiency for an electromagnetic trigger tower energy to be
associated with the correct bunch crossing and pass a  noise cut of 
around 1.2~\GeV as a function of
the sum of raw cell \ET\ within that tower}
  \label{fig:l1calobcidperformance}
\end{figure}

In order to remove the majority 
of fake triggers due to small energy deposits, signals are 
processed by an optimized filter and a noise cut of around 1.2~GeV is applied 
to the trigger tower energy. The efficiency for an electromagnetic tower energy to 
be associated to the correct bunch crossing and pass
this noise cut is shown in
Fig.~\ref{fig:l1calobcidperformance} as a function of
the sum of raw cell \ET\ within that tower, for different regions of
the electromagnetic calorimeter. The efficiency turn-on is consistent with the optimal 
performance expected from a simulation of the signals and the full efficiency
in the plateau region indicates the successful association of these
small energy deposits to the correct bunch crossing. 

Special treatment, using additional bunch crossing identification logic,
is needed for saturated pulses with \et\ above about 250~\GeV.  
It was shown that BCID logic performance was more than adequate for 2010
LHC energies, working for most trigger towers up to transverse energies of 3.5~\TeV\ and
beyond. Further tuning of timing and algorithm parameters will 
ensure that the full LHC energy range is covered.

\begin{figure}[!ht]
  \centering
  \subfigure[]{
  \includegraphics[width=0.45\textwidth]{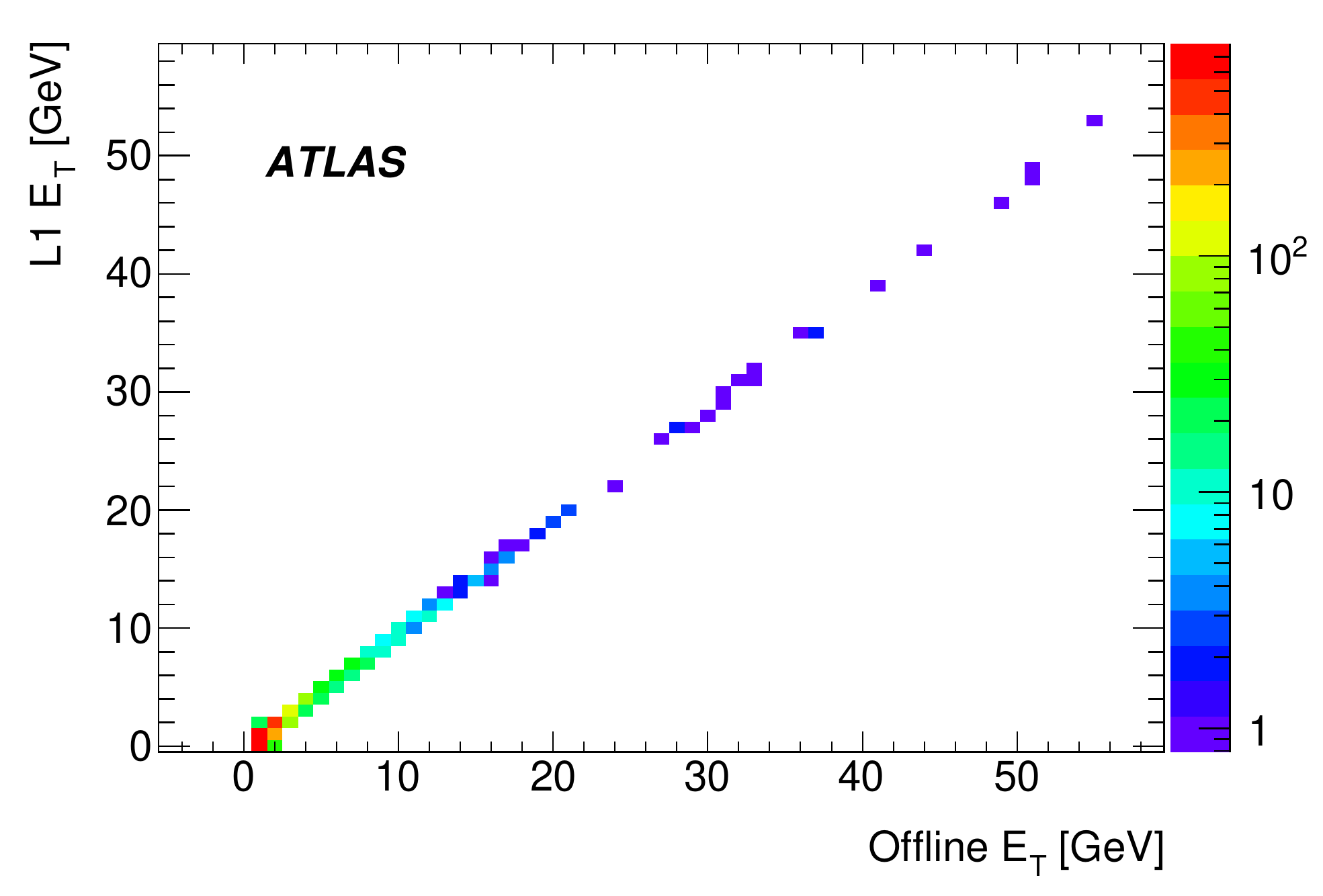}
  \label{fig:l1caloenergycorrelation_a}
   }
   \subfigure[]{
  \includegraphics[width=0.45\textwidth]{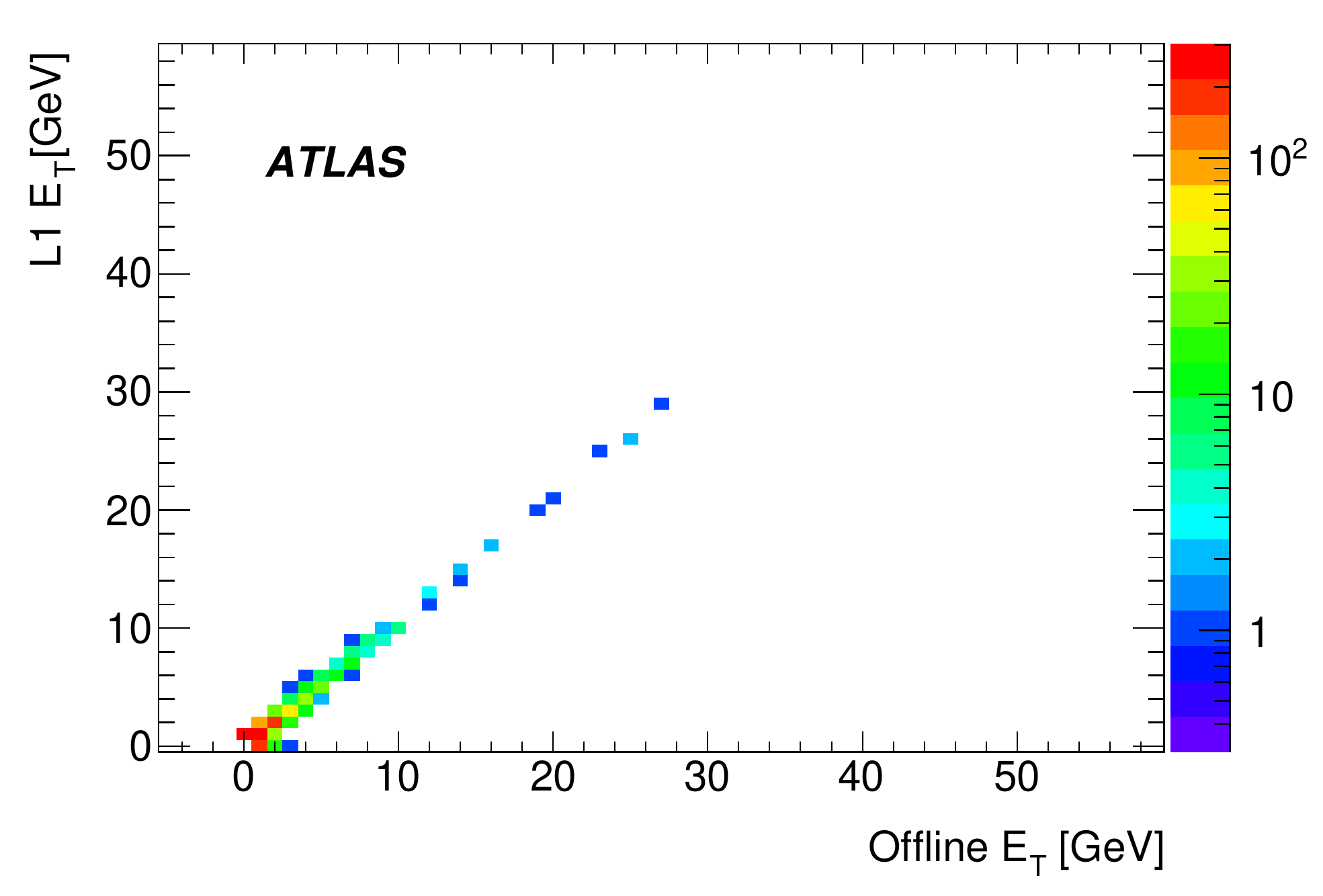}
  \label{fig:l1caloenergycorrelation_b}
  }
  \caption{Typical transverse energy correlation plots for two individual central calorimeter towers,
           \subref{fig:l1caloenergycorrelation_a}  electromagnetic and \subref{fig:l1caloenergycorrelation_b}   hadronic}
  \label{fig:l1caloenergycorrelation}
\end{figure}

\begin{sloppypar}
In order to obtain the most precise transverse energy measurements, 
a transverse energy calibration must be applied to all trigger
towers.
The initial transverse energy calibration was produced by
calibration pulser runs.  In
these runs signals of a controlled size are injected into the 
calorimeters.  Subsequently, with sufficient data, the gains were recalibrated
by comparing the transverse energies from the trigger with those calculated offline
from the full calorimeter information.
By the end of the 2010 data-taking this analysis 
had been extended to provide a more precise calibration
on a tower-by-tower basis.  In most cases, the transverse energies derived from the
updated calibration differed by less than 3\%
from those obtained from the
original pulser-run based calibration.
Examples of correlation plots between trigger
and offline calorimeter transverse energies can be seen in 
Fig.~\ref{fig:l1caloenergycorrelation}.
In the future, with even larger datasets, the
tower-by-tower calibration will be further refined
based on physics objects with precisely
known energies, for example, electrons from $Z$ boson decays.
\end{sloppypar}


\subsection{L1 Muon Trigger}\label{sec:L1muon}

\begin{figure}[!ht]
	\centering
  	\includegraphics[width=0.45\textwidth]{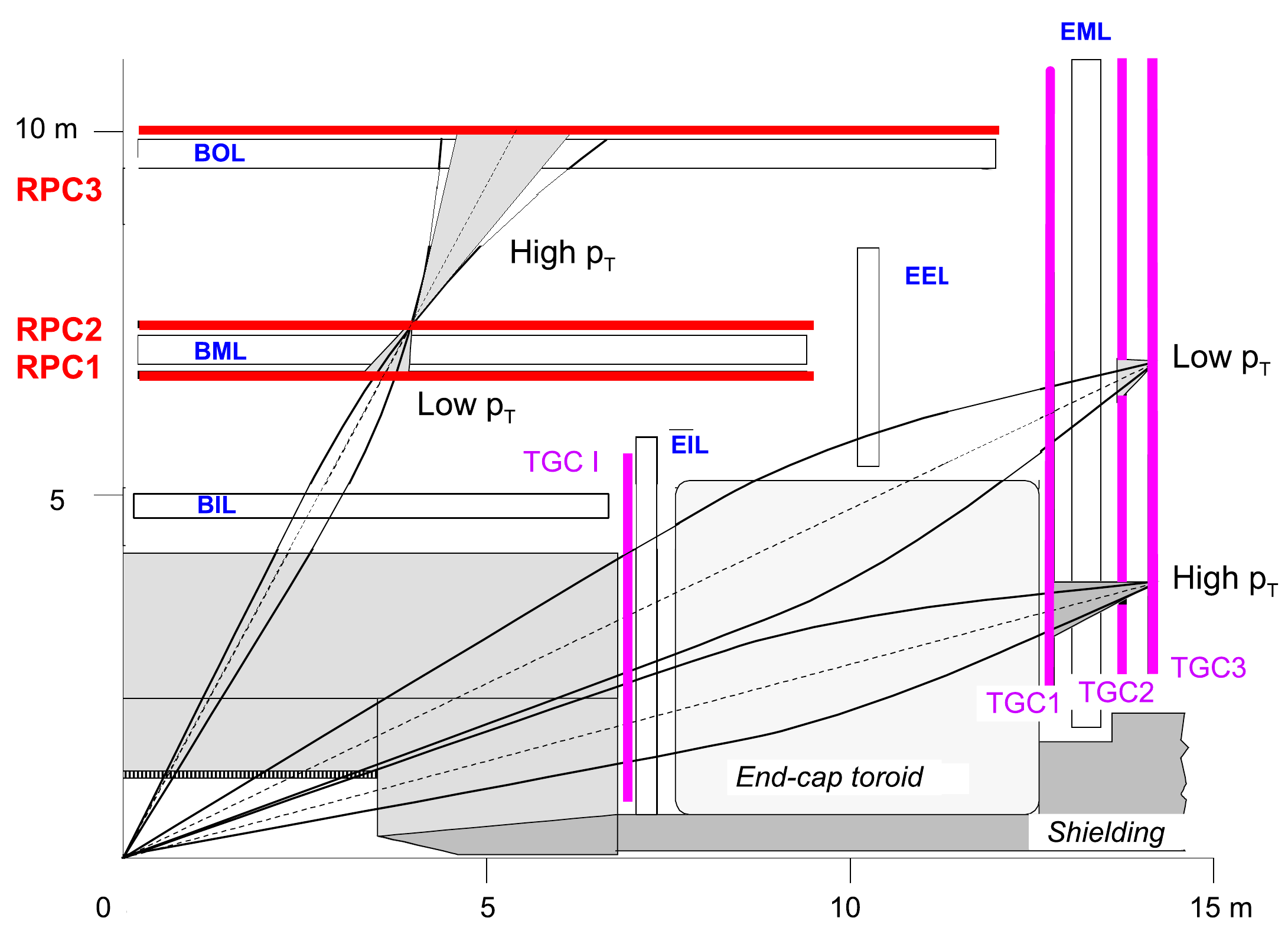}
  	\caption{A section view of the L1 muon trigger chambers. TCG~I was not used in the trigger
in 2010}
    	\label{fig:L1MuonTrigger}
\end{figure}

The L1 muon trigger system \cite{DetectorPaper},\cite{CSCBook} is a hardware-based system 
to process input data from fast muon trigger detectors. 
The system's main task is to select muon candidates and identify the 
bunch crossing in which they were produced.  The primary performance requirement is to be 
efficient for muon \pt\ thresholds above 6 GeV.  
A brief overview of the L1 muon trigger is given here; the performance of the muon trigger
is presented in Section~\ref{sec:muon}.

\subsubsection{L1 Muon Trigger Architecture}

Muons are triggered at L1 using the RPC system in the barrel region ($|\eta|< 1.05$) and the 
TGC system in the end-cap regions
($1.05 <|\eta|< 2.4$), as shown in Fig.~\ref{fig:L1MuonTrigger}.
The RPC and TGC systems provide rough measurements of muon candidate \pt, $\eta$, and $\phi$. 
The trigger chambers are arranged in three planes in the barrel and three in each 
endcap (TCG~I shown in Fig.~\ref{fig:L1MuonTrigger} did not participate in the 2010 trigger).
Each plane is composed of two to four layers. 
Muon candidates are identified by forming coincidences between the muon planes. 
The geometrical coverage of the trigger in the end-caps is  $\approx99$\%. 
In the barrel the coverage is reduced to $\approx80\%$ due to a crack around $\eta=0$, the 
feet and rib support structures for the ATLAS detector and two small elevators in the 
bottom part of the spectrometer.

The L1 muon trigger logic 
is implemented in similar ways for both the RPC and TCG systems, but with the 
following differences:
\begin{itemize}
\item The planes of the RPC system each consist of a doublet of independent detector layers, 
each read out in the $\eta$ ($z$) and $\phi$ coordinates.
A low-\pT\ trigger is generated by requiring a coincidence of hits in at least 3 of 
the 4 layers of the inner two planes, labelled as RPC1 and RPC2 in Fig.~\ref{fig:L1MuonTrigger}).
The high-\pT\  logic starts from a low-\pT\  trigger, 
 then looks for hits in one of the two layers of the high-\pT\ confirmation plane (RPC3). 
\item The two outermost planes of the TGC system (TGC2 and TGC3) each consist of a doublet of independent detectors
read out 
by {\it strips} to measure the $\phi$ coordinate and {\it wires} to measure the $\eta$ coordinate.
A low-\pT\ trigger is generated by a coincidence of hits in at least 3 of 
the 4 layers of the outer two planes.
The inner plane (TGC1) contains 3 detector layers, the wires are read out from all of these, but the
strips from only 2 of the layers.
The high-\pT\ trigger requires at least one of two $\phi$-strip layers
and 2 out of 3 wire layers from the innermost plane
in coincidence with the low-\pT\ trigger.
\end{itemize}
In both the RPC and TGC systems, coincidences are generated separately for $\eta$ and $\phi$ and  
can then be combined with programmable logic to form the final trigger result.  
The configuration for the 2010 data-taking period required a logical AND between the 
 $\eta$ and $\phi$ coincidences in order
to have a muon trigger.

In order to form coincidences, hits are required to lie within parametrized geometrical 
muon \emph{roads}. A road represents an envelope containing the trajectories, 
from the nominal interaction point, of
muons of either charge with a \pT\ above a given threshold. Example roads
are shown in Fig.~\ref{fig:L1MuonTrigger}. 
There are six programmable \pT\ thresholds at L1 (see Table~\ref{tab:ExampleMenu}) which are 
divided into two sets: three low-\pT\ thresholds to cover values up to 10~GeV, and
three high-\pT\ thresholds to cover \pT\ greater than 10 GeV.

To enable the commissioning and validation of the performance of the system for 2010 running, 
two triggers were defined which did not require coincidences within roads and thus 
gave maximum acceptance and minimum trigger bias. One (MU0)
based on low-\pT\ logic and the other \linebreak (MU0\_COMM) based on the high-\pT\ logic. For these
triggers the only requirement was that hits were in the same trigger tower
($\eta~\times~\phi~\sim~$0.1$~\times~$0.1).

\begin{figure}[!htb]
	\centering
	\includegraphics[width=0.45\textwidth]{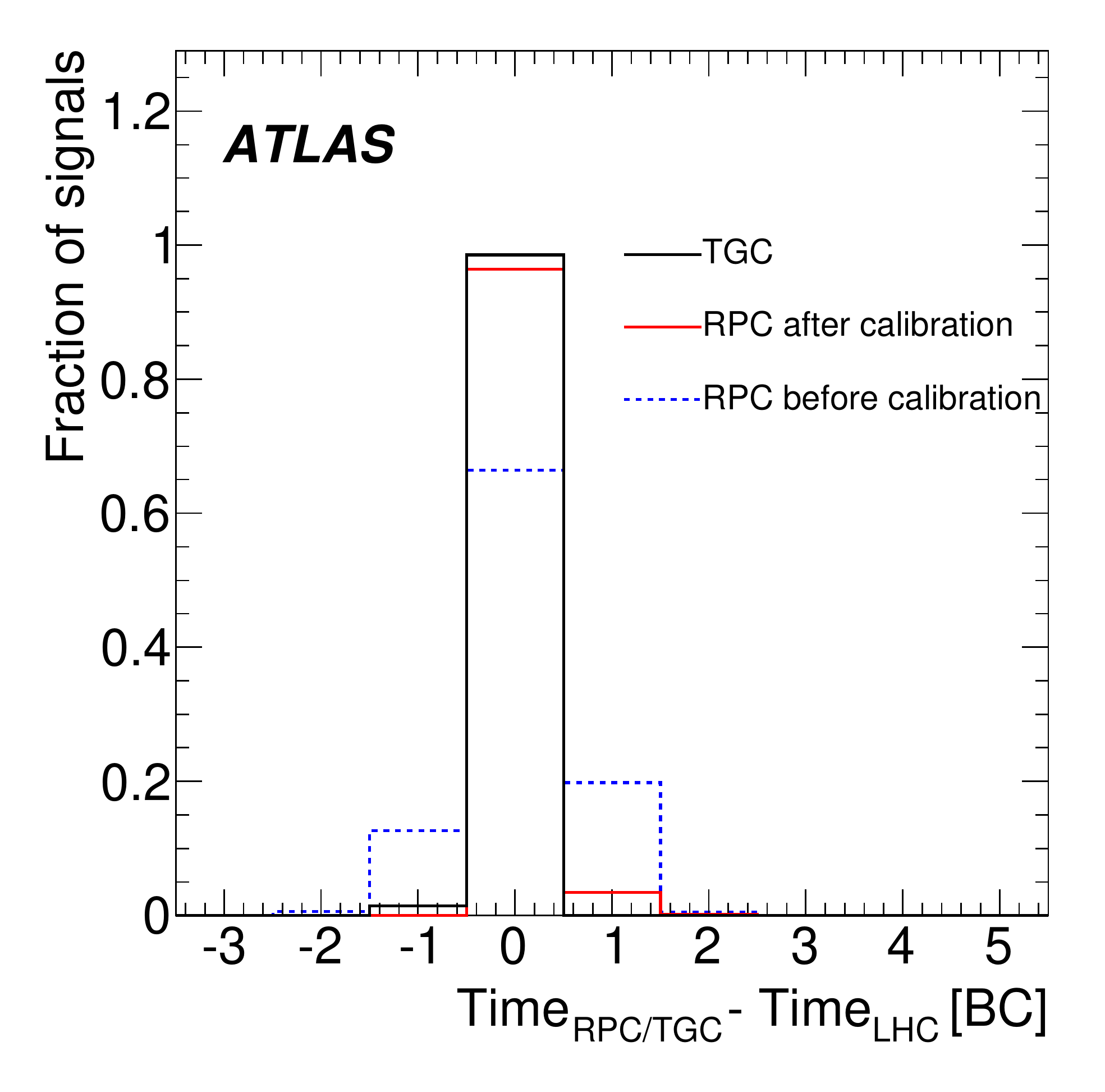}
  	\caption{The timing alignment with respect to the LHC bunch clock (25~ns units) for the RPC system (before and after the timing calibration) and the TGC system}
    	\label{fig:MuonTiming}
\end{figure}

\subsubsection{L1 Muon Trigger Timing Calibration}
In order to assign the hit information to the correct bunch crossing, a precise 
alignment of RPC and TGC signals, or timing calibration, was performed to
take into account signal delays in all components of the read out and trigger chain.
Test pulses were used to calibrate the TGC timing to within 25~ns (one bunch crossing) 
before the start of data-taking.
Tracks from cosmic ray and collision data were used to calibrate the timing of the RPC system.
This calibration required a sizable data sample to be collected before 
a time alignment of better than 25~ns was reached.
As described in section~\ref{sec:L1ctp}, the CTP imposes a 25~ns window about the nominal bunch crossing time
during which signals must arrive in order to contribute to the trigger decision.
In the first phase of the data-taking, while the timing calibration of the RPC system was on-going, 
a special CTP configuration was used to increase
the window for muon triggers to 75~ns. 
The majority of 2010 data were collected with both systems aligned to within one bunch crossing for 
both high-\pt\ and low-\pT\ triggers.
In Fig.~\ref{fig:MuonTiming}~ the timing alignment of the RPC and TGC systems is shown
with respect to the LHC bunch clock in units of the 25~ns bunch crossings (BC).


\section{High Level Trigger Reconstruction}\label{sec:reconstruction}
The HLT has additional information available, compared to L1, including
inner detector hits, full information from the calorimeter 
and data from the precision muon detectors.
The HLT trigger selection is based on features reconstructed in these
systems. The reconstruction is performed, for the most part, inside RoIs 
in order to minimize execution times and reduce data
requests across the network at L2.
The sections below give a brief description of the algorithms for
inner detector tracking, beamspot measurement, calorimeter 
clustering and muon reconstruction. The performance of the algorithms
is presented, including
measurements of execution times which meet the timing constraints 
outlined in Section~\ref{sec:overview}.

\subsection{Inner Detector Tracking}\label{sec:idReco}
\def \figurepath{.}

The track reconstruction in the Inner Detector is an essential
component of the trigger decision in the HLT. A robust and efficient
reconstruction of particle trajectories is a prerequisite
for triggering on electrons, muons, $B$-physics, taus, and $b$-jets.
It is also used for triggering on inclusive \pp\ interactions and for the online
determination of the beamspot (Section~\ref{sec:beamspotReco}), 
where the
reconstructed tracks provide the input to reconstruction of
vertices. This section gives a short description of the reconstruction algorithms and 
an overview of the performance of the
track reconstruction with a focus on tracking efficiencies in the ATLAS trigger system.

\begin{figure*}[!htb]
  \centering
  \subfigure[]{
    \includegraphics[width=0.45\textwidth]{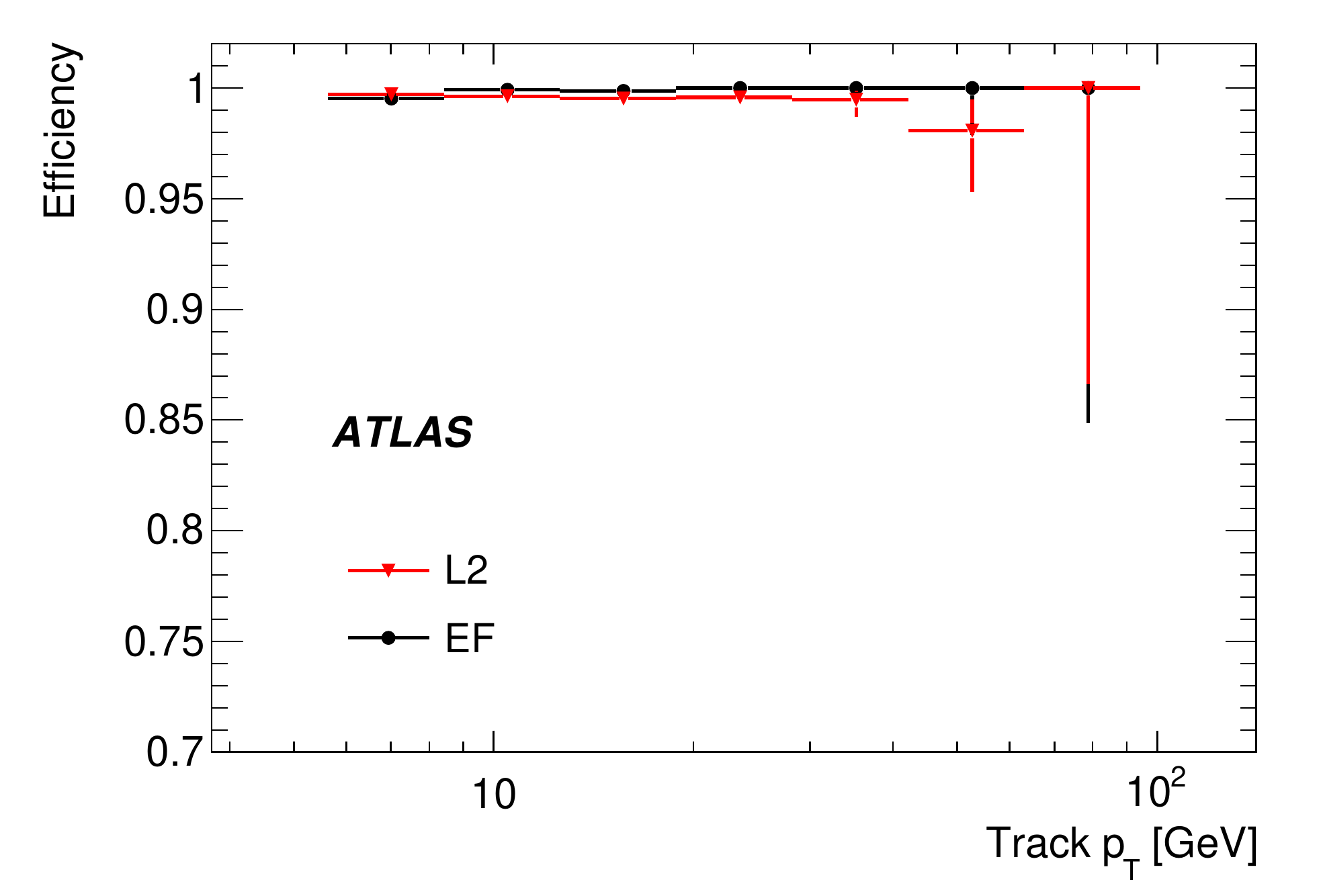}
    \label{fig:eff-mu-staco-pT}
  }
  \subfigure[]{
    \includegraphics[width=0.45\textwidth]{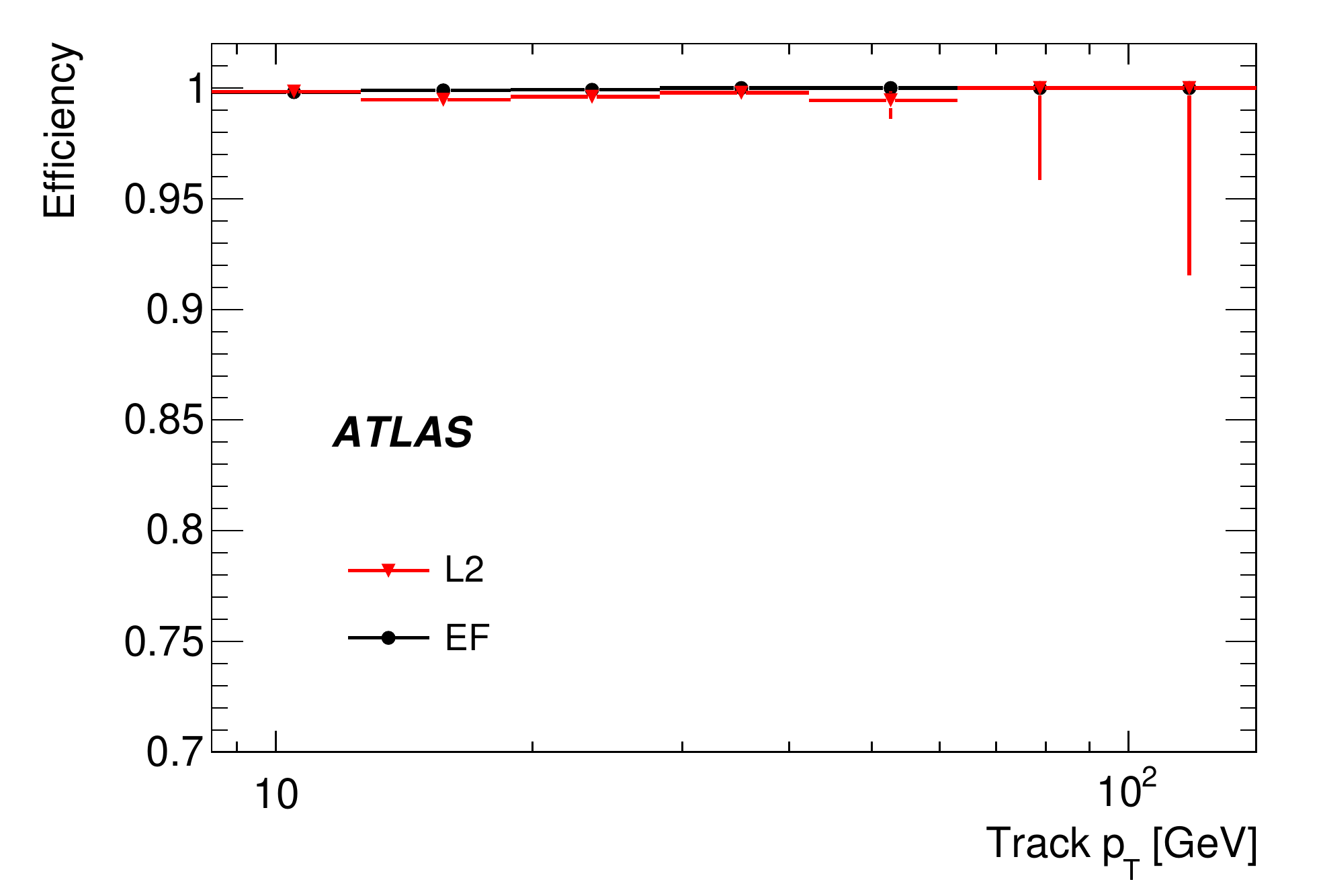}
    \label{fig:elec-trk-eff}
  }
  \caption{L2 and EF tracking reconstruction efficiency with respect to offline \subref{fig:eff-mu-staco-pT} muon candidates
 and \subref{fig:elec-trk-eff} electron candidates} 
  \label{fig:muon-elec-effic}
\end{figure*}
\begin{figure*}[!htb]
  \centering
  \subfigure[]{
    \includegraphics[width=0.45\textwidth]{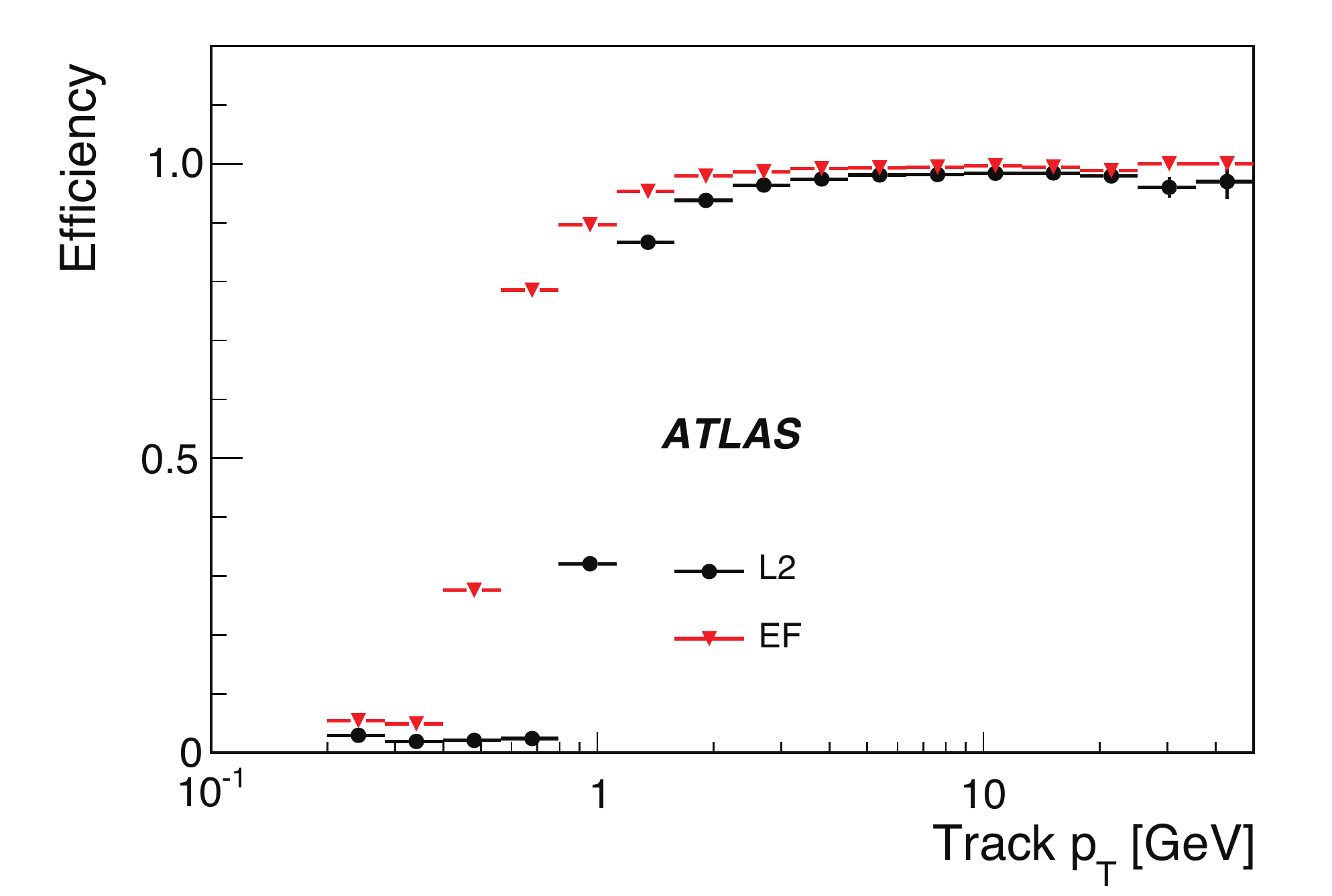}
    \label{fig:tau-eff-pT}
  }
  \subfigure[]{
    \includegraphics[width=0.45\textwidth]{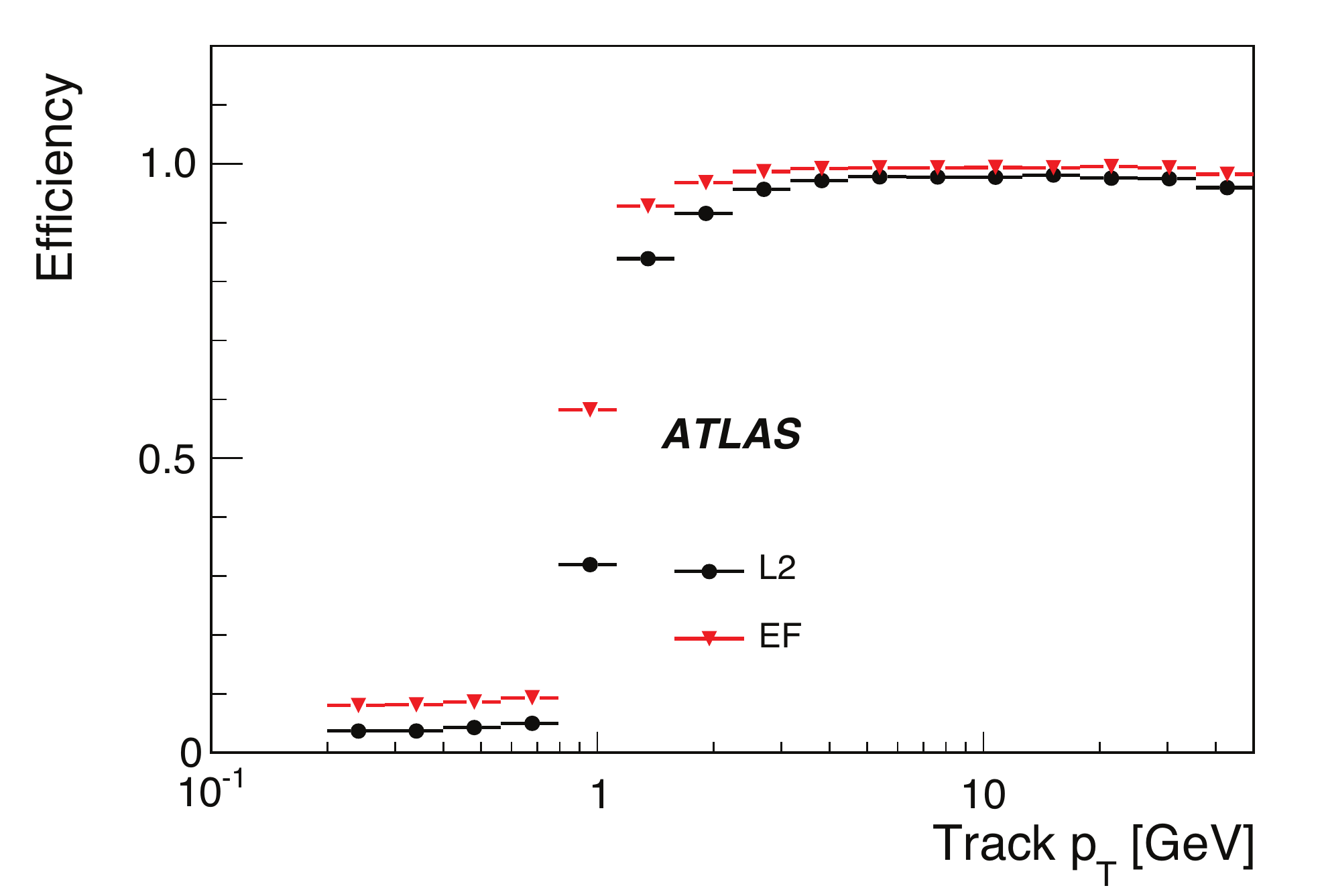}
    \label{fig:jet-eff-pT}
  }
  \caption{L2 and EF tracking reconstruction efficiency with respect to offline reference
tracks inside
\subref{fig:tau-eff-pT} tau RoIs and  \subref{fig:jet-eff-pT} jet RoIs}
  \label{fig:taujet-trk}
\end{figure*}

\subsubsection{Inner Detector Tracking Algorithms}

The L2 reconstruction algorithms are specifically designed to meet the strict timing
requirements for event processing at L2.
The track reconstruction at the EF is less time constrained and can use,
to a large extent, software components from the offline reconstruction.  In
both L2 and EF the track finding is preceded by a data
preparation step in which detector data are decoded and transformed to
a set of hit positions in the ATLAS coordinate system.
Clusters are first formed from adjacent signals on the SCT strips or 
in the Pixel detector.  
Two-dimensional Pixel clusters and pairs of one-dimensional SCT clusters (from back-to-back detectors 
rotated by a small stereo angle with respect to one another) are combined with geometrical information to provide three-dimensional hit information, called \emph{space-points}.
Clusters and space-points provide the input to the HLT   
 pattern recognition algorithms.

The primary track reconstruction strategy is \emph{inside-out} tracking
which starts with pattern recognition in the SCT and Pixel detectors;
 track candidates are then extended to the TRT volume.  In addition, 
the L2 has an algorithm that reconstructs tracks in the  
TRT only and the EF has an additional track reconstruction strategy
that is \emph{outside-in}, starting from the TRT and 
extending the tracks to the SCT and Pixel detectors.  

Track reconstruction at both L2 and EF
is run in an RoI-based mode for electron, muon, tau and
$b$-jet signatures. $B$-physics signatures  are based either on a 
\nomenclature{\bf FS}{FullScan, track reconstruction within the entire volume of the  Inner Detector}
\emph{FullScan} (FS) mode (using the entire volume of the Inner Detector) or a large RoI.
The tracking algorithms can be configured differently for each
signature in order to provide the best performance.

L2 uses two different pattern recognition strategies:
\begin{itemize}
  \setlength{\itemsep}{1pt}
  \setlength{\parskip}{0pt}
  \setlength{\parsep}{0pt}
\item A three-step histogramming technique, called \emph{IdScan}. First, the $z$-position of
the primary vertex, $z_v$, is determined as follows. The RoI is divided into $\phi$-slices and $z$-intercept values
are calculated and histogrammed for lines through all possible pairs 
of space-points in each $phi$-slice; $z_v$ is determined from peaks in this histogram. 
The second step is to fill a $(\eta,\phi)$ histogram with values calculated with respect to $z_v$
for each space-point in the RoI; groups of space-points to be passed on to the third step are
identified from histogram bins containing at least four space-points from different detector layers.
In the third
step, a $(1/\pT, \phi)$ histogram is filled from values calculated for
all possible triplets of space-points from different detector layers; track candidates are formed from
bins containing at least four space-points from different layers. 
This technique is the approach used for electron, muon and $B$-physics triggers due to the slightly higher efficiency of \emph{IdScan} relative to \emph{SiTrack}.

\item A combinatorial technique, called \emph{SiTrack}. First, pairs of hits consistent with a 
beamline constraint are found within a subset of the
inner detector layers. Next, triplets are formed by associating additional hits in the
remaining detector layers consistent with a track from the beamline. In the final step, triplets 
consistent with the same track trajectory are merged, duplicate or 
outlying hits are removed and the remaining hits are passed to the track fitter. \emph{SiTrack}
is the approach used for tau and jet triggers as well as the beamspot measurement as it
has a slightly lower fake-track fraction.

\end{itemize}
In both cases, track candidates are further processed by
a common Kalman~\cite{bib:KalmanFitter} filter track fitter and extended to the TRT for an
improved \pt\ resolution and to benefit from the electron identification capability of the
TRT.  

The EF track reconstruction is based on software shared with the
offline reconstruction~\cite{bib:newtracking}. The offline software was
extended to run in the trigger environment by adding support for
reconstruction in an RoI-based mode. The pattern recognition in the
EF starts from seeds built from triplets of space-points in the Pixel and SCT detectors. 
Triplets consist of space-points from different layers, all in the pixel detector, all in the the SCT or two space-points in the pixel detector and one in the SCT.
Seeds are preselected by imposing
a minimum requirement on the momentum and a maximum requirement on the impact parameters.
The seeds
define a road in which a track candidate can be formed
by adding additional clusters using a combinatorial Kalman filter technique. 
In a subsequent step, the
quality of the track candidates is evaluated and low quality candidates 
are
rejected. The tracks are then extended into the TRT and a final fit 
is performed to extract the track parameters.

\subsubsection{Inner Detector Tracking Algorithms Performance}

\begin{figure}[!htb]
  \centering
    \includegraphics[width=0.45\textwidth]{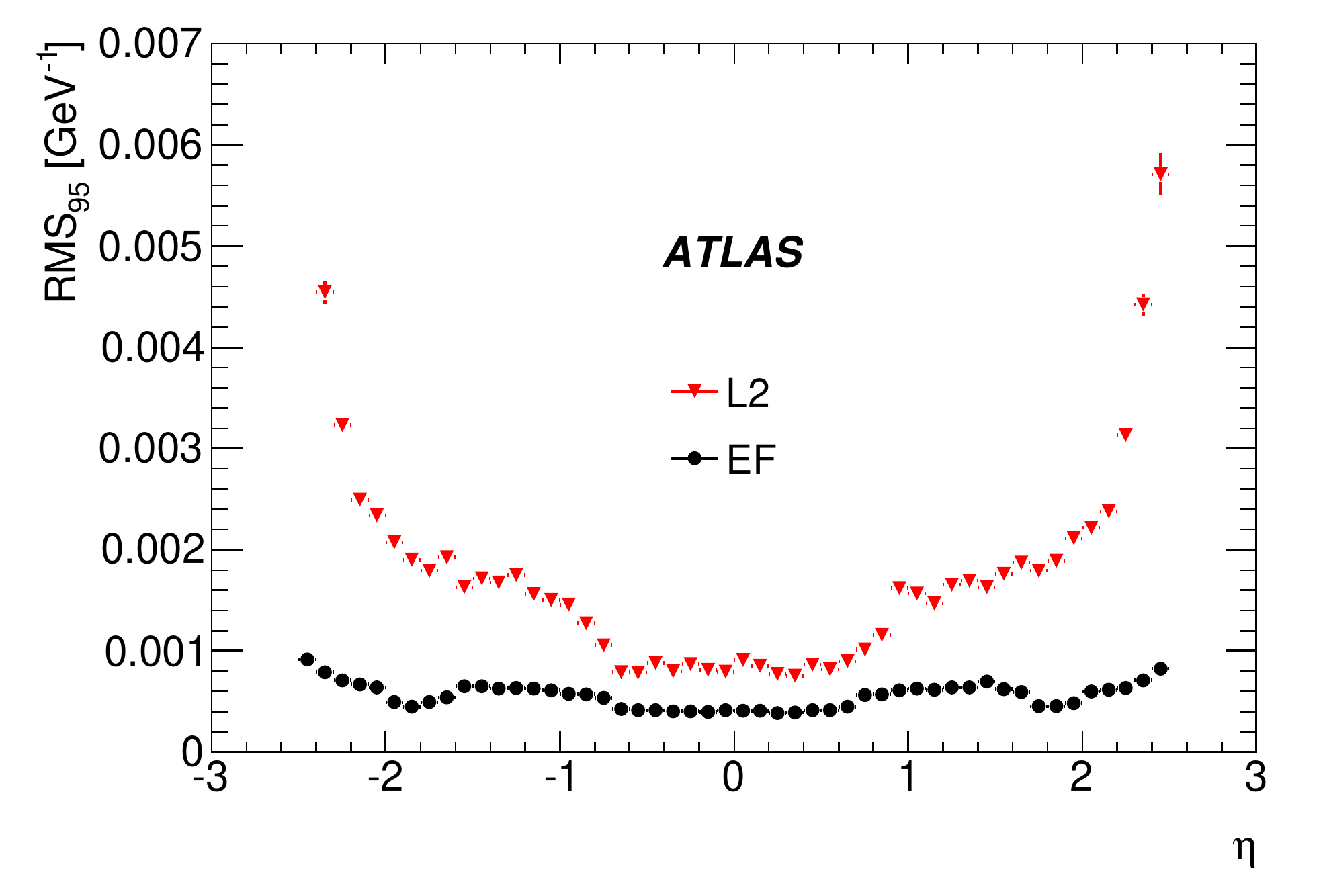}
  \caption{The RMS of the core 95\% (RMS$_{95}$) of the inverse-\pT\ residual as a function of offline track $\eta$}
    \label{fig:id-pt-res}
\end{figure}
\begin{sloppypar}
The efficiency of the tracking algorithms is studied using specific monitoring
triggers, which do not require a track to be present for the event to be accepted,
and are thus unbiased for track efficiency measurements. The efficiency is defined
as the fraction of offline reference tracks that are matched to a trigger track (with matching requirement \linebreak $\dR=\sqrt{\Delta\phi^2+\Delta\eta^2}<0.1$).
Offline reference tracks are required to have 
$|\eta|<2.5$, $|d_0|<1.5$\,mm, $|z_0|<200$\,mm and $|(z_0-z_V)\sin{\theta}|<1.5$\,mm, where
$d_0$ and $z_0$ are the transverse and longitudinal impact parameters,
 and $z_V$ is the position of the primary vertex along the beamline as reconstructed 
offline. 
The reference tracks are also required to have one Pixel hit and at
least six SCT clusters. For tau and jet RoIs, the
reference tracks are additionally required to have $\chi^2$ probability of the track fit higher than 1\%,
two Pixel hits, one in the innermost layer, and a total of at
least seven SCT clusters.
\end{sloppypar}

The L2 and EF tracking efficiencies are shown as a function of \pT\ for offline 
muon candidates in Fig.~\ref{fig:eff-mu-staco-pT} and for offline electron candidates 
in Fig.~\ref{fig:elec-trk-eff}.  Tracking efficiencies in tau and jet RoIs are shown in 
Fig.~\ref{fig:taujet-trk}, determined with respect to all offline reference tracks lying within the RoI.
In all cases, the efficiency is close to 100\% in the \pT\ range important for triggering.

\begin{figure}[!ht]
  \centering
  \subfigure[]{
    \includegraphics[width=0.45\textwidth]{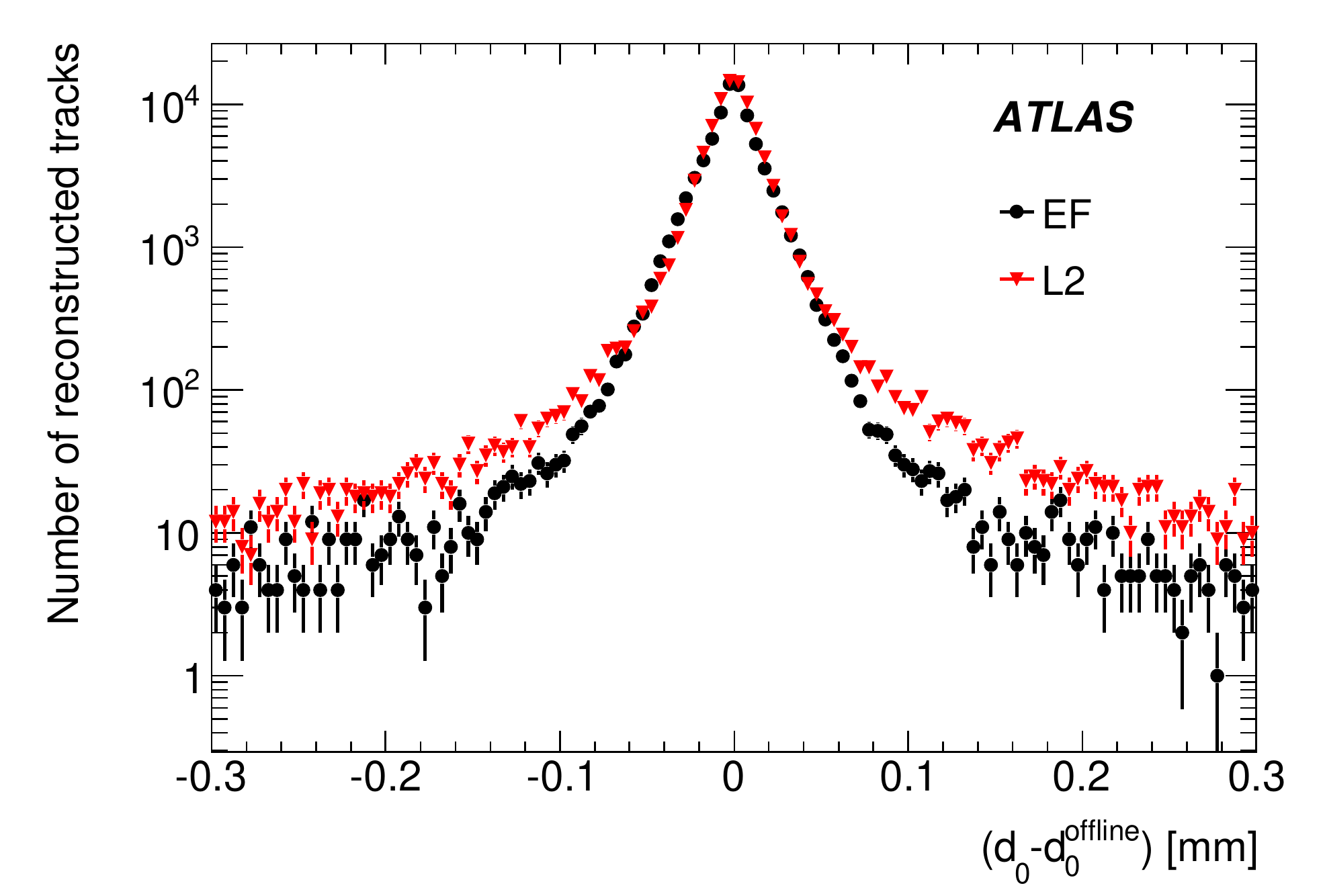}
    \label{fig:id-d0-res}
  }
  \subfigure[]{
    \includegraphics[width=0.45\textwidth]{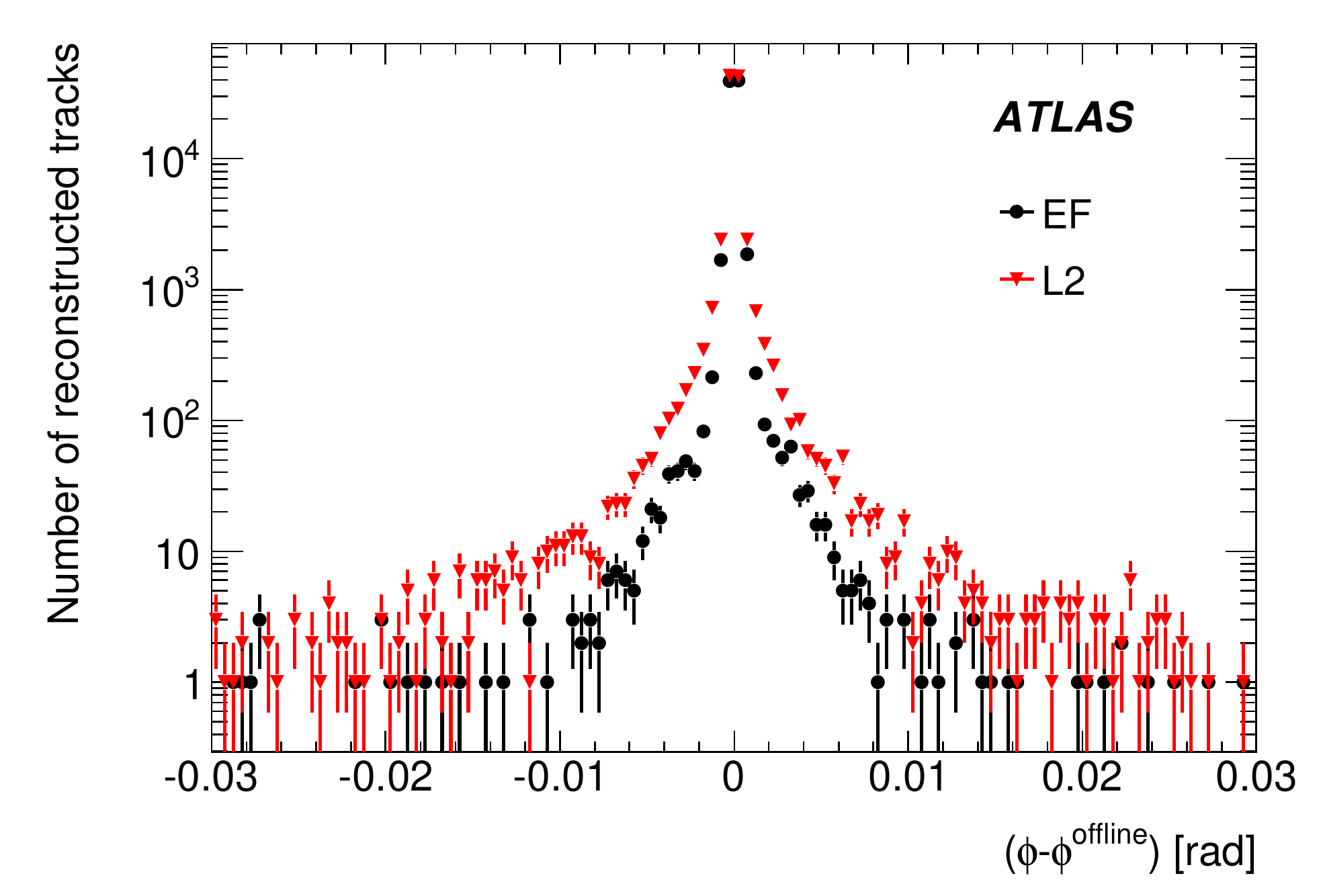}
    \label{fig:id-phi-res}
  }
  \subfigure[]{
    \includegraphics[width=0.45\textwidth]{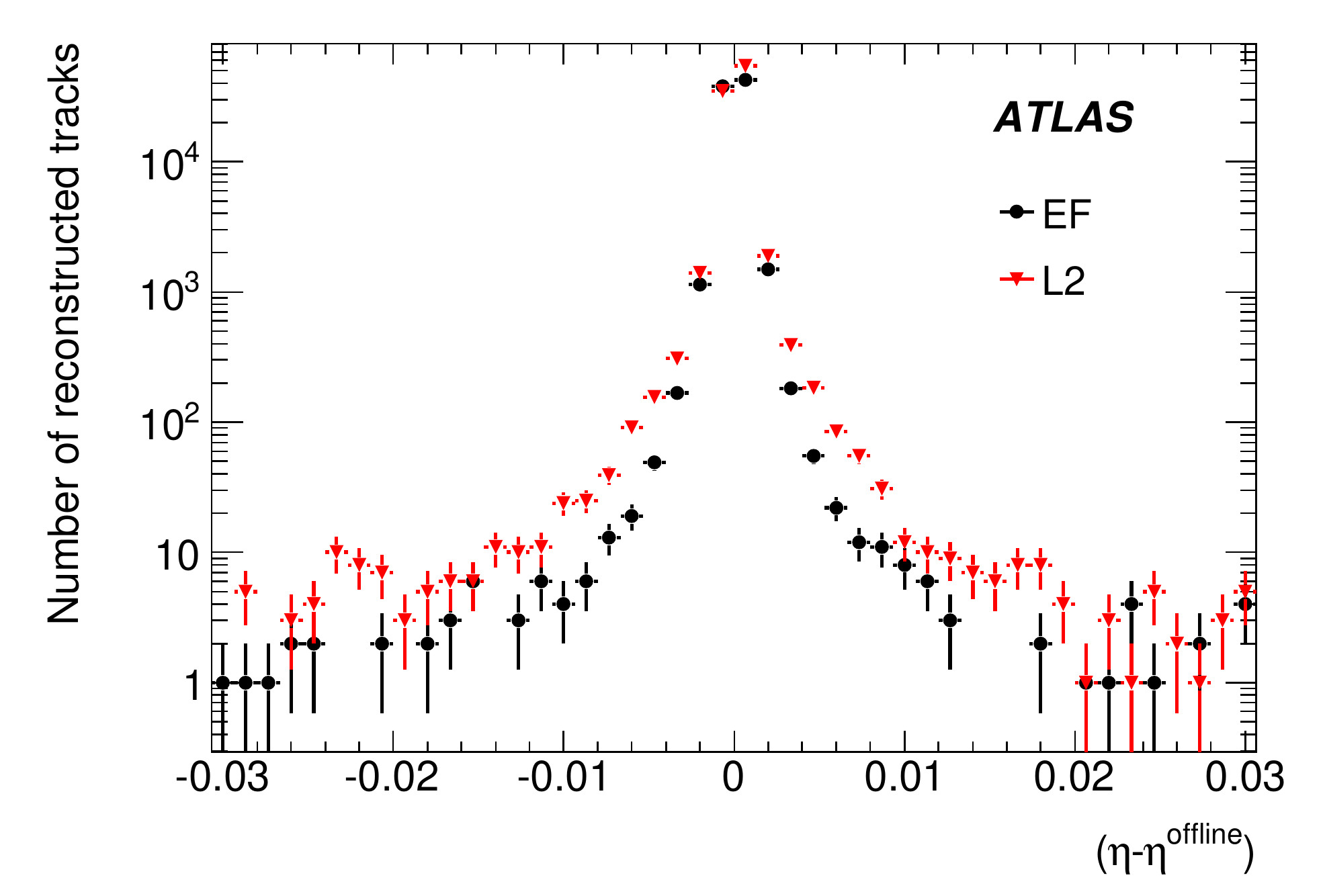}
    \label{fig:id-eta-res}
  }

  \caption{Residuals with respect to offline for track parameters \subref{fig:id-d0-res}
$d_0$, \subref{fig:id-phi-res}  $\phi$ and \subref{fig:id-eta-res} $\eta$}
  \label{fig:trk-residuals}
\end{figure}

\begin{sloppypar}
The RMS of the core 95\% (RMS$_{95}$) of the inverse-\pT\ residual $((\frac{1}{\pt})^{\mathrm{trigger}}-(\frac{1}{\pt})^{\mathrm{offline}})$ distribution is shown as a function of $\eta$ in
Fig.~\ref{fig:id-pt-res}.  Both L2 and EF show good agreement with offline, although
the residuals between L2 and offline are larger, particularly at high \abseta\, as
a consequence of the speed-optimizations made at L2. 
Figure~\ref{fig:trk-residuals} shows the residuals in 
$d_0$, $\phi$ and $\eta$. 
Since it uses offline software, EF tracking performance is close to that of the
offline reconstruction. Performance is not identical, however, due to an
online-specific 
configuration of offline software designed to increase speed and
be more robust to 
compensate for the more
limited calibration and detector status information available in the online environment. 
\end{sloppypar}

\subsubsection{Inner Detector Tracking Algorithms Timing}

\begin{figure}[!htb]
  \centering
  \subfigure[]{
    \includegraphics[width=0.45\textwidth]{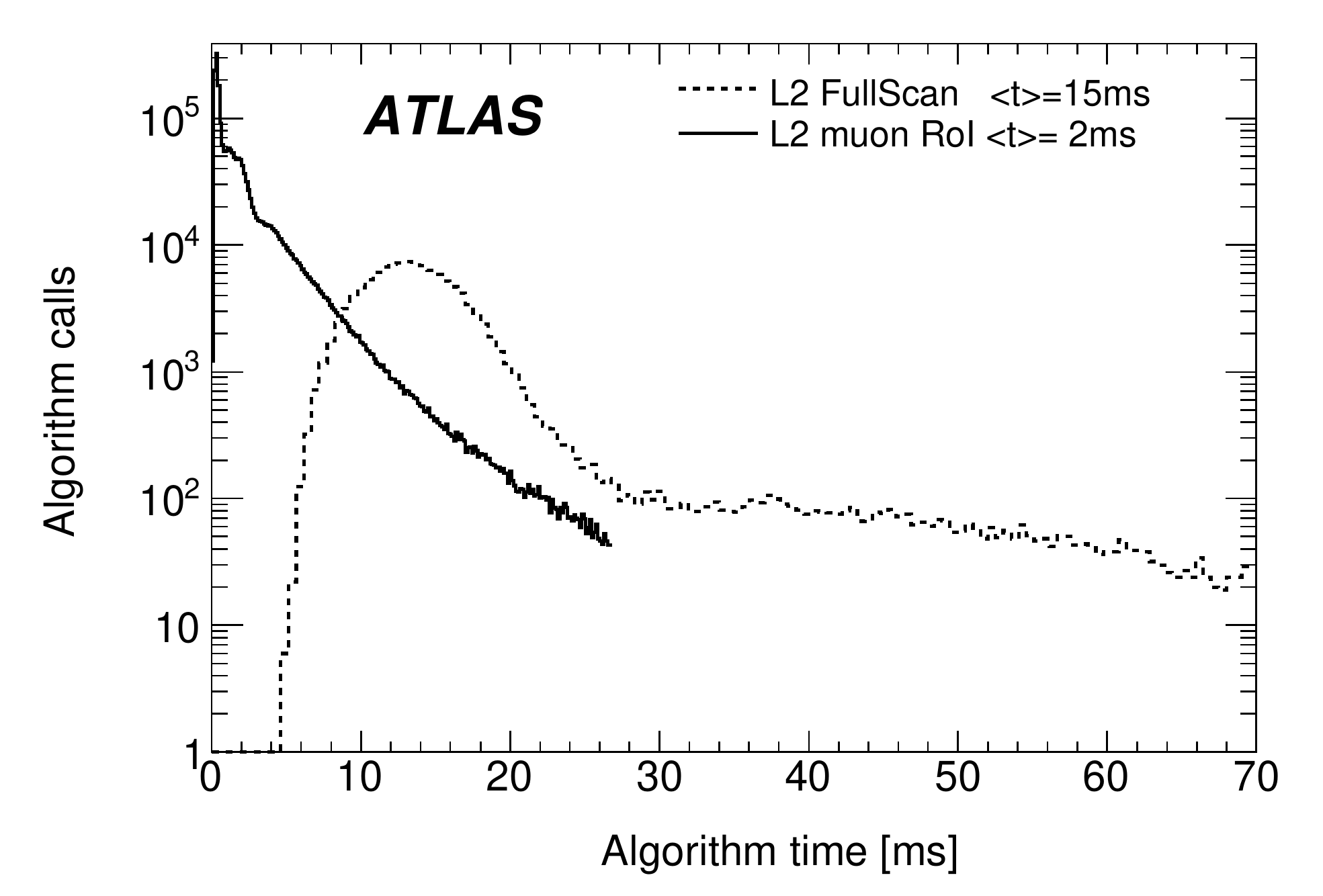}
    \label{fig:id-L2-time}
  }
  \subfigure[]{
    \includegraphics[width=0.45\textwidth]{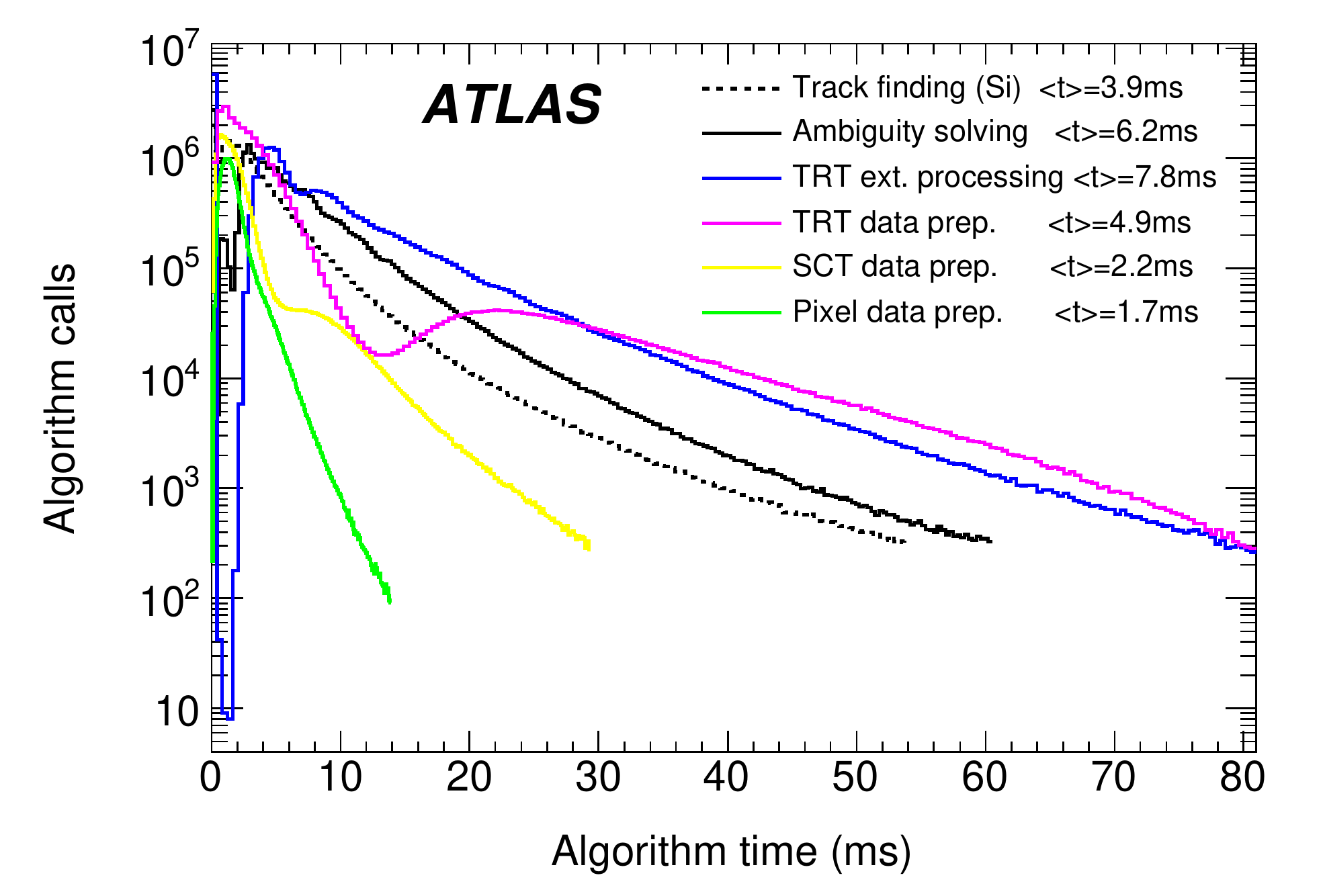}
    \label{fig:id-EF-time}
  }
  \subfigure[]{
    \includegraphics[width=0.45\textwidth]{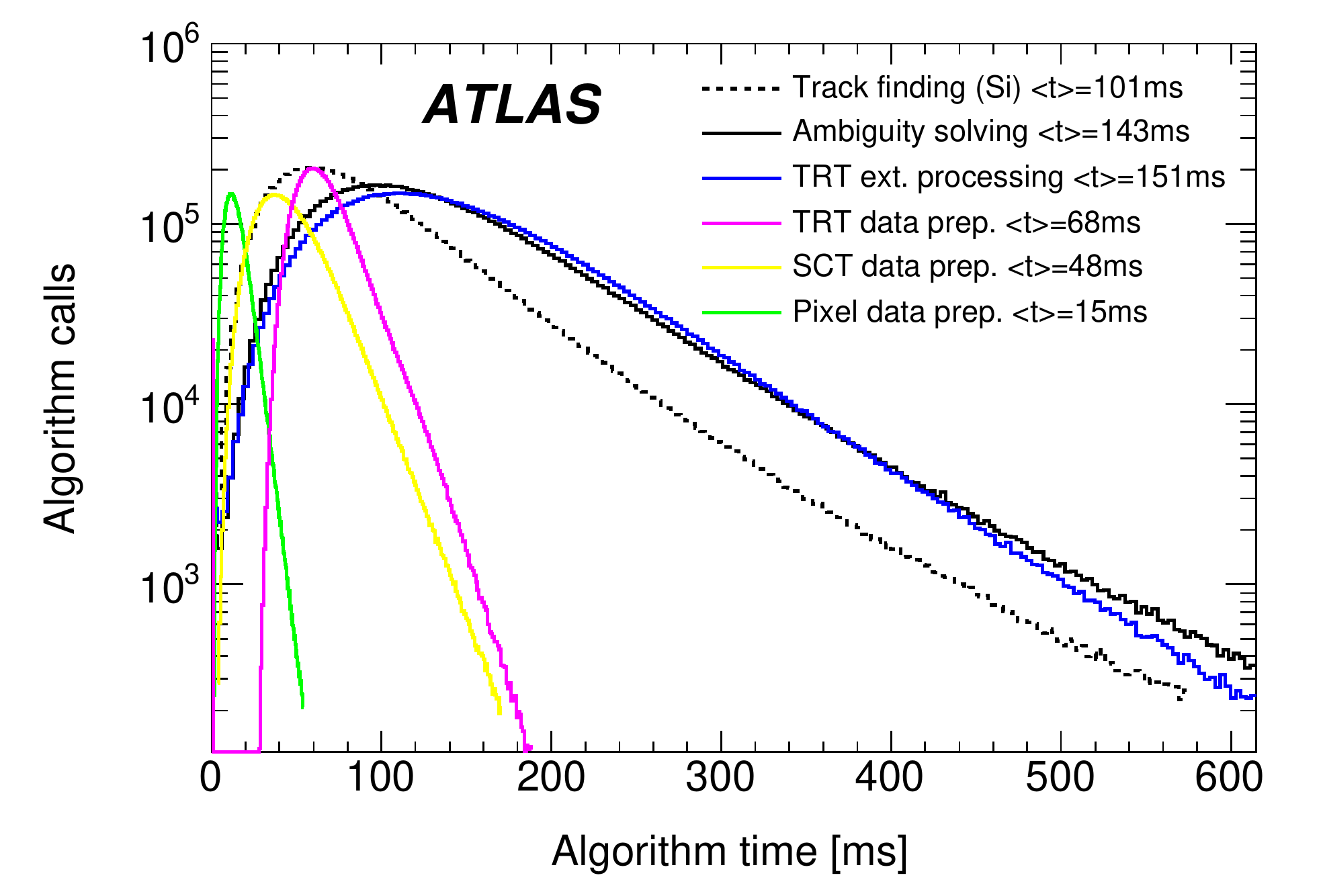}
    \label{fig:id-EF-FS-time}
  }
  \caption{Execution times for \subref{fig:id-L2-time} FullScan reconstruction and 
reconstruction within muon RoIs at L2, \subref{fig:id-EF-time} the different EF 
reconstruction steps in muon RoIs and \subref{fig:id-EF-FS-time} the different steps
of the EF FullScan.  The mean time of each algorithm is marked in the legend.  The structure in the TRT data preparation time in \subref{fig:id-EF-time} is due to caching}
  \label{fig:id-time} 
\end{figure}

Distributions of the algorithm execution time at L2 and EF are shown in Fig.~\ref{fig:id-time}.
The total time for L2 reconstruction is shown in  Fig.~\ref{fig:id-L2-time} for a muon algorithm in RoI 
and FullScan mode. The times of the different reconstruction steps at the EF are shown in 
Fig.~\ref{fig:id-EF-time} for
muon RoIs and in Fig.~\ref{fig:id-EF-FS-time} for FullScan mode. The execution times are shown for all instances of the algorithm execution, whether the trigger was passed or not.  The execution times are
well within the online constraints.  

\subsection{Beamspot}\label{sec:beamspotReco}
\def \figurepath{.}
The online beamspot measurement uses L2 ID tracks from the \emph{SiTrack} 
algorithm (Section~\ref{sec:idReco}) to reconstruct
primary vertices on an event-by-event basis~\cite{bib:beamspotconfnote}.  
The vertex position
distributions collected over short time intervals are used to measure
the position and shape of the luminous region, \emph{beamspot},
parametrized by a three-dimensional Gaussian. The coordinates of the
centroids of reconstructed vertices determine the average position of the 
collision point in the
ATLAS coordinate system as well as the size and orientation of the
ellipsoid representing the luminous region in the horizontal ($x$-$z$) and
vertical ($y$-$z$) planes.

These observables are continuously reconstructed and monitored online
in the HLT, and communicated, for each luminosity block, to displays in 
the control room. In addition, the instantaneous rate of reconstructed
vertices can be used online as a luminosity monitor. Following these
online measurements, a system for applying real-time configuration
changes to the HLT farm distributes updates for use by trigger
algorithms that depend on the precise knowledge of the luminous
region, such as $b$-jet tagging (Section~\ref{sec:bjet}).

\begin{figure}[!ht]
  \centering
  \includegraphics[width=\columnwidth]{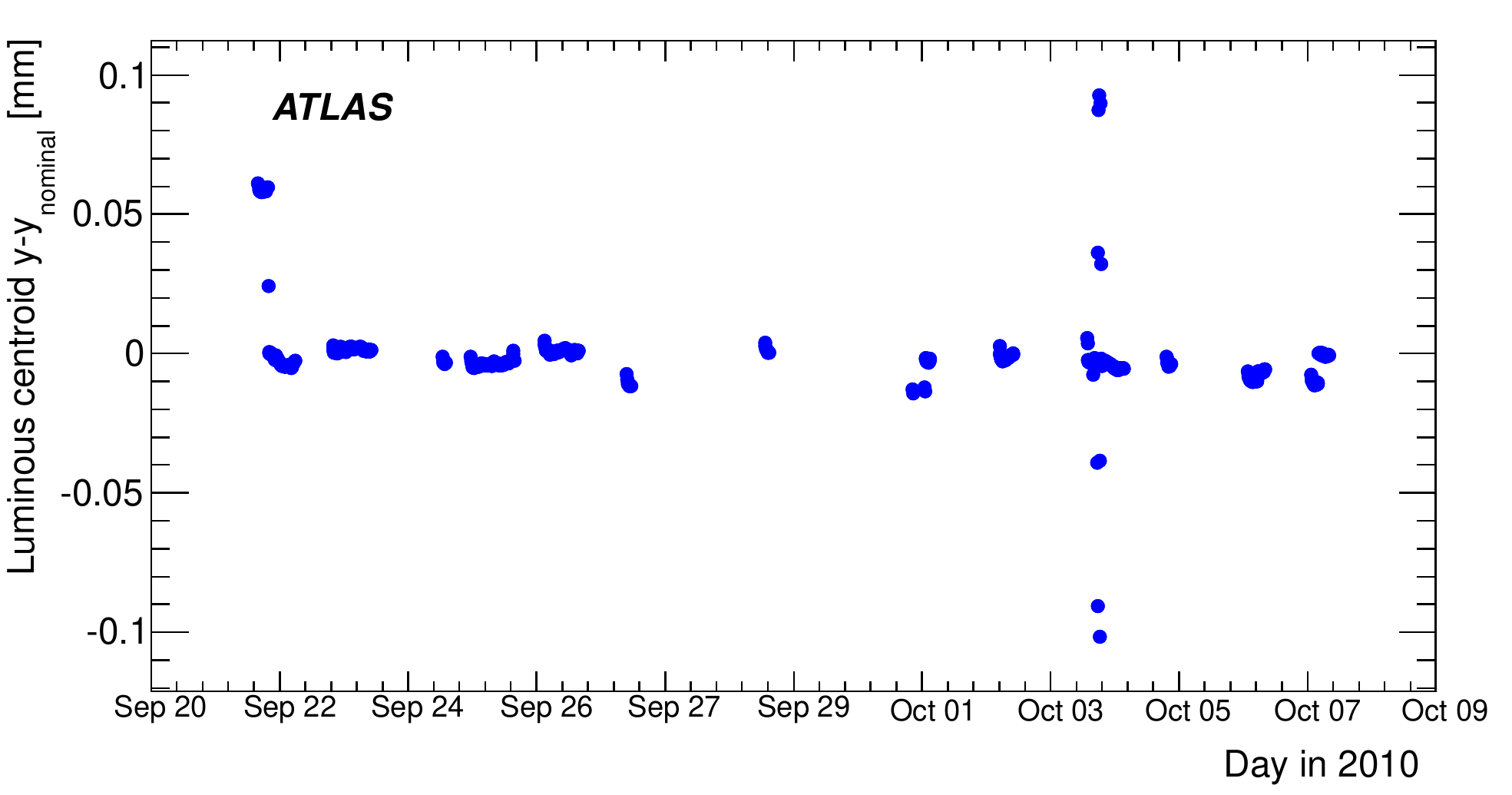}
  \caption{A timeline of the observed collision point centroid position relative to the nominal beam position}
  \label{fig:BeamspotTimeline}
\end{figure}
    
Figure~\ref{fig:BeamspotTimeline} shows the variation of the collision point 
centroid around the 
nominal beam position in the transverse plane
($y_{nominal}$) over a period of a few weeks.   
The nominal beam position, which is typically up to several
hundred microns from the centre of the ATLAS coordinate system,
is defined by a time average of previous measured centroid positions.
The figure shows that
updates distributed to the online system as a part
of the feedback mechanism take account of the measured beam
position within a narrow band of only a few microns.  The large deviations on 
Oct 4 and Sept 22 are from
beam-separation scans.  

\begin{sloppypar}
During 2010 data-taking, beamspot measurements were  
averaged over the entire period of stable beam during 
a run and updates applied, 
for subsequent runs, in the case of significant shifts.  
For 2011 running, when triggers that are sensitive to 
the beamspot position, such as the $b$-jet trigger~(Section~\ref{sec:bjet}), 
are activated, updates will be made more frequently.
\end{sloppypar}

\begin{figure*}[!htb]
  \centering
  \subfigure[]{
    \includegraphics[width=0.45\textwidth]{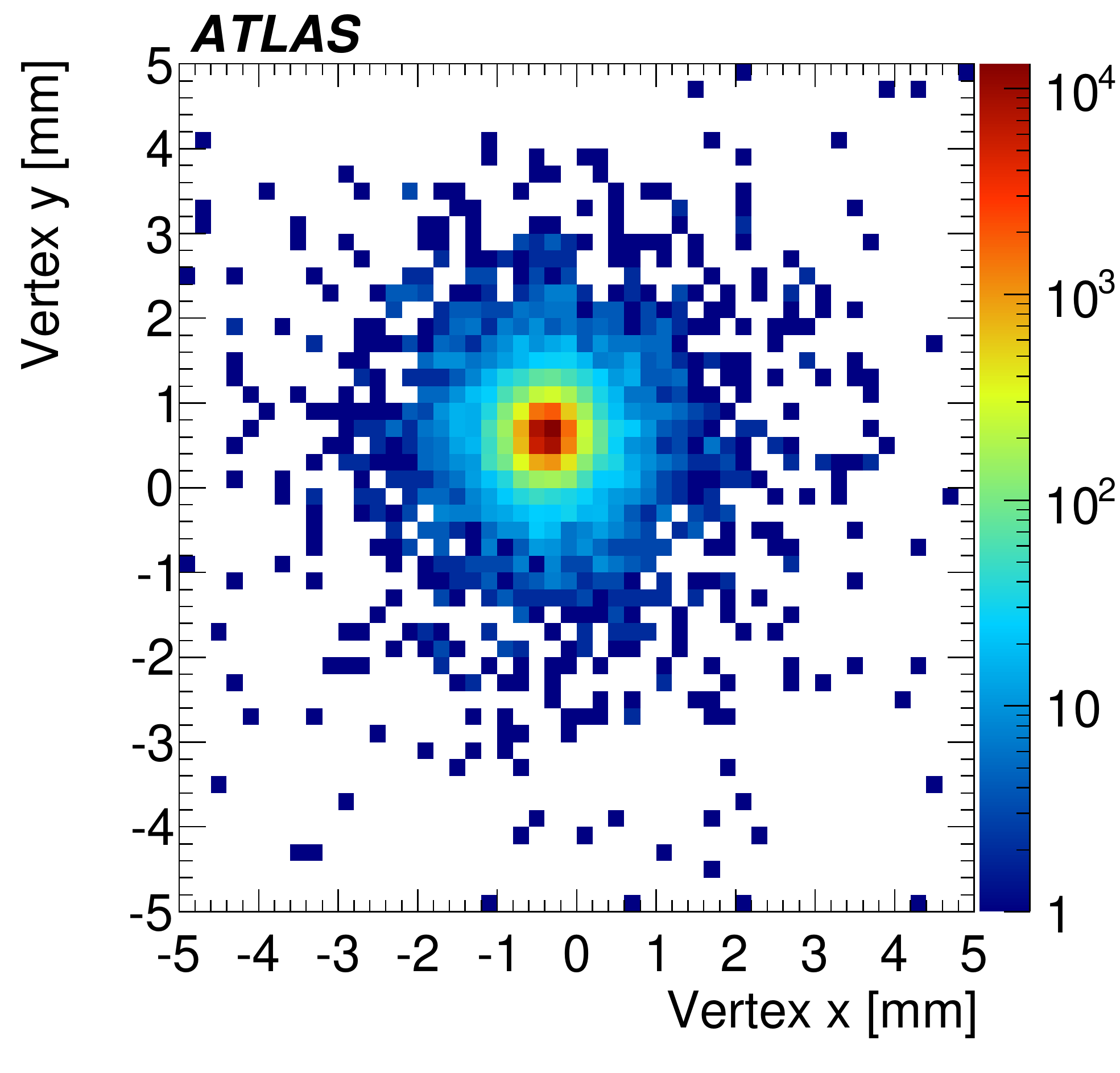}
    \label{fig:BeamspotProfileXY}
  }
  \subfigure[]{
    \includegraphics[width=0.45\textwidth]{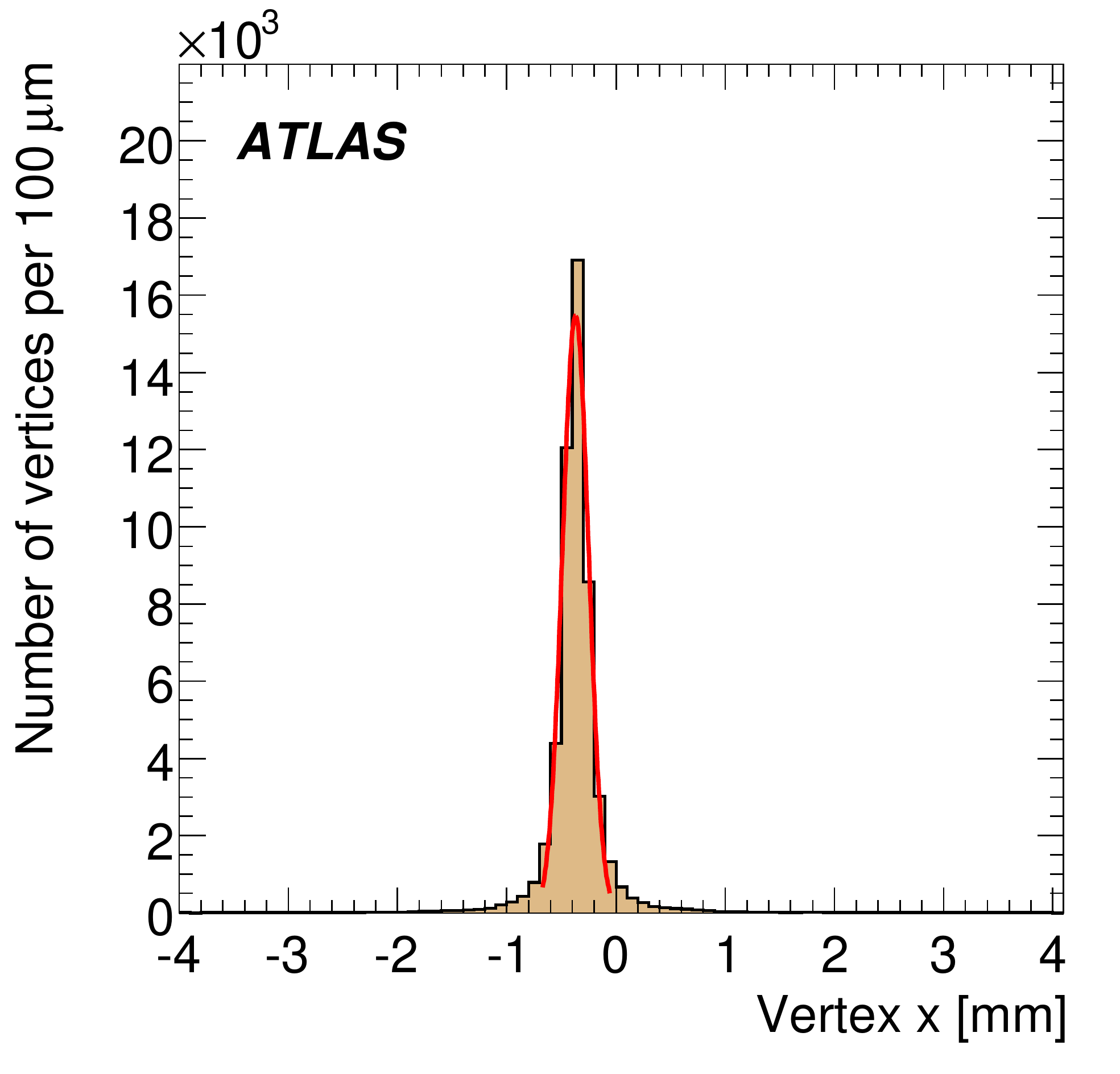}
    \label{fig:BeamspotDistX}
  }
  \subfigure[]{
    \includegraphics[width=0.45\textwidth]{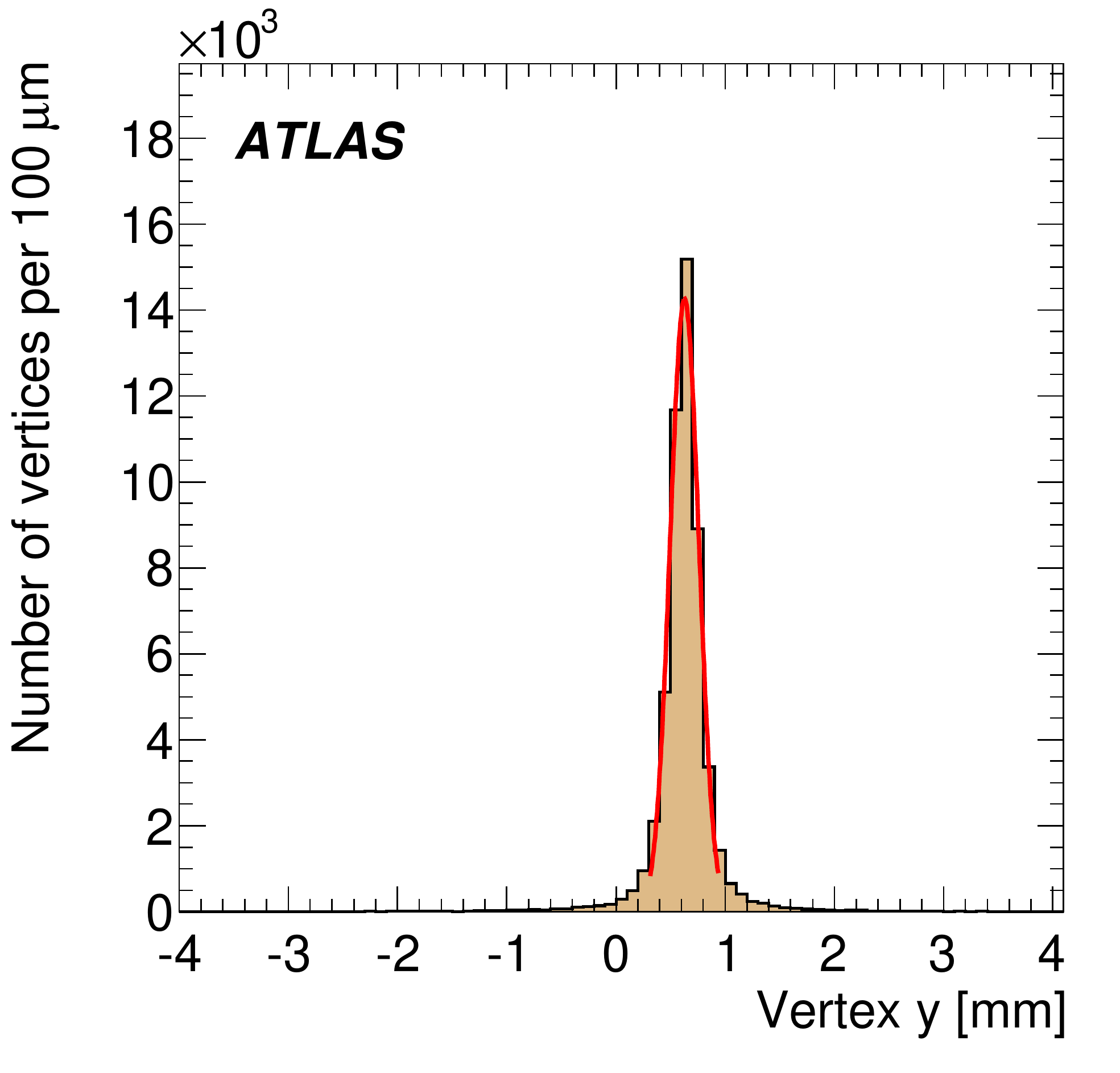}
    \label{fig:BeamspotDistY}
  }
  \subfigure[]{
    \includegraphics[width=0.45\textwidth]{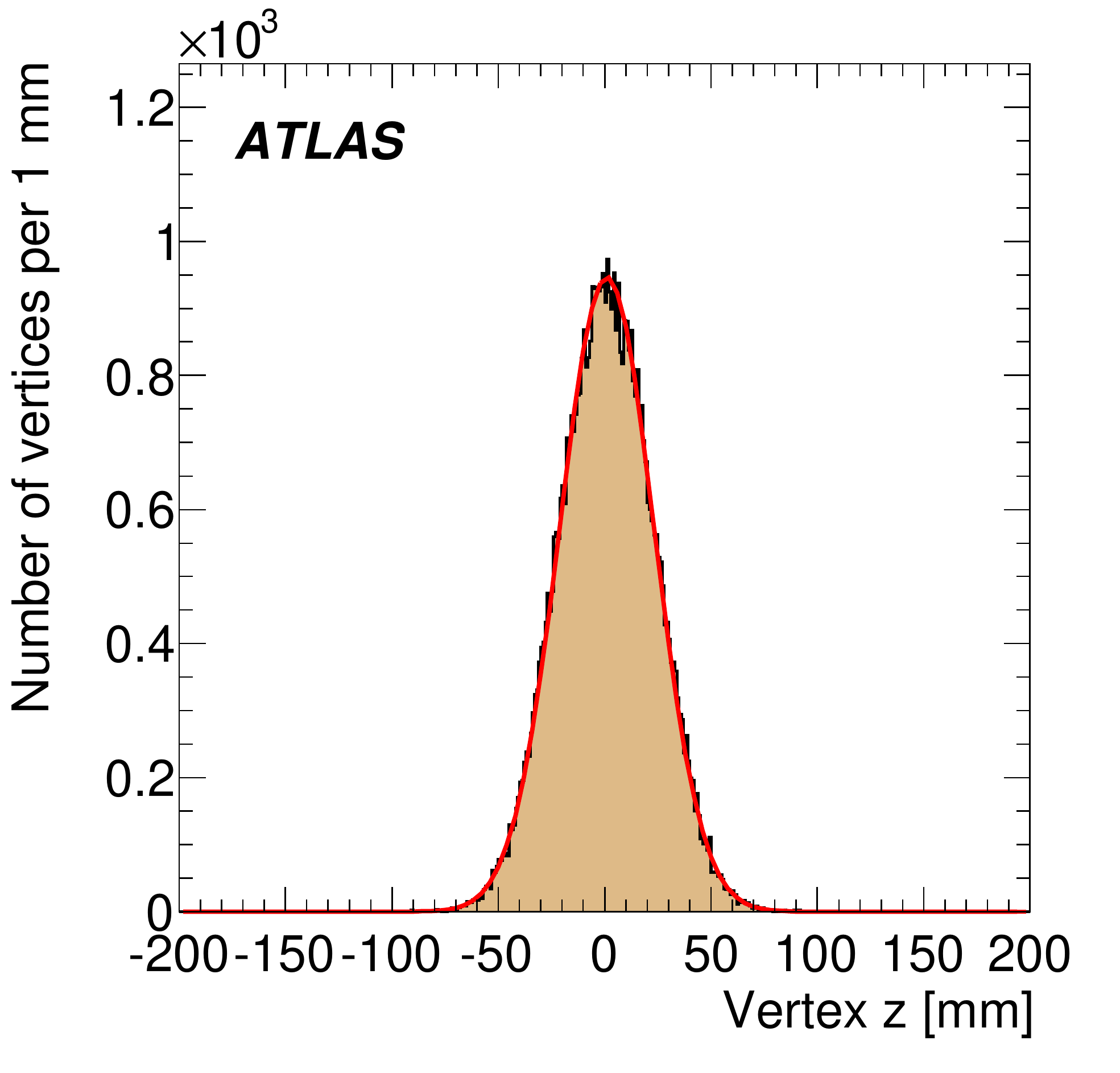}
    \label{fig:BeamspotDistZ}
  }
  \caption{ The distribution of primary vertices reconstructed by the 
online beamspot algorithm in an example run for 
vertices with at least two tracks ($\pt > 0.5 \GeV$ and $|\eta|<2.5$) in 
  \subref{fig:BeamspotProfileXY}  the transverse plane, 
   \subref{fig:BeamspotDistX} $x$,  \subref{fig:BeamspotDistY}  $y$, and 
 \subref{fig:BeamspotDistZ} $z$.
   The mean beam position and observed widths, before correction 
for the intrinsic vertex position resolution; $\mu_x = (-0.370\pm0.001)$mm, $\sigma_x = (0.120\pm0.001)$mm, $\mu_y = (0.628\pm0.001)$mm, $\sigma_y = (0.132\pm0.001)$mm, $\mu_z = (1.03\pm0.10)$mm, $\sigma_z = (22.14\pm0.07)$mm}
 \label{fig:beamspot}
\end{figure*}

\subsubsection{Beamspot Algorithm}

The online beamspot algorithm employs a fast vertex fitter able to
efficiently fit the L2 tracks emerging from the interaction region to
common vertices within a fraction of the L2 time budget.
The 
tracks used for the vertex fits are
required to have at least one Pixel space-point and three 
SCT space-points and a transverse impact parameter with respect to the 
nominal beamline of $|d_0|<1$\,cm. Clusters
of tracks with similar impact parameter ($z_0$) along the
nominal beamline form the input to the vertex fits.
The tracks are ordered in
\pt\ and the highest-\pt\ track above
0.7\,\GeV\ is taken as a seed. The seed track is grouped with all other tracks
with $\pt>0.5\GeV$ within $\Delta z_0<1$\,cm. The average
$z_0$ value of the tracks in the group provides the initial 
estimate of the vertex
position in the longitudinal direction, 
used as a starting point for the vertex fitter. In order to
find additional vertices in the event, the process is repeated
taking the next highest \pt\ track above 0.7\,\GeV\ as the seed.

\subsubsection{Beamspot Algorithm Performance}

Using the event-by-event vertex distribution computed in real-time by
the HLT and accumulated in intervals of typically two minutes,
the position, size and tilt angles of the luminous region
within the ATLAS coordinate system are measured. A view of the transverse
distribution of vertices reconstructed by the HLT is shown in
Fig.~\ref{fig:beamspot} along with the transverse ($x$ and $y$) and
longitudinal ($z$) profiles.

The measurement of the true size of the beam relies on an
unfolding of the intrinsic resolution of the vertex position measurement. 
A correction for the 
intrinsic resolution is determined, in real-time, by measuring
the distance between two daughter vertices constructed from a primary
vertex when its tracks are split into two random sets for
re-fitting. This correction method has the benefit that it allows the
determination of the beam width to be relatively independent of
variations in detector resolution, by explicitly taking the variation
into account.

\begin{figure}[!ht]
  \centering
  \includegraphics[width=0.45\textwidth]{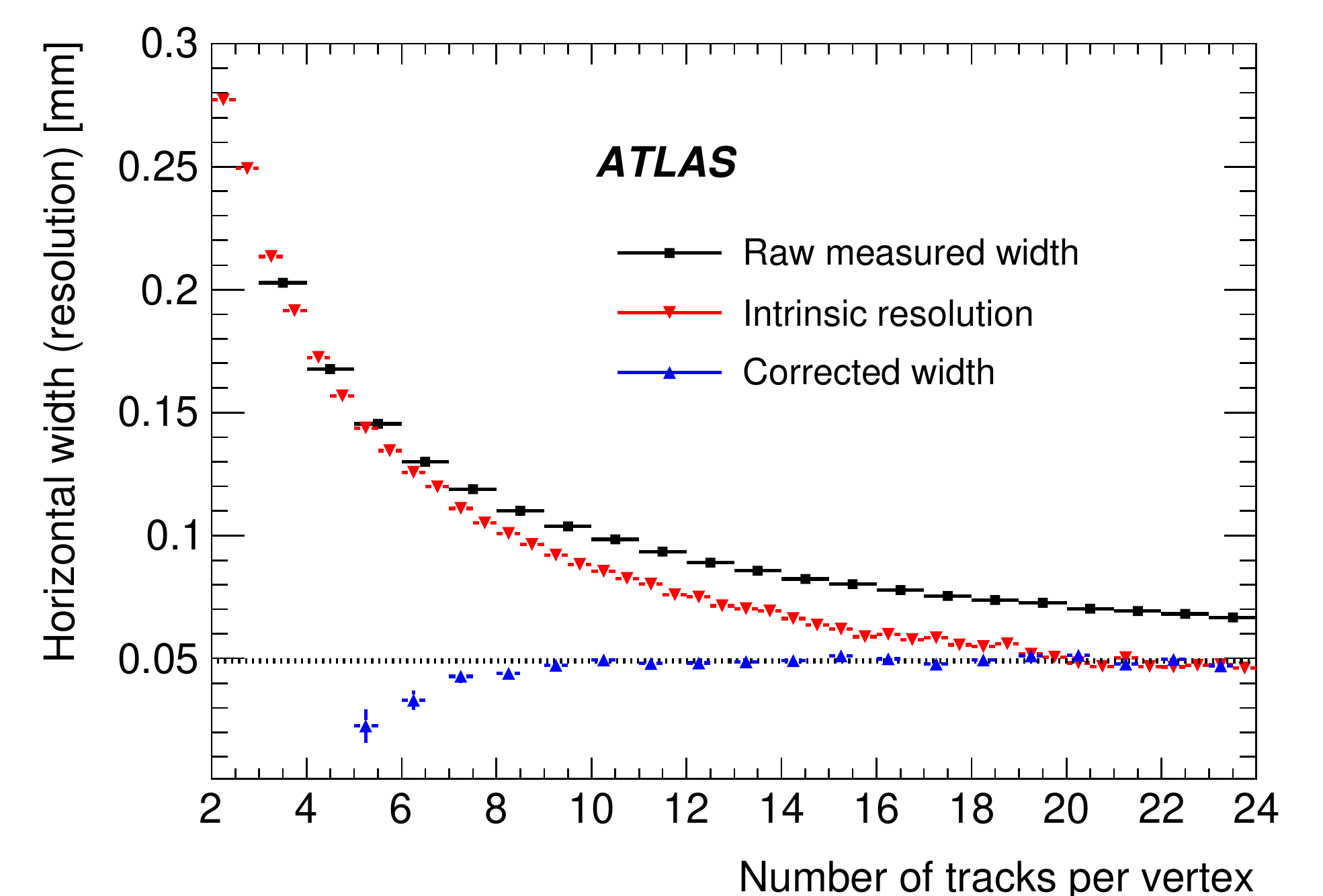}
  \caption{The corrected width of the measured vertex position, in $x$, 
along with the measured intrinsic resolution and the raw measured width before 
correction for the resolution.
The asymptotic value of the corrected width
provides a measurement of the true beam width}
  \label{fig:BeamspotResCorr}
\end{figure}

Figure~\ref{fig:BeamspotResCorr} shows the measured beam width, 
in x, as a function of the number of tracks per vertex.
The raw measured width is shown as well as the width after
correction for the intrinsic resolution of the vertex position 
measurement. 
The measured intrinsic resolution is also shown. 
The intrinsic resolution is overestimated, and hence the
corrected width is underestimated, for vertices with a small number of 
tracks. The true beam width ($50\mu$m) is, therefore, given by the 
asymptotic value of the corrected width. For
this reason vertices used for the beam width measurement 
are required to have more than a minimum number of tracks.
The value of this cut depends
on the beamspot size. Data and MC  studies have 
shown that intrinsic resolution must be less than about two times the 
beamspot size to be measured. For the example fill shown in 
Fig.~\ref{fig:BeamspotResCorr}, this requirement corresponds to 
$\gtrsim10$ tracks per vertex. To resolve smaller beam sizes, the 
multiplicity requirement can be raised accordingly.


\subsection{Calorimeter}\label{sec:caloReco}
\def \figurepath{.}
The calorimeter reconstruction algorithms are designed to 
reconstruct clusters of energy from electrons, photons, taus and jet objects
using calorimeter cell information. At the EF, global \MET\ 
is also calculated.
Calorimeter information is also used to provide information to the
muon isolation algorithms. 

At L2, custom algorithms are used to confirm the results of
the L1 trigger and provide cluster information as input to the signature-specific
selection algorithms. The detailed calorimeter cell information available at the HLT allows
the position and
transverse energy of clusters to be calculated with higher precision than at L1. 
In addition, shower shape variables useful for particle identification are calculated. 
At the EF, offline algorithms with custom interfaces for online running
are used to reproduce offline clustering performance
as closely as possible, using similar calibration procedures.
More details on the HLT and offline clustering algorithms 
can be found in Ref.~\cite{CSCBook,ATL-LARG-PUB-2008-002}.

\begin{figure}[!ht]
\centering
	  \subfigure[]{
	    \includegraphics[width=0.45\textwidth]{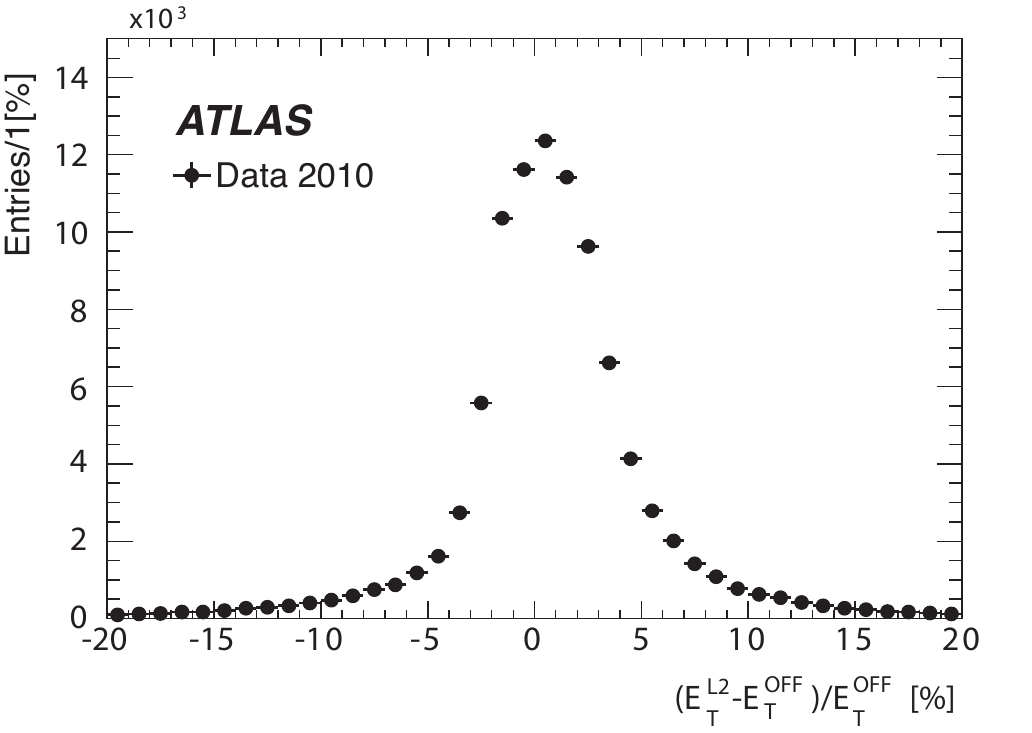}   
	    \label{fig:L2EMEtResolution}
	  }
	  \subfigure[]{
	    \includegraphics[width=0.45\textwidth]{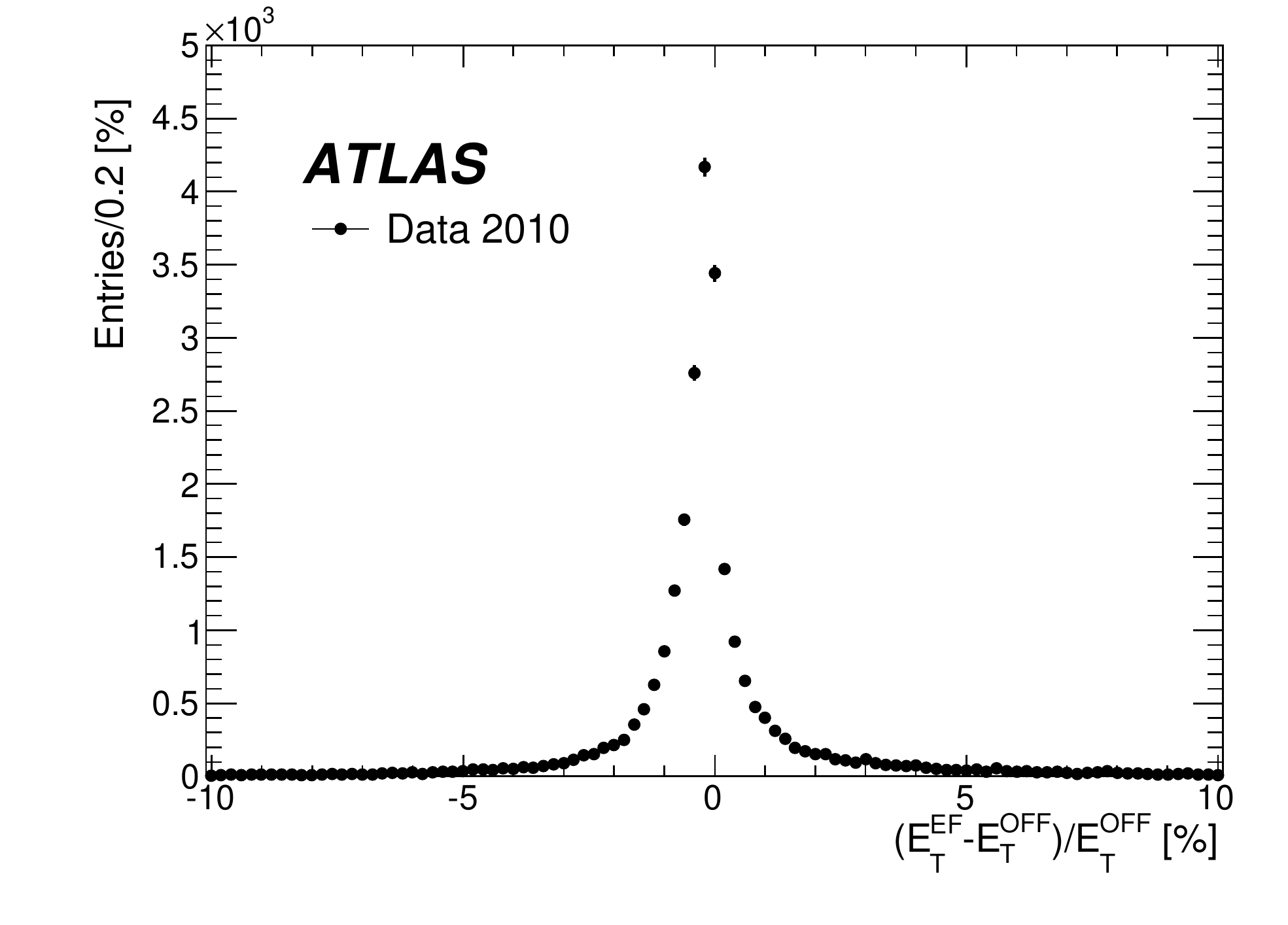}
	    \label{fig:EFEMEtResolution}
	  }
	  \caption{Residuals between online and offline \et\ values for EM clustering at \subref{fig:L2EMEtResolution} L2  and \subref{fig:EFEMEtResolution} EF}
	  \label{fig:EM_Et_Res}
\end{figure}

\subsubsection{Calorimeter Algorithms}
While the clustering tools used in the trigger 
are customized for the different signatures, they take their
input from
a common data preparation software layer. This layer, which is common to L2 and the EF, 
requests data using the general
trigger framework tools  and 
drives sub-detector specific code to
convert the digital information into the input objects \linebreak
(calorimeter cells with energy and geometry)
used by the algorithms. This code is optimized 
 to guarantee fast unpacking of detector data. The data is organized  
so as to allow efficient access by the algorithms. At the EF the  
calorimeter cell information is arranged using projective regions called \emph{towers},
of size $\Delta\eta\times\Delta\phi = 0.025 \times 0.025$ for EM clustering and 
$\Delta\eta\times\Delta\phi = 0.1 \times 0.1$ for jet algorithms.

The L2 electron and photon (\egamma) algorithm performs clustering withing an 
RoI of dimension $\Delta\eta\times\Delta\phi = 0.4\times0.4$.
The algorithm relies on the fact that most of
the energy from an electron or photon is deposited in the
second layer of the electromagnetic (EM) calorimeter. The cell with the most
energy in this layer provides the seed to the clustering process. This cell 
defines the centre of a $\Delta\eta\times\Delta\phi=0.075\times0.125$
window within this layer. The cluster position is calculated by taking an 
energy-weighted average
of cell positions within this window and the cluster transverse energy is calculated by 
summing the cell transverse energies within equivalent windows 
in all layers. Subsequently, a correction for the upstream energy loss and 
for lateral and longitudinal leakage is applied.

At the EF a clustering algorithm similar to the offline algorithm is used. Cluster 
finding is performed
using a sliding window algorithm acting on the towers formed in the data preparation 
step. Fixed window clusters in regions of 
$\Delta\eta\times\Delta\phi=0.075\times0.175~(0.125\times0.125)$ are built in the 
barrel (end-caps). The cluster transverse energy and position are calculated in the same way 
as at L2. 
Distributions of \et\ residuals, defined as the fractional difference between 
online and offline \et\ values, are shown in Fig.~\ref{fig:EM_Et_Res} for L2 and EF. 
The broader L2 distribution is a consequence of 
the specialized fast algorithm used at L2.

\begin{figure}[!htb]
\centering
    \includegraphics[width=0.45\textwidth]{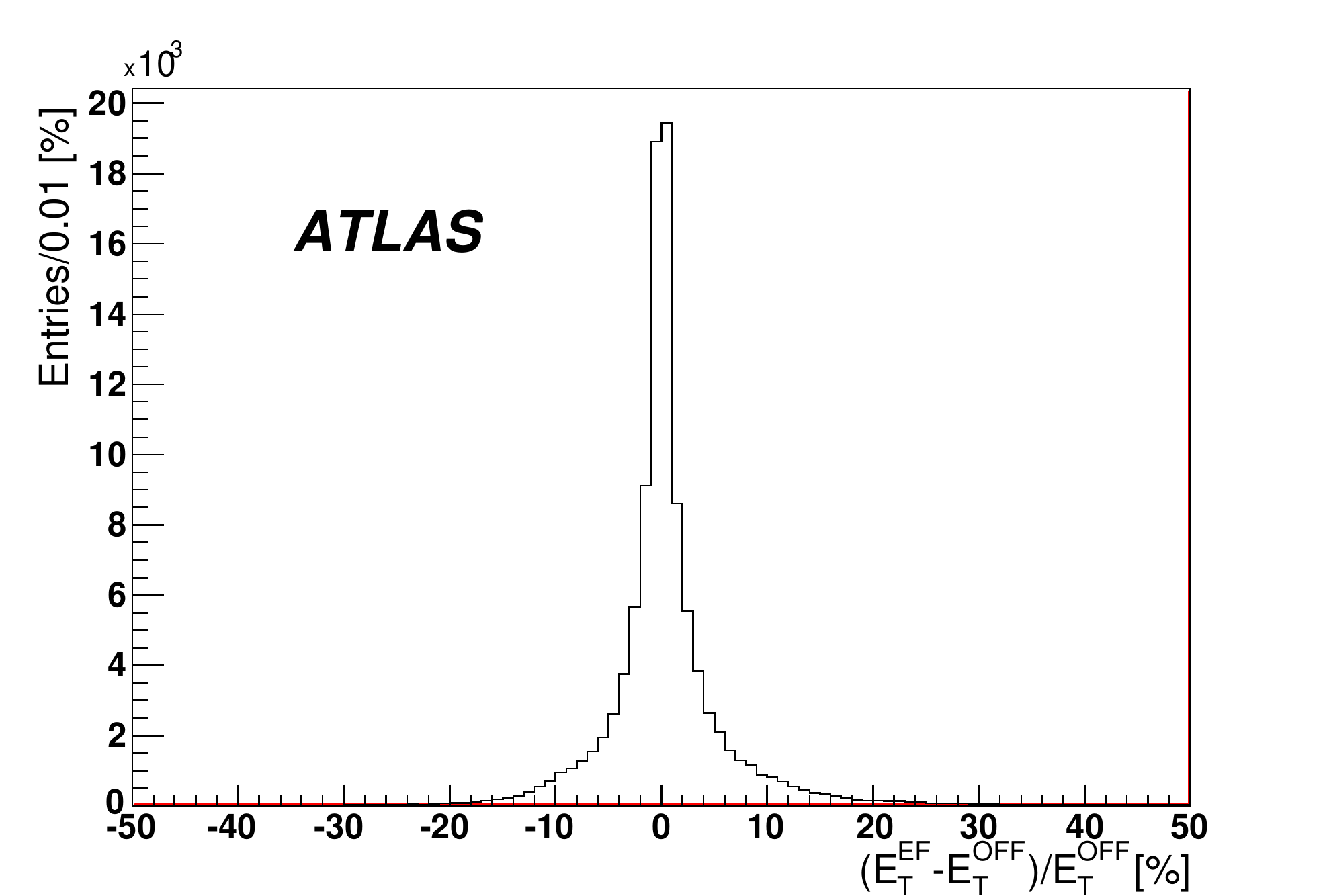}
  \caption{Tau EF \et\ residuals with respect to offline}
  \label{fig:EFTauEtResolution}
\end{figure}

The L2 tau clustering algorithm searches for a seed in all EM and 
hadronic calorimeter layers and within an 
RoI of $\Delta\eta\times\Delta\phi=0.6\times0.6$.  
At the EF
the calorimeter cells within a $\Delta\eta\times\Delta\phi=0.8\times0.8$ region
are used directly as input to a topological clustering algorithm 
that builds clusters of any shape by adding neighbouring cells that have energy above a 
given number (0-4) of standard deviations of the noise distribution. The 
large RoI size is motivated by the cluster 
size used in offline tau reconstruction.
The EF tau \et\ residual with respect to the offline clustering algorithm is 
shown in Fig.~\ref{fig:EFTauEtResolution}.

\begin{figure}[!htb]
\centering
	  \subfigure[]{
	    \includegraphics[width=0.43\textwidth, height=0.37\textwidth]{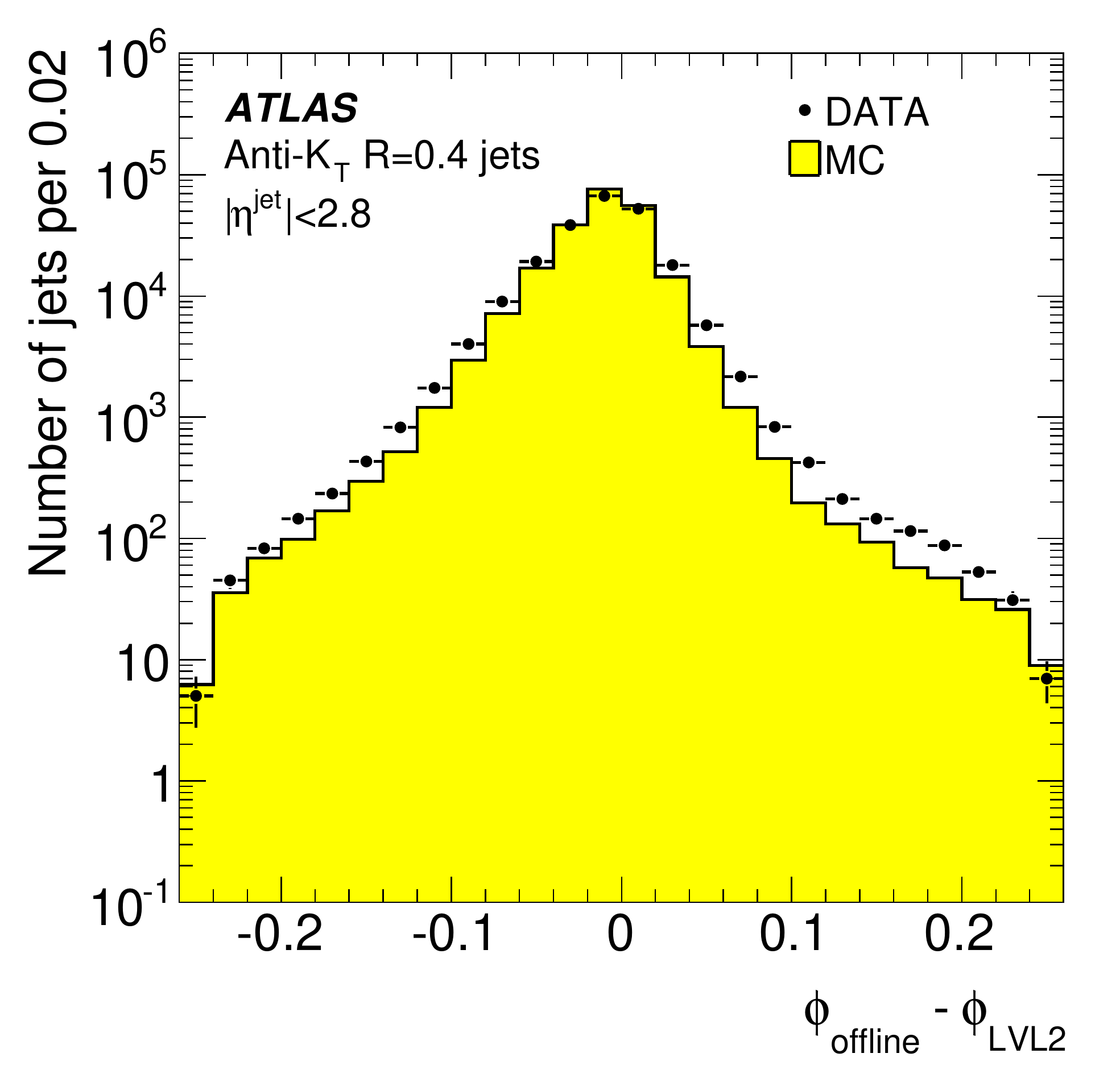}   
	    \label{fig:L2JetPhiResolution}
	  }
	  \subfigure[]{
	    \includegraphics[width=0.43\textwidth, height=0.37\textwidth]{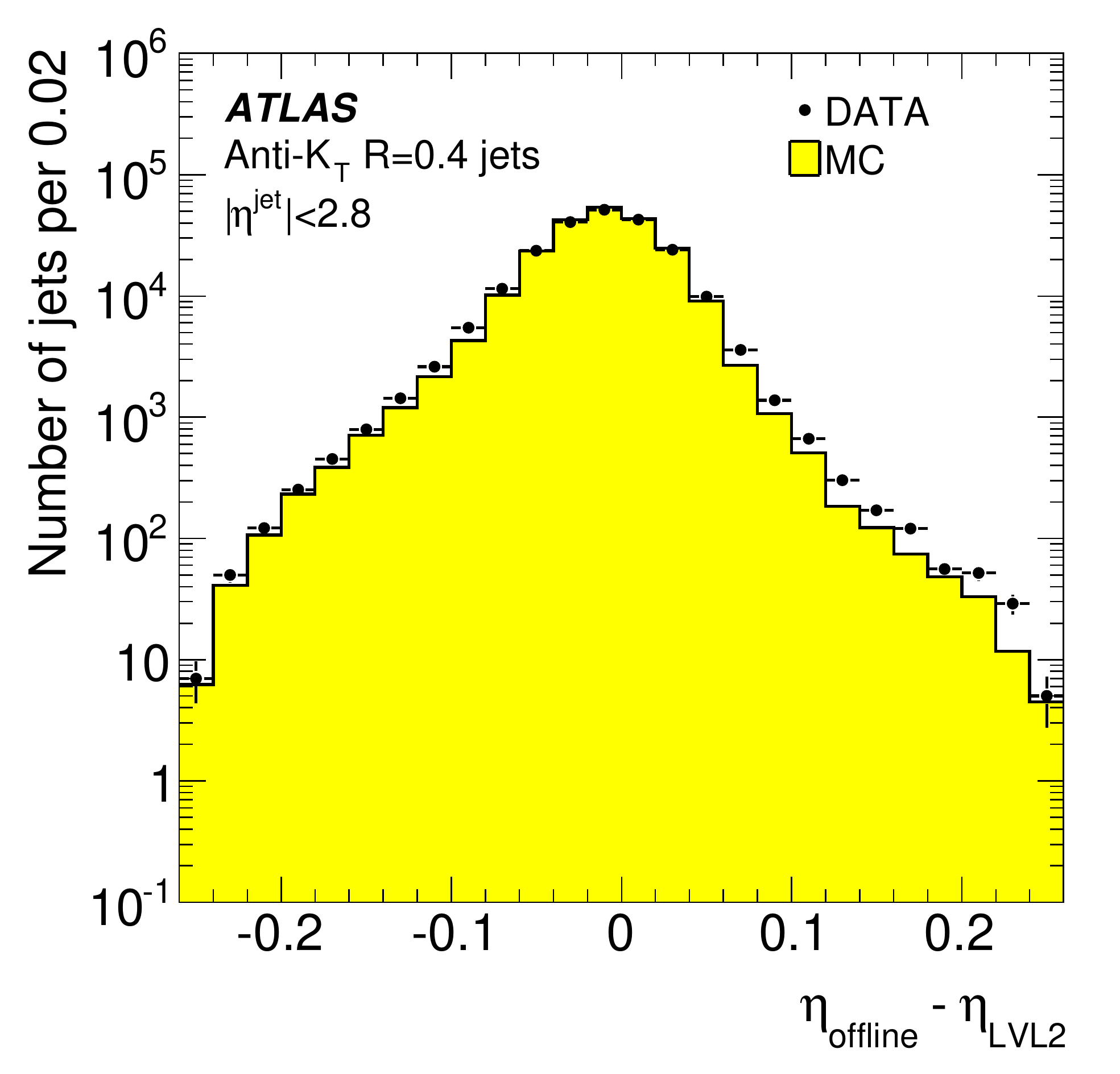}
	    \label{fig:L2JetEtaResolution}
	  }
	  \caption{Residuals between L2 and offline  values of jet cluster \subref{fig:L2JetPhiResolution} $\phi$ and \subref{fig:L2JetEtaResolution} $\eta$ shown for data and MC simulation. The anti-\kt\ algorithm with R=0.4 was used for offline clustering}
	  \label{fig:L2_Jet_Eta_Phi_Res}
	\end{figure}

The L2 jet reconstruction uses a cone algorithm iterating over cells in 
a relatively large RoI ($\Delta\eta\times\Delta\phi=1.0\times1.0$). 
Figure~\ref{fig:L2_Jet_Eta_Phi_Res} shows L2 $\phi$ and $\eta$ residuals with 
respect to offline, 
showing reasonable agreement with simulation.  
The asymmetry, which is reproduced by the simulation, is due to the fact that 
L2 jet reconstruction, unlike offline, is performed
within an RoI whose position is defined with the  granularity of the L1 jet 
element size (Section~\ref{sec:L1calo}).
The L2 jet \et\ reconstruction and jet energy scale are discussed further in 
Section~\ref{sec:jet}.
During 2010, EF jet trigger algorithms ran online in monitoring mode i.e. without rejection.  
In 2011, the EF jet selection will be activated based on EF clustering
within all layers of the calorimeter using the offline anti-\kt\ jet 
algorithm~\cite{antikt}.

Recalculation of \MET\ at the HLT requires data from the whole calorimeter, and so
was only performed at the EF where data from the whole event is available. 
Corrections to account for muons were calculated at L2, but these corrections
were not applied during 2010 data-taking. Future improvements will allow \MET\ 
to be recalculated at L2 based
on transverse energy sums calculated in the calorimeter front-end boards.  
The \MET\ reconstruction, which uses the common
calorimeter data preparation tools, is described in Section~\ref{sec:met}.

\subsubsection{Calorimeter Algorithms Timing}

\begin{figure}[!ht]
  \centering
  \subfigure[]{
  \includegraphics[width=0.45\textwidth]{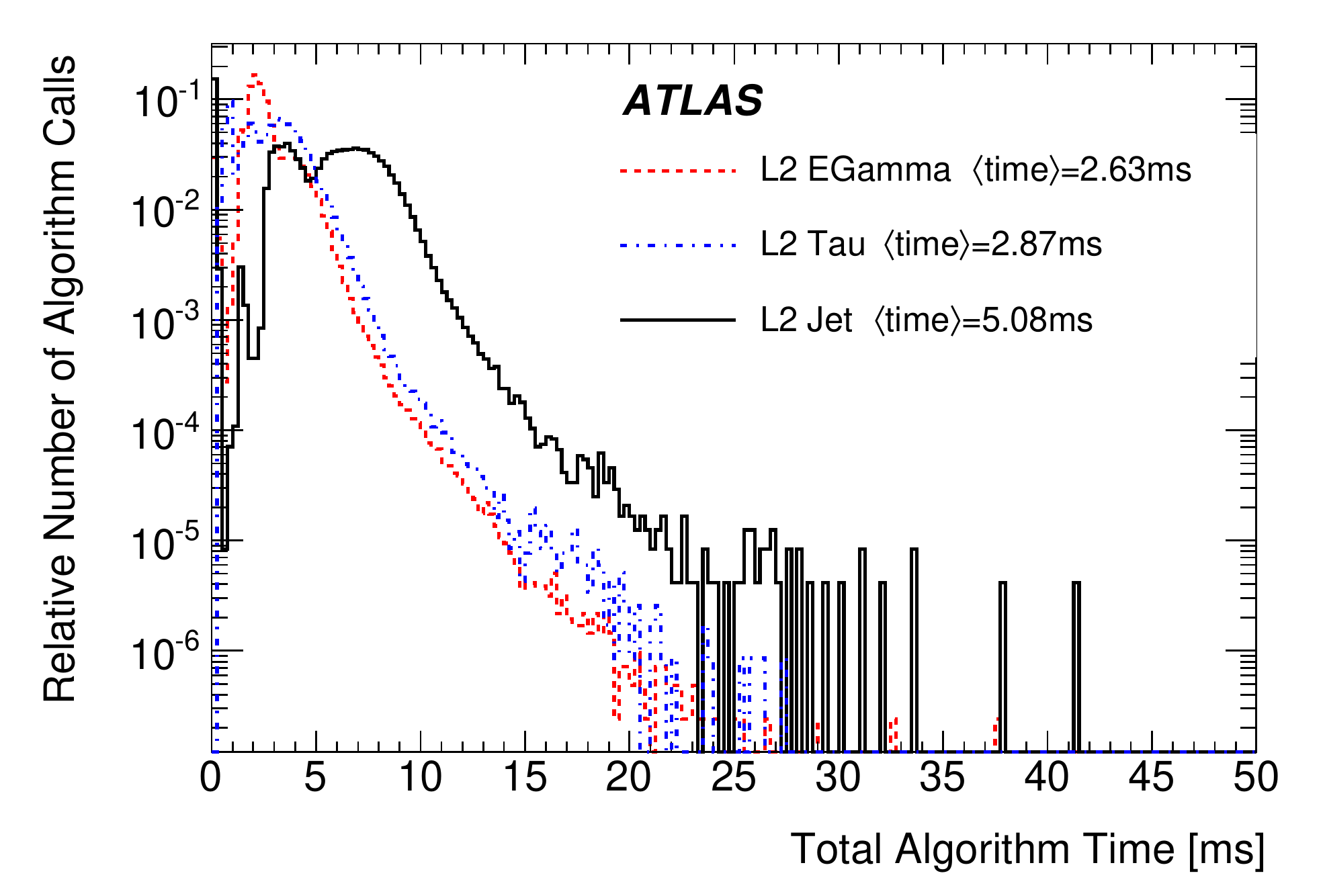}
  \label{fig:TimePerfL2}
  }
  \subfigure[]{
  \includegraphics[width=0.45\textwidth]{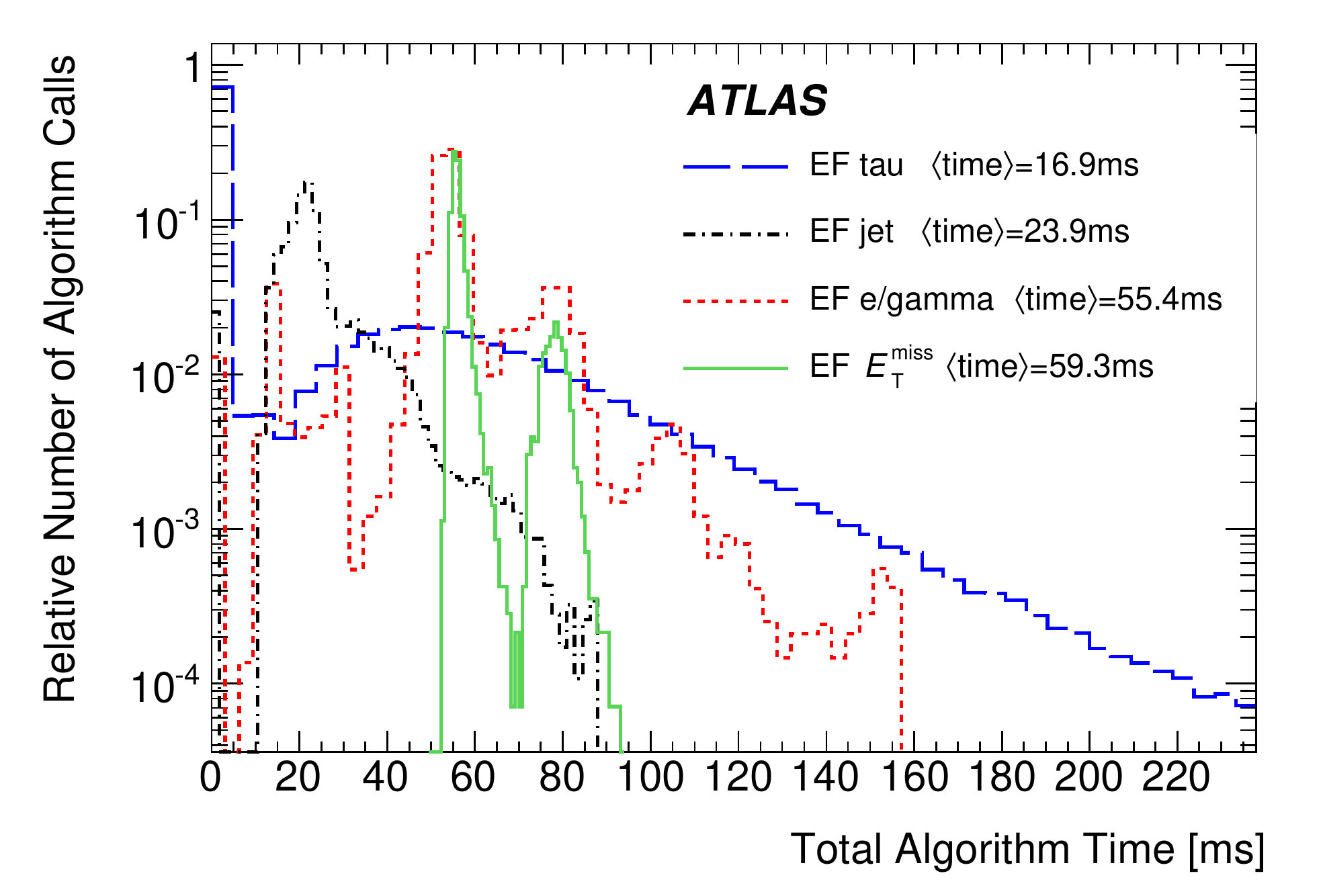}
  \label{fig:TimePerfEF}
  }
  \caption{Execution times per RoI for calorimeter clustering algorithms at (a) L2 and (b) EF.
The mean execution time for each algorithm is given in the legend}
  \label{fig:TimePerf}
\end{figure}

Figure~\ref{fig:TimePerfL2} shows the processing time per RoI 
for the L2 e/gamma, tau and jet clustering algorithms, including data preparation. 
The processing 
time increases
with the RoI size. The tau algorithm has a longer processing time than the \egamma\ algorithm 
due to the larger 
RoI size as well as the seed search in all layers.
The distributions have multiple peaks due to 
caching of results in the HLT, which leads to shorter times when
overlap of RoIs allows cached information to be used.  
Caching of L2 results occurs in two places: 
first, at the
level of data requests from the readout buffers; second, in the data preparation
step, where raw data is unpacked
into calorimeter cell information.  
Most of the L2 time is consumed in requesting data from the detector buffers. 

Figure~\ref{fig:TimePerfEF} shows the processing time per RoI 
for the EF e/gamma, tau, jet and \MET\ clustering algorithms. 
Since more complex offline algorithms are used at the EF,
the processing times are longer and the distributions have more features than for L2.
The mean execution times do not show the same dependence on RoI size as at L2, since 
algorithm differences are more significant than RoI size at the EF. 
The multiple peaks due to caching of data preparation results
are clearly visible. 
The measured L2 and EF algorithm times are well within the requirements
given in Section~\ref{sec:overview}.


\subsection{Muon Tracking}\label{sec:muonReco}
\def \figurepath{.}Muons are triggered in the ATLAS experiment within a rapidity range of 
$|\eta|<2.4$~\cite{DetectorPaper}. In addition to the L1 trigger chambers 
(RPC and TGC), the HLT makes use of information from the MDT chambers, which 
provide precision hits in the $\eta$ coordinate.
The CSC, that form the innermost muon layer in the region $2<|\eta|<2.7$, 
were not used in the HLT during 2010 data-taking period, but will be used in 2011.

\subsubsection{Muon Tracking Algorithms}

The HLT includes L2 muon algorithms that are specifically designed to be fast and EF algorithms 
that rely on offline muon reconstruction software~\cite{CSCBook}.

At L2, each L1 muon candidate is refined by including the precision data from the MDTs in the RoI 
defined by the L1 candidate.  There are three algorithms used sequentially at L2, each building 
on the results of the previous step.
\begin{description}
  \setlength{\itemsep}{1pt}
  \setlength{\parskip}{0pt}
  \setlength{\parsep}{0pt}
\item \emph{L2 MS-only:} The MS-only algorithm uses only the Muon Spectrometer information. 
The algorithm uses L1 trigger chamber hits to define the trajectory of the L1 muon and
opens a narrow road around this to select MDT hits. A track fit
is then performed using the MDT drift times and positions and a \pt\ measurement is assigned
using look-up tables.
\item \emph{L2 Muon Combined:} This algorithm combines the MS-only tracks with tracks
reconstructed in the inner detector (Section~\ref{sec:idReco}) to form a muon candidate
with refined track parameter resolution. 
\begin{sloppypar}
\item \emph{L2 Isolated Muon:} The isolated muon algorithm starts from the 
result of the combined algorithm and incorporates tracking and calorimetric information to 
find isolated muon candidates. The algorithm sums the $|\pt|$ of
inner detector tracks and evaluates 
the electromagnetic and hadronic energy deposits, as
measured by the calorimeters, in cones centred around the muon direction.  
For the calorimeter, two different concentric cones are defined: an internal cone chosen to contain the energy deposited 
by the muon itself; and an external cone, containing energy from detector noise and other 
particles. 
\end{sloppypar}
\end{description}

At the EF, the muon reconstruction starts from the RoI identified
by L1 and L2, reconstructing segments and tracks using information from 
the trigger and precision chambers.
There are three different reconstruction strategies used in the EF:
\begin{description}
  \setlength{\itemsep}{1pt}
  \setlength{\parskip}{0pt}
  \setlength{\parsep}{0pt}
\item \emph{EF MS-only:} Tracks are reconstructed using Muon Spectrometer information and
extrapolated to determine track parameters at the interaction point and form MS-only muon candidates.
\item \emph{EF Combined:} Using an outside-in strategy, MS-only muon candidates are combined with 
inner detector tracks to form combined muon candidates. 
\item \emph{EF Inside-Out:} The inside-out strategy starts with inner detector
tracks and extrapolates them to the Muon Spectrometer to search for MS-only candidates in order to 
form combined muon candidates.
\end{description}

EF Combined and Inside-out are both used for the trigger and offline reconstruction; MS-only is an alternative strategy for specialized triggers.  For the EF MS-only and EF Combined strategies, the reconstruction is performed in the
following steps: 

\begin{description}
  \setlength{\itemsep}{1pt}
  \setlength{\parskip}{0pt}
  \setlength{\parsep}{0pt}
\item \emph{SegmentFinder:} Segments are formed from hits in the trigger and precision chambers within each of the three layers of the muon detector.
\item \emph{TrackBuilder:} The segments are combined to form tracks. 
\item \emph{Extrapolator:} The tracks are extrapolated to the interaction point, track parameters are corrected for energy loss in the traversed material, producing EF MS-only muon candidates.
\item \emph{Combiner:} The tracks from the muon spectrometer are combined with inner detector tracks to form combined tracks, resulting in EF Combined muon candidates.
\end{description}

\subsubsection{Muon Tracking Performance}

\begin{figure}[!htb]
  \begin{center}
  \includegraphics[width=0.45\textwidth]{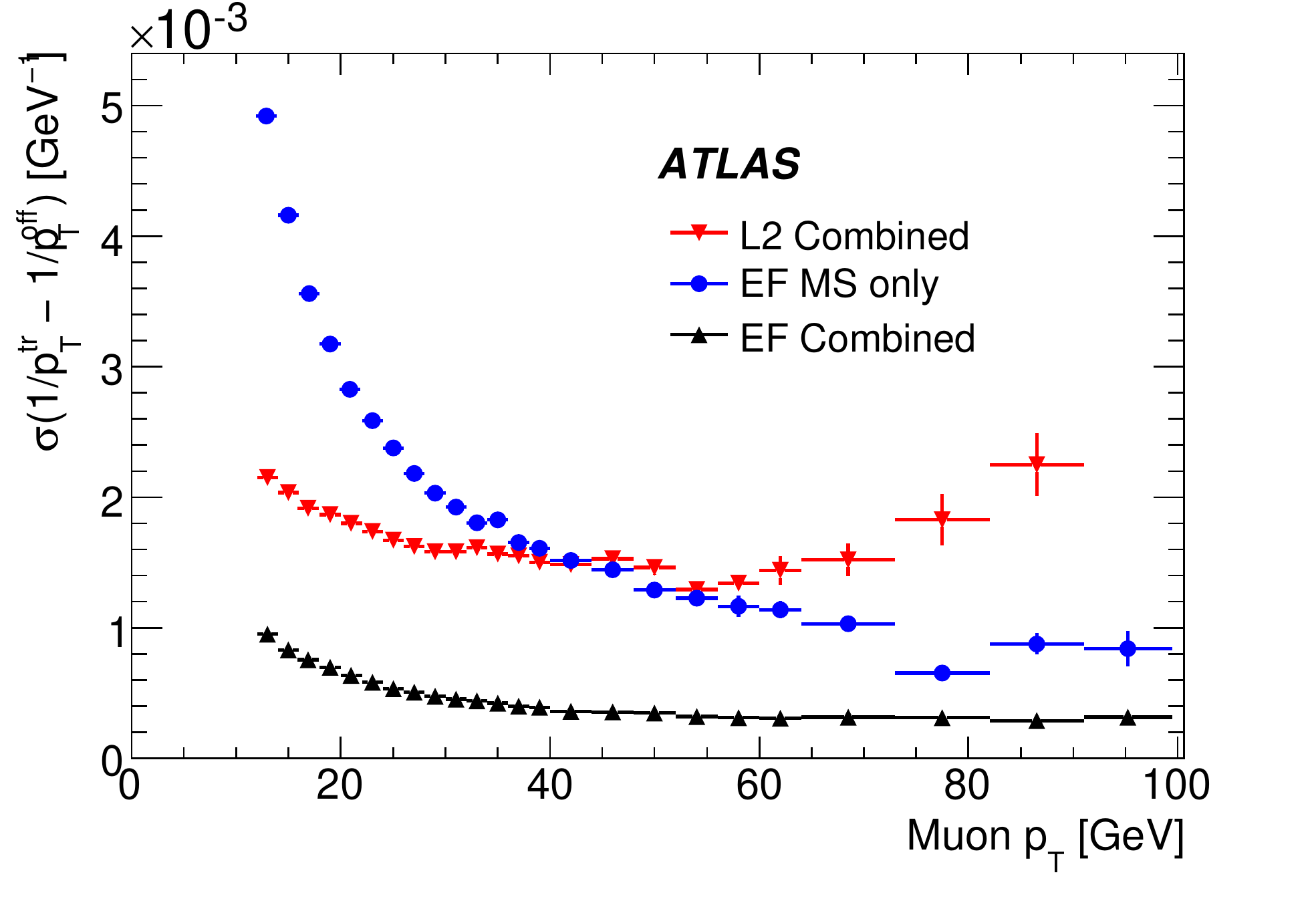}
  \caption{Inverse-\pT\ residual widths as a function of offline muon \pT\ ($\pT>13\GeV$)
for the L2 combined, EF MS-only and EF Combined reconstruction} 
\label{fig-muon-pt-res}
  \end{center}

\end{figure}

\begin{figure*}[!htb]
    \begin{center}
\subfigure[]
{
\includegraphics[width=0.45\textwidth]{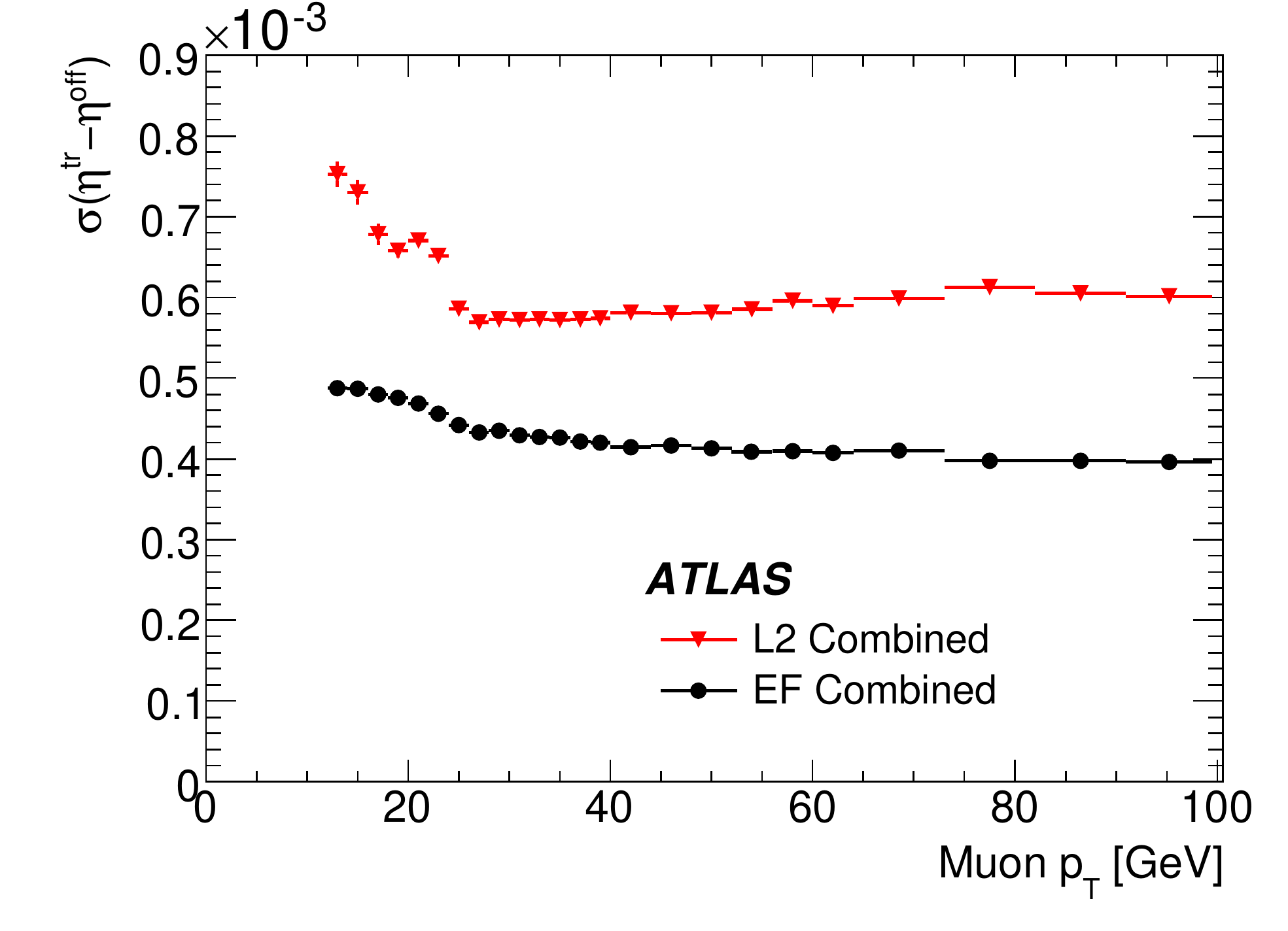}
    \label{fig-muon-eta-res}
}
\subfigure[]
{
\includegraphics[width=0.45\textwidth]{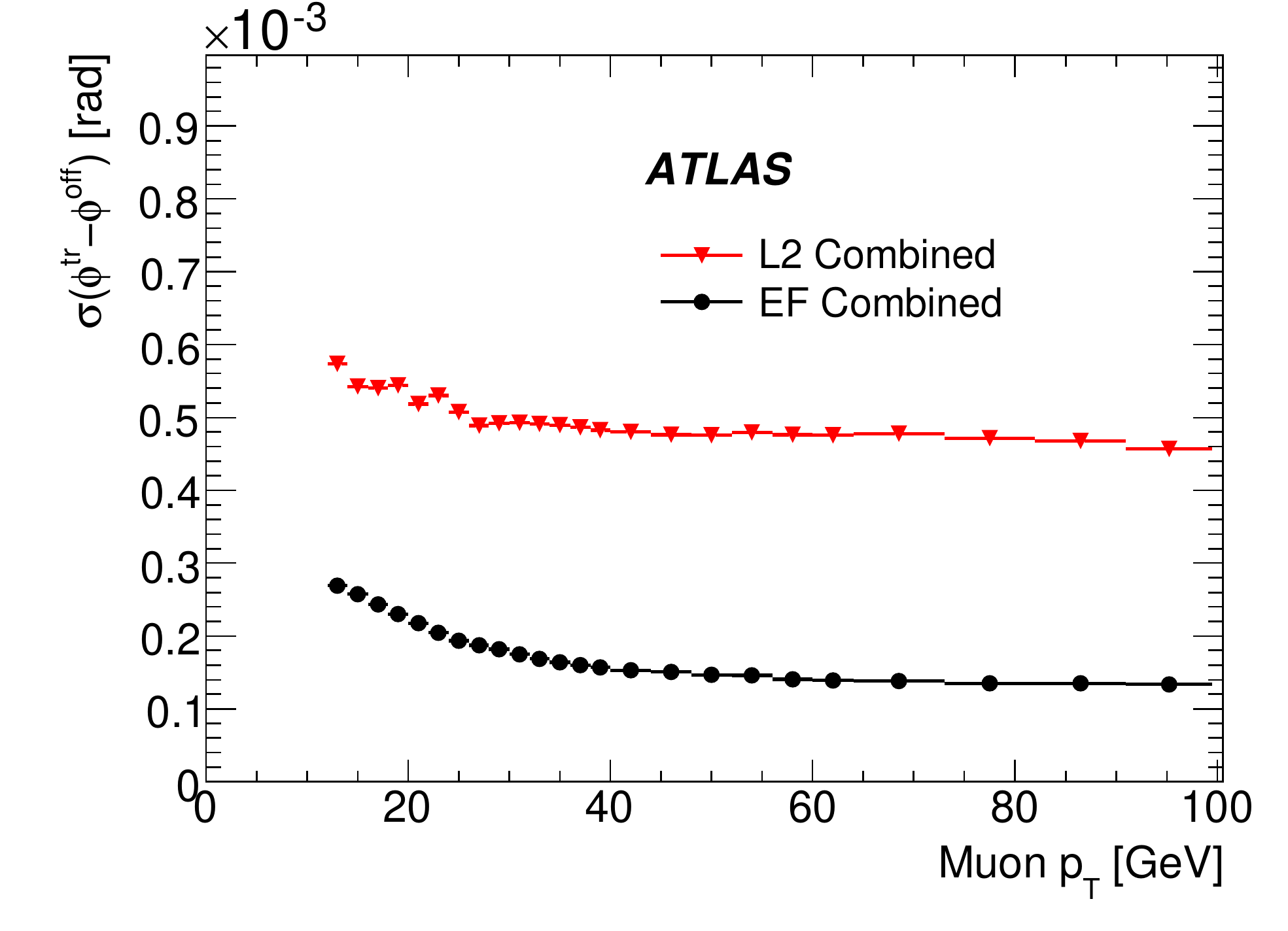}
    \label{fig-muon-phi-res}
}
    \end{center}
    \caption{Residual widths as a function of the offline muon \pT\ ($\pT>13\GeV$) for \subref{fig-muon-eta-res} $\eta$ and  
\subref{fig-muon-phi-res} $\phi$ calculated by L2 and EF Muon Combined 
algorithms}
    \label{fig-muon-eta-phi-res}
\end{figure*}

Comparisons between online and offline muon track parameters are presented in this section; muon
trigger efficiencies are presented in Section~\ref{sec:muon}. 
Distributions of the residuals between online and offline track parameters 
($\frac{1}{\pT}$, $\eta$ and $\phi$) were constructed in bins of \pT\ and Gaussian fits 
were performed to extract the
widths, $\sigma$, of the residual distributions as a function of $\pT$.
The inverse-\pT\ residual widths, $\sigma((\frac{1}{\pT})^{\mathrm{trigger}}-(\frac{1}{\pt})^{\mathrm{offline}})$, are shown in Fig.~\ref{fig-muon-pt-res} as a function of the offline muon \pT\
for the L2 Muon Combined, EF MS-only and EF Combined reconstruction. As a consequence of 
the optimisations made for algorithm speed, the L2 has worse 
track parameter resolution than the EF.
The increase in the L2 inverse-\pT\ widths at high \pT\ is due to the finite granularity of the
look-up table used in the L2 MS-only algorithm;
at lower values of \pT\ the inner detector \pT\ resolution dominates. The improvement
in \pT\ resolution, particularly at lower \pT\,  resulting from the inclusion of inner detector 
information is also evident from a comparison  of the \pT\ resolution of the
EF MS-only and combined algorithms. 
The $\eta$ residual widths, \mbox{$\sigma(\eta^{tr}-\eta^{off})$}, and $\phi$ residual widths, \mbox{$\sigma(\eta^{tr}-\eta^{off})$}, are shown as a function of \pT\ in 
Fig.~\ref{fig-muon-eta-res} and
Fig.~\ref{fig-muon-phi-res} respectively. These figures show the residual widths for L2 and
EF combined reconstruction and illustrate the good agreement between track parameters 
calculated online and offline. 
 

\begin{figure*}[!htb]
\begin{center}
\subfigure[]{
\includegraphics[width=0.45\textwidth]{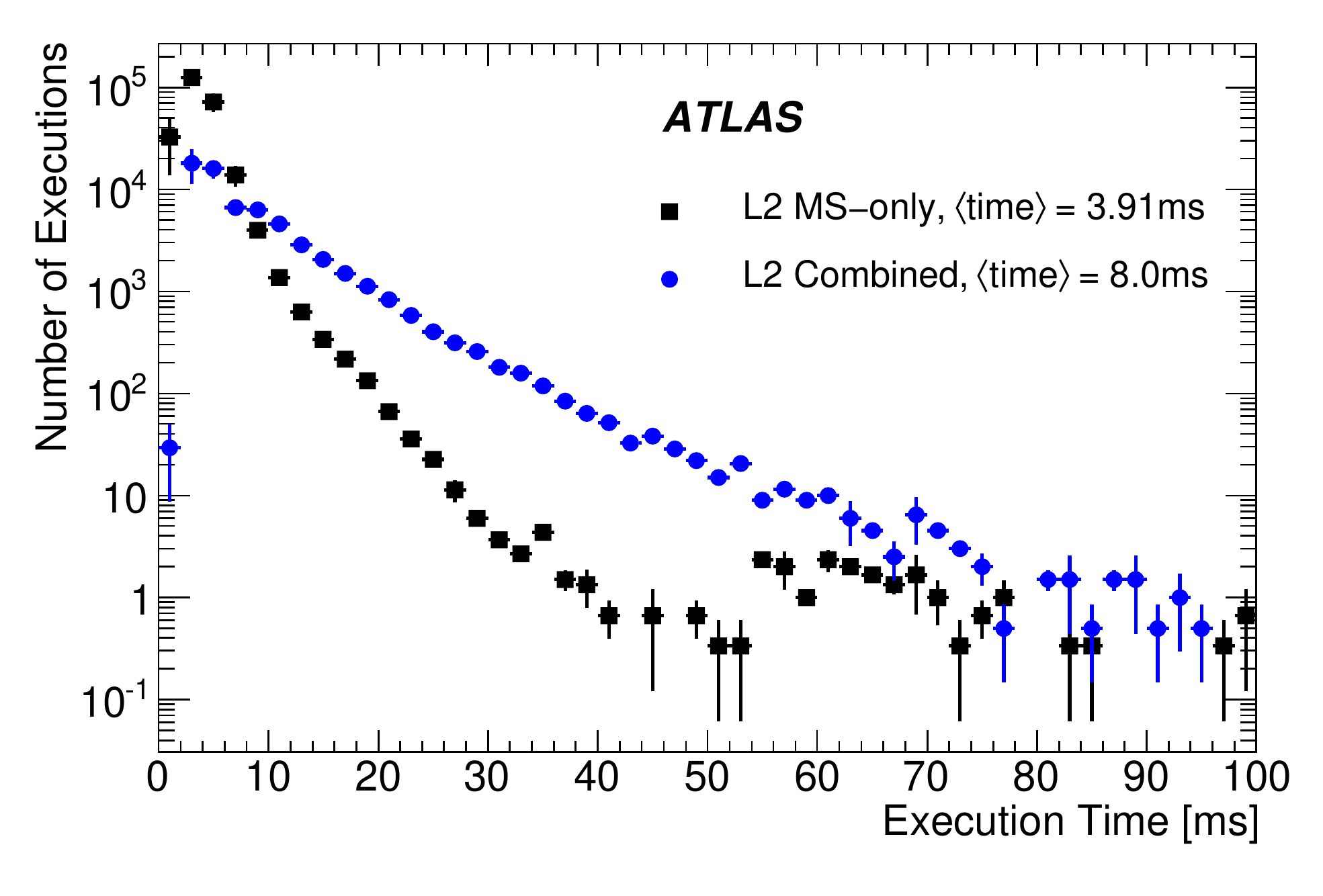}
\label{fig-muon-l2-muFast-time}
}
\subfigure[]{
\includegraphics[width=0.45\textwidth]{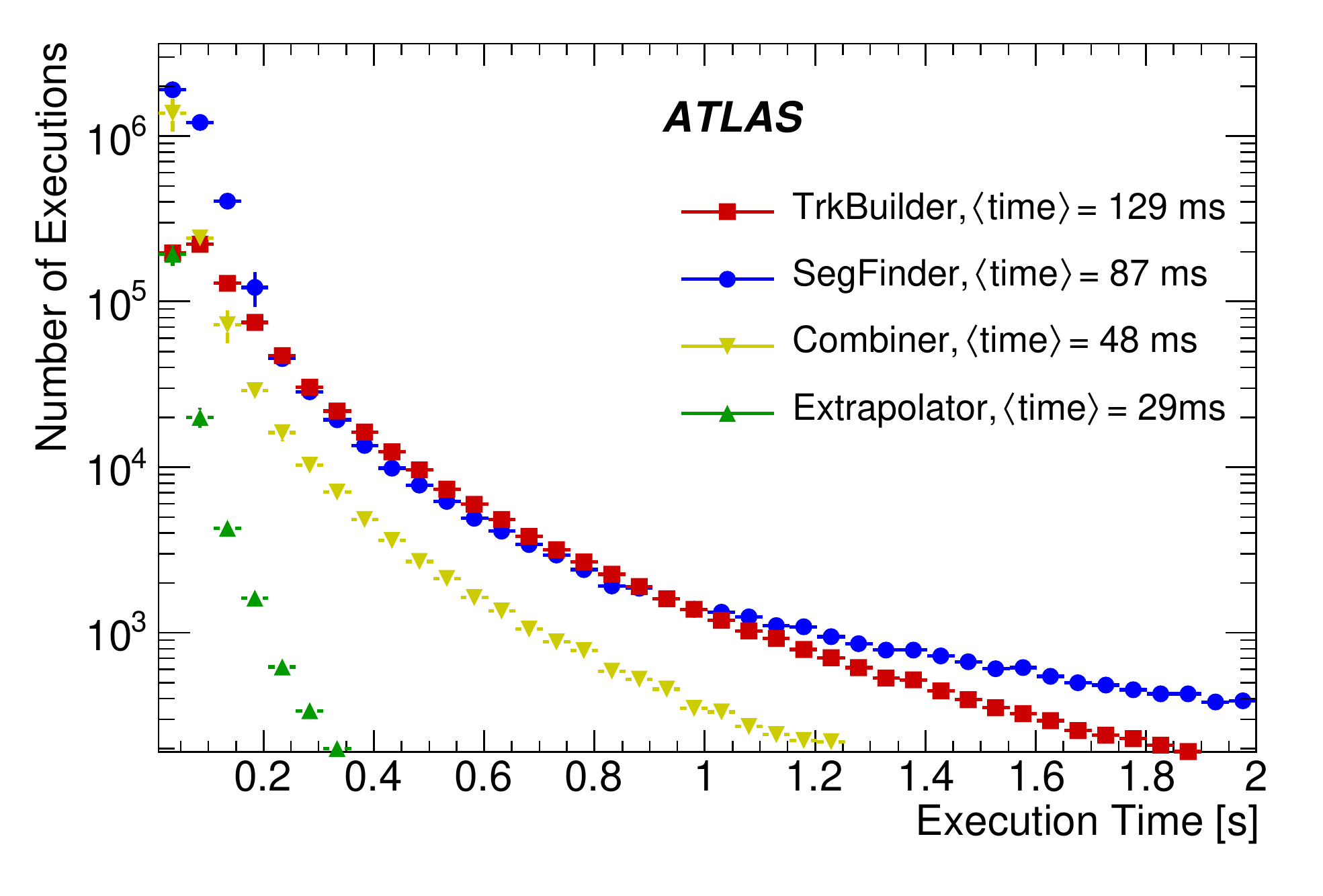}
\label{fig-muon-ef-MSonly-time}
}
\end{center}
\caption{Measured execution times per RoI 
for the \subref{fig-muon-l2-muFast-time} L2 MS-only algorithm and L2 Combined chain  and
\subref{fig-muon-ef-MSonly-time} EF SegmentFinder, TrackBuilder, Extrapolator and Combiner algorithms.  
The mean time of each algorithm is indicated in the legend}
\label{fig-muon-l2-ef-timing}
\end{figure*}

\subsubsection{Muon Tracking Timing}

The processing times for the L2 muon reconstruction algorithms are shown in 
Fig.~\ref{fig-muon-l2-muFast-time} for the MS-only algorithm and for the 
combined reconstruction chain, which includes the ID track reconstruction time.  Figure~\ref{fig-muon-ef-MSonly-time} shows the corresponding times for the EF algorithms.  The execution times are measured for each invocation of the algorithm, and are well within the time restrictions for both L2 and EF given 
in Section~\ref{sec:overview}.



\section{Trigger Signature Performance}\label{sec:signatures}

In this section the different trigger signature selection criteria are described. 
The principal triggers used in 2010 are listed, 
their performance is presented and compared with Monte Carlo simulation and 
some references are given as examples of published results that rely on these triggers.

Efficiencies have been measured using the following\linebreak methods:
\begin{description}
  \setlength{\itemsep}{1pt}
  \setlength{\parskip}{0pt}
  \setlength{\parsep}{0pt}
\item \emph{Tag and probe method}, where the event contains a pair of related objects reconstructed offline, such as electrons from a \Zee\ decay, one that triggered the event and the other that can be used to measure trigger efficiency;
\item \emph{Orthogonal triggers method}, where the event is triggered by a different and independent trigger from the one for which the efficiency is being determined;
\item \emph{Bootstrap method}, where the efficiency of a higher threshold is determined using a lower threshold to 
trigger the event.
\end{description}
An example of the tag and probe method is the determination of low-\pt\ muon trigger efficiencies using \Jmumu\ events.
In this method, \mumu\ pairs are selected from \Jmumu\ decays reconstructed offline in events 
triggered by a single muon trigger.  The \emph{tag} is selected by matching (in $\dR$) one of the offline muons with a trigger muon that passed the trigger selection.  The other muon in the \mumu\ pair is defined as the \emph{probe}.  
The efficiency is then defined as the fraction of probe muons that match (in $\dR$) a trigger muon that passes the trigger selection.
An efficiency determined in this way must be corrected for background due to fake \Jmumu\ decays reconstructed offline.  
The background subtraction uses a variable that discriminates the 
signal from the background, in this case, the invariant mass of \mumu\ candidates. 
By fitting this variable with an exponential background shape in the side bands
and with a Gaussian signal shape in the \Jpsi\ mass region, the background content in the \Jpsi\ mass region can be determined
and subtracted. The subtracted 
distribution is then used to determine the trigger efficiency.  Biases due to, for example, topological correlations, are determined by MC.

\subsection{Minimum Bias, High Multiplicity and Luminosity Triggers}\label{sec:minbias}
\def \figurepath{.}
Triggers were designed for inclusive inelastic event selection with minimal bias, for use in inclusive physics studies as well as luminosity measurements.
Events selected by the \glossary{name={minbias triggers},description={triggers for inelastic \pp\ collisions}}minimum bias (\emph{minbias}) trigger are used directly for physics analyses of inelastic $pp$ interactions~\cite{Atlas:UE, Atlas:ChargedPartMult}, $PbPb$ interactions~\cite{Atlas:DiJetAsymHI2010}, as well as indirectly as control samples for other physics analyses.  A high multiplicity trigger is also implemented for studies of two-particle correlations in high-multiplicity events.

\subsubsection{Reconstruction and Selection Criteria}

The minbias and luminosity triggers are primarily hardware-based L1 triggers, defined using signals from the Minimum Bias Trigger Scintillators (MBTS), a Cherenkov light detector (LUCID), the Zero Degree Calorimeter (ZDC), and the random clock from the CTP.  In addition to these L1 triggers, HLT algorithms are defined using inner detector and MBTS information (Section~\ref{sec:overview}).

In 2010, inelastic $pp$ events were primarily selected with the L1\_MBTS\_1 trigger requirement, defined as having at least one of the 32 MBTS counters on either side of the detector above threshold.   Several supporting MBTS requirements were also defined in case of higher beam-induced backgrounds and for online luminosity measurements. For some of these triggers (e.g. L1\_MBTS\_1\_1) a coincidence was required between the signals from the counters on either side of the detector.   In all cases, a coincidence with colliding bunches was required.  During the $PbPb$ running the beam backgrounds were found to be significantly higher and selections requiring more MBTS counters above threshold on both sides of the detector were used.

The \emph{mbSpTrk} trigger~\cite{Atlas:minbCONFnote}, used for minbias trigger efficiency measurements,
selects events using the random clock of the CTP at L1 and inner detector tracker silicon space-points (Section~\ref{sec:idReco}) at the HLT.

\begin{sloppypar}
The LUCID triggers were used to select events for comparison with real-time luminosity measurements.      LUCID trigger items required a LUCID signal above threshold on one side
\footnote{The $\pm z$ sides of the ATLAS detector are named ``A" and ``C"}
, either side, or both sides of the detector.  In all cases a coincidence with colliding proton bunches was required.
\end{sloppypar}

The ZDC detector was included in the ATLAS experiment primarily for selection of $PbPb$ interactions with minimal bias.   Due to the ejection of neutrons from colliding ions, the ZDC covers most of the inelastic $PbPb$ cross-section, but not the inelastic $pp$ cross-section.  Like the LUCID triggers, the ZDC triggers included a one-sided, either side, and two-sided trigger.

The high multiplicity trigger was based on a L1 total energy trigger and includes requirements on the number of L2 SCT space-points and the number of EF inner detector tracks associated to a single vertex.

The Beam Conditions Monitor (BCM) detectors were used to trigger on events with higher than nominal beam background conditions and were also used to monitor the  luminosity.

\subsubsection{Menu and Rates}

The main minbias, high multiplicity and luminosity triggers used in the 2010 run are shown in Table~\ref{tab:mbts_trigger_items}.  These triggers were prescaled for the majority of the 2010 data-taking to keep the rates around a few Hz.

\begin{table*}[!ht]
\begin{center}
\caption[Minbias trigger items]{
Minimum bias, high multiplicity and luminosity trigger items defined during the 2010 $pp$ and $PbPb$ running periods. In all cases, a filled bunch crossing is also required}
\begin{tabular}{l l l l}\hline \hline 
& Trigger item & Definition \\  \hline 
{\em minbias triggers:} & L1\_MBTS\_1(2) 	    & $\ge$1(2) of the 2$\times$32 counters above threshold \\
& L1\_MBTS\_1\_1  & $\ge$1 counter above threshold in each side, in coincidence\\
& L1\_MBTS\_4\_4  & $\ge$4 counters above threshold in each side, in coincidence\\
& mbSpTrk & Random L1 trigger and space-points in the ID at HLT\\ 
\hline
{\em luminosity triggers:} & L1\_LUCID\_A(C) 	      & At least one A(C)-side signal above threshold \\
& L1\_LUCID\_A\_C  & $\ge$1 signal above threshold in each side, in coincidence\\
& L1\_LUCID 	       & $\ge$1 signal above threshold \\
& L1\_ZDC\_A(C) 	  & $\ge$1 A(C)-side signal above threshold \\
& L1\_ZDC\_A\_C  &  $\ge$1 signal above threshold in each side, in coincidence\\ 
& L1\_ZDC 	            & $\ge$1 signal above threshold \\ 
\hline
{\em high luminosity trigger:} & mbSpTrkVtxMh & High Multiplicity trigger starting from L1\_TE20 \\
\hline
\end{tabular}
\label{tab:mbts_trigger_items}
\end{center}
\end{table*}

\subsubsection{Minimum Bias Trigger Efficiency}

The efficiency of the L1\_MBTS\_1 trigger was studied in the context of the charged particle 
multiplicity analysis~\cite{Atlas:ChargedPartMult} which used the L1\_MBTS\_1 trigger to
select its dataset.  The efficiency of the L1\_MBTS\_1 trigger was determined using the 
\emph{mbSpTrk} trigger as an orthogonal trigger. The efficiency was defined as 
the fraction of events triggered by mbSpTrk passing the offline 
selection of an inelastic $pp$ interaction that also passed the L1\_MBTS\_1\ trigger. 
This efficiency was determined with respect to offline-selected events containing at least two good tracks with $\pT > 100 \MeV$, $|\eta|<2.5$, and transverse impact parameter with respect to the beamspot satisfying $d_0^{BS} < 1.8$~mm.  Events with more than one interaction were vetoed.

\begin{figure}[!ht]
  \centering
  \includegraphics[width=0.45\textwidth]{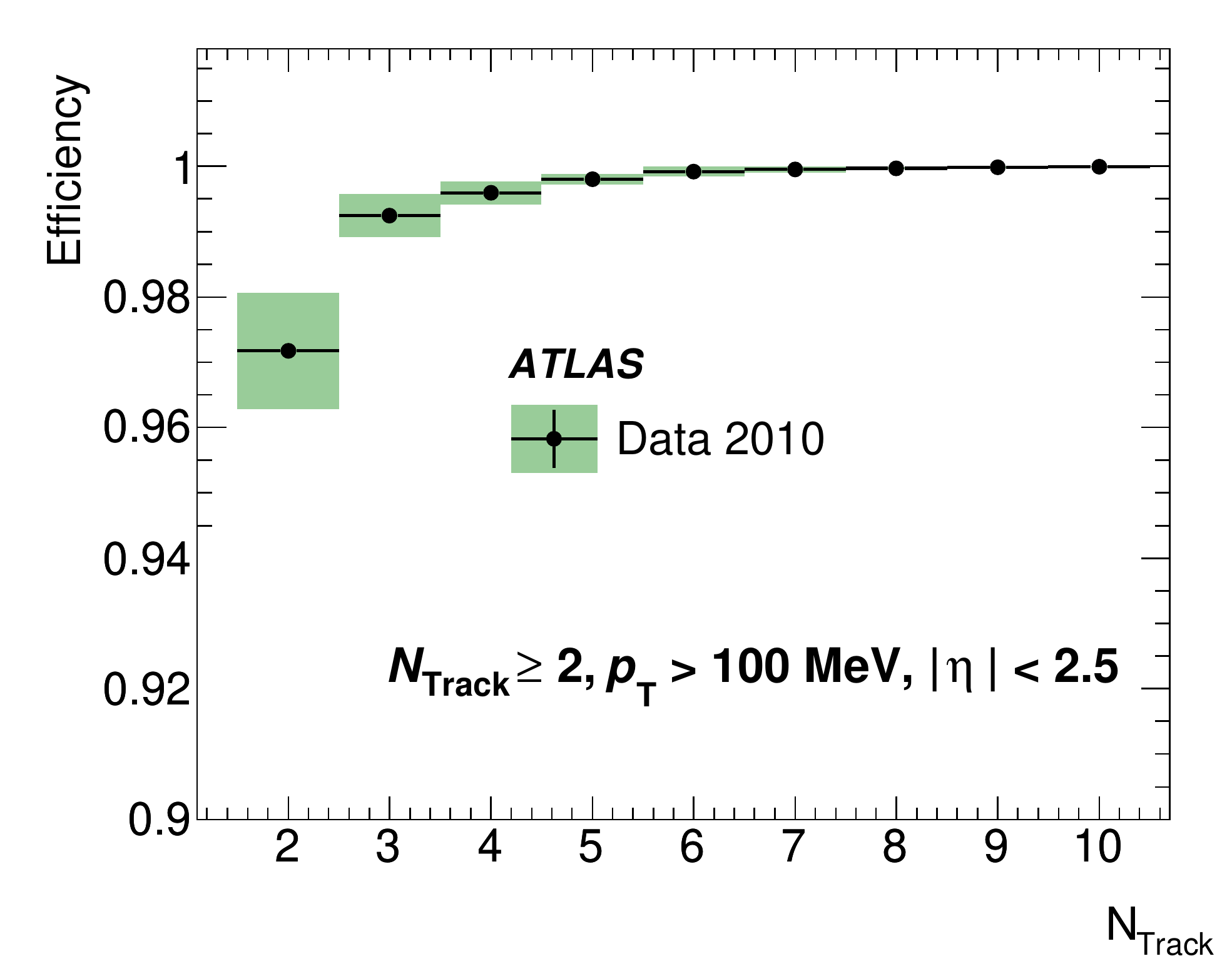}
	\label{fig:eff}
\caption{The L1\_MBTS\_1 trigger efficiency for inelastic $pp$ collisions at \sTev{7}. The shaded areas represent the statistical and systematic uncertainties added in quadrature. The statistical
uncertainty is negligible compared to the systematic uncertainty}
\label{fig:mbts_1_trigger_eff}
\end{figure}

Figure~\ref{fig:mbts_1_trigger_eff} shows the L1\_MBTS\_1\ efficiency as a function of the number of selected offline tracks per event, $N_\mathrm{Track}$, in the data sample. 
 The inefficiency in the low $N_\mathrm{Track}$ region is small but visible.

One source of systematic uncertainty in the measured efficiency is a possible correlation between the control trigger (mbSpTrk) and L1\_MBTS\_1. The trigger efficiency of L1\_MBTS\_1\ in the MC inelastic sample was calculated with and without the control trigger. The difference was found to be negligible.  A second source investigated was the different impact parameter requirements from those in the offline selection. The trigger efficiency was studied with various sets of these requirements and the largest difference among these sets in each bin was taken as the systematic uncertainty for that bin. This variation provides a very conservative estimate of the effect of beam-induced background and secondary tracks on the trigger efficiency.

\begin{figure}[!ht]
\centering
\includegraphics[width=0.45\textwidth]{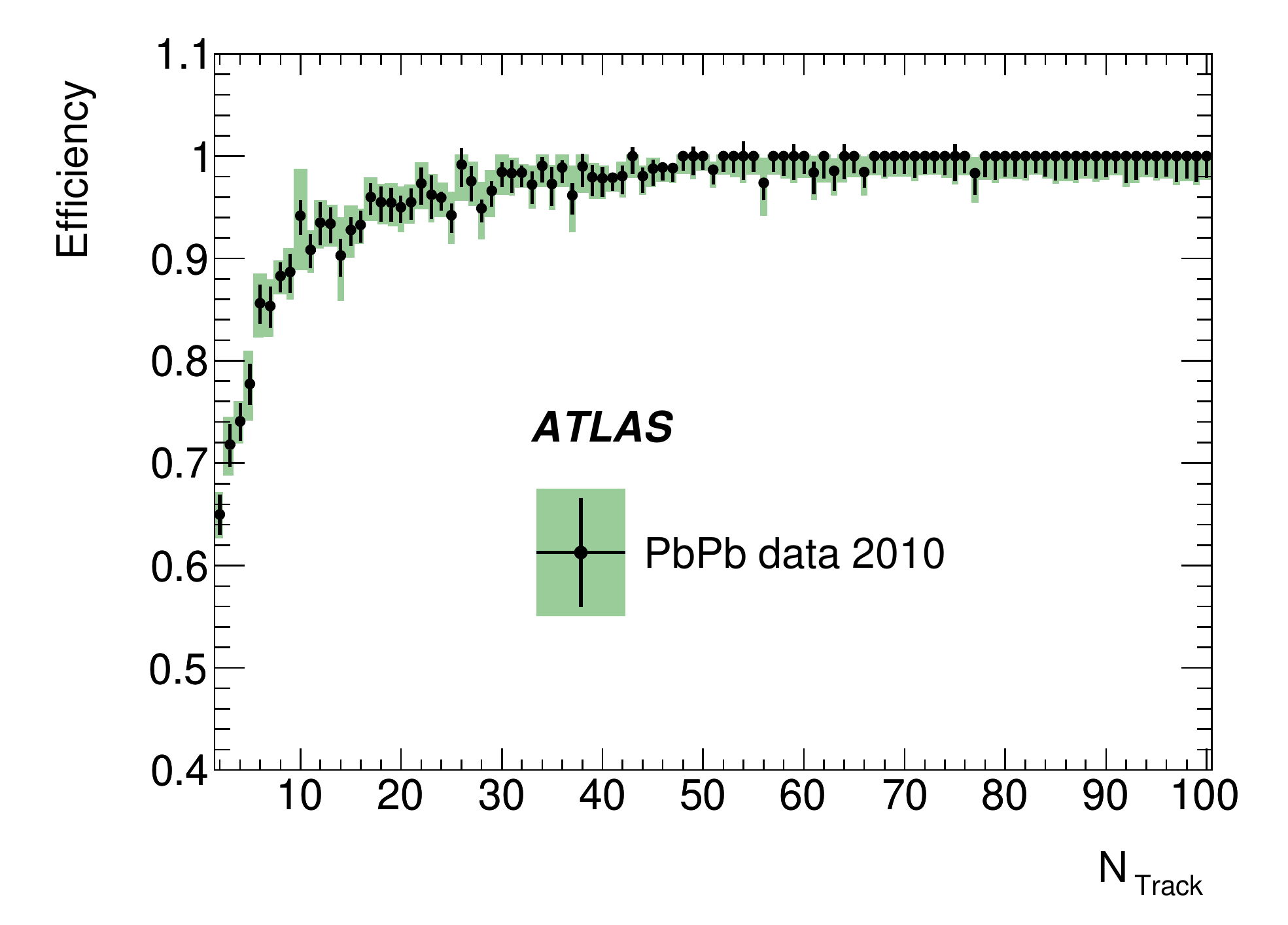}
\caption{The L1\_ZDC\_A\_C trigger efficiency as a function of the number of tracks for inelastic $PbPb$ collisions at $\sHi$. The vertical bars represent the statistical uncertainty, while the shaded areas represent the statistical and systematic uncertainties added in quadrature}
\label{fig:zdc_trigger_eff}
\end{figure}
The efficiency of the ZDC trigger was measured in \pbpb\ collisions using a procedure similar to that used for the initial L1\_MBTS\_1 efficiency measurement.  The efficiency is shown as a function of the number of tracks in the event in Fig.~\ref{fig:zdc_trigger_eff}.


\subsection{Electrons and Photons}\label{sec:egamma}
\def \figurepath{.}

Events with electrons and photons (\egamma) in the final state are important signatures for many ATLAS physics analyses, from SM precision physics, such as top quark or \Wboson\ boson mass measurement, to searches for new physics. Various triggers cover the energy range between a few GeV and several TeV.  In the low-\ET\ range (5-15\,\GeV), the data collected are used for measuring the cross sections and properties of standard candle processes, such as \Jee, di-photon, low mass Drell-Yan, and \Ztau\ production.  
The data collected in the higher \ET\ range ($> 15\GeV$)
are used to measure the production cross-sections for top quark pairs, direct photons and for the \Zee\ and \Wen\ channels~\cite{Atlas:WZleptons, Atlas:topCS2010, InclPromptPhotonPaper, Atlas:Wjets2010}, as well as searches for new physics such as Higgs bosons, SUSY and exotic particles as in extra-dimension models~\cite{Atlas:diphotonMet2010, Atlas:susyOneLepton2011}. Some of these channels, such as \Jee, \Zee, \Wen ~and $\gamma$+jet, are valuable benchmarks to extract the calibration and alignment constants, and to establish the detector performance.  

\subsubsection{Electron and Photon Reconstruction and Selection Criteria}
\label{sec:egreco}

Electrons and photons are reconstructed in the trigger system in the region $|\eta|<2.5$. At L1, photons and electrons are selected using calorimeter information with reduced granularity. For each identified electromagnetic object, RoIs are formed containing the $\eta$ and $\phi$ directions and the transverse energy thresholds that have been passed, e.g. EM5, EM10, as specified by the L1 trigger menu (Table~\ref{tab:ExampleMenu}). Seeded by the position of the L1 cluster, the L2 photon and electron selections employ a fast calorimeter reconstruction algorithm (Section~\ref{sec:caloReco}), and in the case of electrons also fast track reconstruction (Section~\ref{sec:idReco}). The EF also performs calorimeter cluster and track reconstruction, but uses the offline reconstruction algorithms~\cite{CSCBook}. 

\begin{figure}[!ht]
  \centering
  \subfigure[]{
    \includegraphics[width=0.45\textwidth]{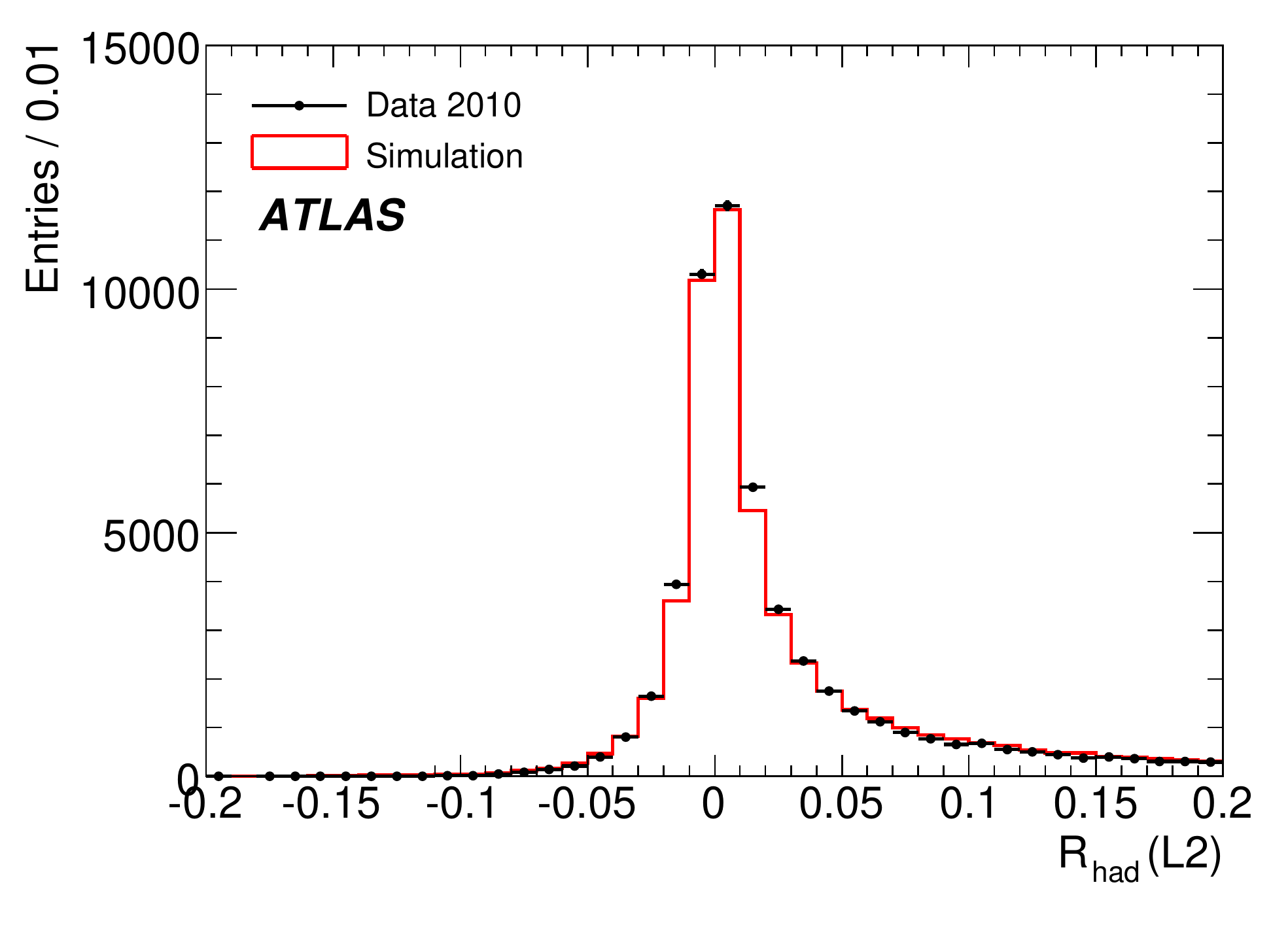}
    \label{fig:ethad}
  }
  \subfigure[]{
    \includegraphics[width= 0.45\textwidth]{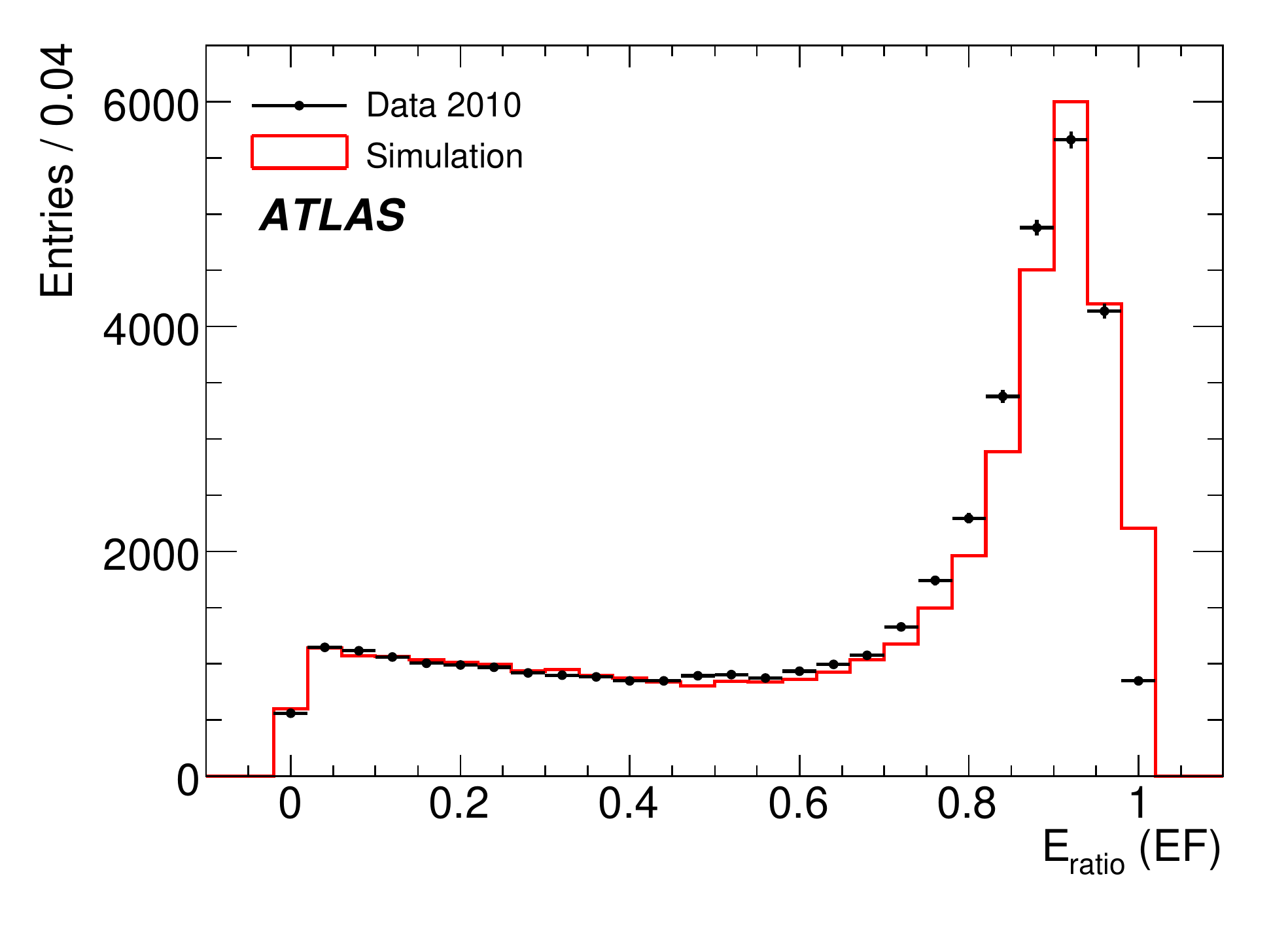}
    \label{fig:eratio}
  }
  \caption{Distributions of the \egamma\ cluster shape variables \subref{fig:ethad} $R_{had}$ at L2 and
\subref{fig:eratio} $E_{ratio}$ at the EF for offline electrons passing the L1 EM trigger with 
a nominal 3~GeV threshold}
  \label{fig:eg_shapes}
\end{figure}

\begin{sloppypar}
At L2 and the EF a calorimeter-based selection is made, for both electrons and photons, based on cluster \et\ and cluster shape parameters.  Distributions of two important parameters are shown in Fig.~\ref{fig:eg_shapes}. The hadronic leakage parameter, $R_{had}=E^{had}_{\textrm{T}} / E^{EM}_{\textrm{T}} $, is the ratio of the cluster transverse energy in the hadronic calorimeter to that in the electromagnetic calorimeter; the distribution for offline reconstructed electrons is shown in Fig.~\ref{fig:ethad} for L2.
Figure~\ref{fig:eratio} shows the distribution, at the EF, of the parameter 
$E_{ratio}=(E_T^{(1)}-E_T^{(2)})/(E_T^{(1)}+E_T^{(2)})$ where 
$E_T^{(1)}$ and $E_T^{(2)}$ are the transverse energies of the two most energetic cells in the first layer of 
the electromagnetic calorimeter in a region of $\Delta\eta\times×\Delta\phi=0.125\times0.2$. The distribution of
this parameter peaks at one for showers with no substructure and so distinguishes clusters due to
single electrons and photons from hadrons and $\pi^0\rightarrow\gamma\gamma$ decays. 
Another important
parameter, $R_\eta$, is based on the cluster shape in the second layer of the 
electromagnetic calorimeter; it is defined as the ratio of transverse energy in a core region of
$3\times7$ cells in $\eta\times\phi$ to that in a $7\times7$ region, expanded in $\eta$ 
from the $3\times7$ core.    
In addition, the electron selection requires that a track be matched to the calorimeter cluster.
\end{sloppypar}

For electrons, three sets of reference cuts are defined with increasing power to reject background: {\it loose}, {\it medium}, and {\it tight}. All selections include the same cuts on the shower shape parameter, $R_\eta$, and hadronic leakage parameter, $R_{had}$. The medium selection adds cuts on the shower shape in the first calorimeter layer, $E_{ratio}$, track quality requirements and stricter cluster-track matching.The tight selection adds, on top of the medium selection, requirements on the ratio, $E_T/p_T$,
of calorimeter cluster $E_T$ to inner detector track $p_T$, a requirement for a hit on the 
innermost tracking layer, and particle identification by the TRT. 

For photons, two reference sets of cuts, {\it loose} and {\it tight}, are defined.  Only the loose selections were used for triggering in 2010. The loose photon selection is the same as the calorimeter-based part of the loose electron selection. The tight selection, in addition, applies cuts on cluster shape in the first calorimeter layer, $E_{ratio}$, and further requirements on cluster shape in the second calorimeter layer.  For more detailed information on \egamma\ triggers in 2010, see Ref.~\cite{Atlas:egammaCONFnote}.

\begin{table*}[hbtp]
   \begin{center}
   \caption{Principal \egamma\ triggers and approximate HLT rates at a luminosity of \Lumi{32}.
The rates given are inclusive; there are significant overlaps between triggers}
   \small{
      \begin{tabular}{llc}
         \hline
         Trigger  & Motivation &  Rate [Hz]\\
         \hline
         \hline
         2e5\_tight      & \Jee, Drell-Yan  & 1 \\
         2e10\_loose     & \Zee, Drell-Yan & 1 \\
         e10\_loose\_mu6 & SM physics, di-lepton searches e.g. for Higgs searches & 1 \\
         2g15\_loose     & di-photon cross-section, di-photon searches  & 1 \\
         e15\_medium     & high-\pt\ physics  & 20   \\
         e20\_loose      & high-\pt\ physics   & 22      \\
         g40\_loose      & direct photons and searches for new particles  & 5 \\
         \hline 
      \end{tabular}
      }
   \normalsize
   \label{tab:eg_rates32}
   \end{center}
\end{table*}

\subsubsection{Electron and Photon Trigger Menu and Rates}
\label{sec:egmenu}
\begin{sloppypar}
Table~\ref{tab:eg_rates32} gives an overview of the rates of the main \egamma\ triggers used in the
2010 menu for instantaneous luminosities around \Lumi{32}. 
The \ET\ thresholds of the electron and photon triggers range from $5\GeV$ to $40\GeV$. 
In addition, supporting triggers were deployed, which were used for efficiency extraction, 
monitoring, commissioning and cross-checks. The overall rate of the \egamma\ trigger 
stream was $\sim$70~Hz at \Lum=\Lumi{32}, constituting $\sim$25\% of the total bandwidth 
written to mass storage at the end of 2010.
\end{sloppypar}

\begin{figure}[!ht]
  \centering
  \subfigure[]{
    \includegraphics[width=0.45\textwidth]{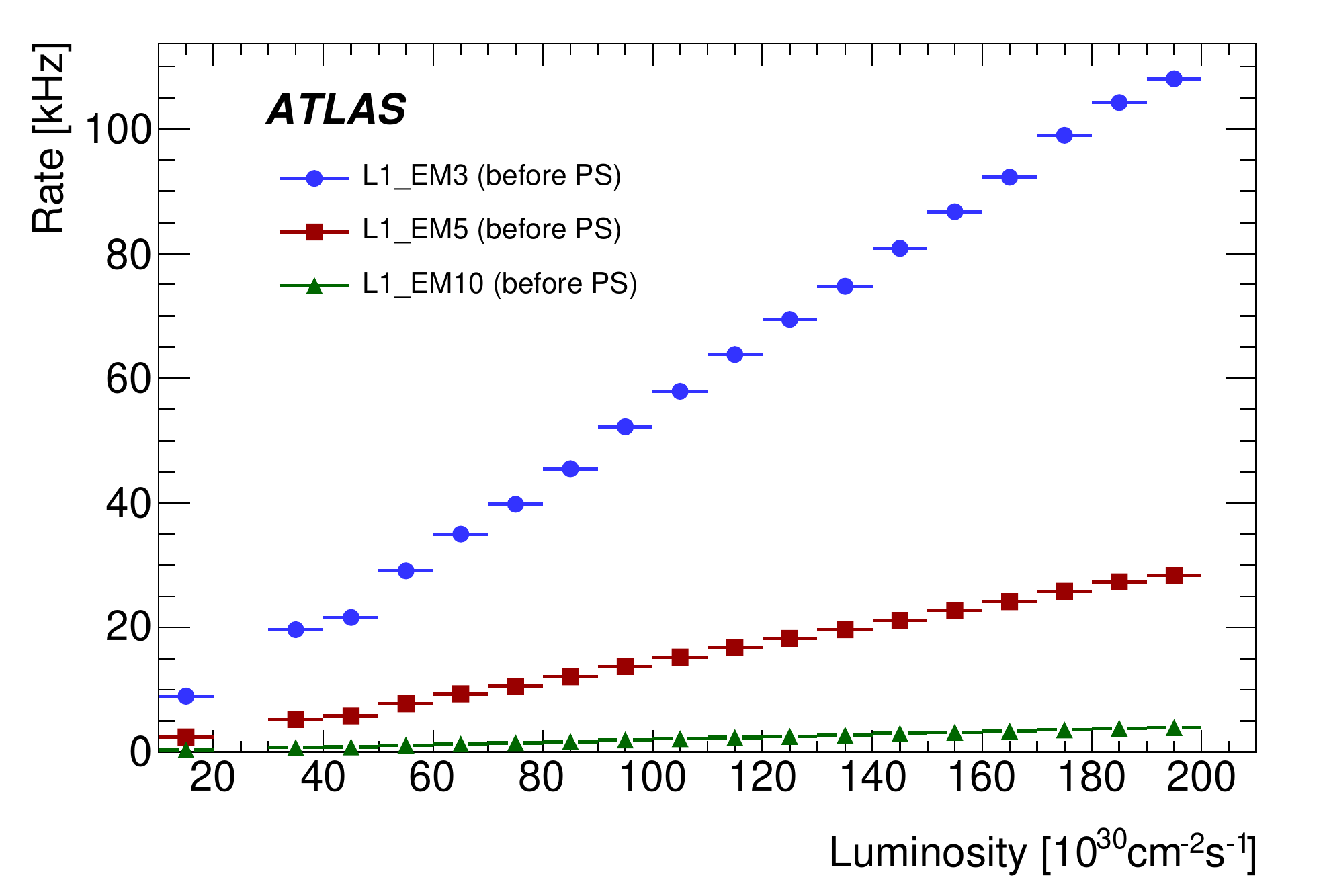}
    \label{fig:L1Rates}
  }
  \subfigure[]{
    \includegraphics[width= 0.45\textwidth]{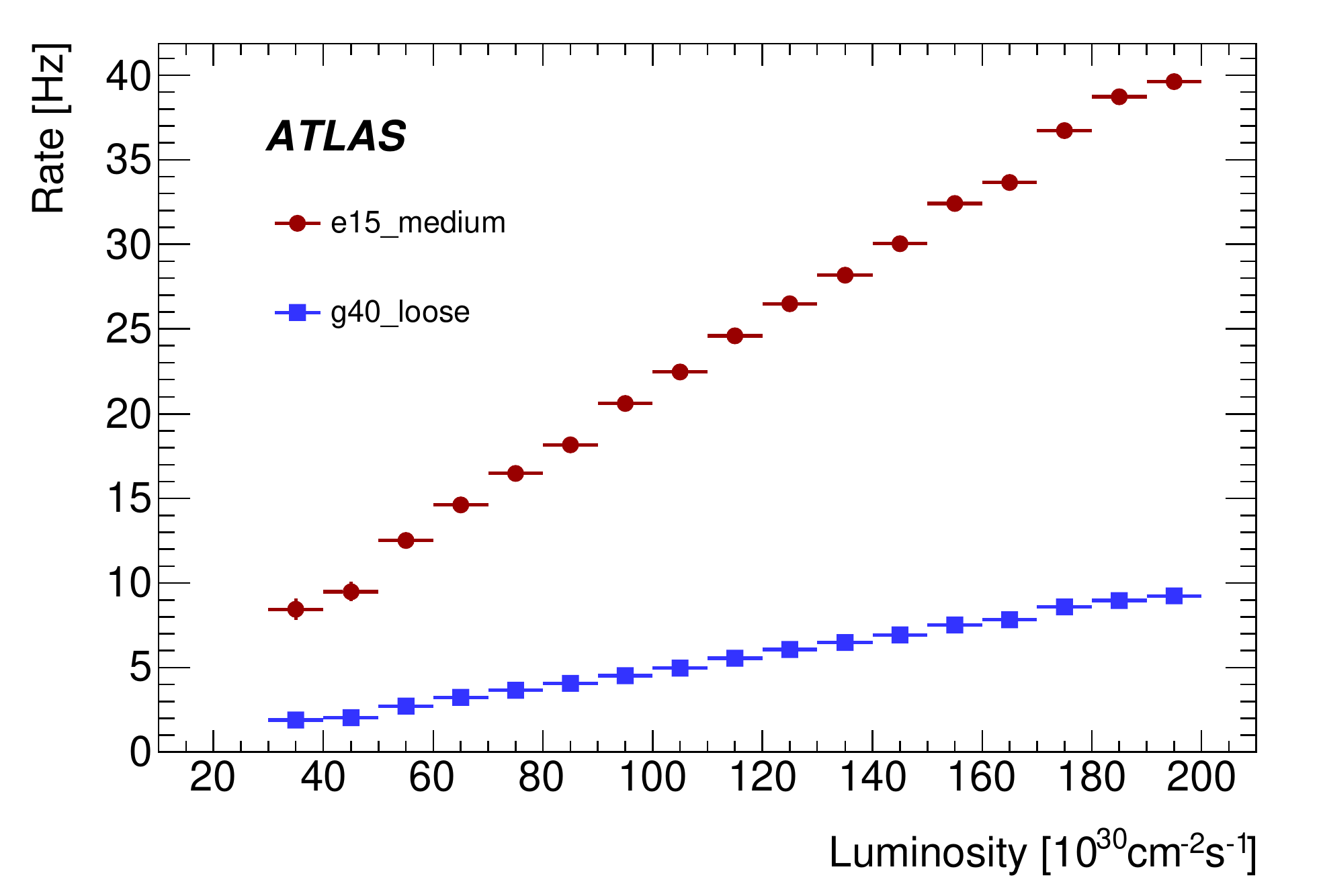}
    \label{fig:HLTRates}
  }
  \caption{Observed rates for primary \egamma\ triggers at (a) L1, before pre-scaling (PS), and (b) EF, after pre-scaling}
  \label{fig:egRates}
\end{figure}

The L1 and HLT trigger rates of \egamma\ triggers are shown in Fig.~\ref{fig:egRates} as a function of  luminosity. No significant
deviation from linearity was observed during 2010 running. 
It should be noted that during the course of 2010,  no deterioration in performance of \egamma\ triggers or effect on rates was observed due to in-time or out-of-time pile-up.

\subsubsection{Electron and Photon Trigger Efficiencies}
\label{sec:egPerf}

\begin{sloppypar}
Trigger efficiencies are presented for electrons and photons identified by the offline reconstruction. More details are given in Ref.~\cite{Atlas:egammaCONFnote}, including a full study of the systematic uncertainties in the plateau efficiencies which amount to~$\sim$~0.4\% for the electron trigger and ~$\sim$~1\% for the photon trigger. The EF selection of electrons and photons is very similar to the  offline identification: the same criteria are used for loose, medium and tight selections in offline reconstruction as detailed in Section~\ref{sec:egreco}. 
\end{sloppypar}

The determination of the efficiencies of electron and photon triggers share the following common selection criteria. Collision event candidates are selected by requiring a primary vertex with at least three tracks. 
Rare events that contain very localised high-energy calorimeter deposits not originating from proton-proton collisions, for example from sporadic discharges in the calorimeter or cosmic ray muons undergoing a hard 
bremsstrahlung, are removed, resulting in predicted losses of less than 0.1\% of minimum-bias events and 0.004\% 
of \Wen\ events~\cite{WinclPaper}.
In addition, events are rejected if the candidate electromagnetic cluster is located in a problematic region of the EM calorimeter, for example where the effect of inactive cells could be significant. Due to hardware problems~\cite{LArPaper}, the signal could not be read out from $\sim$2\% of the EM calorimeter cells in 2010.  Offline electrons are selected if they are within the region $|\eta|<2.47$ and outside the transition between the barrel and end-caps of the EM calorimeter, $1.37<|\eta|<1.52$.  The acceptance region for photons is limited to $|\eta|<2.37$ due to the geometrical acceptance of the first layer of the EM calorimeter (fine strips in the $\eta$ direction), which is crucial for the rejection of background photons originating from \pizero\ decay. 

\begin{figure}[!ht]
  \centering
  \subfigure[]{
    \includegraphics[width=0.45\textwidth]{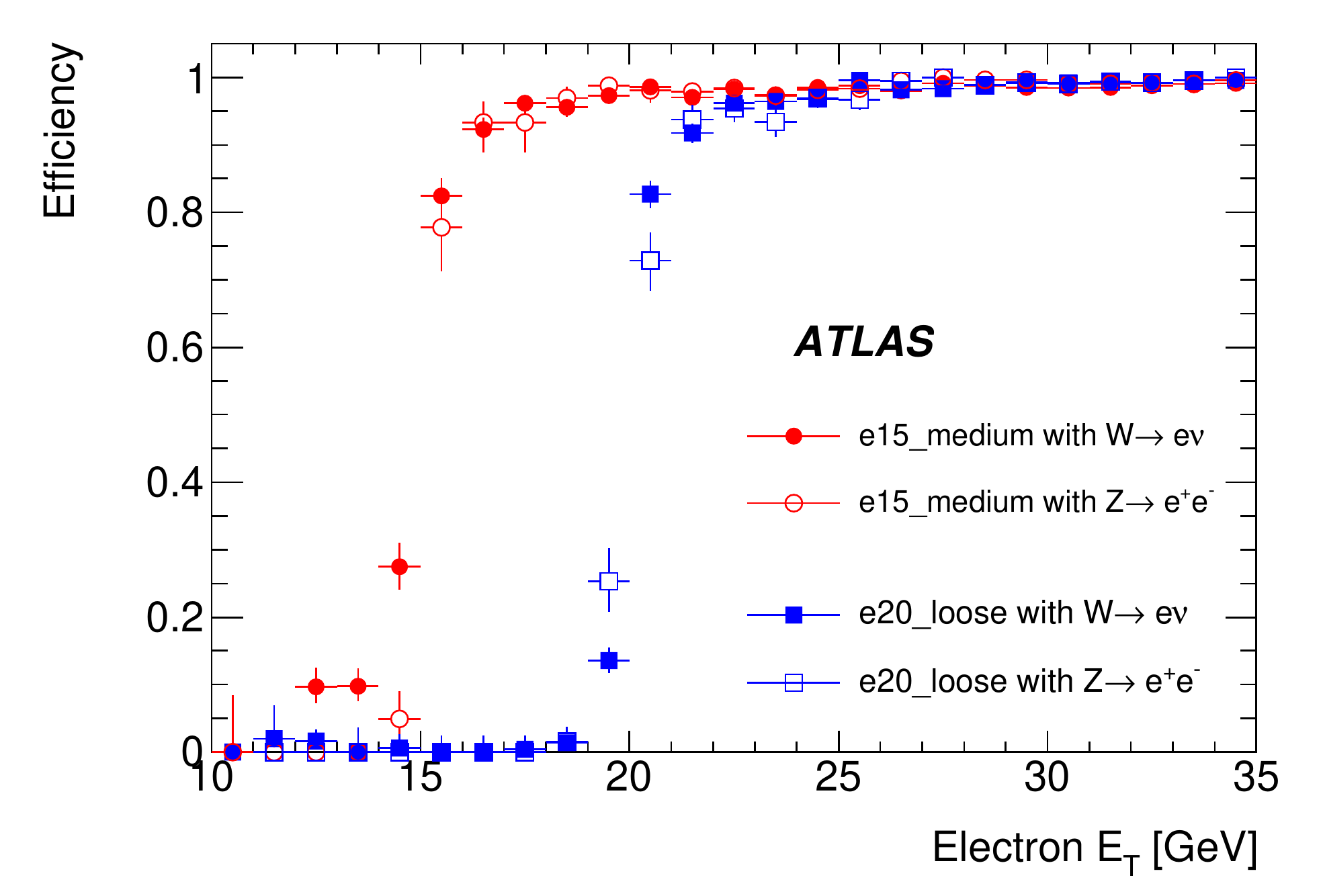}
    \label{fig:WZ_PT}
}
  \subfigure[]{
  
    \includegraphics[width= 0.45\textwidth]{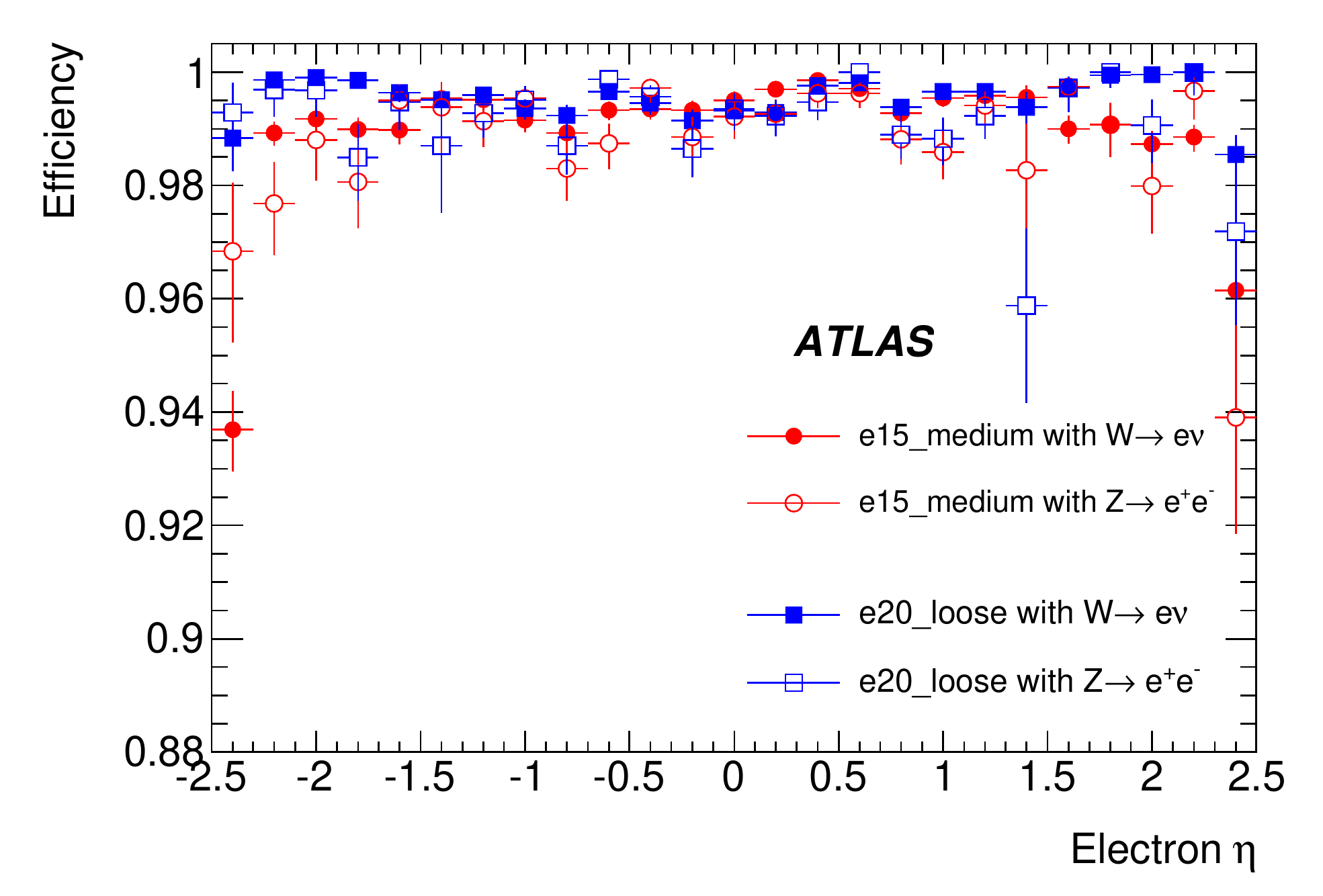}
    \label{fig:WZ_Eta}
}
  \caption{Efficiencies for the e15\_medium and e20\_loose triggers, measured with respect to offline tight electrons in \Wen\ and \Zee\ events, shown as a function of \subref{fig:WZ_PT} \ET\ and \subref{fig:WZ_Eta} $\eta$}
  \label{fig:WZ}
\end{figure}

The decays \Zee\ and \Wen\ provide samples to measure the electron trigger efficiency in the higher-\ET\ range ($>15\GeV$).  
The \Zee\ decays provide a sample of electrons to use with the tag-and-probe method.
In the case of \Wen\ decays, the orthogonal trigger method is employed, using the \MET\ triggers with thresholds between 20 and $40\GeV$ to collect the data sample.  Figure~\ref{fig:WZ}
compares the efficiencies of the e15\_medium and e20\_loose triggers at the EF, measured in \Wboson\ boson events with those measured in \Zboson\ boson events. The efficiencies are measured with respect to offline tight electrons as functions of the offline electron \ET\ and $\eta$.   
The efficiencies measured by the two methods are in excellent agreement, differing by less than about 1\%.
The dominant contribution (0.4\%) to the systematic uncertainty in the plateau efficiency 
comes from an analysis of
the spread of differences in efficiency between data and simulation as a function of \ET\ and 
$\eta$.  
Figure~\ref{fig:WZ_Eta} shows that the response in $\eta$ is flat except at the outer edges of the end-caps. 
Above $20\GeV$ the e15\_medium trigger efficiency for \Wen\ and \Zee\ events is greater than 99\%.

\begin{figure}[!ht]
  \centering
  \subfigure[]{
    \includegraphics[width=0.45\textwidth]{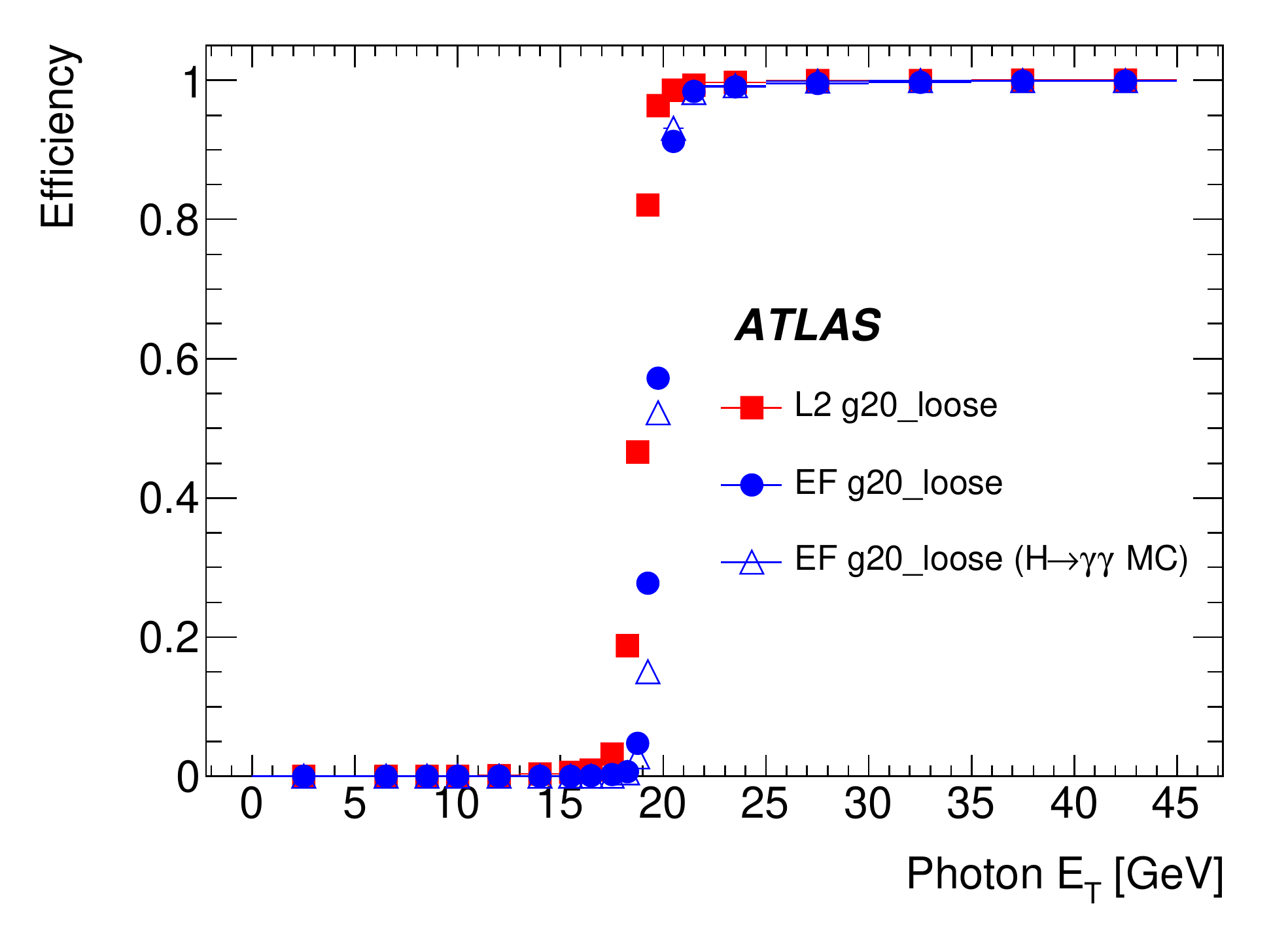}
    \label{fig:g20loose_et}
  }
  \subfigure[]{
    \includegraphics[width= 0.45\textwidth]{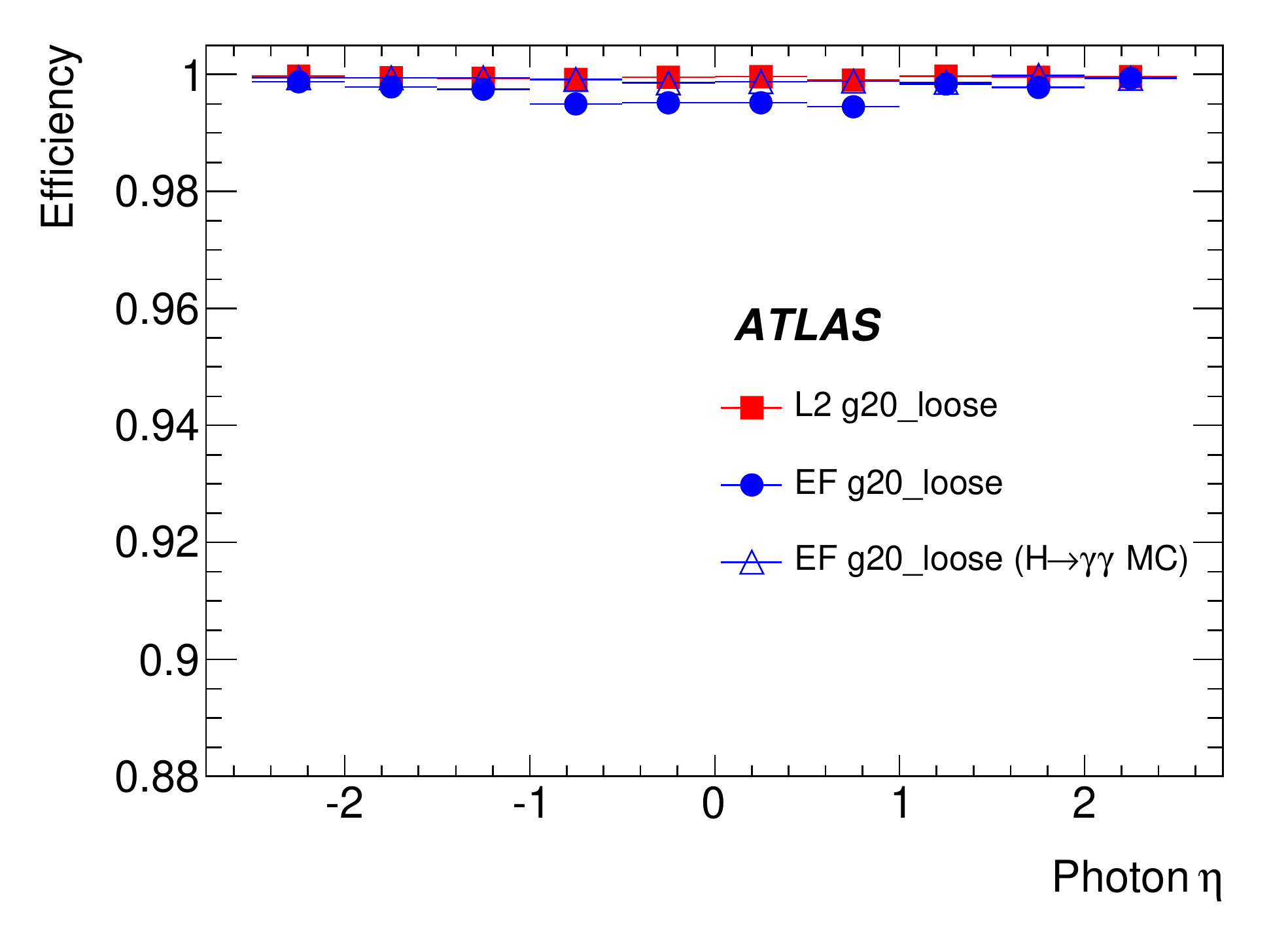}
    \label{fig:g20loose_eta}
  }
  \caption{Photon efficiencies measured with the bootstrap method, as a function of the offline tight photon \subref{fig:g20loose_et} \ET\ and \subref{fig:g20loose_eta} $\eta$ (with $\et>25\GeV$) for events passing the L1\_EM14 threshold}
  \label{fig:g20loose}
\end{figure}

In contrast to electrons, there is no suitable decay channel that would allow the
trigger efficiency to be measured for prompt photons 
in the $\sim10-50 \GeV$ energy range using tag and probe or 
orthogonal triggers.
Therefore, the bootstrap method is used, where the 
HLT efficiency is measured for events that pass a lower L1 \ET\ threshold. For example, the 
g20\_loose efficiency is measured using a sample of events passing the $14\GeV$ \ET\ L1 
threshold (EM14). In most physics analyses, the photons are selected offline with tight 
identification requirements. Thus, the trigger efficiency is shown with respect 
to photons identified with the tight offline requirements. The bootstrap method relies on 
measuring the HLT efficiency in a \pT\ region where the L1 trigger is fully efficient with
respect to offline photons. It has been verified that L1\_EM14 is fully efficient 
for photon clusters with $\ET>20 \GeV$ using a sample of events selected by the L1\_EM5 trigger.  
The bootstrap method suffers from a large contamination of fake photons, such as hadronic jet clusters mis-reconstructed as photons. The bias on the measured efficiency has been estimated to be less than $\sim$0.25\% for photons with 
$\ET>~25$~GeV by comparing the efficiencies from data with those from 
a signal-only simulation.
Figure~\ref{fig:g20loose} shows the L2 and EF efficiencies for the 
g20\_loose trigger, as functions of offline tight photon \ET\ and $\eta$. For the $\eta$ 
distribution, photons were selected with $\et>25\GeV$  in the plateau region 
of the turn-on curve.

\begin{table*}[!htb]
   \begin{center}
   \caption{Principal muon triggers and their approximate HLT rates at a luminosity of \Lumi{32}}
   \small{
      \begin{tabular}{llc}
         \hline
         Trigger & Motivation & Rate [Hz]\\
         \hline
         \hline
         mu13     & \Zmumu, \Wmunu, top, new physics  & 12   \\
         mu40\_MSonly     & new physics   & 20  \\
         2mu6     & \Zmumu, Drell-Yan, $B$ physics, new physics  & 14   \\
         \hline 
      \end{tabular}
      }
   \normalsize
   \label{tab:muon_primary_trigger}
   \end{center}
\end{table*}

The L2 and EF g20\_loose triggers reach the efficiency plateau at about $\ET=25 \GeV$, 
with efficiencies above this threshold of greater than 99\% for both L2 and EF.
The efficiency remains flat, at the plateau value, as far as can be tested in the 2010 data, up to $\sim 500\GeV$.  The agreement between the efficiencies measured in data and simulated events is better than 1\%.


\subsection{Muons}\label{sec:muon}
\def \figurepath{.}
Muons are produced in many final states of interest to the broad physics 
programme being conducted at the LHC, from SM precision physics, such as top quark and \Wboson\ boson mass measurements, to searches for new physics.  Muons are identified with high purity compared to other signatures and cover a wide momentum range between a few GeV and several TeV.  Trigger thresholds in the \pT\ range $4-10\GeV$ are used to collect data for measurements of processes such as \Jmumu, low-\pT\ di-muons, and \Ztau.  Higher \pT\ thresholds are used to collect data used to measure the properties of SM particles such as \Wboson and \Zboson\ bosons, and top quarks, \cite{Atlas:WZleptons, Atlas:topCS2010, Atlas:Wjets2010} as well as to search for new physics, like the Higgs boson, SUSY~\cite{Atlas:susyOneLepton2011} and extra-dimension models.  Some of these channels, such as \Jmumu, \Zmumu, and \Wmunu\ decays are valuable benchmarks to extract calibration and alignment constants, and to establish the detector performance.

\subsubsection{Muon Reconstruction and Selection Criteria}
\label{subsec:muon_reco}

The trigger reconstruction algorithms for muons at L1 and the HLT are described in Sections~\ref{sec:L1muon} and \ref{sec:muonReco} respectively.   The selection criteria applied to reconstructed muon candidates depend on the algorithm with which they were reconstructed.  The MS-only algorithm selects solely on the \pT\ of the muon; the combined algorithm makes selections based on the match between the inner detector and muon spectrometer tracks and their combined \pT; the isolated muon algorithm applies selection criteria based on the amounts of energy found in the isolation cones.

\subsubsection{Muon Trigger Menu and Rates}
\label{ref-ss-trigrates}

Table~\ref{tab:muon_primary_trigger} gives an overview of  the principal muon triggers and their approximate rates at a luminosity of \Lumi{32}.  In addition to these principal physics triggers, a range of supporting triggers were included for commissioning, monitoring, and efficiency measurements. In 2010 running, in order to maximize acceptance, all HLT selections were based on L1 triggers using the low-\pT\ logic 
(described in Section~\ref{sec:L1muon}), including mu13, mu20 and mu40 that were seeded from the L1 MU10 
trigger.

\begin{figure}[!htb]
  \centering
  \subfigure[]{
    \includegraphics[width=0.45\textwidth]{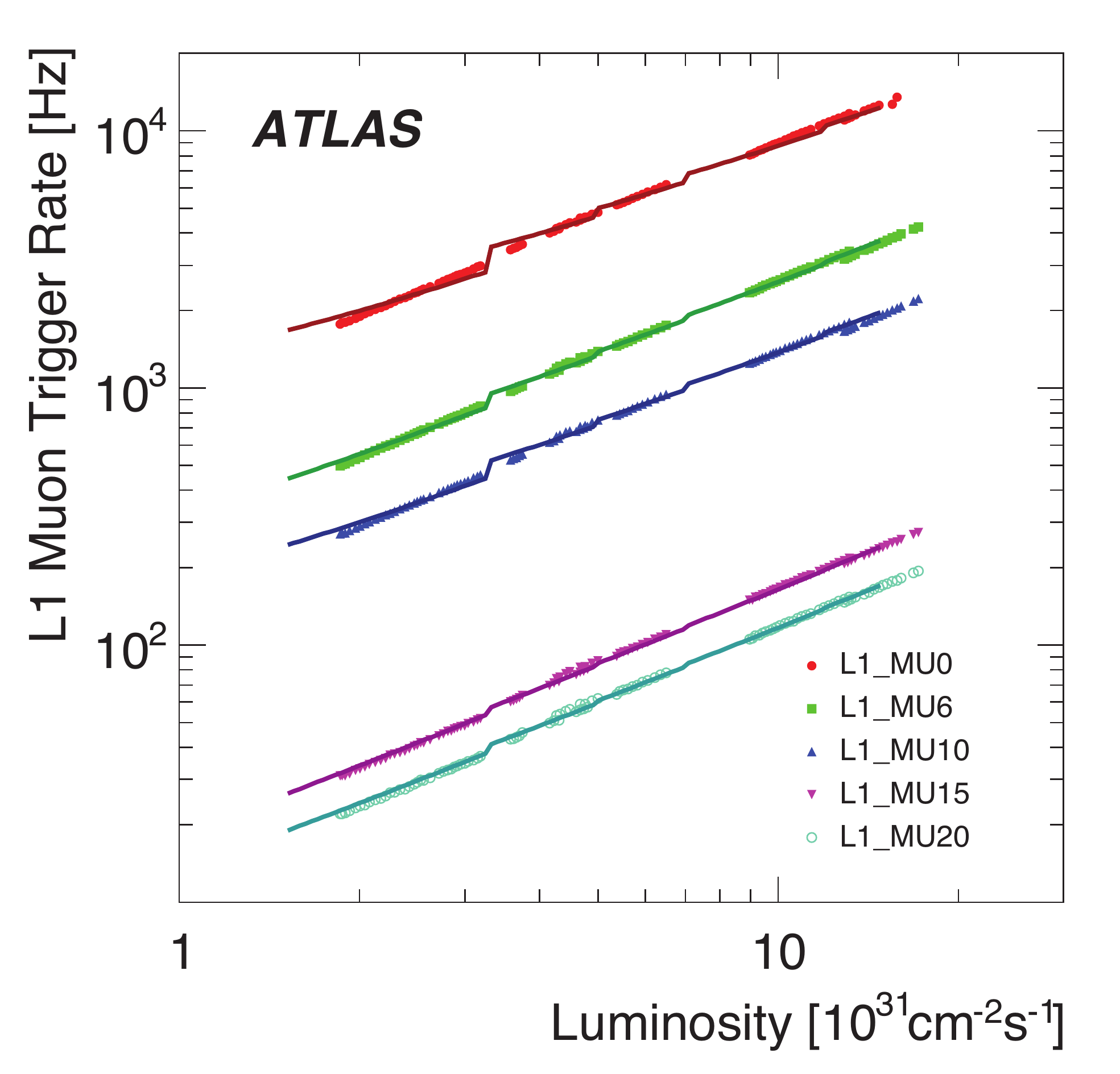}
        \label{fig:rateL1}
	}  
  \subfigure[]{
    \includegraphics[width=0.45\textwidth]{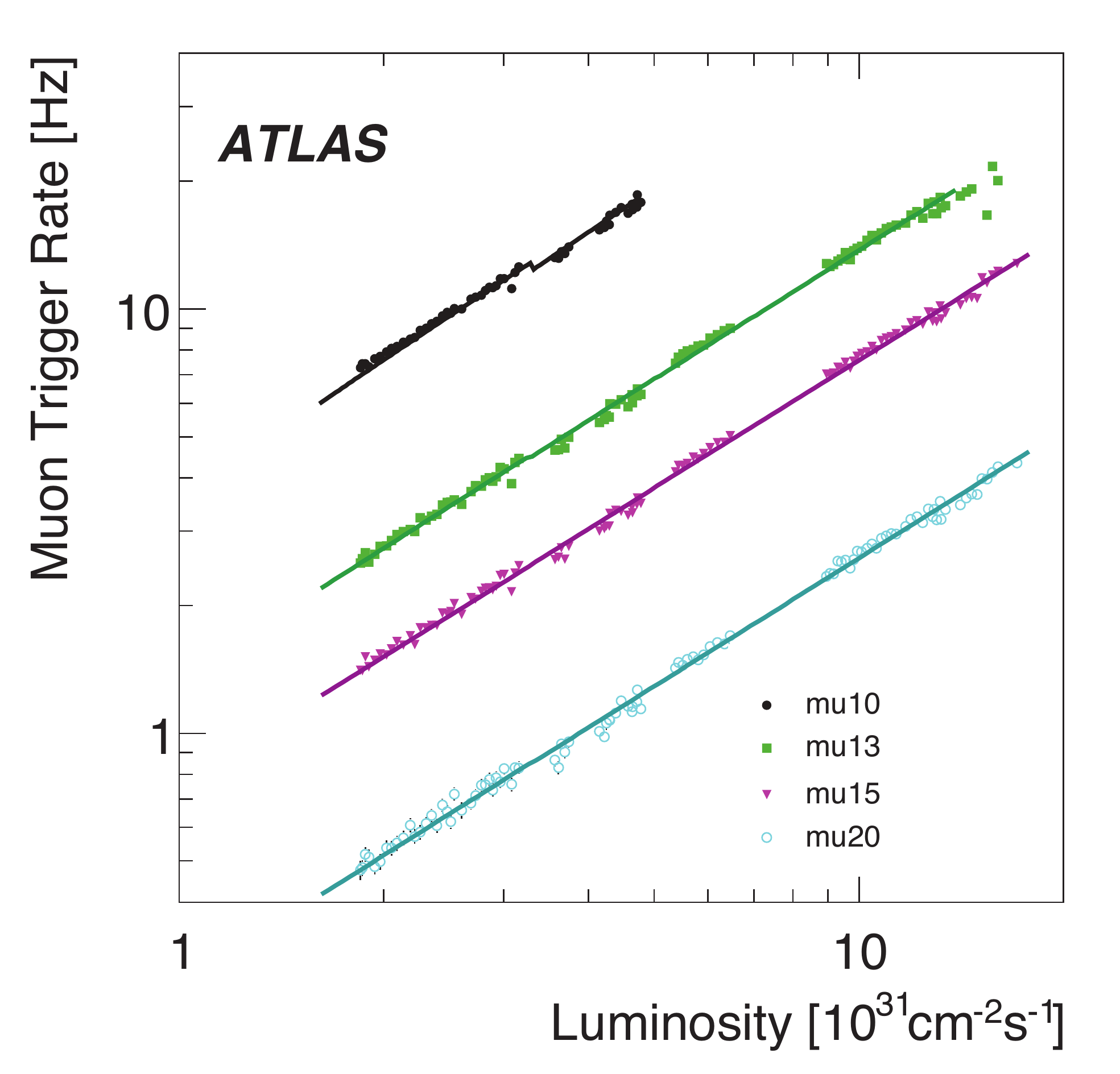}
    \label{fig:rateEF}
    }
  \caption{Observed mean trigger rates as a function of the instantaneous luminosity for \subref{fig:rateL1} L1 and \subref{fig:rateEF} the EF}
  \label{fig:ratePlots}
\end{figure}

The trigger rates at L1, L2, and EF are dependent on thresholds, algorithms (Section \ref{sec:muonReco}) and luminosity.  The trigger rates have been measured as a function of the luminosity
and parametrized with Equation~\eqref{eq:rate}: 
\begin{equation}\label{eq:rate}
r = c_1  \Lum + c_0 N_{BC} ,
\end{equation}
\noindent where $r$ is the rate, \Lum\ the instantaneous luminosity, $N_{BC}$
the number of colliding bunches, and $c_1$, $c_0$ are proportionality constants. The
second term represents the contribution to the trigger rate from cosmic rays: 
as the number of colliding bunches 
increases, so does the amount of time the trigger gate is open to accept cosmic rays. 
The instantaneous luminosity was 
taken from the online measurements averaged over ten successive luminosity
blocks.

The measured muon trigger rates are shown for L1 and EF in 
Fig.~\ref{fig:ratePlots}  together with lines representing 
 the result of fitting Equation~\eqref{eq:rate} to the measurements.  
Steps in the rate are due to the increases in $N_{BC}$, and hence the contribution
to the rate from cosmic rays.
This is significant at L1 and for algorithms using only the muon
spectrometer data at the HLT.  For combined algorithms, the contribution from 
cosmic rays to the rate is negligible (within the
errors of the fit).

\begin{figure*}[!p]
\centering
\subfigure[]{
	\includegraphics[width= 0.45\textwidth,height=5cm]{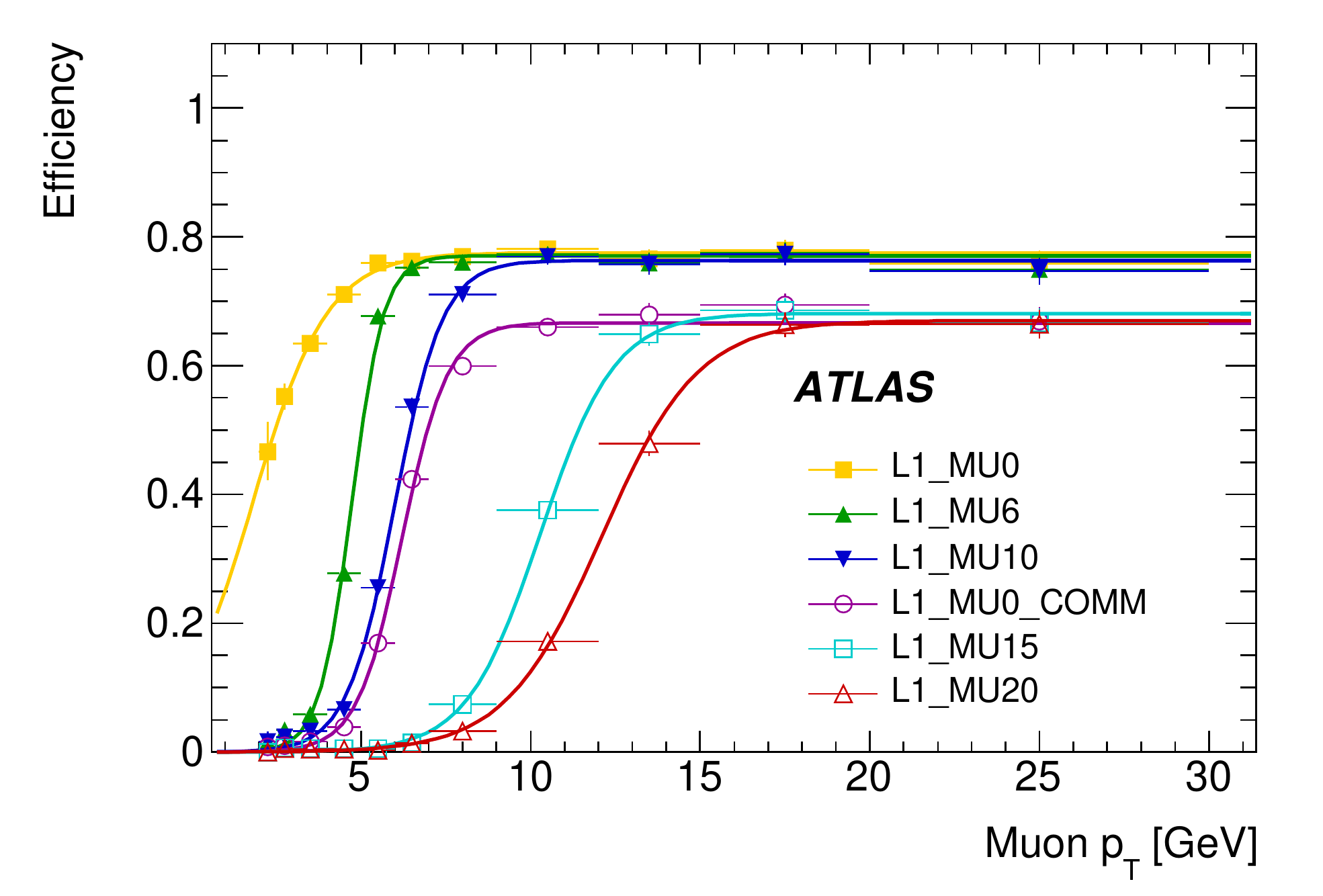}
	\label{fig:RPCL1MU}
}
\subfigure[]{
	\includegraphics[width=0.45\textwidth,height=5cm]{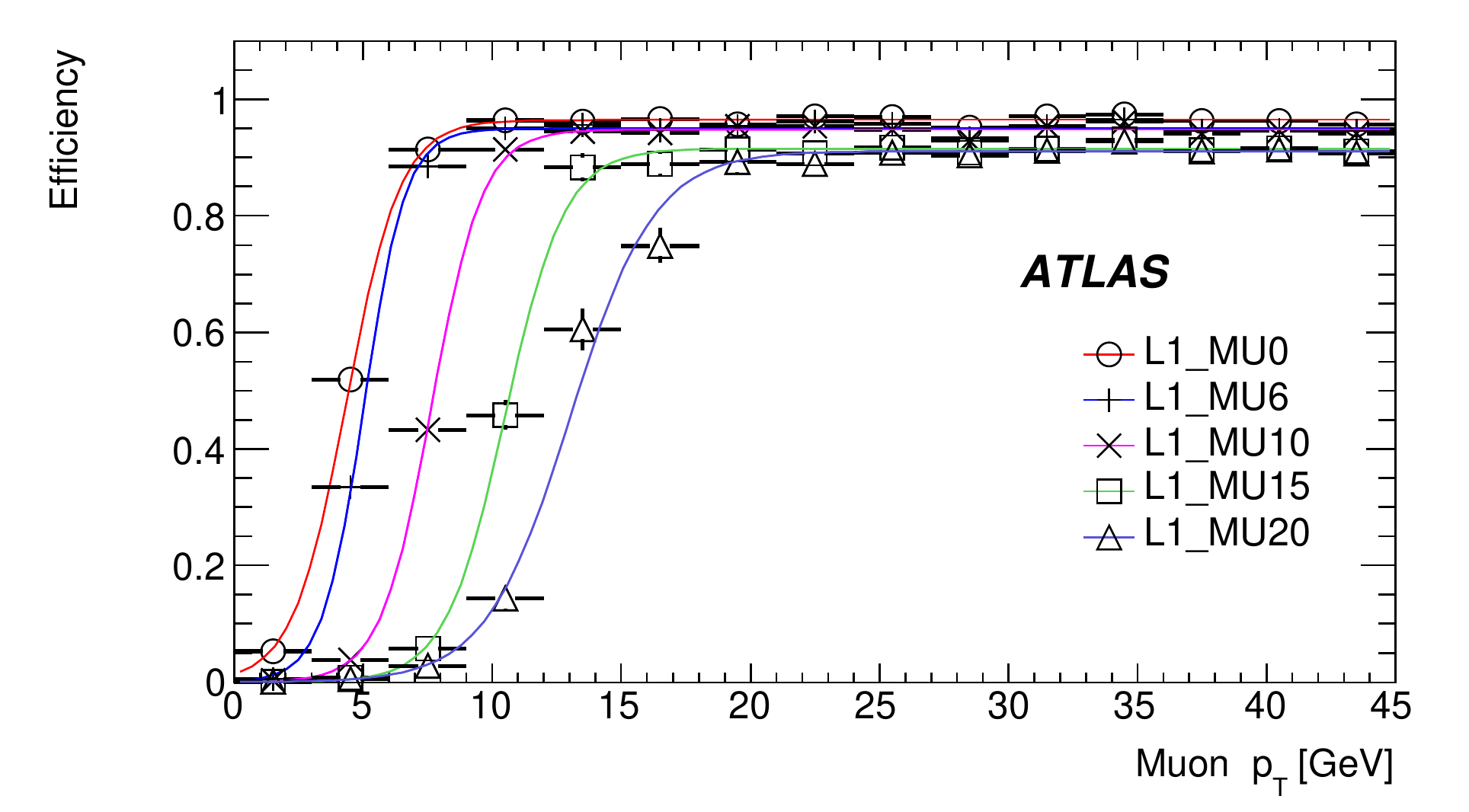}
	\label{fig:TGCL1MU}
}
\caption{L1 trigger efficiency for combined muons reconstructed offline for \subref{fig:RPCL1MU}~RPC and \subref{fig:TGCL1MU}~TGC triggers.  A Fermi function is fitted to the measurements for each trigger.  Statistical uncertainties are represented by vertical bars} 
\label{fig:InclusiveEff}
\end{figure*}
\begin{figure*}[!p]
  \begin{center}
   \subfigure[]{
      \includegraphics[width=0.45\textwidth]{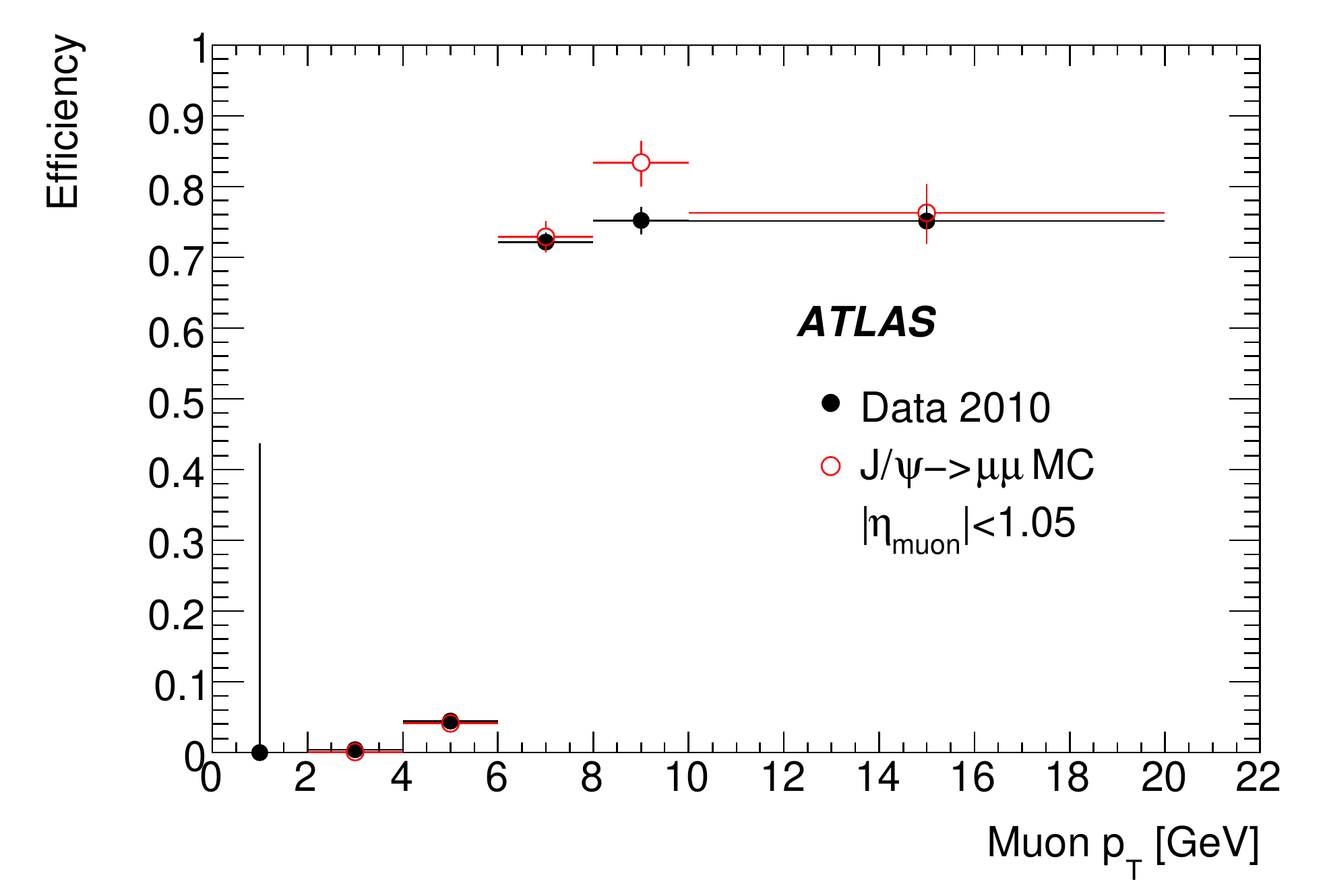} 
      \label{fig:mu_lowpt_hlteff_ba}
   }
    \subfigure[]{
       \includegraphics[width=0.45\textwidth]{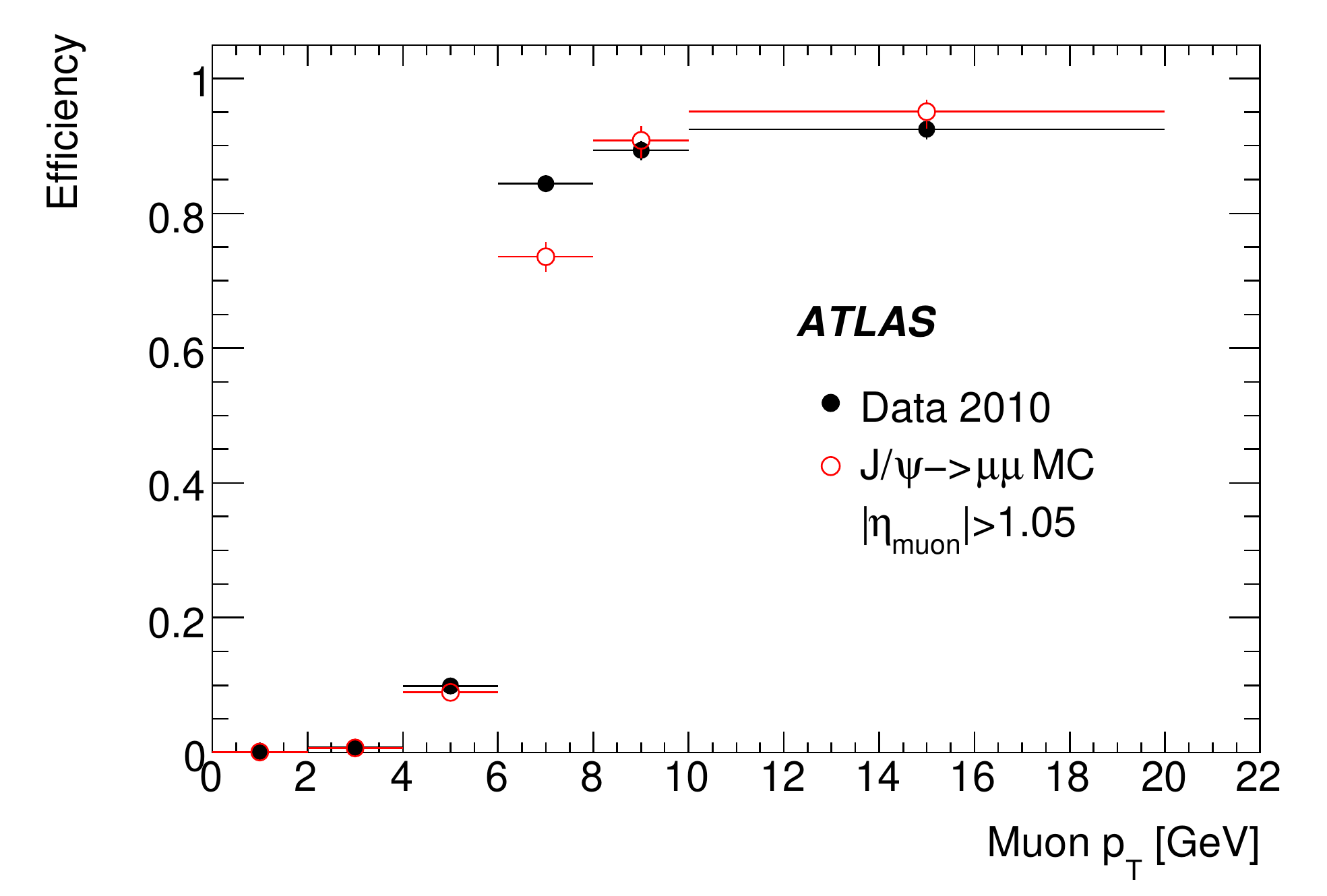} 
       \label{fig:mu_lowpt_hlteff_ec}
    }
    \caption{Efficiencies of the 6 GeV threshold combined muon trigger (mu6) as a function of reconstructed muon \pt\ for \Jmumu\ decays in data and simulation for the \subref{fig:mu_lowpt_hlteff_ba} barrel and \subref{fig:mu_lowpt_hlteff_ec} end-caps.  Statistical uncertainties are represented by vertical bars}
    \label{fig:mu_lowpt_hlteff}
  \end{center}
\end{figure*}
\begin{figure*}[!p]
\centering
\subfigure[]{
	\includegraphics[width=0.45\textwidth, height=4.8cm]{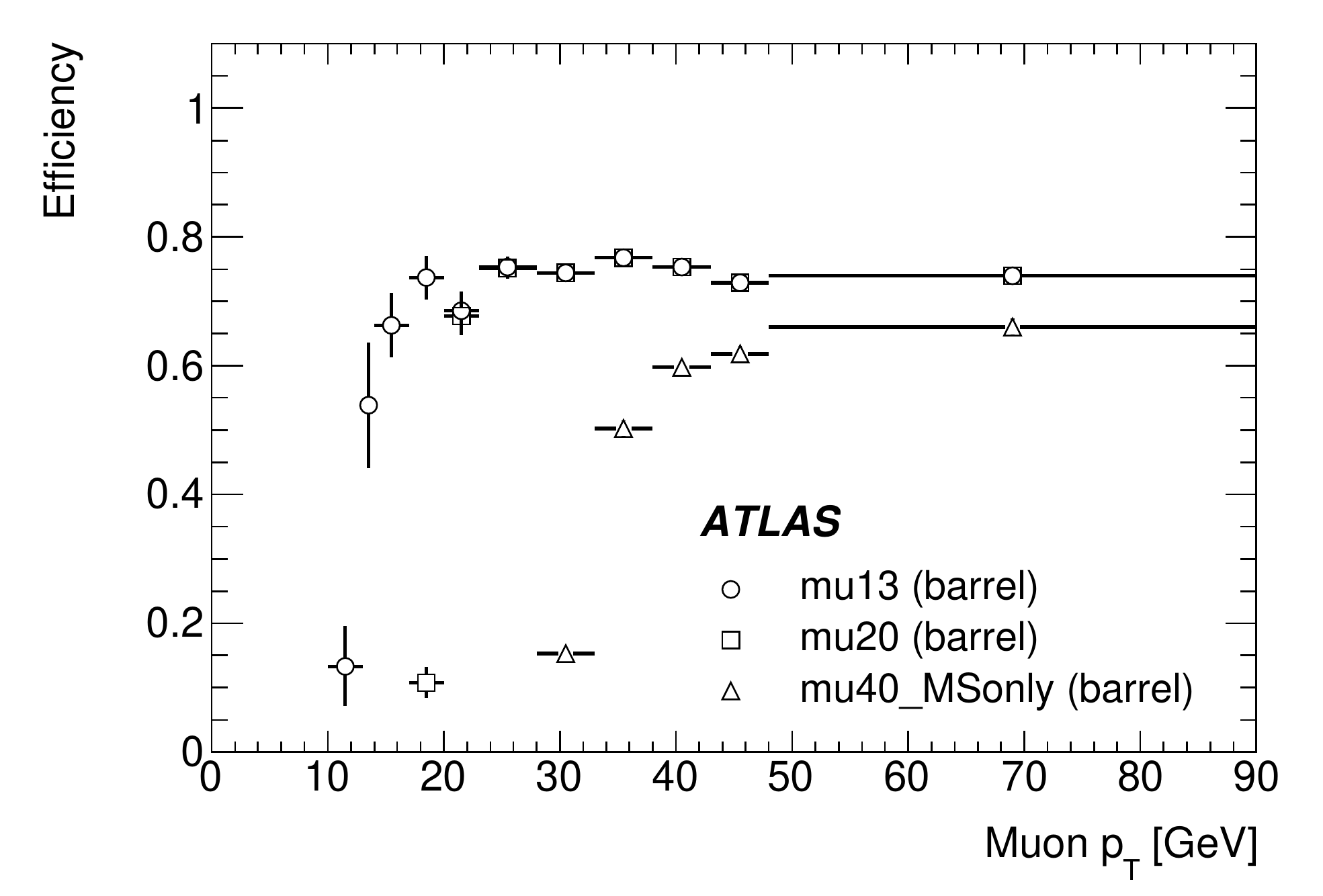}
	\label{fig:mu_hlteff_ba}
}
\subfigure[]{
	\includegraphics[width=0.45\textwidth, height=4.8cm]{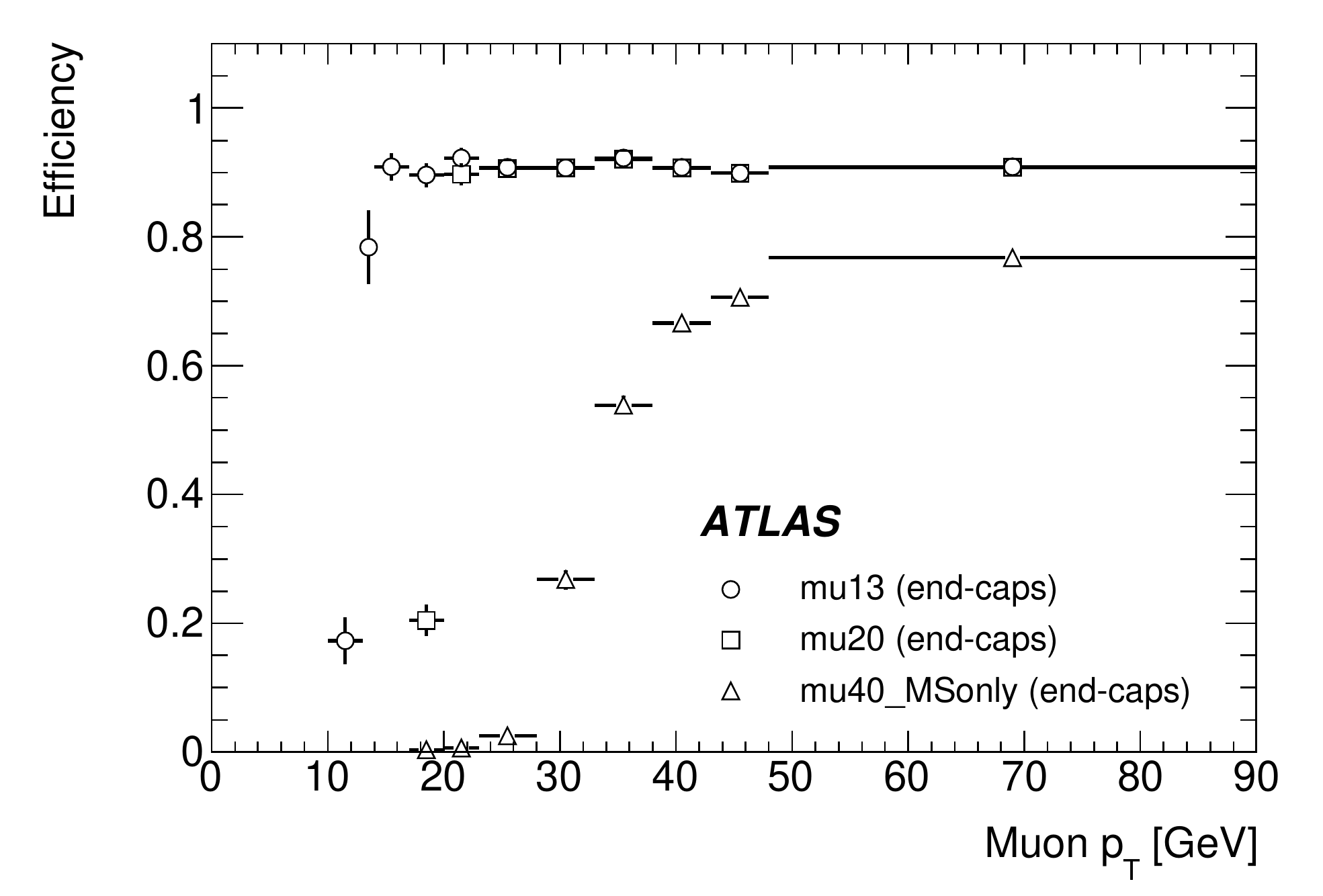}
	\label{fig:mu_hlteff_ec}
}
\caption{Efficiencies in the \subref{fig:mu_hlteff_ba} barrel and 
\subref{fig:mu_hlteff_ec} end-caps of the combined muon triggers with 13 GeV and 20 GeV thresholds and the MS-only trigger with a 40 GeV threshold.   Statistical uncertainties are represented by vertical bars}
\label{fig:mu_hlteff}
\end{figure*}

\subsubsection{Muon Trigger Efficiency}

The muon trigger efficiencies have been measured for offline muons~\cite{Atlas:muonCONFnote}.   The L1 RPC trigger efficiencies measured using an orthogonal 
L1 calorimeter trigger are shown in Fig.~\ref{fig:RPCL1MU} for various thresholds. The efficiencies measured using the tag and probe method with  \Jmumu\ and \Zmumu\ decays are shown for the L1 TGC trigger in  Fig.~\ref{fig:TGCL1MU}.
The geometrical acceptance of the RPC low-\pT\ trigger is about 80\% which 
explains the lower efficiency compared to the TGC trigger, 
which has a geometrical acceptance close to 95\%. 
For the RPC trigger, a further reduction in plateau efficiency is evident for the high-\pT\ ($\pT>$10~GeV) 
triggers compared
to the low-\pT\ triggers ($\pT\le$10~GeV).
About half (6\%) of this difference is due to a smaller geometrical coverage of the high-\pT\ triggers.  Part of this inefficiency will be recuperated in the muon spectrometer upgrade planned for 2013.
The remaining difference is largely due to detector inefficiency which affects the high-\pT\ trigger more than the low-\pT\ trigger
due to the additional coincidence requirements.  Improved efficiency is expected for 2011 running.

The efficiency in the HLT was determined using the tag and probe method with \Jmumu\ samples  
for low \pT\ \linebreak ($6\GeV$) triggers and \Zmumu\ for high \pT\ ($13\GeV$) triggers.  In both 
studies, collision events were selected by requiring that the
 event has at least three tracks associated with the same reconstructed primary vertex.   
Reference muons reconstructed offline using both ID and MS information
 were required to be inside the fiducial volume of the muon triggers  
($|\eta|<2.4$) and the associated 
ID track was required to have at least one Pixel hit and at least six SCT hits. 
Events were required to contain
a pair of reference muons with opposite charge and an 
invariant mass lying within a window around the  mass of the relevant resonance: \\ \mbox{$2.86 \GeV < m_{\mu\mu} < 3.34 \GeV$} for \Jmumu\ decays and \linebreak \mbox{$77 \GeV < m_{\mu\mu} < 106 \GeV$} for \Zmumu\ decays. 
The resulting efficiency in the low-\pT\ region for the mu6 trigger is shown in Fig.~\ref{fig:mu_lowpt_hlteff}.  
For the high-\pt\ region, Fig.~\ref{fig:mu_hlteff} shows the efficiency as a function of \pT\ for the mu13, mu20 and mu40\_MSonly triggers in the TGC and RPC regions derived from the weighted average of the efficiency measured from the \jpsi\ and \Zboson\ samples.  Note that the 40~GeV threshold trigger has not yet reached its plateau efficiency in the highest \pT\ bin in the figure; extending the figure to higher \pt\ is limited by the small number of probe muons above 90\GeV.
The efficiencies are seen to have a sharp turn-on with a plateau efficiency  ($\pt > 13$~GeV) for
the mu13 trigger of
74\%  for the barrel region (dominated by the RPC geometrical acceptance), Fig.~\ref{fig:mu_hlteff_ba}, and 91\% for the end-cap region, Fig.~\ref{fig:mu_hlteff_ec}.  The systematic uncertainty on the plateau efficiency has been evaluated to be $\sim$1\%.


\begin{table*}[htbp]
   \begin{center}
   \caption{The primary triggers in each of the jet trigger categories with their L1 threshold and approximate prescale factor for an instantaneous luminosity of $\sim \Lumi{32}$ (a prescale value of 1 means unprescaled). The trigger name contains the EF threshold value; the L2 threshold is 5~GeV lower}
      \begin{tabular}{lllll}
\hline
Category & Name & L1 Threshold & Prescale & Motivation\\ 
\hline \hline
\emph{ Inclusive Jets:}  & j20  & J5 &  $\sim 10^{5}$ & jets in central region ($|\eta|<3.2$) \\
 & j30 & J10 & $\sim 10^{4}$ &\\
 & j35 & J15 & $\sim 10^{4}$ &\\
 & j50 & J30 & $\sim 10^{3}$ &\\
 & j75 & J55 & $\sim 10^{2}$ &\\
 & j95 & J75 & $\sim 10$ &\\
 & none &  J95 & 1 &\\
 \hline
\emph{ Forward Jets:} & fj30 & FJ10 & $\sim 10^{3}$ & jets in the forward region ($|\eta|>3.2$) \\
 & fj50 & FJ30 & $\sim 50$ &\\
 & fj75 & FJ55 & $\sim 10$ &\\
 & none &  FJ95 & 1 &\\
 \hline
\emph{ Multi-jets:} & 2j75 & 2J70 & $\sim 30$ & two or more central jets above threshold \\
 & 3j30 & 3J10 & $\sim 200$ &\\
 & 4j30 & 4J10 & $\sim 5$ &\\ 
 & 5j30 & 5J10 & 1 &\\
 \hline
\emph{ Total Jet \et:}  & je195 & JE100 & $\sim 70$ & total \et\ of all jets above threshold \\
 & je255 & JE140 & 1 &\\
\hline
\end{tabular}
   \normalsize
   \label{tab:jet_physics}
   \end{center}
\end{table*}

\subsection{Jets}\label{sec:jet}
\def \figurepath{.}
Jet signatures are important for QCD measurements \cite{Atlas:JetDiJetInc2010, Atlas:JetShapes2010}, top quark measurements, and searches for new particles decaying into jets \cite{Atlas:diJetSearch2010, Atlas:ContIntSearch2010}.  Data collected with jet triggers also provide important control samples for many other physics analyses.  Jet triggers select events containing high \pt\ clusters, and can be separated into four categories: \emph{inclusive jets}~(J), \emph{forward jets}~(FJ), \emph{multi-jets} (nJ, n=2,3..), and \emph{total jet \et}~(JE).

\subsubsection{Jet Reconstruction and Selection Criteria}

For a large part of 2010 data-taking, only L1 jet triggers (Section~\ref{sec:L1calo}) were used for 
selection.  L2 rejection was enabled late in 2010, while EF rejection was not enabled during 2010 running
as it was not needed~\cite{Atlas:jetCONFnote}.

\par
Calibration constants that correct for the hadron response of
the non-compensating calorimeters in ATLAS (hadronic energy scale)
were not applied in the trigger during 2010 data-taking. As a result, the jet trigger 
algorithms applied cuts to energy variables at the electromagnetic scale, the scale for 
energy deposited by electrons and photons in the calorimeter.

\begin{figure}[!hb]
  \centering
  \includegraphics[width=0.4\textwidth,trim=10 10 1 10,clip=true]{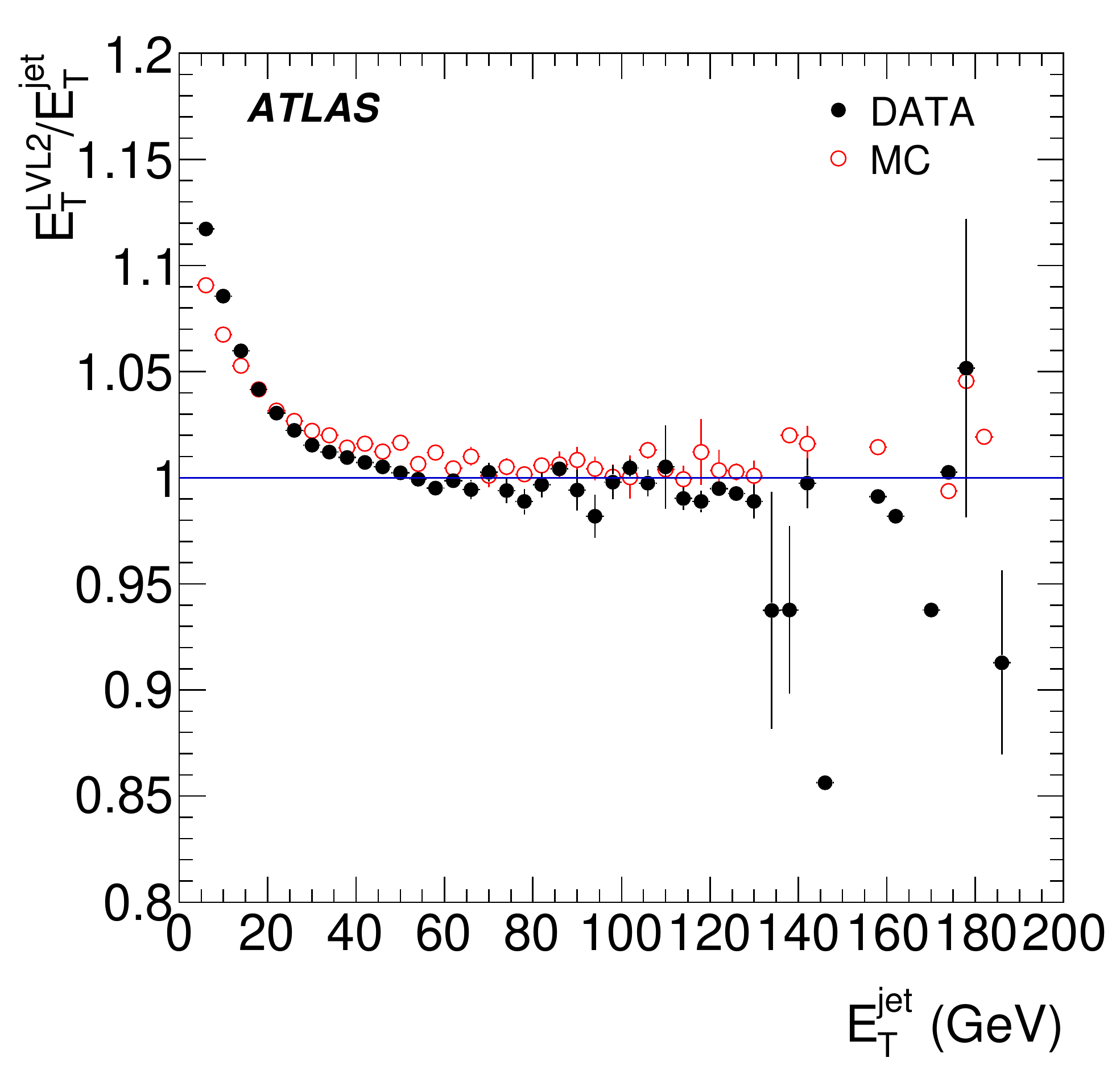}
  \caption{Ratios of transverse momenta of jets in $|\eta|<2.8$ reconstructed
at L2 and jets reconstructed offline with the anti-\kt\ algorithm with parameter $R=0.4$,
as a function of the offline jet \et.  For this comparison, both 
online and offline jet energies have been
  calibrated to the electromagnetic energy scale. The errors shown are statistical only}
  \label{fig:L2JetEtResolution}
\end{figure}

Figure~\ref{fig:L2JetEtResolution} shows the ratio of the L2 jet \et\ to 
the offline jet \et\  as a function of the offline jet \et .  
Data and MC simulation agree well.

\begin{figure}[!h]
\begin{center}
\subfigure[]{
\includegraphics[width=0.4\textwidth,trim=10 10 10 10,clip=true]{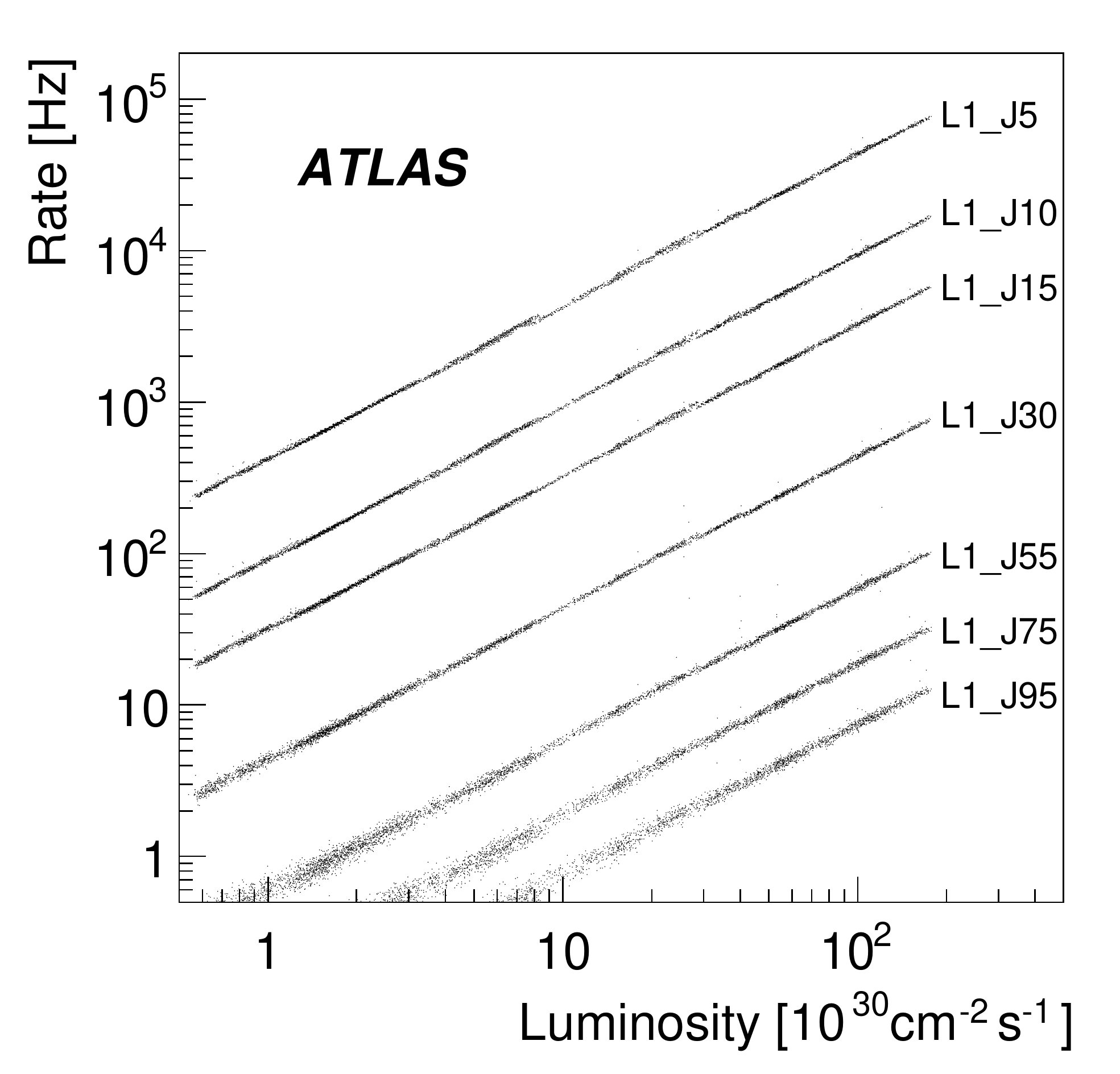}
\label{fig:jetrates_single}
}
\subfigure[]{
\includegraphics[width=0.4\textwidth,trim=10 10 10 10,clip=true]{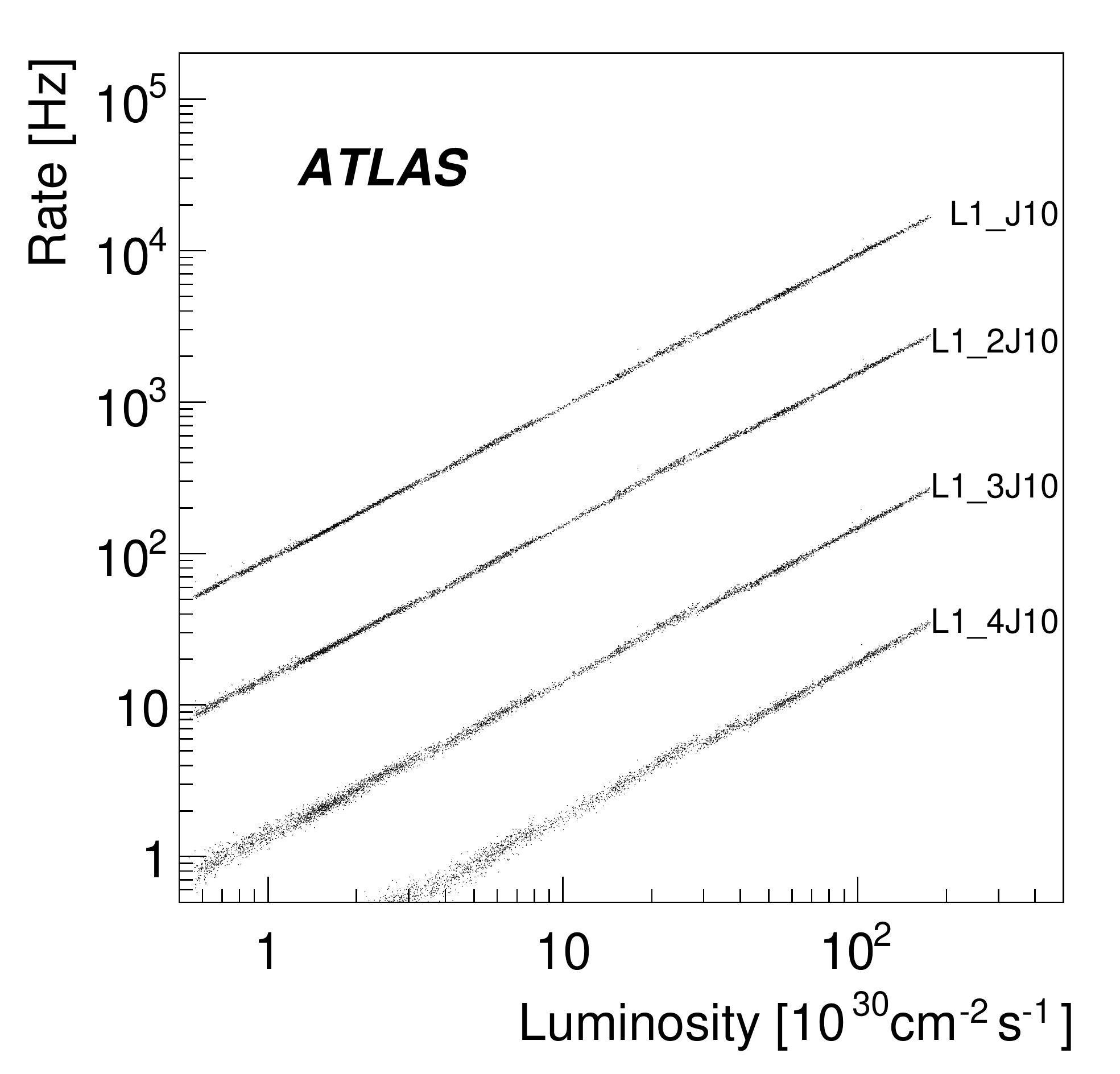}
\label{fig:jetrates_multi}
}
\end{center}
\vglue-0.5cm
 \caption{\label{fig:jetrates} Rates for various L1 jet triggers
as a function of the instantaneous luminosity during the 2010 run for \subref{fig:jetrates_single}
inclusive triggers and \subref{fig:jetrates_multi} multi-jet triggers (with the inclusive trigger L1\_J10 for reference)}
\vglue-0cm
\end{figure}

\subsubsection{Jet Trigger Menu and Rates}

The principal jet triggers for an instantaneous luminosity of $\sim \Lumi{32}$ are listed in 
Table~\ref{tab:jet_physics} for inclusive jets, forward jets, multi-jets, and total jet \et. 
The set of L1 prescales applied provided an 
approximately flat event yield as a function of jet \pt.  The L1 rates of the inclusive and 
multi-jet triggers are shown in Fig.~\ref{fig:jetrates}.  During 2010 running, the level of
pileup was small enough not to have a visible effect on the
rates, which were observed to rise linearly with instantaneous luminosity.

\subsubsection{Jet Trigger Efficiency}

\begin{sloppypar}
The jet trigger efficiency was measured using the orthogonal trigger and bootstrap methods.  
For the lowest-threshold chains, the jet trigger efficiency was calculated using the orthogonal trigger method with
 events selected by the  L1\_MBTS\_1 trigger (Section~\ref{sec:minbias}). For the higher thresholds, the bootstrap method was used. The systematic uncertainty in the plateau 
efficiencies is less than $\sim$1\%.\par
\end{sloppypar}

This efficiency determination~\cite{Atlas:JetDiJetInc2010} used jets that were
reconstructed offline from calorimeter clusters at the electromagnetic
scale, using the anti-\kt\ jet algorithm~\cite{antikt} with $R=0.4$
or $R=0.6$, in the region $|\eta|<2.8$.  These jets were calibrated for calorimeter 
response to hadrons using parameters taken from the simulation, 
after comparison with the data~\cite{JES}.  Cleaning cuts were applied to suppress fake jets
from noise, cosmic rays, and other sources.  These cleaning cuts were designed to
reject pathological jets with almost all energy coming from a very small number of cells, out-of-time cell signals,
or abnormal electromagnetic components.  These cuts are explained in
detail in Ref.~\cite{JetCleaning}.

\begin{figure}[h!]\begin{center}
\subfigure[]{
\includegraphics[height=0.25\textheight,trim=5 5 5 5,clip=true]{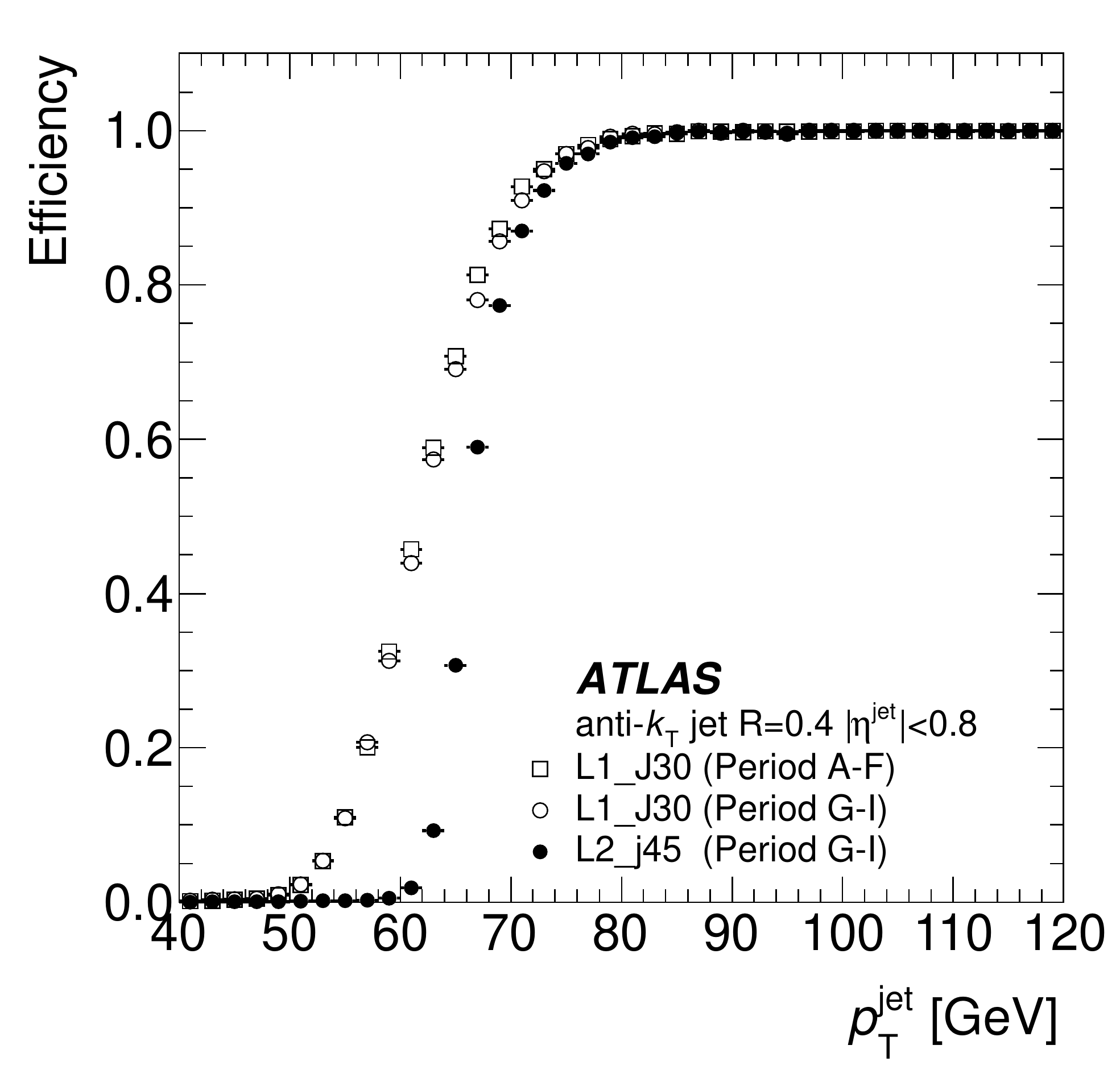}
\label{fig:l1jeteff-a}
}
\subfigure[]{
\includegraphics[height=0.25\textheight,trim=5 5 5 5,clip=true]{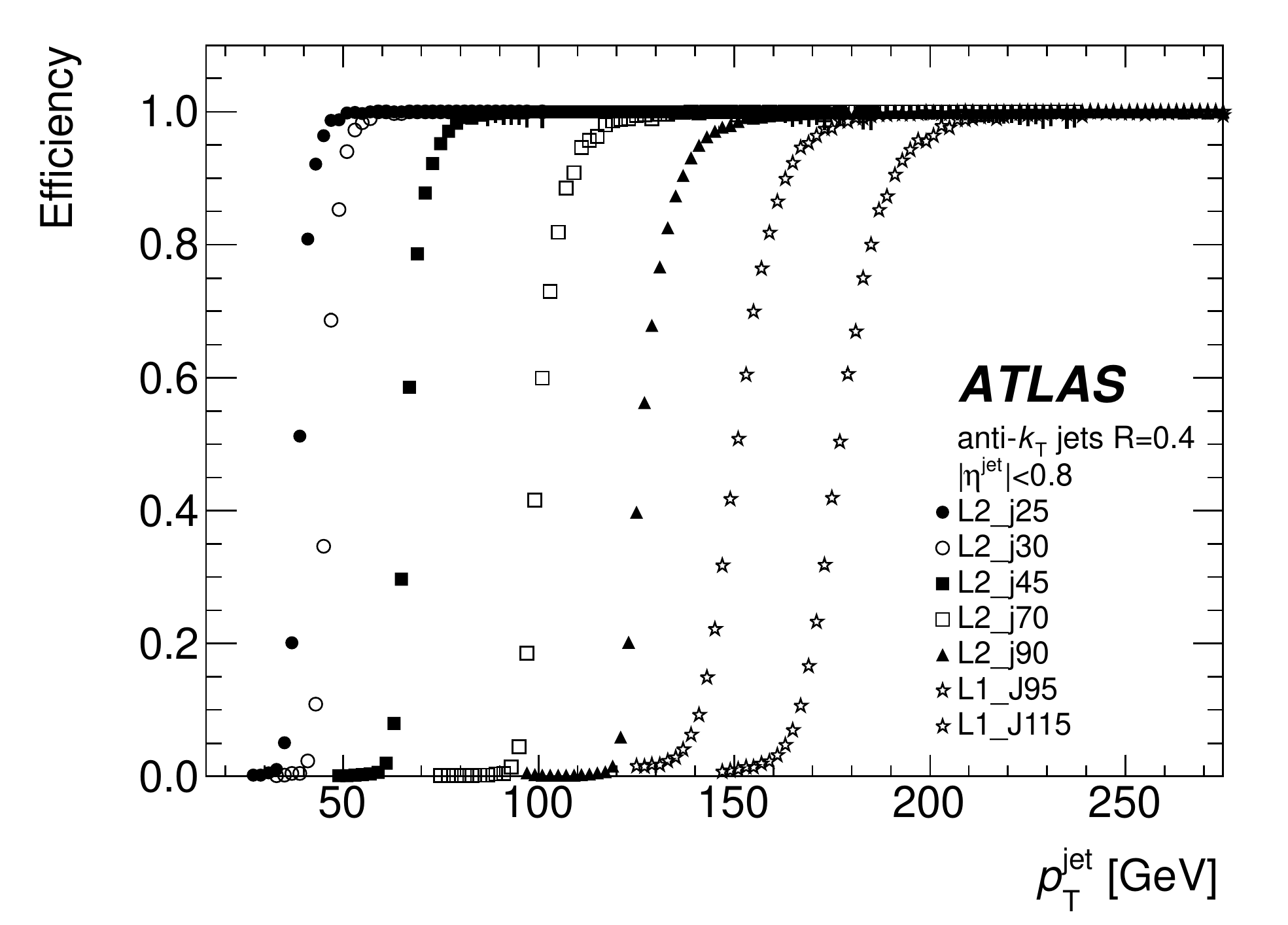}
 \label{fig:l1jeteff-b}
}
\end{center}
\caption{\label{fig:l1jeteff} \subref{fig:l1jeteff-a} Efficiency of the L1\_J30 trigger as a function of offline jet 
transverse momentum (after applying hadronic calibration) for two different data-taking
periods. For the second period the efficiency of the L2\_j45 trigger is also shown. \subref{fig:l1jeteff-b} Efficiency for several triggers, integrated over 2010}
  \vglue-0cm
\end{figure}

\begin{figure}[h!]
\begin{center}
 \includegraphics[height=0.25\textheight,trim=5 5 5 5,clip=true]{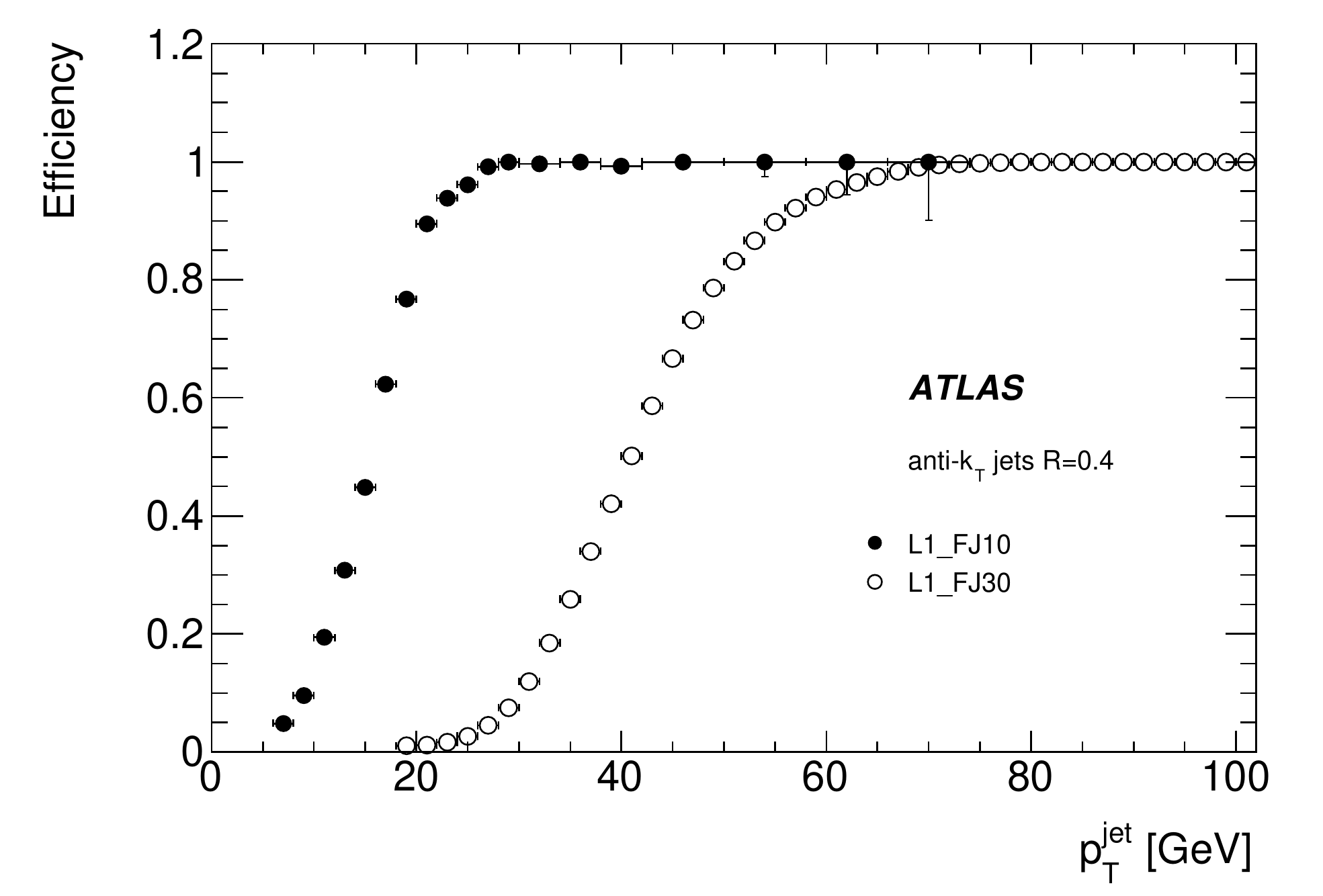}
\end{center}
\caption{\label{fig:JE-a} Efficiency for two L1 thresholds of the forward jet trigger}
\end{figure}

\begin{figure}[h!]
\begin{center}
 \includegraphics[height=0.25\textheight,trim=5 5 5 5,clip=true]{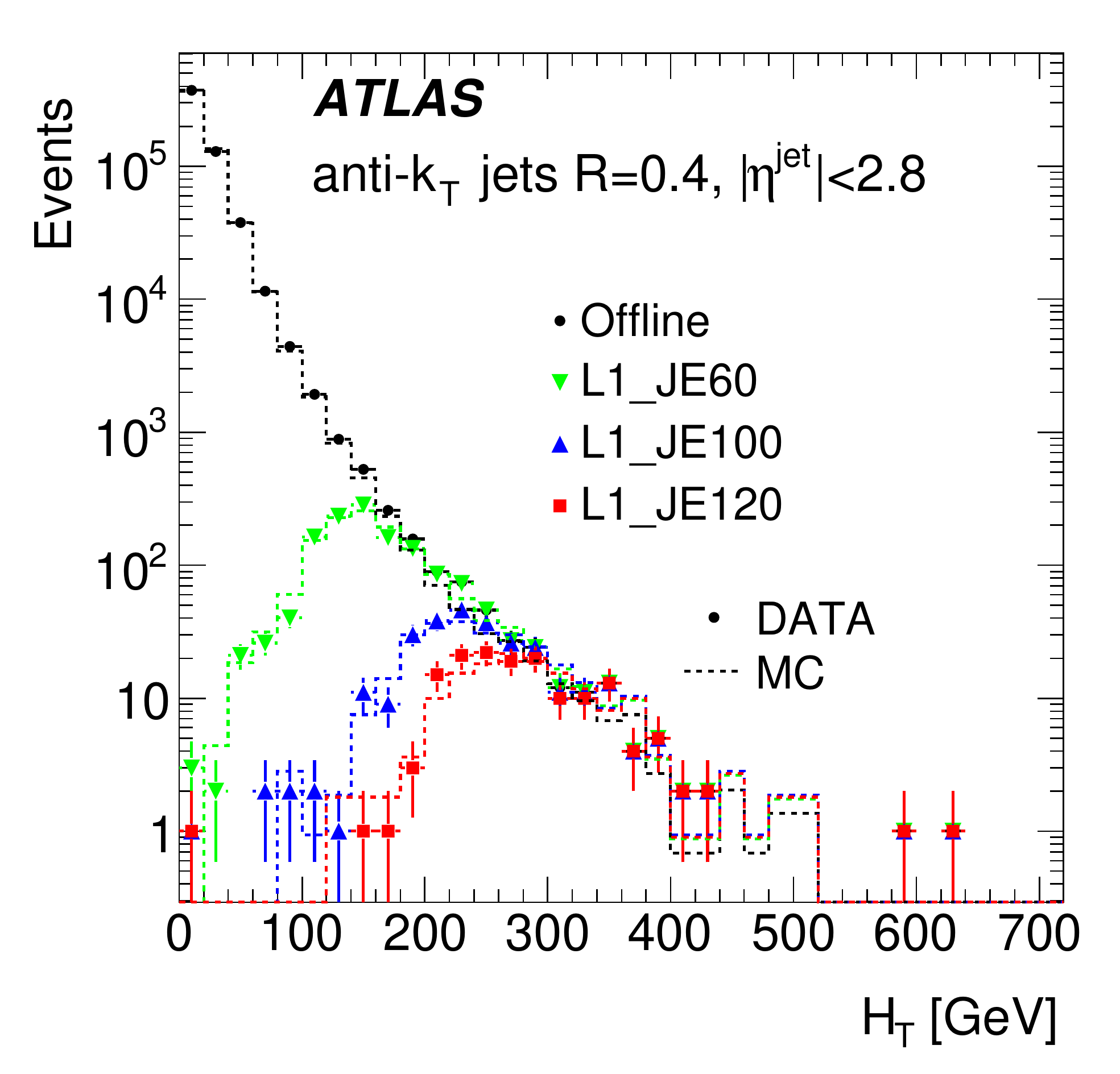}
\end{center}
\caption{\label{fig:JE-b} Offline transverse
momentum distribution for all jets, and for those passing each of three thresholds of
the L1 total jet \ET\ trigger}
\end{figure}

The efficiency of the L1\_J30 jet trigger in the central
region, $\abseta<0.8$, of the detector is shown in Fig.~\ref{fig:l1jeteff-a}
as a function of offline jet \pt\ 
for two different data-taking periods, the 
difference between the periods being that in periods G to I the LHC beam had a
bunch train structure. The change in bunch structure had a small effect on the efficiency 
turn-on curve 
and a negligible effect on the efficiency in the plateau region. The efficiency of the
L2\_j45 trigger chain, which includes the L1\_J30 trigger, is also shown in
Fig.~\ref{fig:l1jeteff-a} for periods G to I, for which L2 rejection was enabled. 
Since the efficiency turn-on is significantly sharper for
L2 than L1, the L2 thresholds were set 15~GeV higher than 
the L1 values, reducing the overall trigger rate while ensuring that the 
L2 trigger reached full efficiency at the same \pt\ value as the corresponding L1 trigger. 
Jet trigger efficiencies integrated over the whole year are shown in
Fig.~\ref{fig:l1jeteff-b} for several 
chains as a function of the calibrated offline jet \pt.  
Figure~\ref{fig:JE-a} shows the efficiency for two
thresholds of the inclusive forward trigger.  The efficiency plateaus at
a lower \pt\ than for central jet triggers due to different energy resolutions 
and different contributions from noise and pile-up. After reaching the plateau, 
the jet and forward jet triggers remain fully efficient to within $\sim$1\%. 

The total jet \et\ triggers require the \et\ sum of all jets in the event (defined as \HT) 
to be higher than a given threshold and have the effect of selecting events with 
high jet multiplicity.
Figure~\ref{fig:JE-b} shows the 
 distribution of \HT\ for events, triggered by an orthogonal muon trigger,
that pass three different JE
trigger thresholds, compared to predictions from the MC.
The MC distributions are in agreement with the data.

\begin{figure}[b!]
\begin{center}
 \includegraphics[height=0.25\textheight,trim=10 10 10 10,clip=true]{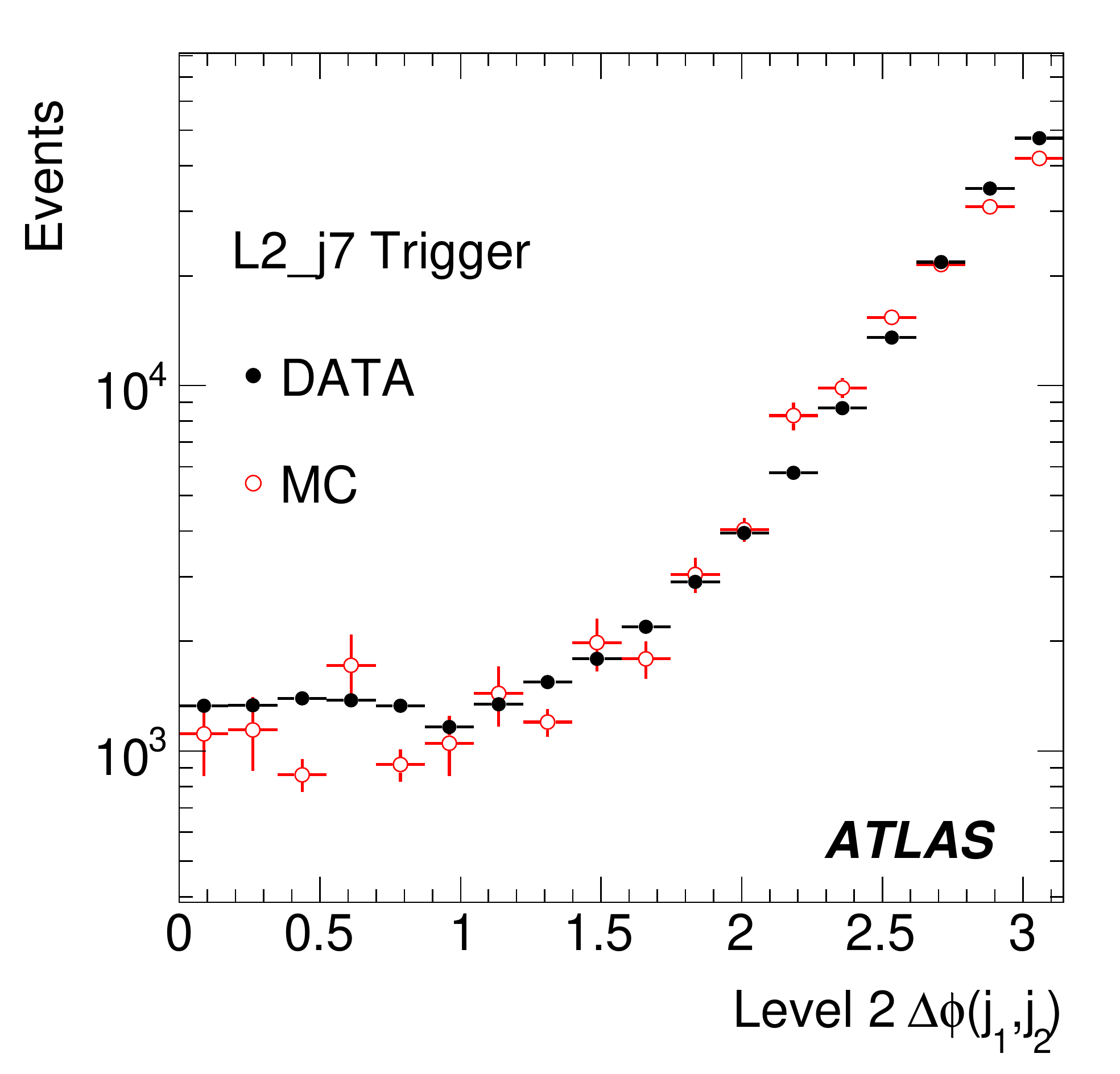}
\end{center}
\caption{\label{fig:hltres} The $\Delta\phi$ between the highest \pt\ and second highest \pt\
 jet in the event for jets reconstructed at L2}
\end{figure}

In the initial phase of data-taking the jet triggers were limited 
to inclusive and multi-jet topologies, with no cuts on the relative
directions of the jets.  Near the end of the 2010 data-taking,  additional 
triggers that require di-jets with 
large rapidity differences or small differences in azimuthal angle
were implemented at L2.  Figure~\ref{fig:hltres}
shows the $\Delta\phi$ distributions for di-jets at L2,
indicating that these distributions are well described by the
simulation.


\subsection{Taus}\label{sec:tau}
\def \figurepath{.}
\begin{sloppypar}
The ATLAS physics programme uses tau leptons for SM measurements and new physics searches.  Being able to trigger on 
hadronic tau signatures is important for this part of the ATLAS physics programme.  Dedicated trigger algorithms have been designed and implemented based on the features of hadronic tau decays: 
narrow calorimeter clusters and a small number of associated tracks.   Due to the high production rate of jets
with very similar features to hadronic tau decays,  keeping the rate of tau triggers under control is particularly challenging.
\end{sloppypar}

\begin{figure}[!htb]
  \centering
  \subfigure[]{
\includegraphics[width=0.45\textwidth]{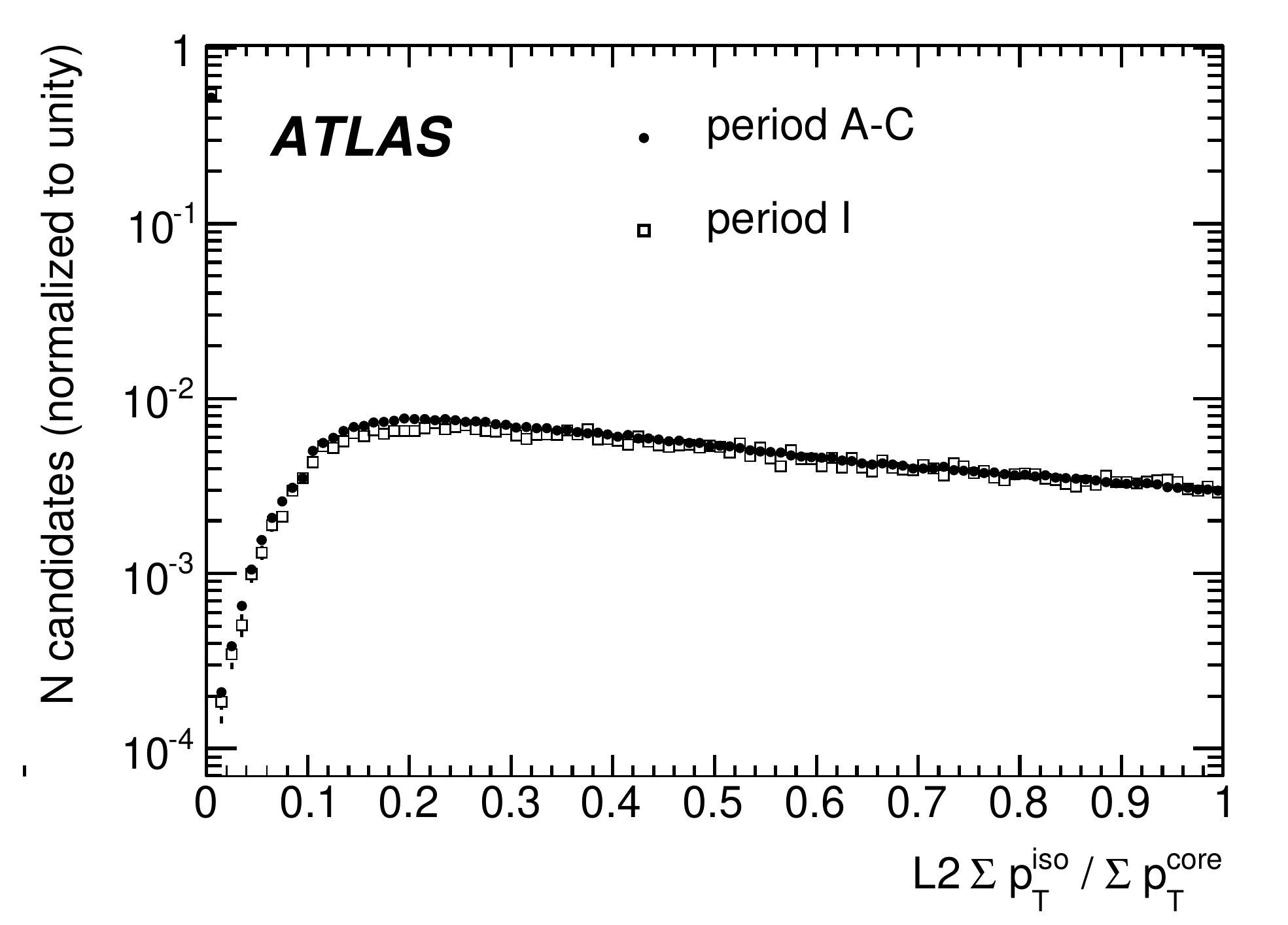}
\label{fig:tau_pileup_PtIsoCore}
}
  \subfigure[]{
\includegraphics[width=0.45\textwidth]{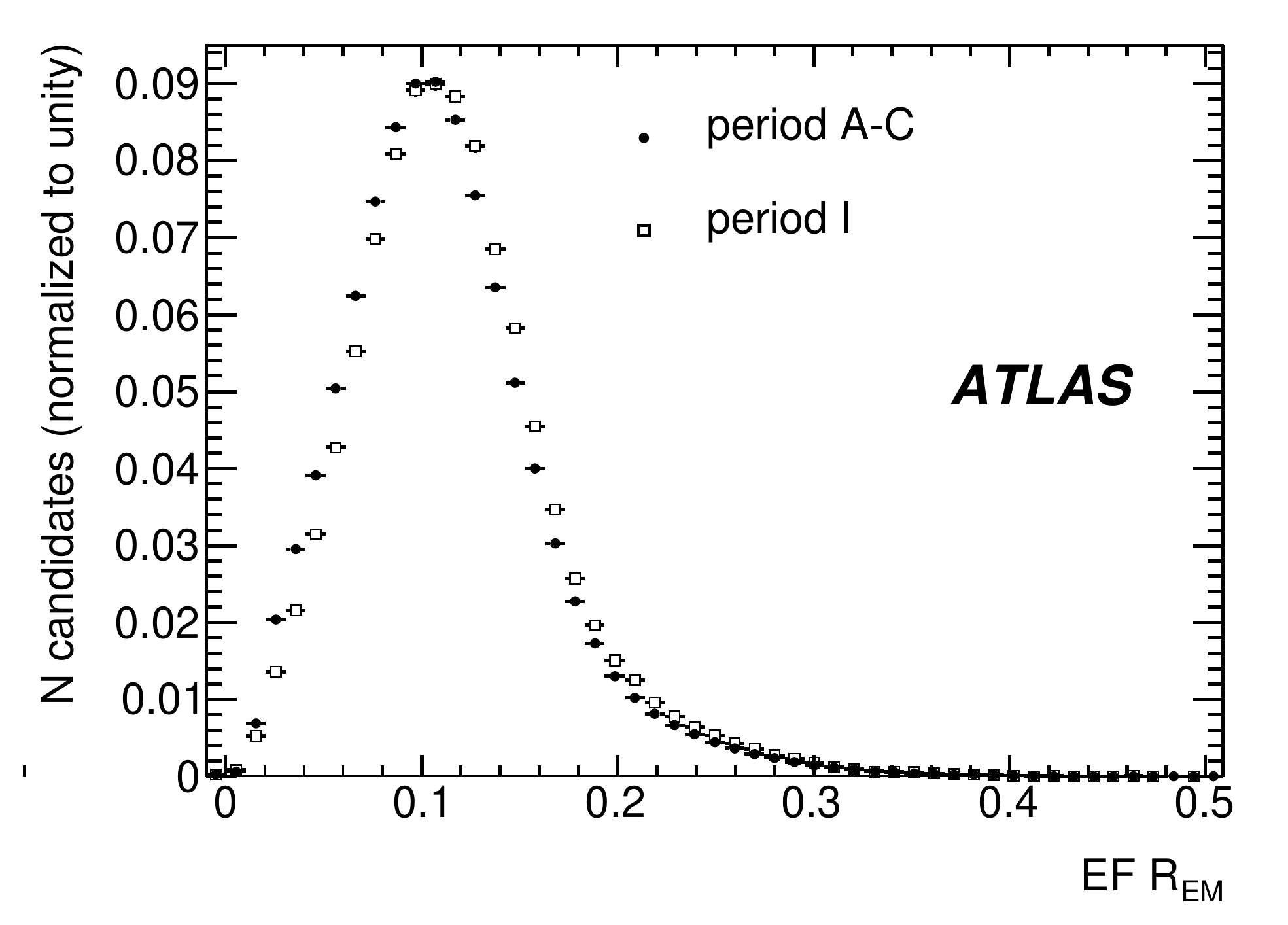}
\label{fig:tau_pileup_EFEMRad}
}
    \caption{Comparison of  (a) the variable $\Sigma \pT^{\mathrm{iso}} / \Sigma \pT^{\mathrm{core}}$ at L2 and (b) $\mathcal{R}_{\mathrm{EM}}$ at the EF in periods A-C (no pile-up) and period I (in-time pile-up of between two and three collisions per bunch crossing and 150 ns bunch trains)}\label{fig:tau_pileupData}
\end{figure}

\subsubsection{Tau Reconstruction and selection criteria}

\label{sec:l1hlt}
At L1 the tau trigger uses EM and hadronic calorimeter
information within regions of 4$\times$4 trigger towers
($\Delta \eta \times \Delta \phi \approx 0.4 \times 0.4$)
to calculate the energy in a core and an isolation
region (Section~\ref{sec:L1calo}).  

At L2 selection criteria are applied using tracking and
calorimeter information, taking advantage of narrowness and
low track multiplicity to discriminate taus from jets.  
The L2 tau candidate is reconstructed from cells in a rectangular L2 RoI of 
size $\Delta \eta \times \Delta
\phi = 0.6 \times 0.6$ centred at the L1 RoI position. The L2 calorimeter
algorithm first refines the L1 RoI position using the second layer of the
EM calorimeter.  It then selects narrow jets in the detector by means
of a calorimeter shape variable determined only from the second layer
of the EM calorimeter.   The shape variable, $\mathcal{R}_{EM}$, is an
energy-weighted radius squared within the L2 RoI, i.e.

\begin{equation}
  \mathcal{R}_{\mathrm{EM}} = \frac {\displaystyle\sum_{\mathrm{cell}} E_{\mathrm{cell}} \cdot (\Delta
    R_{\mathrm{cell}})^n} {\displaystyle\sum_{\mathrm{cell}} E_{\mathrm{cell}}}
  \label{eq:L2_EMRadius}
\end{equation}

\noindent where $E_{\mathrm{cell}}$ is the energy of the calorimeter cell and
$\dR_{\mathrm{cell}}$ is the radius \dR\ (defined in Section~\ref{sec:idReco}) 
of the cell from the centre of the L2 RoI, which is squared ($n=2$).  
Track reconstruction at L2 uses the SiTrack algorithm  (Section~\ref{sec:idReco}), but to
minimize the execution time, tracks are not extended to the TRT.  
Tracks with $\pT > 1.5$ GeV are reconstructed in the L2 RoI.

Exploiting the same characteristics of narrowness and
low track multiplicity, the EF selects 1-prong and multi-prong decays, with different selection 
criteria,
using algorithms that are similar to the offline
reconstruction algorithms~\cite{DetectorPaper}.
The EF tau candidate is reconstructed from cells in a rectangular region of size $\Delta \eta \times \Delta
\phi = 0.8 \times 0.8$ centred at the L1 RoI position.  The position, transverse energy, and calorimeter shower shape
variables of the EF tau candidate are calculated from cells of all calorimeter layers 
within this $0.8 \times 0.8$ region.
An overall hadronic calibration~\cite{JetCalibration} is
applied to all cells, and a tau-specific calibration is applied to
the tau trigger candidate.  The EM radius shape variable used at the EF is defined by
Equation~\ref{eq:L2_EMRadius} with $n=1$. 
Additional quality criteria are applied to tracks reconstructed in the RoI, and if more
than one track is found a secondary vertex reconstruction is attempted. 

The stability of the tau trigger selection variables against pile-up was evaluated by comparing the distributions of these variables for events passing the L1\_TAU5 trigger from data-taking periods A-C with those from period I.  Periods A-C contain a negligible amount of pile-up, while events from period I contain the largest amount of pile-up (Section~\ref{sec:overview}) observed in 2010.   The distributions of the two most important variables ($\Sigma \pT^{\mathrm{iso}} / \Sigma \pT^{\mathrm{core}}$ at L2 and $\mathcal{R}_{\mathrm{EM}}$ at EF) are shown in Fig.~\ref{fig:tau_pileupData} for events with and without pile-up.  The variable $\Sigma \pT^{\mathrm{iso}} / \Sigma \pT^{\mathrm{core}}$ describes the ratio of the scalar \pt\ sums of the tracks in an isolation ring (R=0.1 to 0.3) and in the core area (R=0.1).  
The plots show a small shift due to the presence of additional energy and tracks, but these variables
are in general quite stable with respect to the pile-up of two to three collisions per bunch crossing.
The same behaviour was observed for other variables used for making the HLT decision.

\begin{table*}[!hbt]
\centering
\newcommand{\m}{\hphantom{$-$}}
\newcommand{\cc}[1]{\multicolumn{1}{c}{#1}}
\renewcommand{\tabcolsep}{2pc} 
\renewcommand{\arraystretch}{1.2} 
\caption{Tau trigger menu and approximate HLT rates for a luminosity of \Lumi{32}}
\begin{tabular}{llc}
\hline
Trigger   &  Motivation  & Rate [Hz]\\
\hline
tau50\_medium & $H^+$, $Z^{\prime}$ & 8 \\
2tau29\_loose1 & $H$, $Z^{\prime}$  & 1 \\
tau16\_medium\_xe22 & $H^+$, $t$, \W  & 10 \\
tau12\_loose\_e10\_medium & $H$, $t$, \Z & 4  \\
tau12\_loose\_mu10 & $H$, $t$,  \Z  & 3 \\
\hline
\end{tabular}\\[2pt]
\label{table:1}
\end{table*}

\begin{figure}[!htb]
  \centering
   \subfigure[]{
\includegraphics[width=0.45\textwidth]{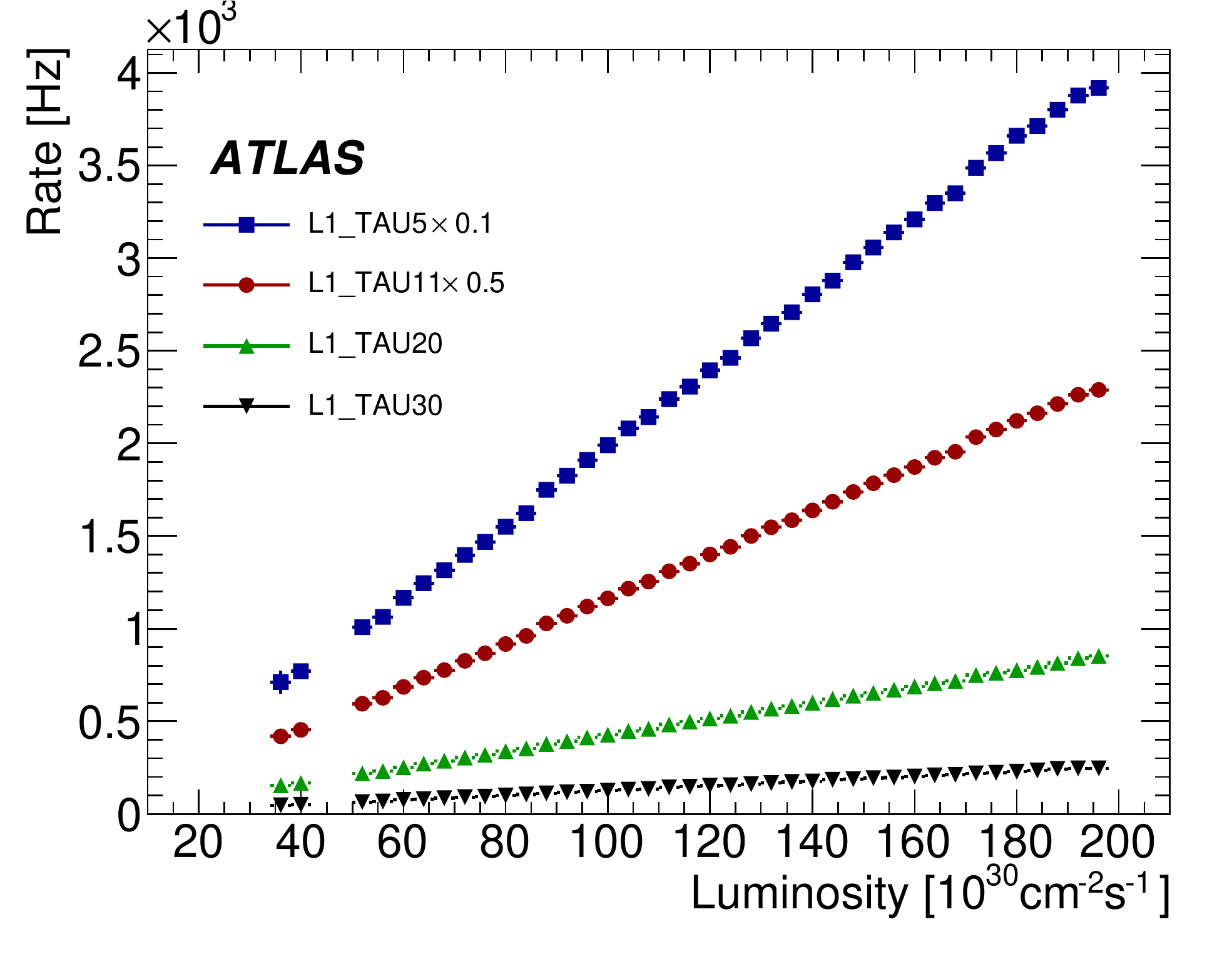}\label{fig:tau_lumis}
}
   \subfigure[]{
\includegraphics[width=0.45\textwidth]{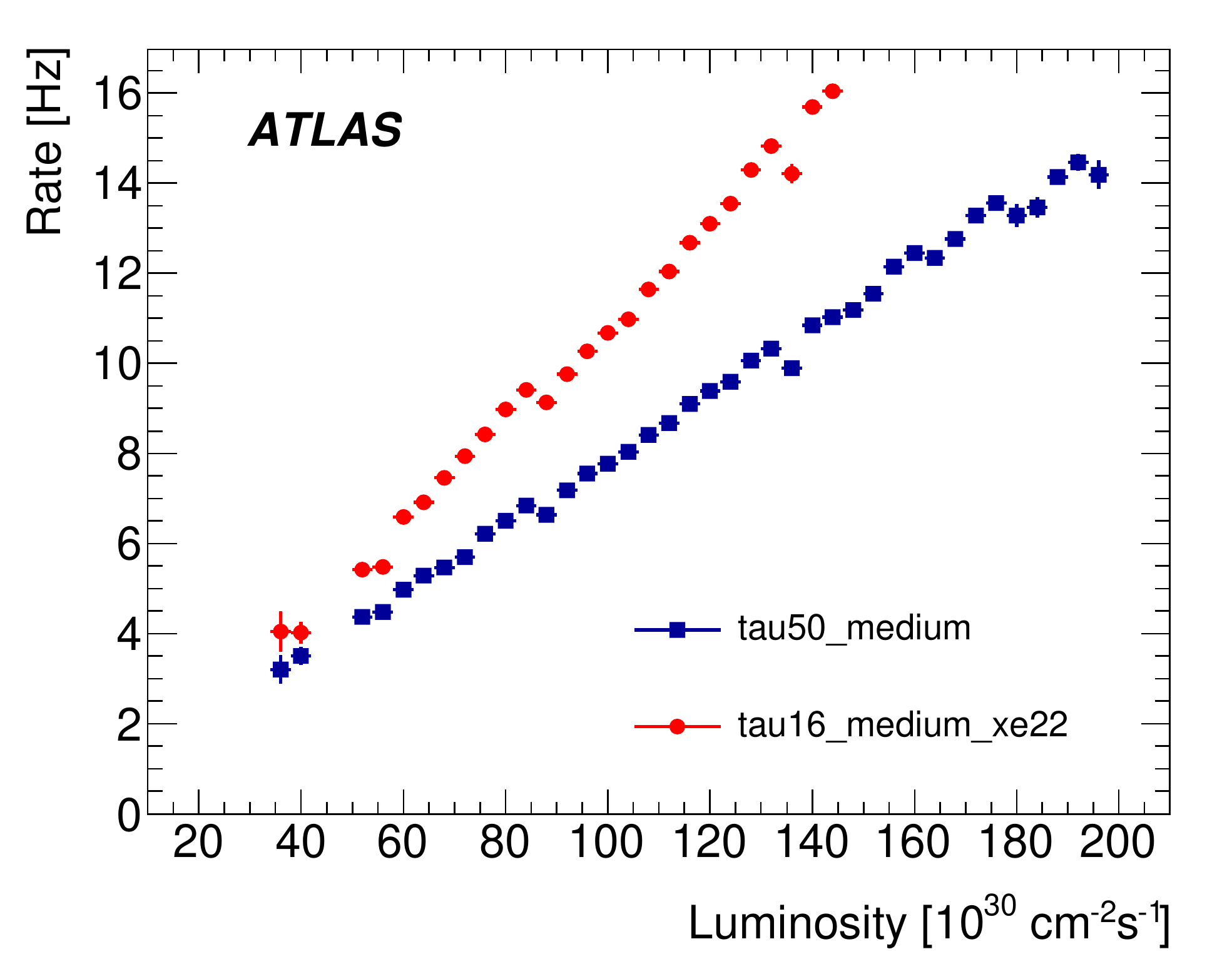}\label{fig:tauHLTrates}
}
    \caption{Trigger rates as a function of instantaneous luminosity for several (a) L1 and (b) HLT tau triggers. The rates for L1\_TAU5 and L1\_TAU11 have been scaled by 0.1 and 0.5 respectively}
    \label{fig:tauRates}
\end{figure}

\subsubsection{Tau Trigger Menu and Rates}
\begin{sloppypar}
Both single tau triggers and tau triggers in combination with electrons, muons, jets and missing energy signatures were present in the 2010 trigger menus.  Tau signatures were used in combination with other triggers to keep rates low enough while maintaining acceptance for the physics processes of interest. Table~\ref{table:1} shows a subset of these items with their rates that represent the lowest threshold triggers that remained unprescaled at a luminosity of \Lumi{32}. 
\end{sloppypar}

Figure \ref{fig:tauRates} shows the trigger rates for various L1 and HLT tau triggers as a function
of instantaneous luminosity showing 
a linear increase of rates during 2010 running.   

\begin{figure}[!htb]
  \centering
  \subfigure[]{
\includegraphics[width=0.45\textwidth]{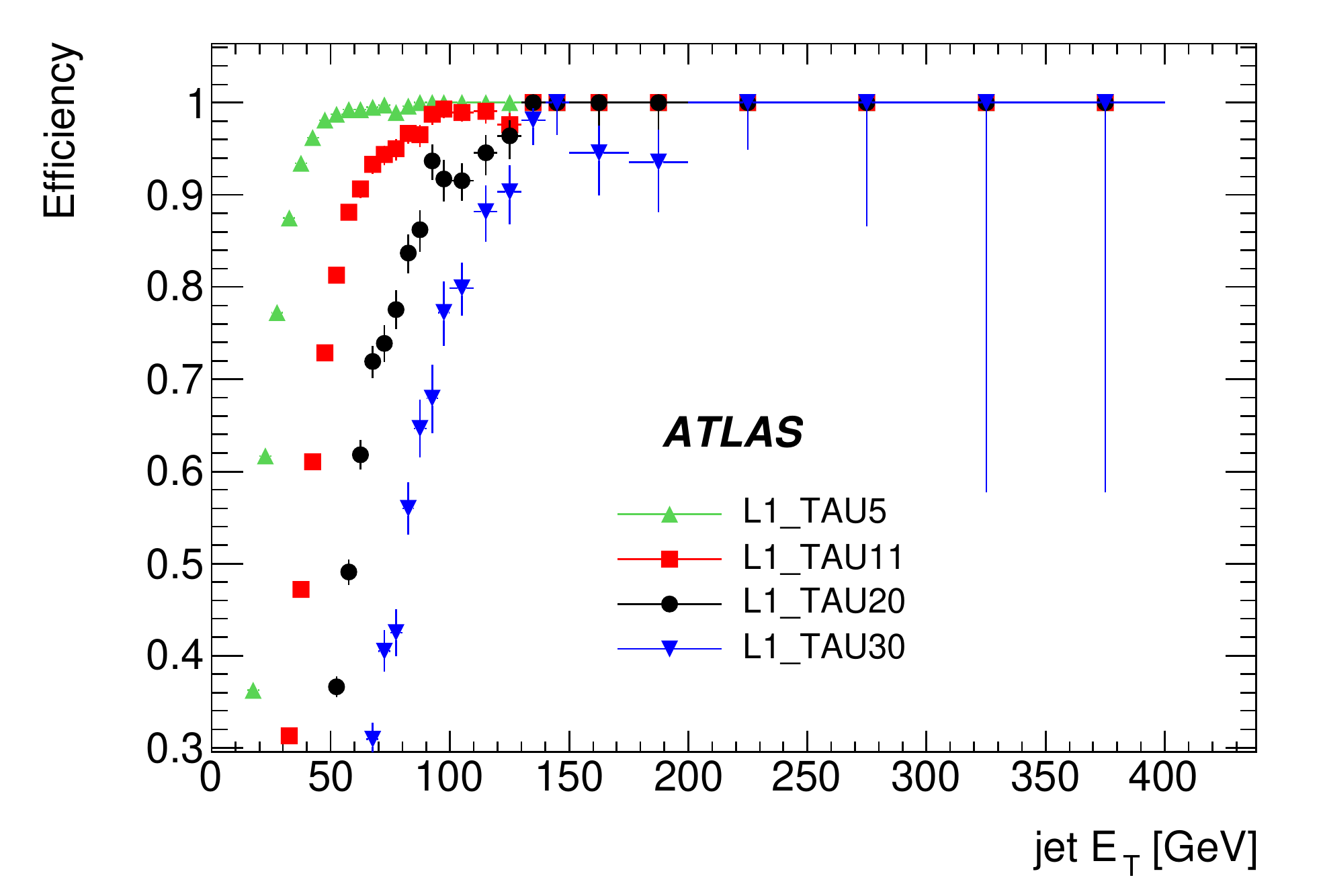}\label{fig:tau_L1TurnOn}
}
    \subfigure[]{
\includegraphics[width=0.48\textwidth]{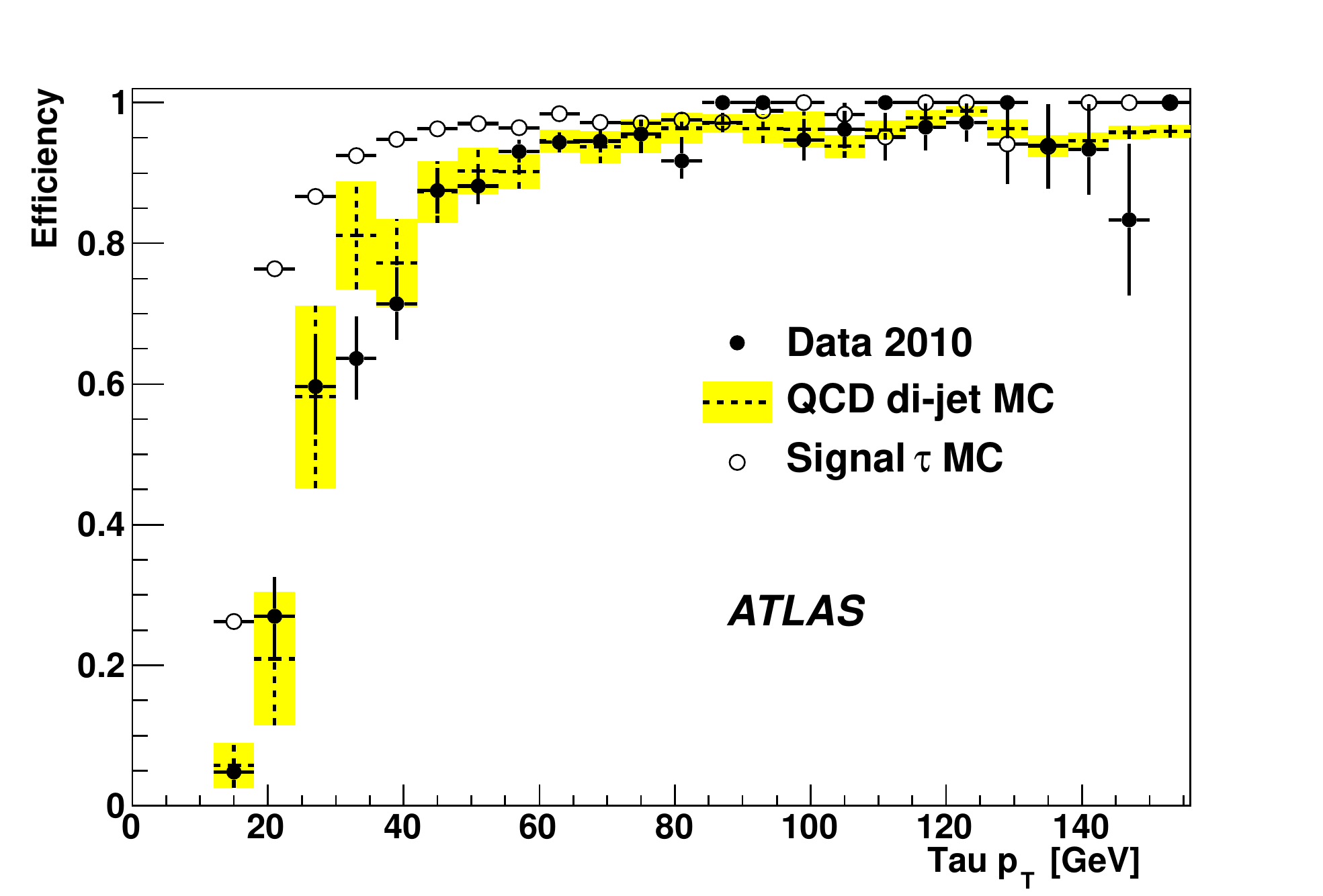}\label{fig:tau_EFTurnOn}
}
    \caption{ Efficiency for \subref{fig:tau_L1TurnOn} jets reconstructed offline
with at least one associated track to pass the L1 tau trigger with thresholds of 5, 11, 20 
and $30\GeV$ as a function of the offline jet \et\ and
\subref{fig:tau_EFTurnOn} offline tau candidates 
to pass the HLT tau16\_loose trigger in a di-jet data sample, simulated QCD di-jets and a simulated
tau signal sample, as a function of the offline tau $\pt$}
\label{fig:tau_HLTTurnOn}
\end{figure}

\subsubsection{Tau Trigger Efficiency}

Tau trigger efficiencies were measured using offline reconstructed tau candidates in 
events containing QCD jets.  
Since QCD jets are the biggest source of fake taus in data, a sample of
jets reconstructed offline provides a useful
reference for tau trigger performance measurements.  For the L1 trigger
efficiency determination, offline jets were reconstructed 
with the anti-\kt\ algorithm (using parameter R = 0.4) and required to have at least
one associated track. Figure~\ref{fig:tau_L1TurnOn}
shows the efficiency of the L1\_TAU trigger for these
jets, as a function of the jet \et.  Although the L1 trigger efficiency has a slower turn-on for 
jets than for true taus, due to the wider shower profile of
QCD jets, above the turn-on region the performance is similar, as confirmed from 
MC simulation studies. 
The L1 trigger efficiency reaches a plateau value of 100\% (to within a systematic uncertainty 
of $\sim$1\%).

Figure~\ref{fig:tau_EFTurnOn} shows the
efficiency of the tau16\_loose trigger for offline tau candidates in data, 
simulated di-jet events, and simulated signal 
$\tau$ events.  Data events were selected by requiring two back-to-back jets (within 
0.3 radians), balanced in \pt\ (within 50\% of the higher \pt\ jet).  The data sample was 
collected with jet triggers (Section~\ref{sec:jet}).  Bias related to the jet trigger 
selection was removed by randomly selecting one of the jets (tag jet) that passed 
the jet trigger 
and using the other jet (probe jet) to match to a reconstructed tau candidate.  Reconstructed 
tau candidates that pass the tight offline identification requirements and match a 
probe jet ($\dR<0.4$) were used as the denominator of the efficiency measurement.  The 
numerator was defined 
as the subset of those candidates that also passed the tau16\_loose trigger.  The efficiencies 
from data agree with those for the simulated di-jets, but have a slower
turn-on than for the simulated signal sample. This is because of the lower L1 efficiency
for jets than taus in the threshold region. 
The trigger efficiency for offline tau candidates with $\pt>30 \GeV$ is 94\% with a 
total uncertainty of 
$\sim$5\%.   Measurements of the tau trigger efficiency from  \Ztau\ and  \Wtau\ decays are 
consistent with 
the QCD jet measurement but, with 2010 data, 
have relatively large statistical uncertainties.

\subsection{Missing Transverse Energy}\label{sec:met}
\def \figurepath{.}

The missing transverse-energy (\MET) signature is exploited  in the measurement of the \Wboson\ 
boson and top quark \cite{Atlas:WZleptons, Atlas:topCS2010, Atlas:Wjets2010} to provide 
information on the kinematics of neutrinos in the events.  It is also extensively used in 
searches for new physics \cite{Atlas:diphotonMet2010, Atlas:susyOneLepton2011} including possibly 
new particles that are not directly detected~\cite{Atlas:squarksandgluinos}.  The \MET\ is 
estimated by calculating the vector sum of all energies deposited in the calorimeters, projected 
onto the transverse plane, corrected for the transverse energies of all reconstructed muons.  
The \MET\ 
triggers~\cite{bib:metpubnote} are designed to select events for which the measured transverse 
energy imbalance is above a given threshold. Triggers based on the scalar sum of the transverse 
energies (\SumET) are also used.

\begin{figure}[!htbp]
  \centering
  \subfigure[]{
    \includegraphics[width=0.45\textwidth]{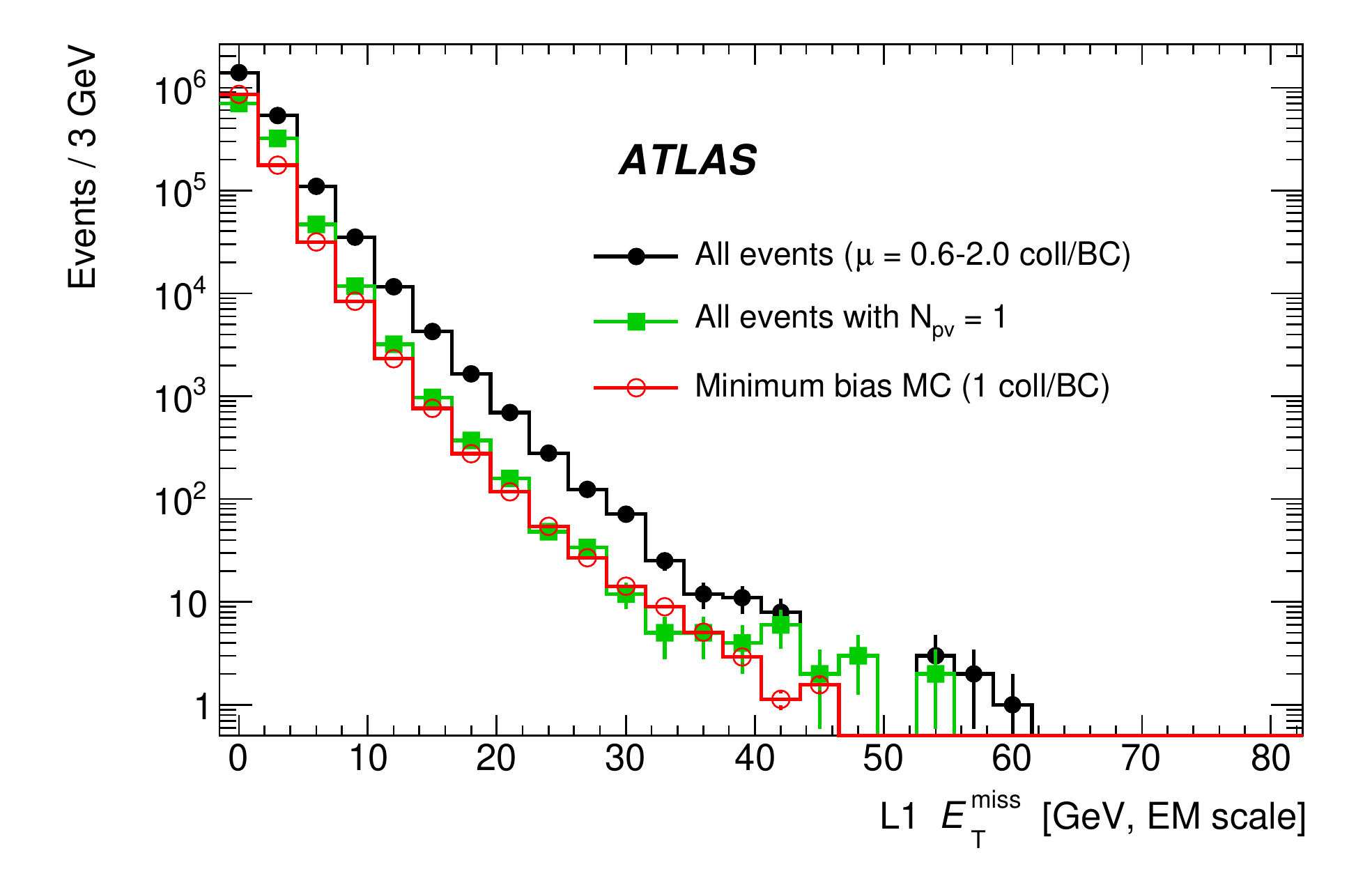}
     \label{fig-L1met-pileup}
  }
  \subfigure[]{
  \includegraphics[width=0.45\textwidth]{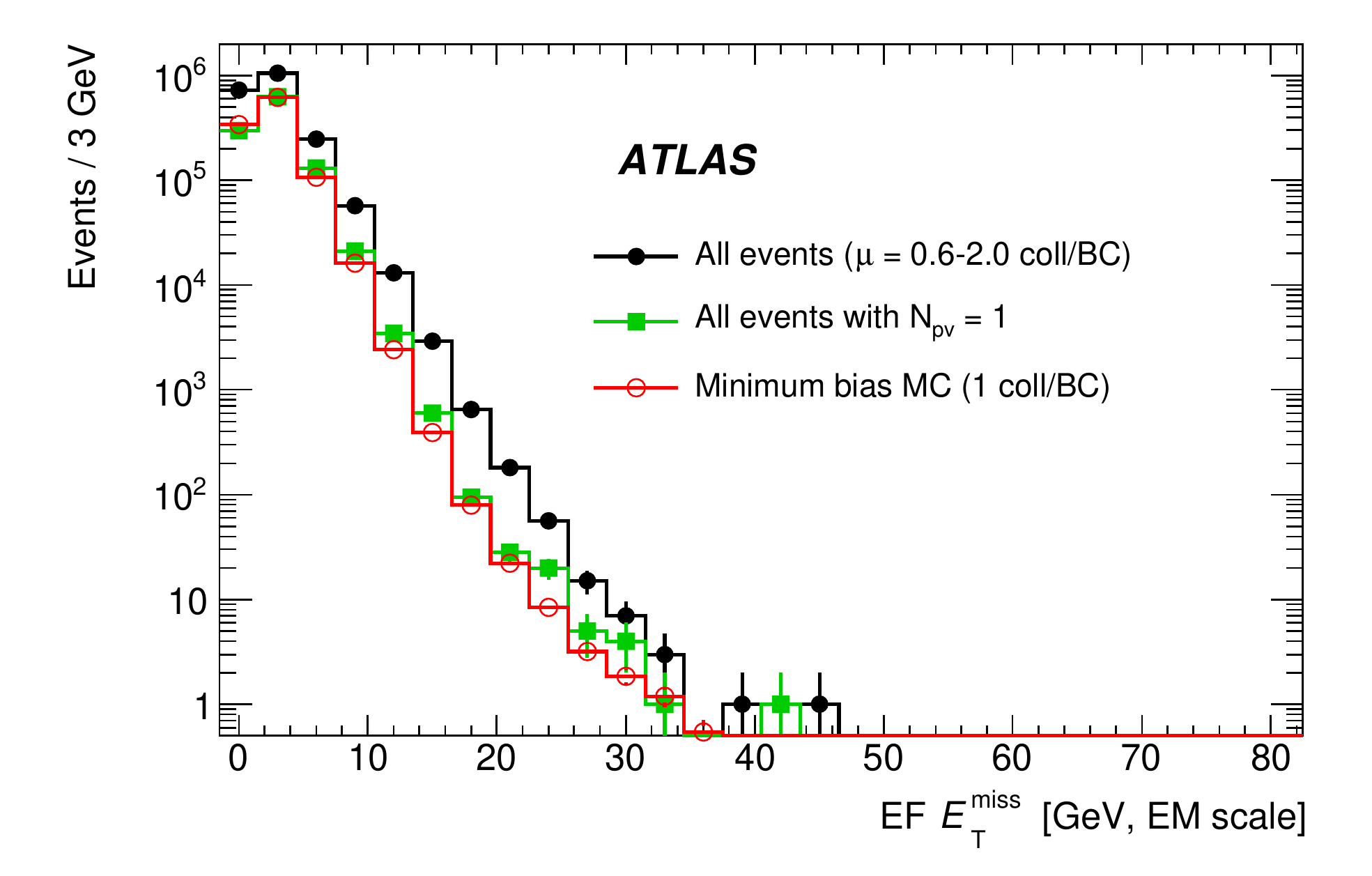}
  \label{fig-EFmet-pileup}
 }
  \caption{Distributions of \MET\ computed at \subref{fig-L1met-pileup} L1  
  and \subref{fig-EFmet-pileup} the EF
    for all 2010 \pp\ collision events (dots) and for the subset obtained by
    rejecting events with multiple primary vertices (squares),
    compared to simulated minimum bias events that do not include
    pile-up effects (circles)}
 \label{fig-met-pileup}
\end{figure}
\begin{figure}[!htbp]
  \centering
  \subfigure[]{
  \includegraphics[height=5cm]{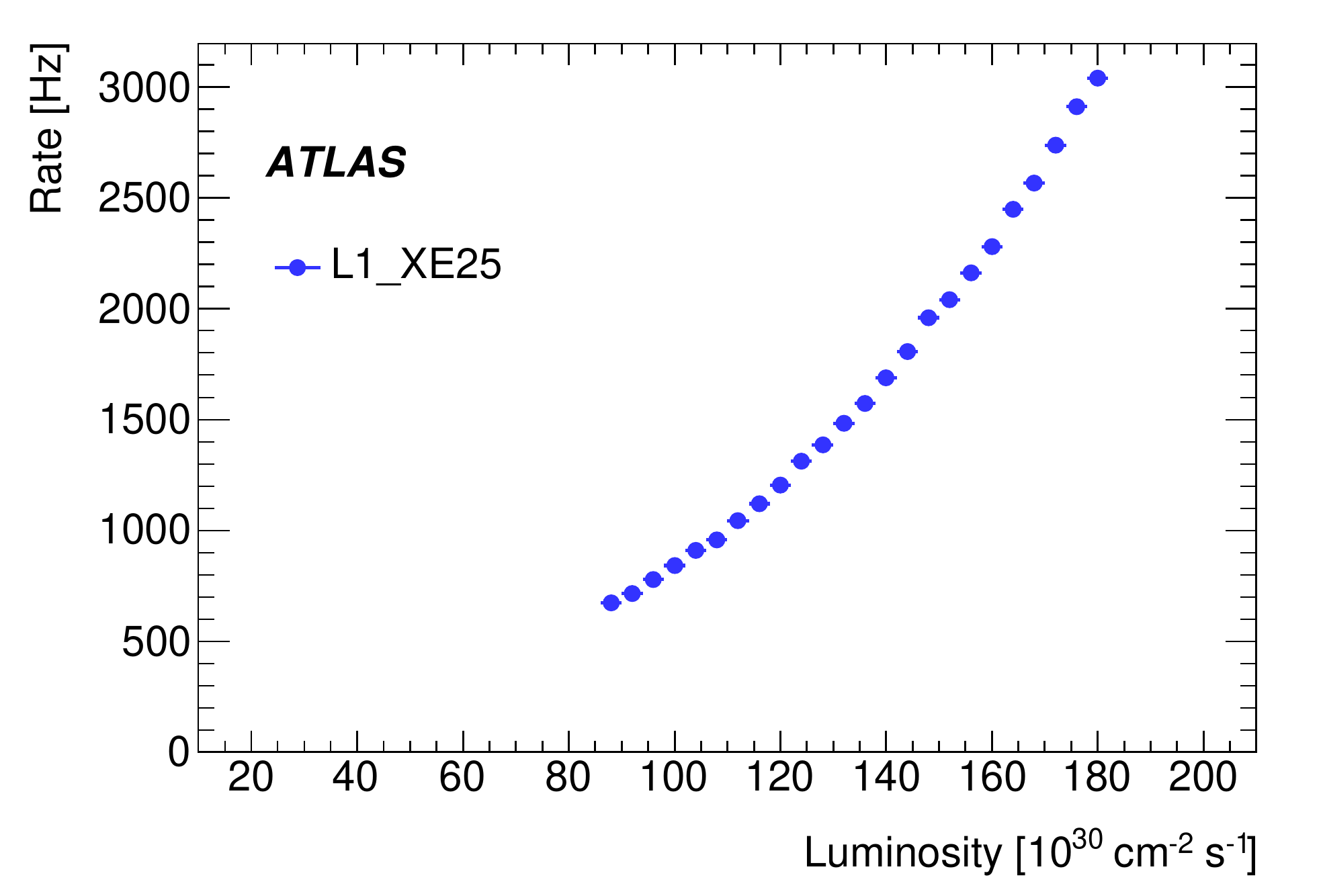}
  \label{fig:XE25rate}
  }
  \subfigure[]{
  \includegraphics[height=5cm]{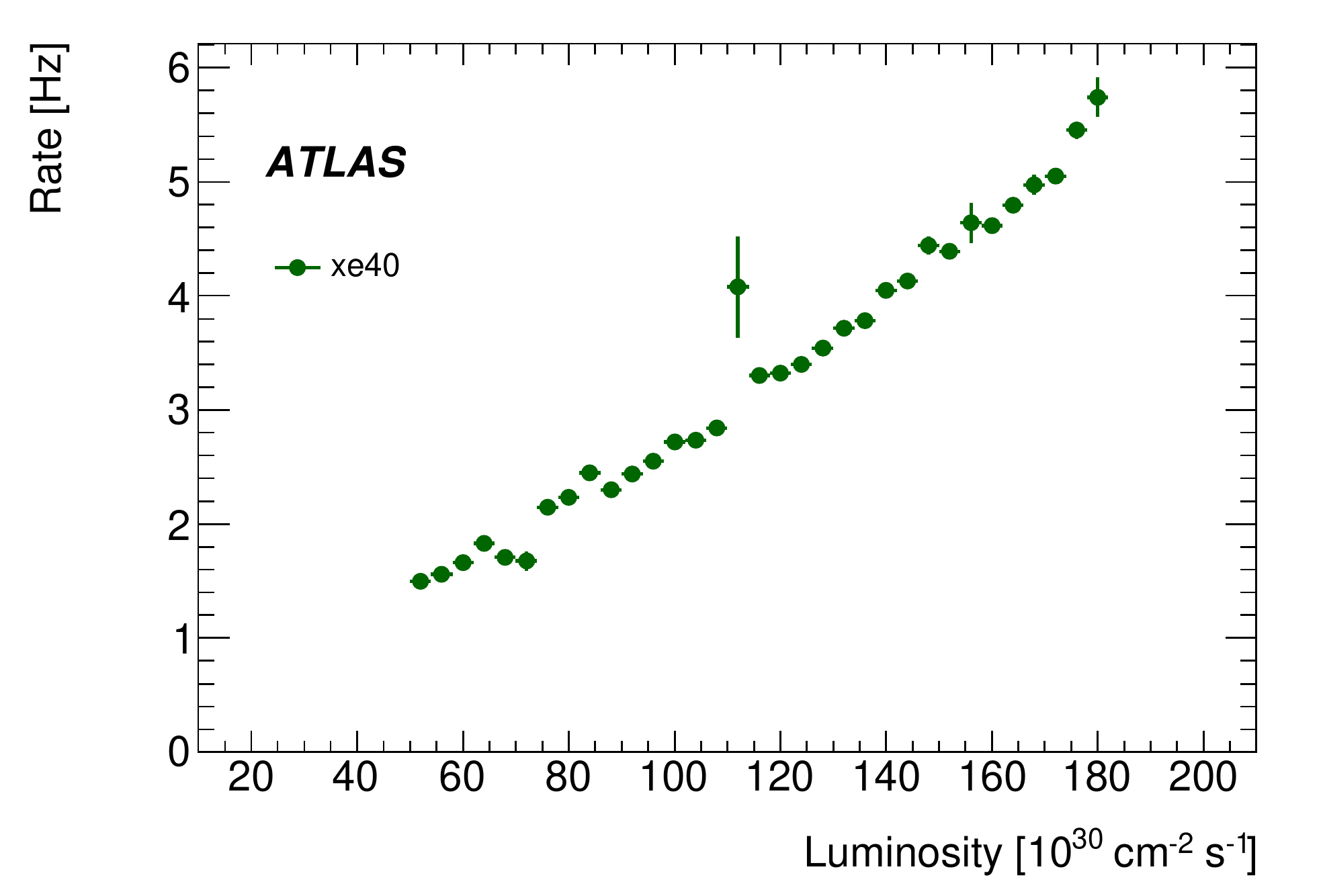}
  \label{fig:metHLTrate}
  }
  \caption{Rate of the  \MET\ triggers at \subref{fig:XE25rate} L1 (L1\_XE25) and \subref{fig:metHLTrate} the EF
(xe40) as a function of instantaneous luminosity for 
a set of runs taken near the end of 2010 running} 
 \label{fig-xe40-rate}
\end{figure}

\subsubsection{Reconstruction and Selection Criteria}\label{sec:metRec}

During 2010, the \MET\ and \SumET\ triggers used calorimetric measurements calibrated at the EM scale. 
In the L1 calorimeter trigger system trigger towers are used to compute both \MET\ and \SumET\
 over the full ATLAS acceptance ($|\eta|<4.9$).
The magnitude of \MET\ is not calculated directly at L1, but rather is derived from a look-up table 
that takes the values of $\Sigma$\Ex\ and $\Sigma$\Ey\
(expressed in integer values in \GeV) as inputs~\cite{bib:metpubnote}.   The resulting resolution 
smearing is $\sim1~\GeV$.
The noise suppression scheme adopted at L1 in 2010 was very conservative with a
rather high \et\ threshold, in the range $1.0-1.3~\GeV$, applied to each
 trigger tower before computing the sums $\Sigma$\Ex, $\Sigma$\Ey\ and \MET.

\begin{figure*}[!htbp]
  \centering
  \subfigure[]{
    \includegraphics[width=0.45\textwidth]{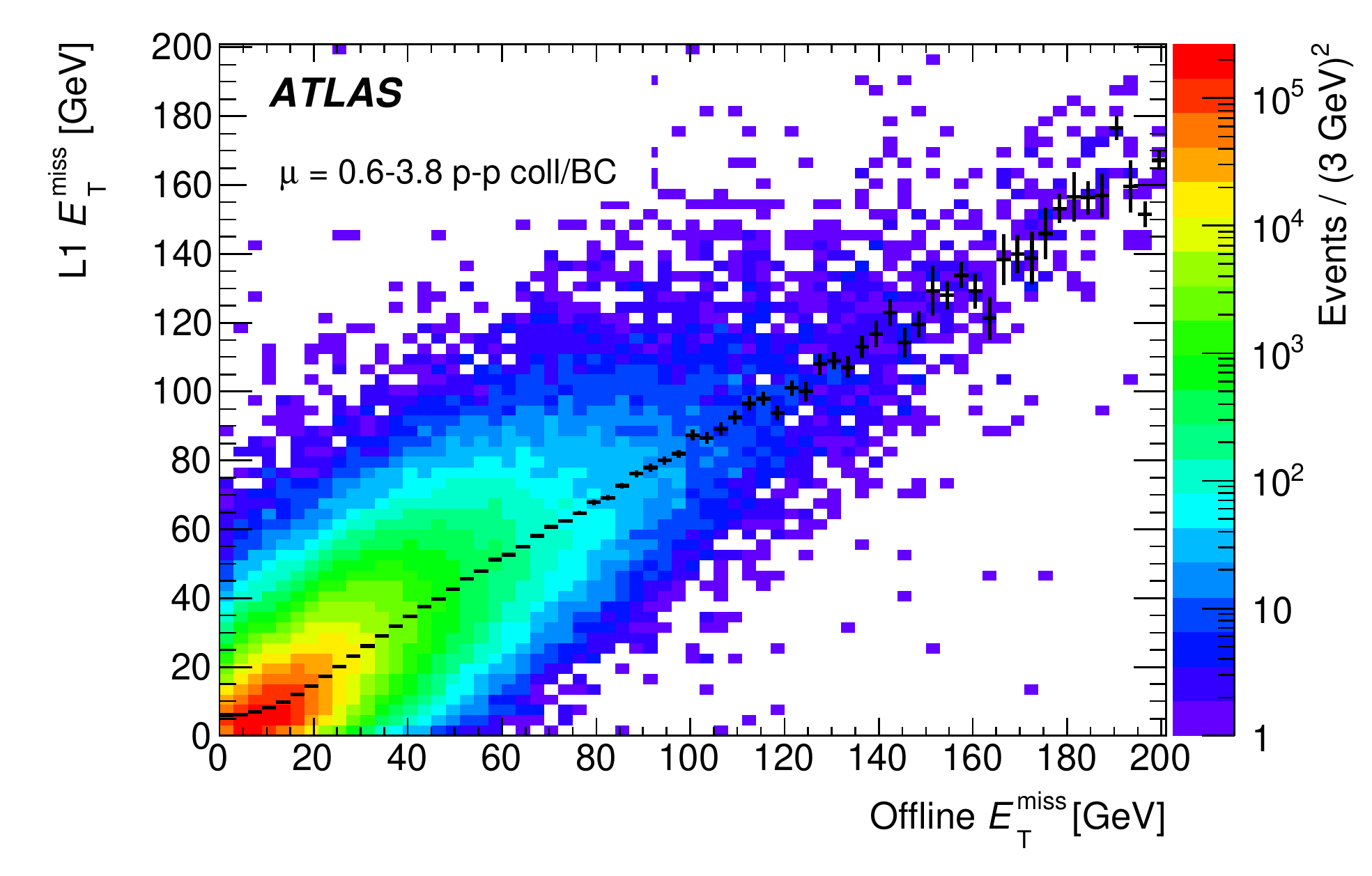}
  \label{fig-L1-met-2D}   
  } 
  \subfigure[]{
    \includegraphics[width=0.45\textwidth]{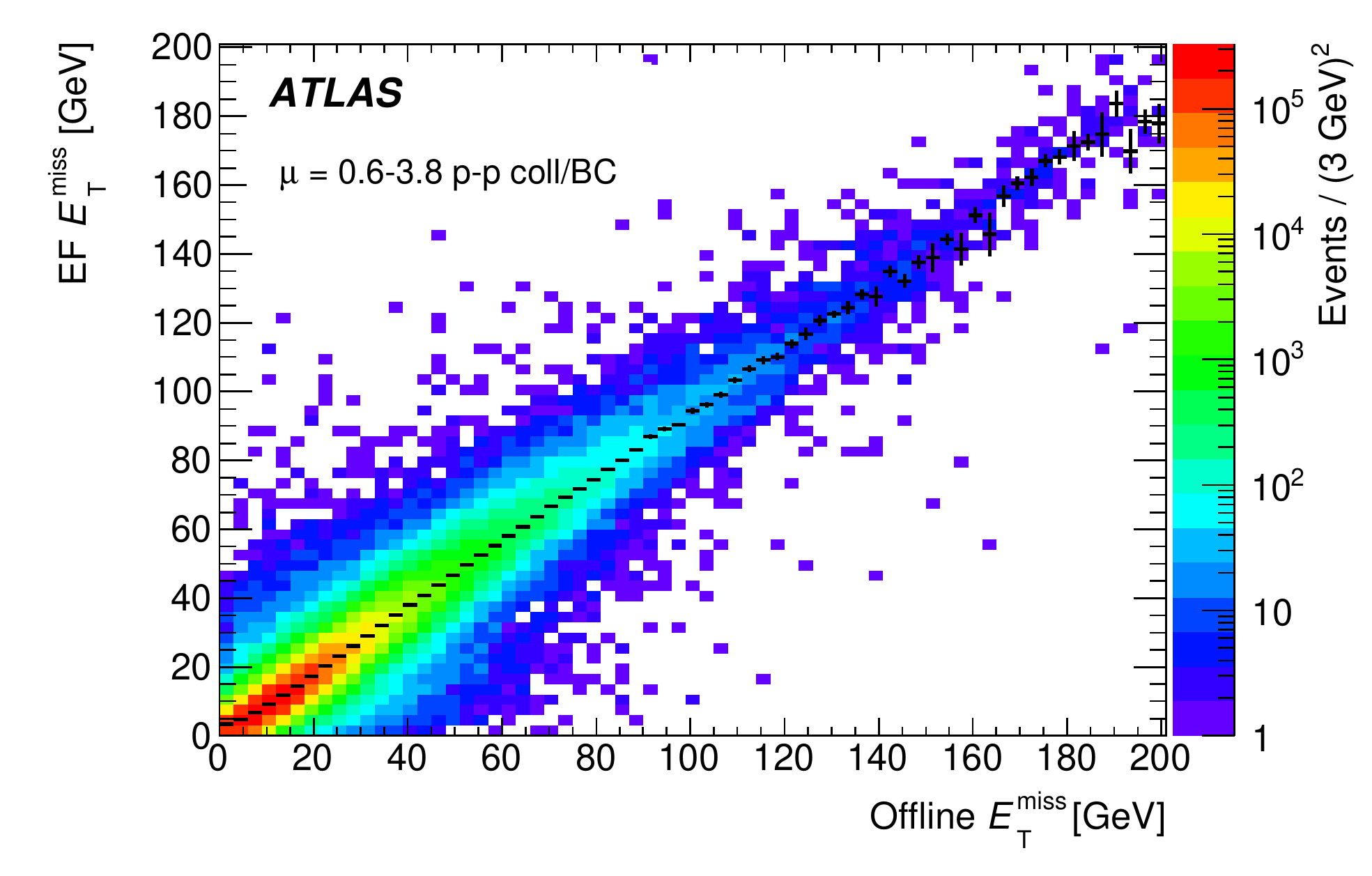}
  \label{fig-EF-met-2D}   
}
  \caption{Correlation between the trigger and offline 
\MET\ shown for \subref{fig-L1-met-2D} L1 and \subref{fig-EF-met-2D} the EF for events
  triggered by the mu13 trigger. The black crosses show
the mean value of online \MET\ calculated in each vertical slice with the vertical length of the cross representing the error on the mean.  Energies are at EM scale}
\label{fig-met-dist-2D}
\end{figure*}
\begin{figure*}[!htbp]
  \centering
  \subfigure[]{
    \includegraphics[width=0.45\textwidth]{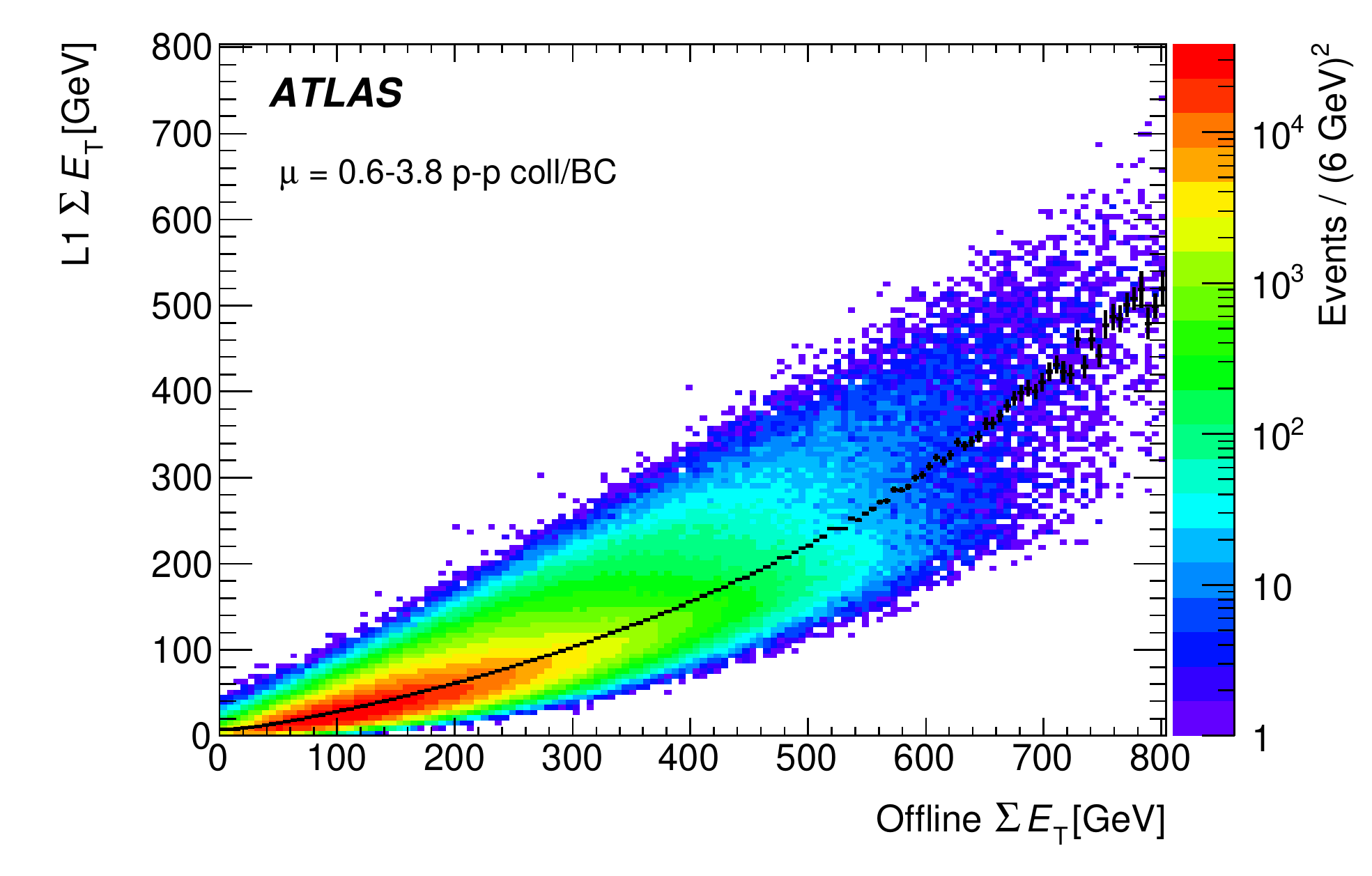}
  \label{fig-L1-set-2D}   
  }
  \subfigure[]{
  \includegraphics[width=0.45\textwidth]{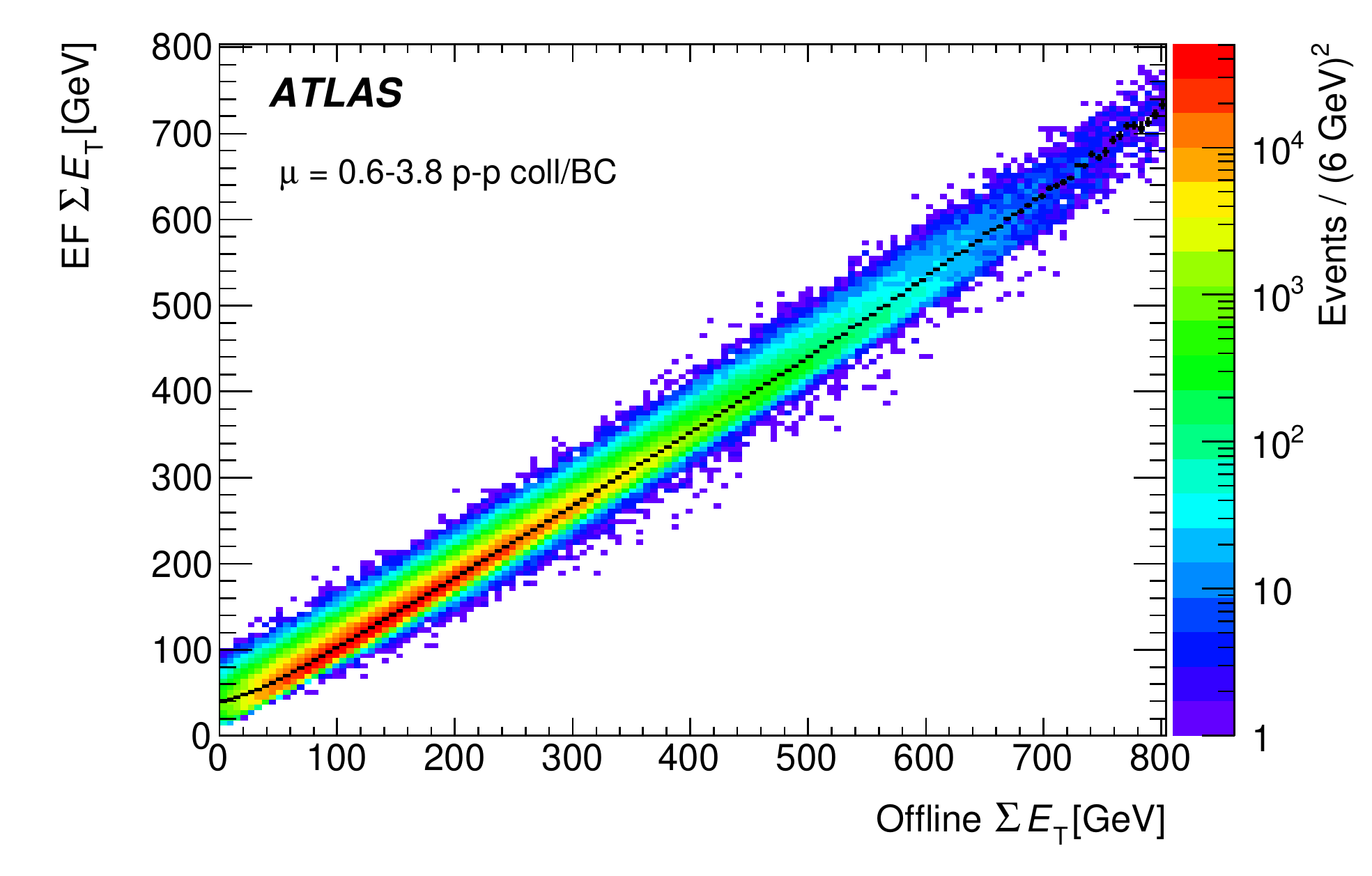}
  \label{fig-EF-set-2D}   
  }
  \caption{Correlation between the trigger and offline 
\SumET\ measurements shown for \subref{fig-L1-set-2D} L1  and \subref{fig-EF-set-2D} the EF 
for events triggered by the mu13 trigger. The black crosses have
the same meaning as in Fig.~\ref{fig-met-dist-2D}.   Energies are at EM scale}
\label{fig-set-dist-2D}
\end{figure*}
\begin{figure*}[!htbp]
 \centering
 \subfigure[]{
   \includegraphics[width=0.45\textwidth]{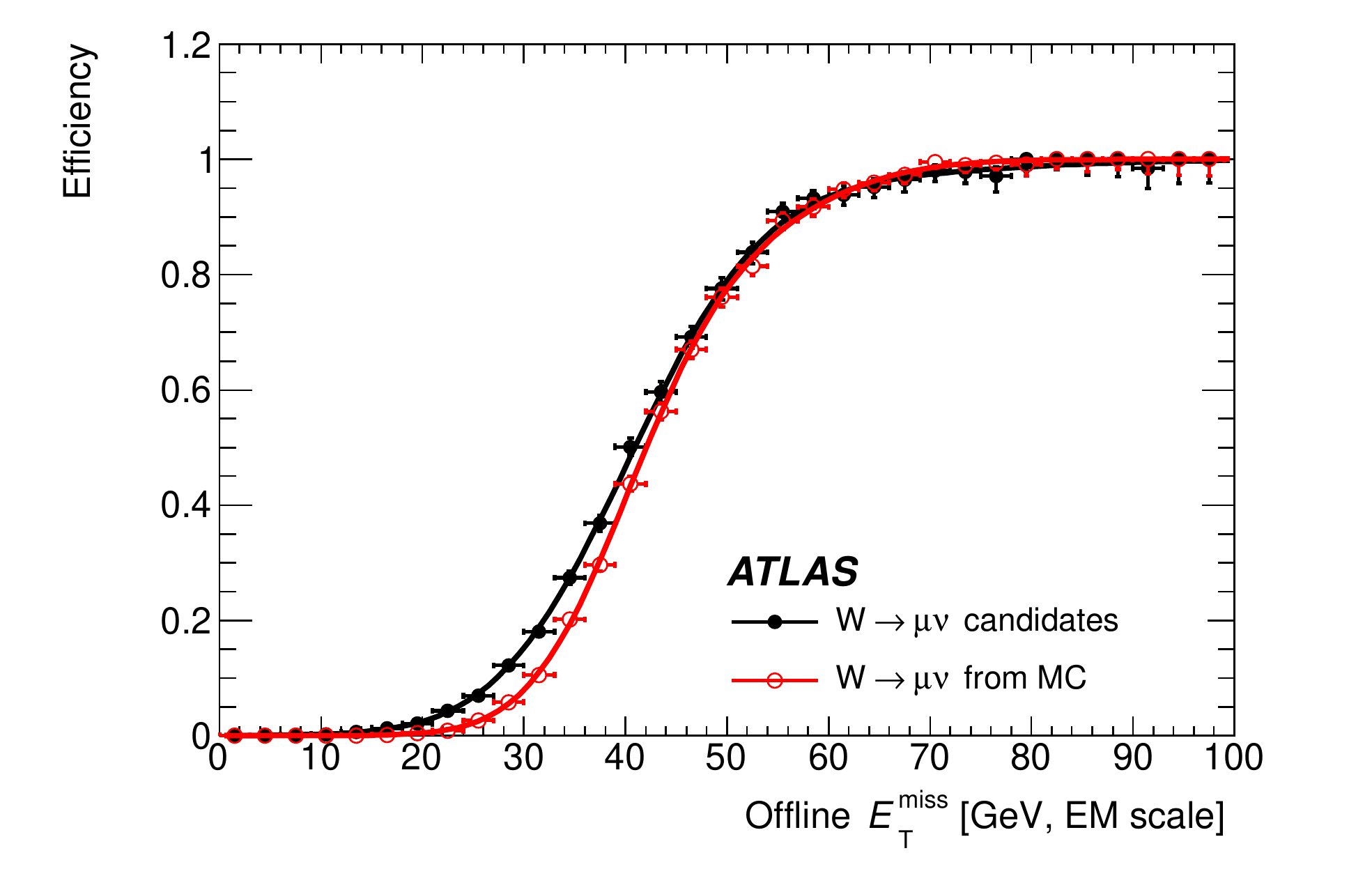}
   \label{fig-Wmunu-turnon-L1-met}
 }
 \subfigure[]{
  \includegraphics[width=0.45\textwidth]{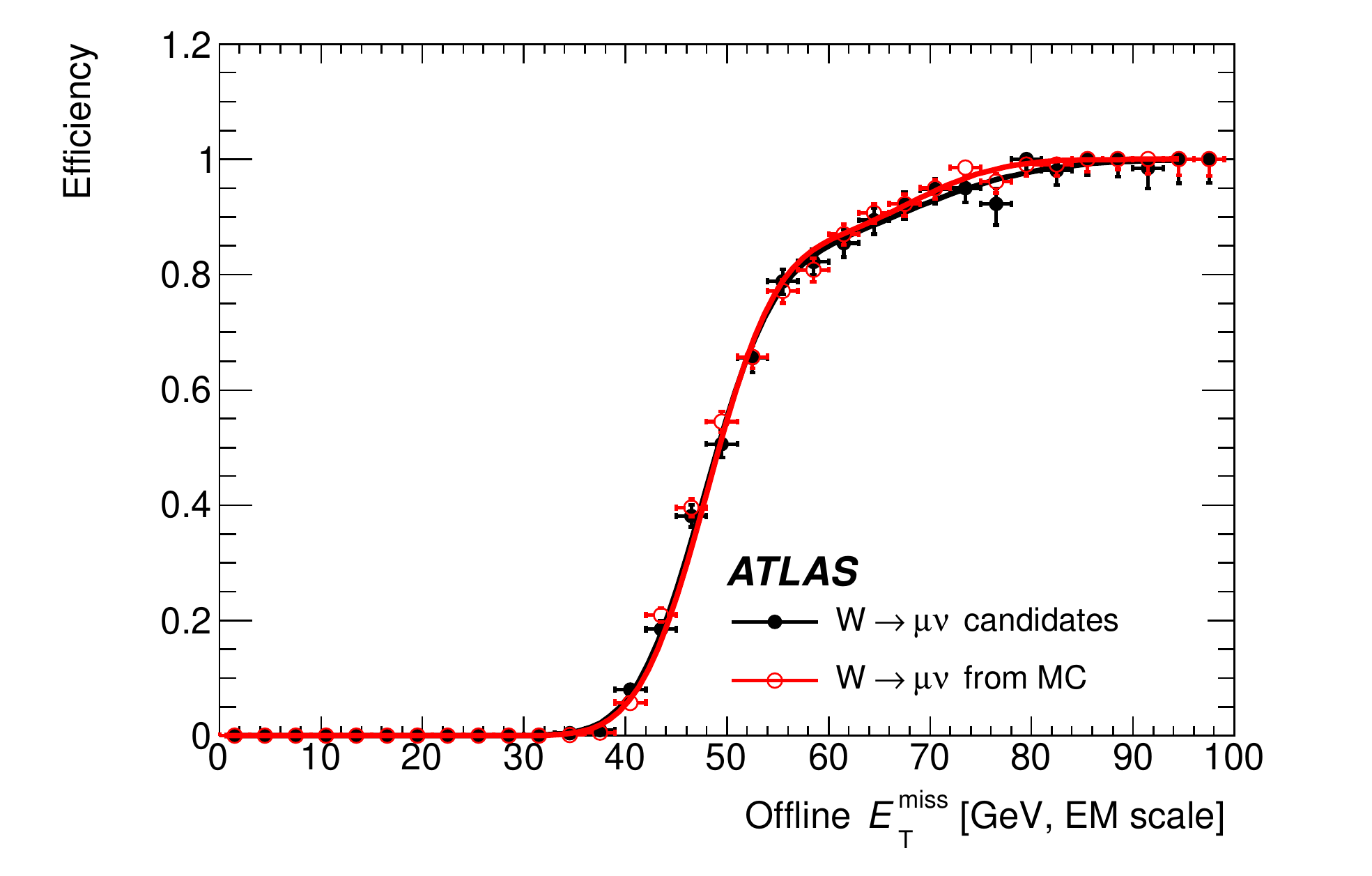}
  \label{fig-Wmunu-turnon-EF-met}
 }
 \caption{Efficiency of the \MET\ trigger as a function of offline \MET\ 
for \subref{fig-Wmunu-turnon-L1-met} the $25 \GeV$ L1 threshold  
and \subref{fig-Wmunu-turnon-EF-met} the $40\GeV$ EF threshold, 
for \Wmunu\ candidates selected offline and a sample of 
simulated \Wmunu\ events}
 \label{fig-Wmunu-turnon-L1-EF-met}
\end{figure*}

The discreteness of the L1 approach is smoothed out at L2, where the $\Sigma$\Ex\ and $\Sigma$\Ey\ values
from L1 are summed in
quadrature and a threshold is placed on the magnitude of \MET=$\sqrt{({\Sigma}E_x)^2+({\Sigma}E_y)^2}$.    
At L2, the L1 energy measurement can also be corrected using the measured
 momenta of detected muons in the event. Since the muon
correction has only a small impact on trigger rates, for 2010 running the
correction was calculated at L2 and the value of the correction stored
in the event.  However, this correction was not applied to the \MET\ value
calculated online, and thus was not used in the trigger decision.

Because recalculation of \MET\ and \SumET\ using the full
granularity of the calorimeters requires access to the whole event, it is only performed at
the EF. 
Both \MET\ and \SumET\
 are estimated by the same algorithm, which loops over all calorimeter
 cells discarding those whose energy is negative or has a value less than three 
 standard deviations of the noise distribution.  For each of the cells with energy above threshold,
an energy
 vector is defined whose direction is given by the unit vector
 starting from the nominal interaction point and pointing to the cell
 centre, with magnitude equal to the measured cell energy.

\subsubsection{Menu and Rates}

\begin{table}[!hbtp]
   \begin{center}
   \caption{Principal \MET\  and \SumET\ triggers and their rates at a luminosity of \Lumi{32}}
   \small{
      \begin{tabular}{llc}
         \hline
         Trigger  & Motivation  & Rate [Hz] \\
         \hline
         \hline
         xe40      &   \W, top, new physics  &  3 \\
         te350      &   new physics & 3  \\
         \hline 
      \end{tabular}
      }
   \normalsize
   \label{tab:met_triggers}
   \end{center}
\end{table}

 There are eight L1 \MET\ thresholds shown in Table~\ref{tab:ExampleMenu}. 
The L2 (EF) thresholds were set at least $2 \GeV$ ($10 \GeV$) higher than the corresponding 
thresholds at L1 to mask
the reduced granularity of the look-up table and the effects of the slowly increasing efficiency 
at L1. For example the xe40 trigger has a 25~GeV threshold at L1 (L1\_XE25) and a 30~GeV threshold 
at L2 (L2\_xe30).
  To control the trigger rate as the instantaneous luminosity increased it
was necessary to reduce the energy difference between the L1 and EF thresholds
for some chains; these chains were suffixed with ``tight'' 
in the trigger menu, e.g. xe30\_tight. For
these triggers, the effect of the L1 efficiency turn-on extends above the
EF threshold.  The principal \MET\ and \SumET\ triggers used in 2010 and their rates 
at a luminosity of \Lumi{32} are shown in Table~\ref{tab:met_triggers}.

 Figure~\ref{fig-met-pileup} shows the impact of in-time pile-up
  on \MET.  The measured L1
 and EF distributions are compared to a MC sample of minimum bias
 events simulated without pile-up. The simulation reproduces the
 \MET\ distributions for the bunch crossings with a single \pp\ collision ($N_{pv}~=~1$).
For data events with multiple
 collisions (0.6-2.0 collisions/BC) there is a visible broadening of the \MET\ distribution 
reflecting an increase in \MET\ due to pile-up.
The \MET\ trigger rates at L1 and the EF are shown in Fig.~\ref{fig-xe40-rate} for the xe40 
trigger which has a 25~GeV threshold
at L1 (L1\_XE25) and a 40~GeV thresholds at the EF. 
The \MET\ rate  rate increase with luminosity
is faster than linear, due to the effects of pile-up.

\subsubsection{Resolution}\label{sec:metRes}

 The correlations between the trigger and offline values of \MET\ and \SumET\ 
using uncalibrated calorimeter 
energies are shown in Figs.~\ref{fig-met-dist-2D} and \ref{fig-set-dist-2D}. The offline 
calculations use an algorithm (MET\_Topo) which sums the energy
 deposited in topological clusters~\cite{ATL-LARG-PUB-2008-002}.
Figure~\ref{fig-L1-met-2D} shows the correlation between L1 and
 offline \MET\ for events triggered by the mu13 trigger (Section~\ref{sec:muon}).  
The L1 \MET\ resolution is worse than offline, as expected, while the EF
 shows a good correlation and improved resolution with respect to L1, as seen in 
Fig.~\ref{fig-EF-met-2D}.
Figure \ref{fig-L1-set-2D} shows the correlation between the L1 \SumET\ and that 
calculated by the offline algorithm
 MET\_Topo for events selected by the mu13 trigger.  
L1 underestimates the \SumET\, particularly at low values, due to the rather conservative noise
suppression (i.e. high trigger tower \et\ thresholds) employed at L1. The effect
is to shift the energy scale at low \SumET\ values, as shown by the non-linear behaviour in 
Figure \ref{fig-L1-set-2D}.

 The plot in
 Fig.~\ref{fig-EF-set-2D} shows the correlation between the
EF and offline values of \SumET. There is an offset of about $10 \GeV$ for the
 values of \SumET\ computed at EF, as the offline \SumET\ approaches
 zero.  The offset arises because of a one-sided noise
 cut applied by the trigger, compared to symmetric cuts applied offline.
The main motivation for the choice made at the EF is to protect
 against large negative energy values, which could arise from read-out
 problems and which would constitute a source of fake \MET.  
The choice of the online noise cut (of three times the r.m.s. noise) 
is a compromise between
minimising the offset (a lower cut of twice the r.m.s. noise would
give a much larger bias of $\sim200 \GeV$) and maintaining sensitivity, since
higher thresholds would cause a greater loss of the real signal~\cite{bib:metpubnote}.

\subsubsection{Efficiency}

\begin{figure}[!b]
 \centering
 \subfigure[]{
    \includegraphics[width=0.45\textwidth]{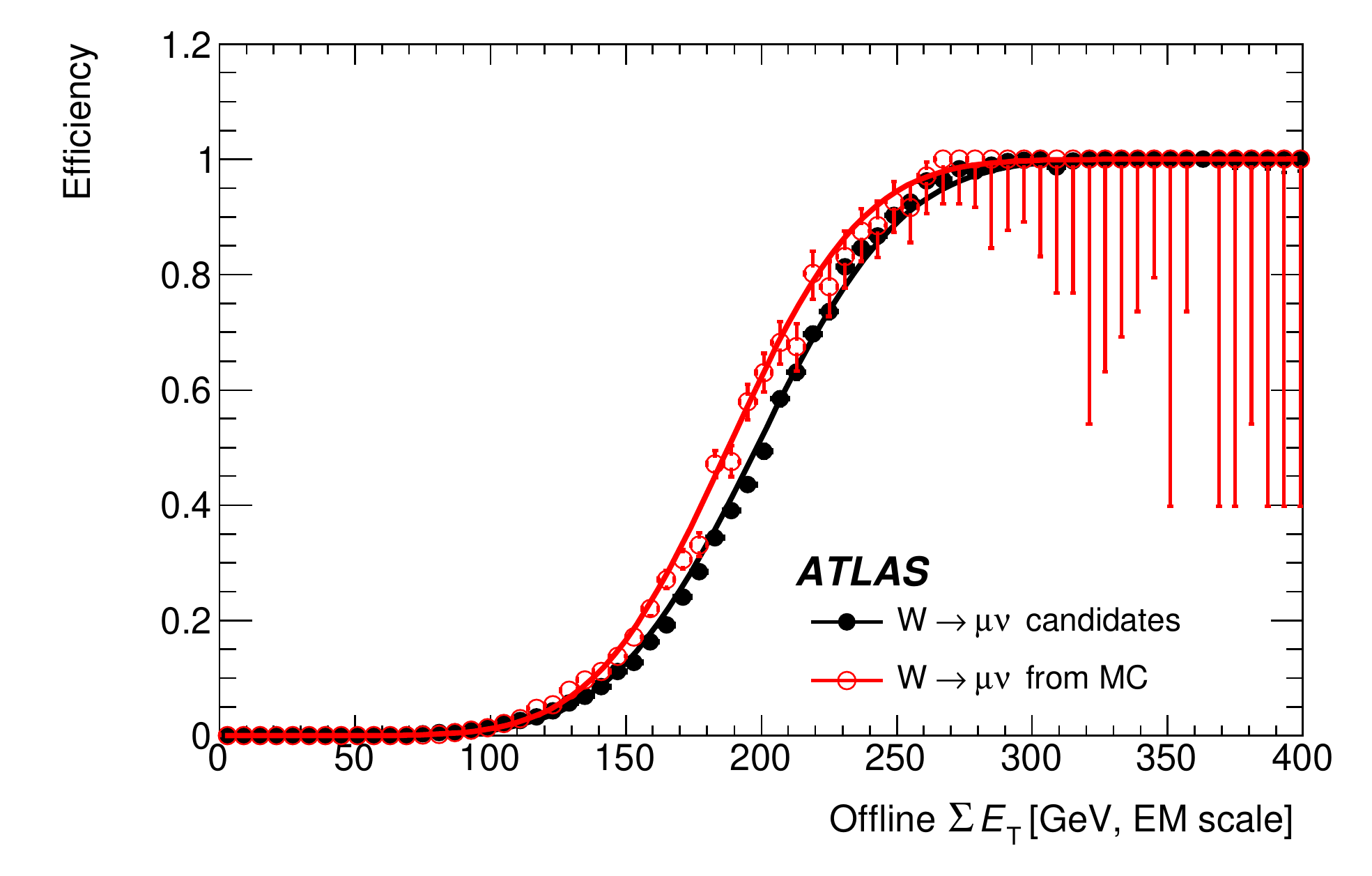}
  \label{fig-L1-MinBias-turnon-TE}
  }
 \subfigure[]{
  \includegraphics[width=0.45\textwidth]{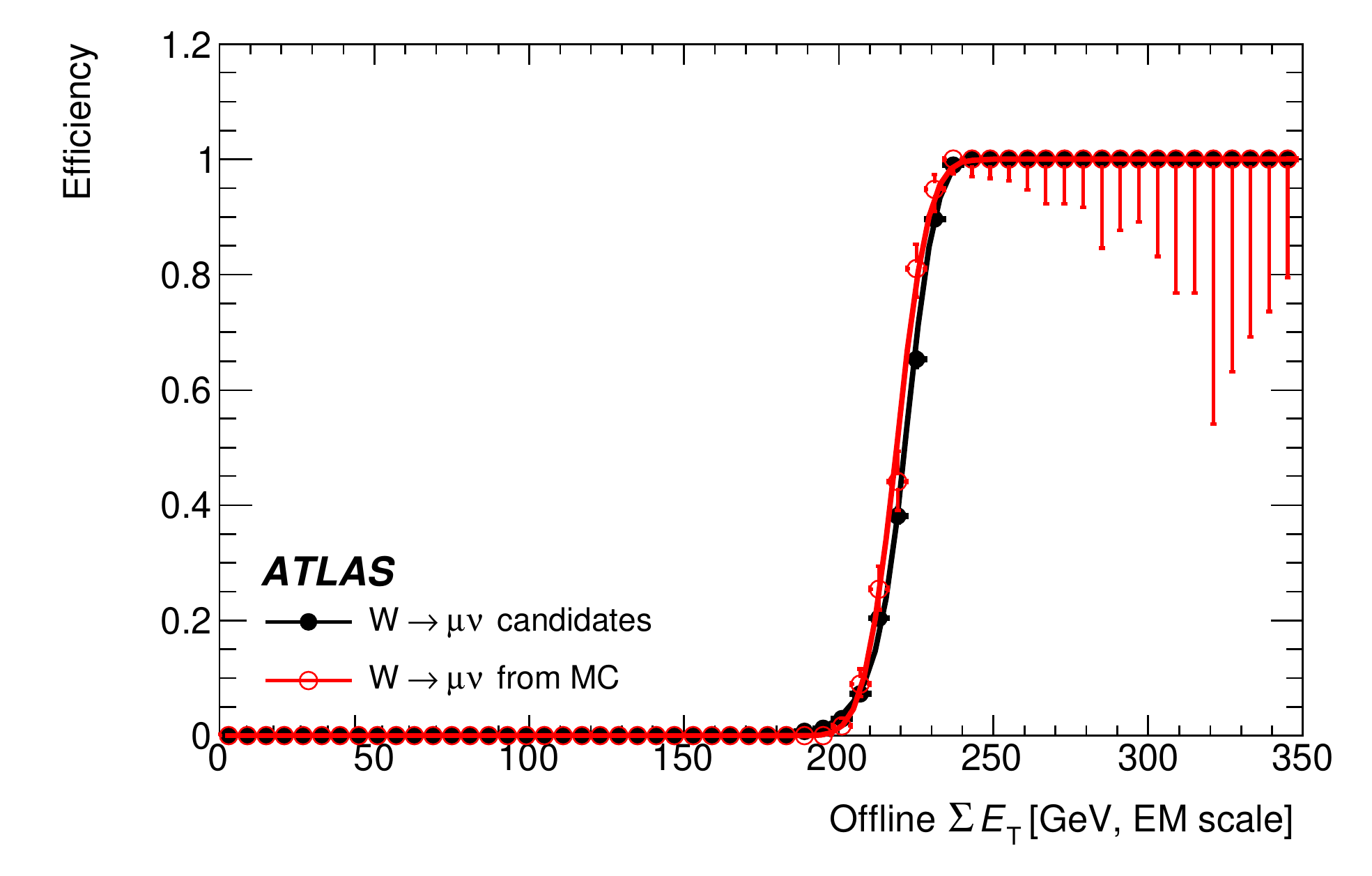}
  \label{fig-EF-MinBias-turnon-TE}
}
 \caption{Efficiency of \subref{fig-L1-MinBias-turnon-TE} the L1 \SumET\ threshold at 50 GeV  and
 \subref{fig-EF-MinBias-turnon-TE}  the EF-only \SumET\ threshold at $200 \GeV$,  as a function of
  \SumET\ calculated by the offline algorithm MET\_Topo, for \Wmunu\ candidates selected offline and a sample of simulated \Wmunu\ events}
\label{fig-MinBias-turnon-TE}
\end{figure}

 Figure \ref{fig-Wmunu-turnon-L1-met} shows the efficiency of the
L1 \MET\ trigger with a $25 \GeV$ threshold 
as a function of \MET\ reconstructed offline for \Wmunu\ candidates selected 
in offline reconstruction, triggered by mu13.  The plateau region is
 described well by the MC. The agreement with the simulation is not
 perfect for low energies; 
background events from QCD processes and \Wboson\ boson decays into taus, which
subsequently decay into muons, are difficult to simulate precisely. 
Figure~\ref{fig-Wmunu-turnon-EF-met} shows the corresponding 
efficiency for the full trigger chain including a $40 \GeV$ \MET\ threshold 
at EF.  The initial faster rise of the efficiency turn-on is
 dominated by the EF \MET\ resolution whereas the slower rise approaching 
the plateau is due to the slower L1 turn-on. This behaviour is modelled well by the
simulation.  Once the plateau has been reached the \MET\ triggers remain fully
efficient within a negligible systematic uncertainty.

 Figure~\ref{fig-L1-MinBias-turnon-TE} shows the L1 efficiency turn-on for a
 nominal \SumET\ threshold of $50 \GeV$. The late turn-on, starting only
at about 150~GeV in offline \SumET, results from an under-estimation of  \SumET\ at L1 due 
to the noise suppression scheme, as described in subsection \ref{sec:metRes}. The efficiency reaches 
90\% at about $260 \GeV$. Data and MC agree reasonably well; the shift in the efficiency turn-on
is due to small errors in the modelling of noise at the individual cell level in the simulation.  
Figure~\ref{fig-EF-MinBias-turnon-TE} shows the efficiency
of the EF selection alone, not including L1 and L2. 
The EF efficiency reaches 90\% at about $230 \GeV$. 
Once the plateau has been reached the \SumET\ triggers remain fully
efficient within a negligible systematic uncertainty.  Data and simulation agree well.
More details can be found in Ref.~\cite{bib:metconfnote}.


\subsection{$b$-Jets}\label{sec:bjet}
\def \figurepath{.}
\begin{sloppypar}
The ability to separate heavy flavour jets from light-quark and gluon jets is an important asset 
for many physics analyses, such as measurements in the top-quark sector 
and searches for Higgs bosons or other new physics signatures.  The ability to 
identify $b$-jets in the ATLAS trigger system gives access to signals that 
would otherwise be lost in the multi-jet background, such as $t\bar{t} \to jets$.  
ATLAS employs two categories of $b$-jet triggers: 
\emph{lifetime} triggers that exploit the $B$-hadron time-of-flight and \emph{muon-jet}
 triggers that exploit the presence of a muon in $B$-hadron decays.

During the 2010 data-taking period, the lifetime triggers were not in active rejection mode and 
the muon-jet triggers were used to collect data to validate the 
lifetime triggers.  The lifetime triggers will be used in 2011 to collect 
data for physics analysis.  In this section a brief description of the 
muon-jet triggers is given, but the main focus is on the performance of the lifetime triggers.
\end{sloppypar}

\subsubsection{$b$-Jets Reconstruction and Selection Criteria}

\emph{Muon-jet} triggers were used to select events containing jets
 associated with a low \pt\ muon. At L1 a combined
muon-jet trigger, L1\_MU0\_JX (X=5,10,15,30,55), required
the lowest threshold
muon trigger in combination with a jet.
No topological matching between muon and jet is possible at L1. The HLT selection introduces
a refinement of the muon selection (L2\_mu4) and requires matching within $\dR <0.4$ 
between the muon and the corresponding L1 jet.
The selected jet sample is 
  enriched in $b$-jets and is used to calibrate 
both trigger and offline $b$-tagging algorithms. 

 \emph{Lifetime} triggers use tracks and
vertices reconstructed at the HLT (in the region $\eta<2.5$) to select a sample enriched in $b$-jets. 
These triggers are based on the impact parameters of tracks with respect to the
reconstructed primary vertex. 
The HLT selection is based on inner detector tracks reconstructed within a L1 jet RoI.
The lowest threshold $b$-jet trigger is b10 which starts from a 
L1 jet with a $10 \GeV$ \ET\ threshold (L1\_J10).

At the HLT, the first step for the lifetime triggers is to find the location of the primary vertex. 
The coordinates of the primary vertex in the transverse plane are
determined by the beamspot information which is part of the configuration data provided
to the algorithm via the online conditions database. The beamspot position
can be updated during a run based on information from the online beamspot 
measurement~(Section~\ref{sec:beamspotReco}).
During 2010 running, when the lifetime triggers were not in active rejection mode,
this update was initiated manually  whenever the beamspot showed a significant displacement.  
The longitudinal coordinate of the primary vertex 
is determined on an event-by-event basis from a histogram of the $z$ positions of all tracks
in the RoI. The $z$ position of the vertex is identified, using a sliding window
algorithm, as the $z$ position at which the window contains the most histogram entries. 
In the case of multiple primary vertices, this algorithm selects the vertex 
with the most tracks.

The transverse and longitudinal impact parameters are determined, for each track, 
as the distances from the primary vertex to 
the point of closest approach of the track, in the appropriate projection. 
The impact parameters are signed with respect to the jet axis determined by a track-based cone jet reconstruction
algorithm.
The impact parameter is positive
if the angle between the jet axis and a line from the primary vertex to the point of closest approach of the 
track is less than $90\degr$.

  Two different methods, \emph{likelihood} and $\chi^2$ taggers, both based on the track impact parameters, 
  are then used to build a variable discriminating between $b$ and light jets:
  \begin{description}
    \item \emph{Likelihood taggers:} longitudinal and transverse impact parameters
       are combined, using a likelihood ratio method, to form a discriminant
       variable.
    \item \emph{$\chi^2$ tagger:} the compatibility of the tracks in the RoI with the beamspot 
is tested using the transverse impact parameter significance (defined as the transverse impact 
parameter divided by the transverse impact parameter resolution)~\cite{aleph:JetProb}. 
The distribution 
of the $\chi^2$ probability of the impact parameter significance for all the tracks reconstructed in an RoI
is expected to be uniform for light jets, as tracks come from the primary vertex,
while it peaks toward 0 for $b$-jets, which contain tracks that are not from the primary vertex. 
The $\chi^2$ probability can, therefore, be used as a discriminant variable. It is set to 1 for RoIs 
that do not contain any reconstructed tracks.

  \end{description}

  Likelihood taggers are more powerful, in principle, but require significant validation 
from data as they rely on
determining probability density functions that give the signal and background
probabilities corresponding to a given impact parameter value.
  The $\chi^2$ tagger, though less powerful,
  can be tuned more easily on data using the negative side of the transverse 
  impact parameter distribution.   This technique is used because the shape of the negative side of the distribution is determined only by resolution effects and there is no significant contribution from highly displaced tracks in this part of the distribution.

\begin{figure}[!ht]

  \centering
    \includegraphics[width=0.45\textwidth]{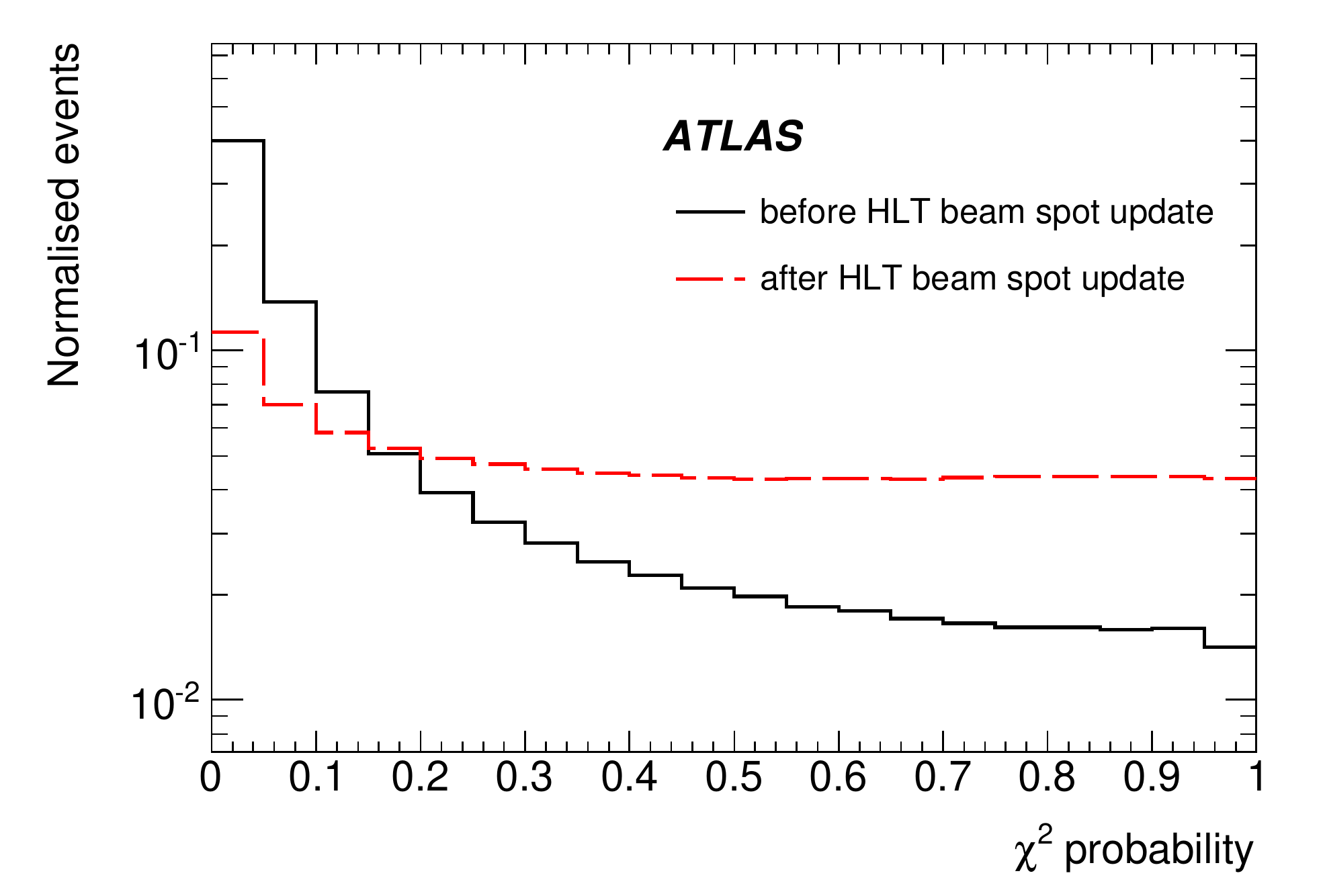}
  \caption{The $\chi^2$ probability distribution before and after the beamspot
  measurement update in a data-taking period 
  when the beamspot was significantly displaced 
  with respect to the reference}
    \label{fig:bs}

\end{figure}

The importance of the online beamspot measurement is demonstrated  
in Fig.~\ref{fig:bs} which shows the $\chi^2$ probability distribution of the  $\chi^2$ tagger 
before and after a beamspot
update in a data-taking period when the beamspot was significantly displaced
  with respect to the initial reference.  In 2011 the beamspot will be updated automatically every 
few minutes because a transverse displacement of the beamspot
can cause tracks in light-quark jets to artificially acquire large impact 
parameters and so resemble the tracks in  $b$-jets.

\subsubsection{$b$-Jets Menu and Rates}

During the 2010 data-taking period the muon-jet triggers were the only $b$-jet triggers 
in active rejection mode, selecting the calibration sample.  The lifetime triggers ran in monitoring mode, allowing
for tuning in preparation for activation in 2011 running. Similar algorithms ran at both L2 and the EF. 

\begin{sloppypar}
The muon-jet triggers were maintained at a rate of about 7~Hz, using prescaling when luminosity exceeded \linebreak \Lumi{31}.   
Prescaling of the triggers with lower jet thresholds was done in such a way as to collect a sample of events with a 
uniform jet transverse momentum distribution in the reconstructed muon-jet pairs. The uniformity of the distribution 
is important for a precise determination of the $b$-jet efficiency in a wide range of 
jet transverse momenta.
\end{sloppypar}

\begin{figure}[h!]
\centering
\subfigure[]{
        \includegraphics[width=0.45\textwidth]{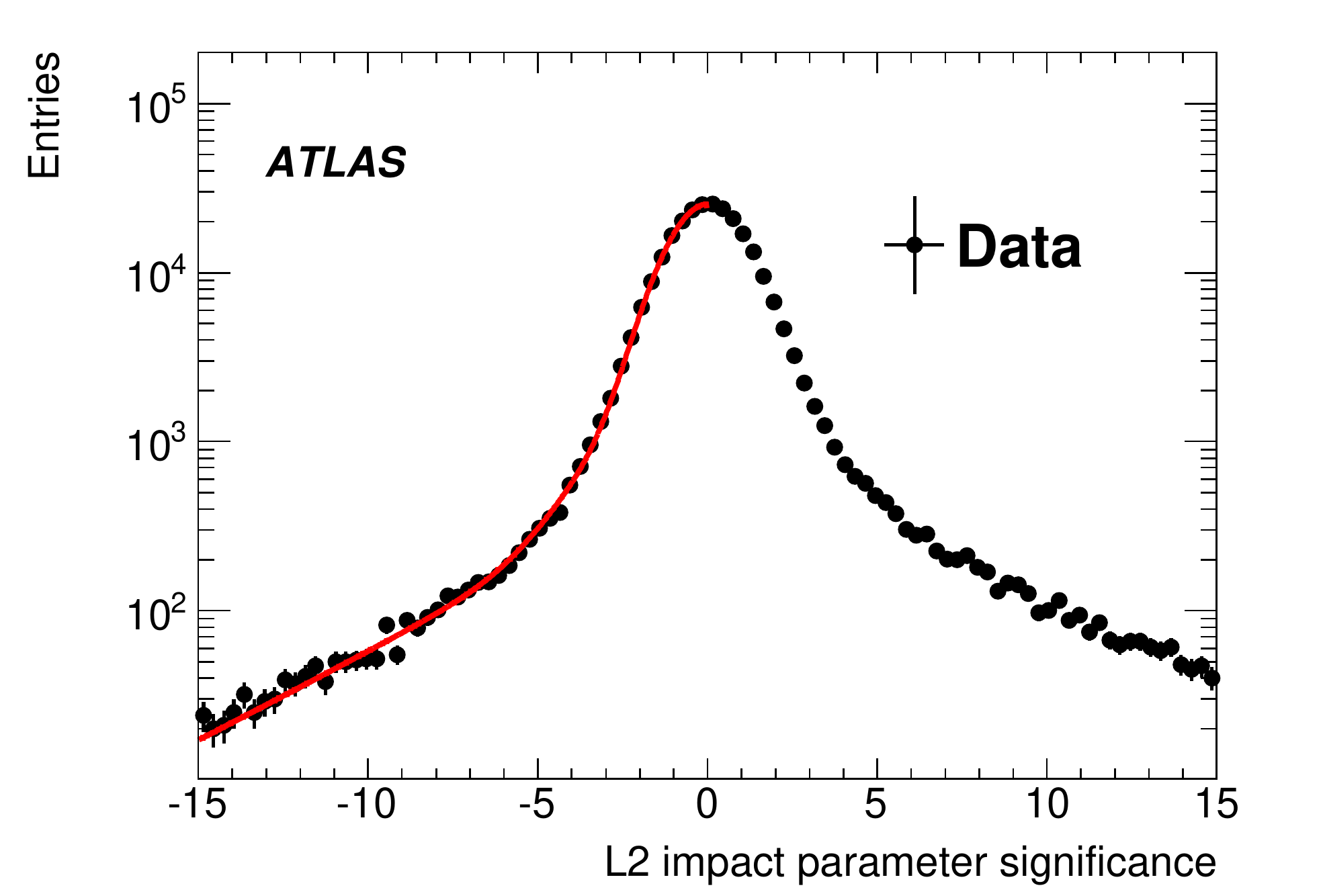}
        \label{fig:tuneJp} 
        }
\subfigure[]{
        \includegraphics[width=0.45\textwidth]{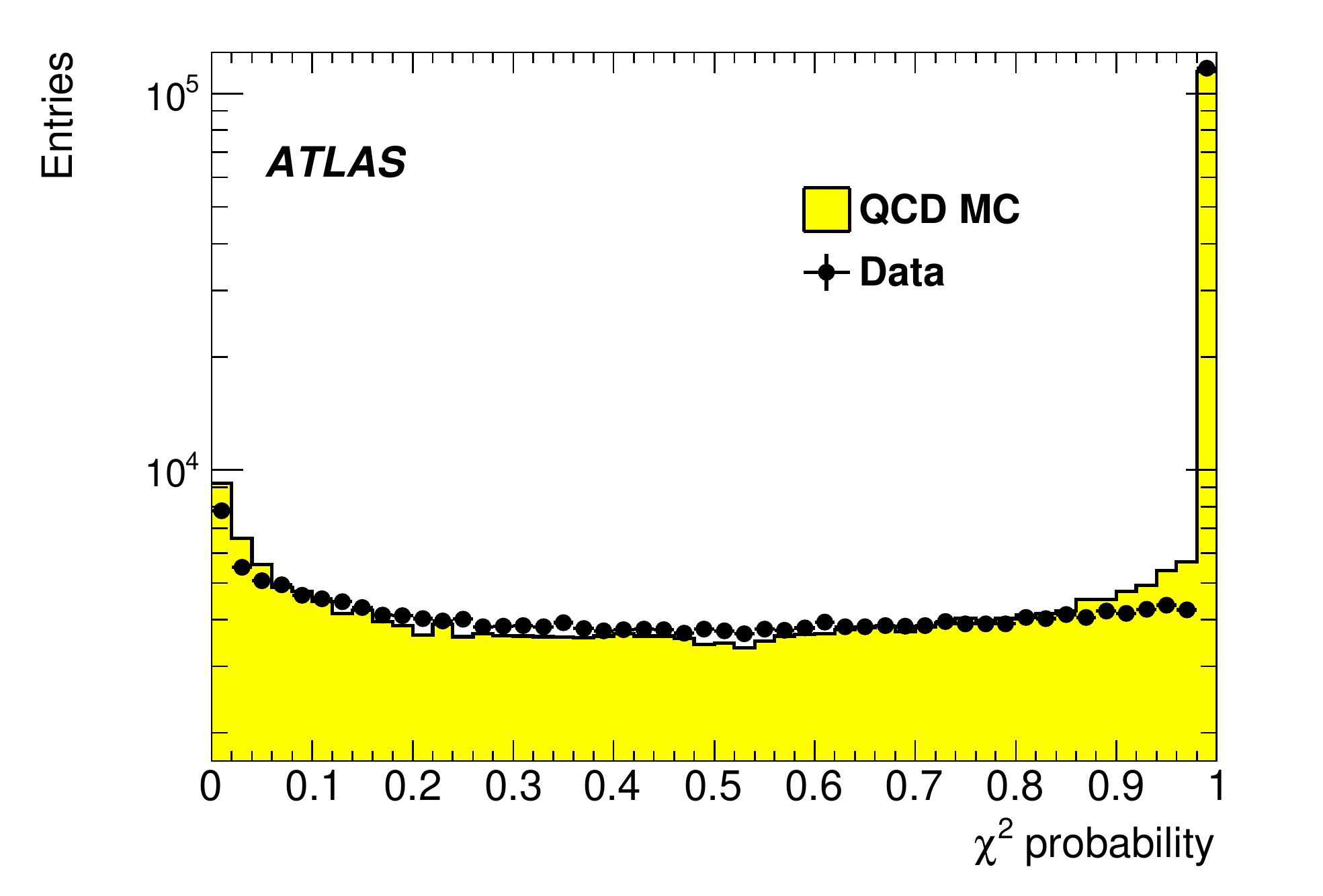}
        \label{fig:tuneChi2} 
}
\caption{
  \subref{fig:tuneJp} 
Transverse impact parameter significance distribution for L2 tracks with a fit to the negative side.  
  \subref{fig:tuneChi2} $\chi^2$ distribution in data and simulation after tuning procedure}
\label{fig:tune}
\end{figure}

\subsubsection{$b$-Jet Trigger Performance}
The performance of the $\chi^2$ tagger
is shown in Table~\ref{singleBjetPerf}, which 
gives the rejection obtained from data collected with the b10 trigger and the efficiency obtained from simulation of $b$-jets with a similar \pt\ distribution to the data.
The efficiency measurement from simulation 
requires a tagged jet RoI matched with an offline jet ($\dR<0.4$).
The offline jet is required to be 
associated with a true $b$ quark ($\dR<0.3$) and identified by an offline tagger
based on the secondary vertex transverse flight length significance. 
\begin{table}[!ht]
  \centering
 \caption{Single $b$-RoI efficiency and rejection for the $\chi^2$ tagger at L2 and for the HLT. 
The efficiency is computed using MC with respect to tagged offline $b$-jets while rejection is computed on data}
 \begin{tabular}{ccc} 
    \hline  
    Efficiency & L2 Rejection & HLT Rejection \\ \hline \hline
     0.7       & 2  &  8 \\ \hline
     0.6       & 3  & 17 \\ \hline
     0.5       & 6  & 28 \\ \hline
 \end{tabular}
  \label{singleBjetPerf}
\end{table}

The data collected with the b10 trigger has been used to tune the 
$\chi^2$ tagger ready for the activation of the $b$-jet
trigger in 2011 data-taking. 
The tuning procedure is identical for L2 and EF and 
consists mainly of a parameterization of
the transverse impact parameter resolution. 
The selection cuts applied at L2 and the EF are chosen to give the optimum overall balance of
efficiency and rejection at each level, taking into account the different 
impact parameter resolutions of the L2 and EF tracking algorithms 
(Section~\ref{sec:idReco}).
Figure~\ref{fig:tuneJp} shows the L2 transverse impact parameter significance distribution for
data, where the impact parameter is signed with respect to the jet axis.
The negative side of this distribution is mainly
due to tracks originating from light quarks decays, allowing the resolution to be studied 
using an almost pure sample of
tracks coming from the primary vertex.
A fit was made to the  negative part of the impact parameter significance distribution
using a double Gaussian function. The result of the fit is shown superimposed on the
data points in Fig.~\ref{fig:tuneJp}. The same tuning procedure was applied separately
to MC simulated data.
The $\chi^2$ probability distributions obtained using the parameterized resolution are shown
in Fig.~\ref{fig:tuneChi2} for data and simulation. Data and MC simulation show reasonable agreement, 
although there are some differences
at values of the $\chi^2$ probability close to 0 and 1.   A typical cut would be to select jets with a 
$\chi^2$ probability less than 0.07. The peak at 1 reflects the choice of setting the $\chi^2$ probability 
to 1 for RoIs that do not contain any reconstructed tracks.


\subsection{$B$-Physics}\label{sec:bphys}
\def \figurepath{.}

The ATLAS $B$-physics programme includes searches for rare $B$ hadron decays and CP violation 
measurements, as well as tests of QCD calculations through production and spin-alignment 
measurements of heavy flavour quarkonia and $B$ baryons~\cite{bib:BphysUpsilon, bib:BphysJpsi}.  
$B$-physics triggers complement 
the low-\pt\ muon triggers by providing invariant mass based 
selections for \jpsi, \Ups, and $B$ mesons.   There are two categories of $B$-physics 
triggers, \emph{topological} and \emph{single RoI seeded}, each one exploiting a different 
characteristic of the ATLAS trigger system to manage the event rates.

\subsubsection{$B$-Physics Reconstruction and Selection Criteria}

\begin{table*}[!ht]
  \begin{center}
  \caption[Bphys menu]{$B$-physics primary trigger menu and rates at a luminosity of \Lumi{32}}
    \begin{tabular}{llc}
      \hline 
      Trigger    & Motivation & Rate [Hz] \\ 
      \hline \hline
      mu4\_DiMu     &  a loose single RoI seeded trigger  & 8 \\
      mu4\_DiMu\_FS     & a loose single RoI seeded trigger using FullScan & 13  \\
      mu4\_Jpsimumu     & a single RoI seeded selection for \jpsi & 3\\
      mu4\_Upsimumu\_FS     & a single RoI seeded selection for \Ups & 3 \\
      mu4\_Bmumu     &   a single RoI seeded selection for $B$ mesons & $<$1 \\
      2mu4\_DiMu     &  a loose topological selection & 4 \\
      2mu4\_Jpsimumu     &  topological selection for \jpsi & 1  \\
      2mu4\_Upsimumu     &   topological selection for \Ups & $<$1\\
      2mu4\_Bmumu    &  topological selection for $B$ mesons & $<$1\\
      \hline \hline
    \end{tabular}
  \label{tab:Bphys1}
  \end{center}
\end{table*}

\begin{description}
\begin{sloppypar}
\item \emph{Topological triggers} require 2 muon RoIs to have been found at L1 and the HLT (see Section~\ref{sec:muon}). The 
$B$-physics algorithms in the HLT 
then combine the information from the two muon RoIs to search for the parent \jpsi, \Ups, or $B$ meson, 
and a vertex fit 
is performed for the two reconstructed ID tracks. 
The requirement for two muons at L1 reduces the rate, but is inefficient for events where the 
second muon does not give rise to a L1 RoI because it has low momentum, or falls outside the L1 acceptance.  
\end{sloppypar}
\item \emph{Single RoI seeded triggers} recover events that have been missed by the topological triggers
by starting
from a single L1 muon and finding the second muon at the HLT. In this approach, tracking is 
performed in a large region  ($\Delta\eta\times\Delta\phi=1.5\times1.5$) around the L1 muon. 
At L2, tracks found in this large RoI are
 extrapolated to the muon system.
The algorithm searches for muon hits within a road around the extrapolated track; if enough hits are found then 
the track is flagged as a muon. At the EF the search for tracks within the large RoI uses the EF Combined 
strategy (Section~\ref{sec:muonReco}) which starts from the Muon Spectrometer and then adds inner detector information.
If a second track is found, it is combined with the first one to search for the parent 
di-muon object in the same way as in the topological trigger. This approach can also be used in \emph{FullScan} (FS) mode 
(Section~\ref{sec:idReco}). The FS mode is particularly useful for triggering \Ups\ events 
where the muons tend to be separated by more than the RoI size, but requires approximately 8 times more CPU time than the RoI approach.

\end{description}

In both approaches, a series of cuts can be made on the muon pair requiring: the two muons are opposite charge; the mass cuts \jpsi: $2.5-4.3 \GeV$, \Ups: $8-12 \GeV$, $B$: $4-7 \GeV$, DiMu $> 0.5 \GeV$;  a cut on the $\chi^2$ of the reconstructed vertex. The mass cuts are very loose compared to the mass resolutions ($\sim$40 MeV and $\sim$100 MeV for \jpsi\ and \Ups\ respectively). In 2010 chains were run both with and without the opposite sign requirement and with and without a requirement on the vertex $\chi^2$.

\begin{figure}[!ht]
  \centering
  \subfigure[]{
    \includegraphics[width=0.45\textwidth]{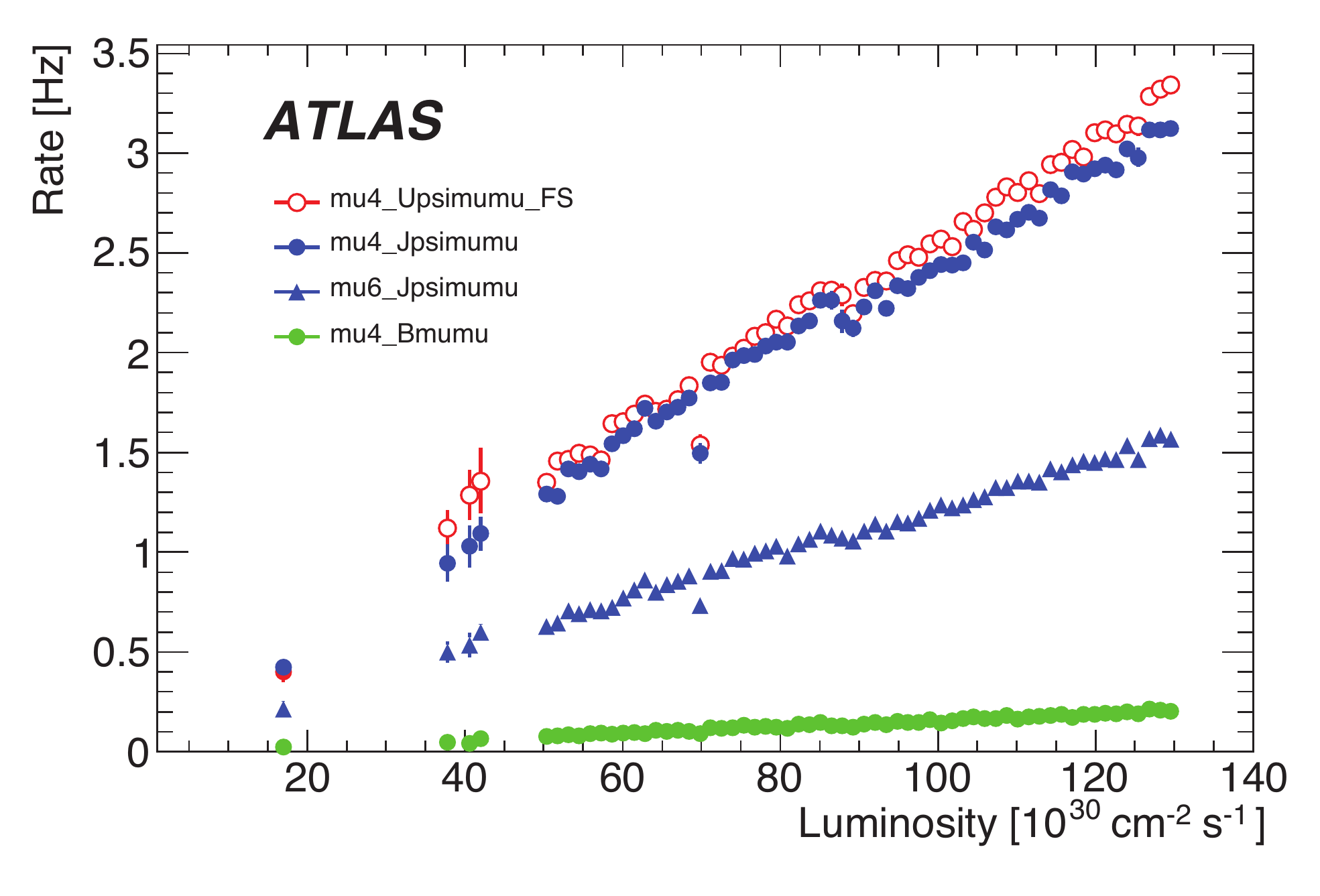}
    \label{fig:BphysRate1}
  }
  \subfigure[]{
    \includegraphics[width= 0.45\textwidth]{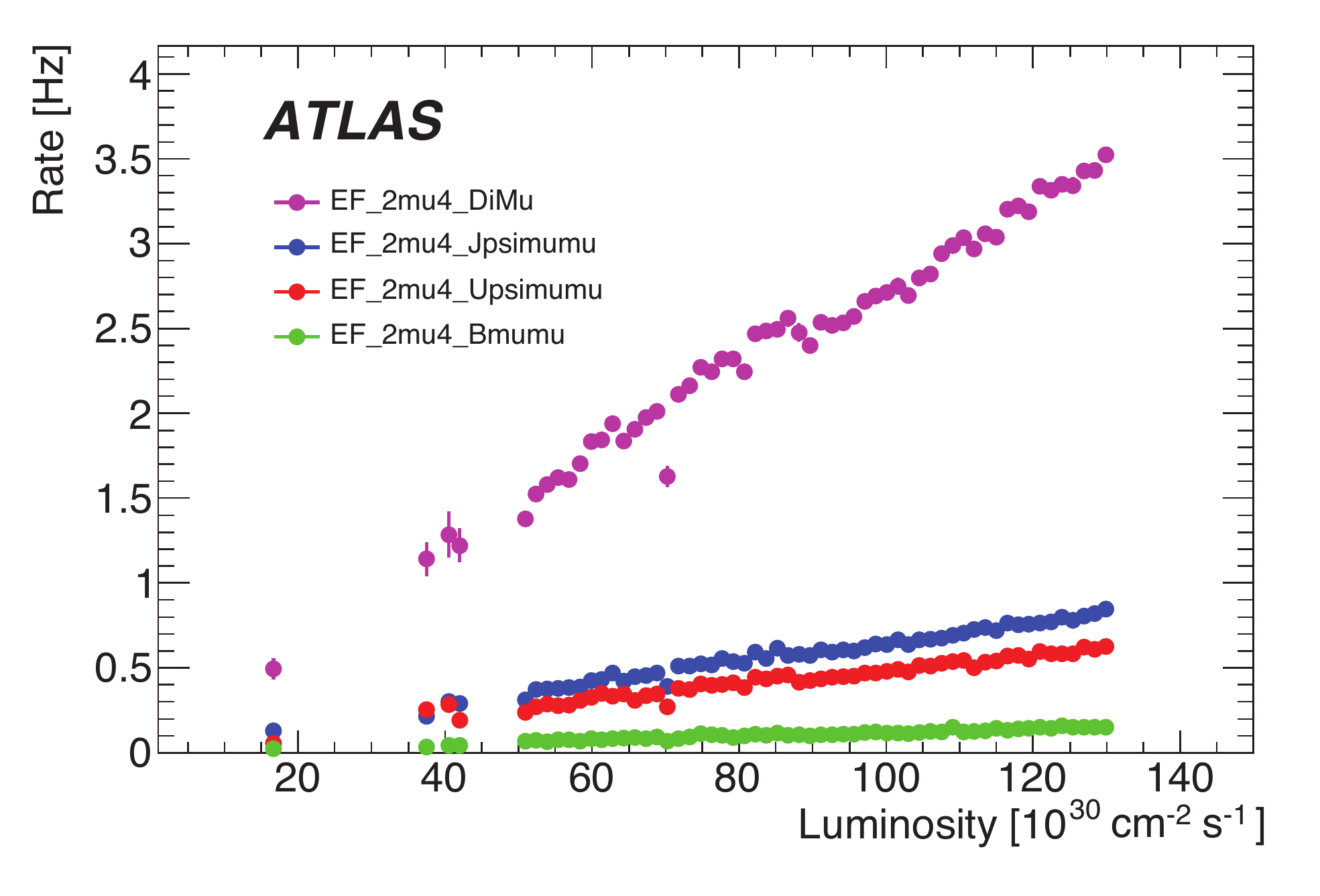}
    \label{fig:BphysRate2}
  }
  \caption{Plot of rates for several triggers for \subref{fig:BphysRate1} single RoI seeded chains and \subref{fig:BphysRate2} topological chains. }
  \label{fig:BphysRate}
\end{figure}

\subsubsection{$B$-Physics Trigger Menu and Rates}

Table~\ref{tab:Bphys1} gives an overview of the main $B$-physics triggers and their rates at a luminosity of \Lumi{32}. 
At this luminosity the mu4 trigger was prescaled by 1500 and the 2mu4 trigger was prescaled by 85.
The single muon-seeded  ``DiMu" triggers needed to be prescaled by $\sim$20; however the
topological triggers ran unprescaled.  Figure~\ref{fig:BphysRate} shows the rates for some of the triggers 
shown in Table~\ref{tab:Bphys1} as a function of instantaneous luminosity.

\subsubsection{$B$-Physics Trigger Efficiency}

\begin{figure}[!ht]
  \centering
  \subfigure[]{
     \includegraphics[width=0.45\textwidth]{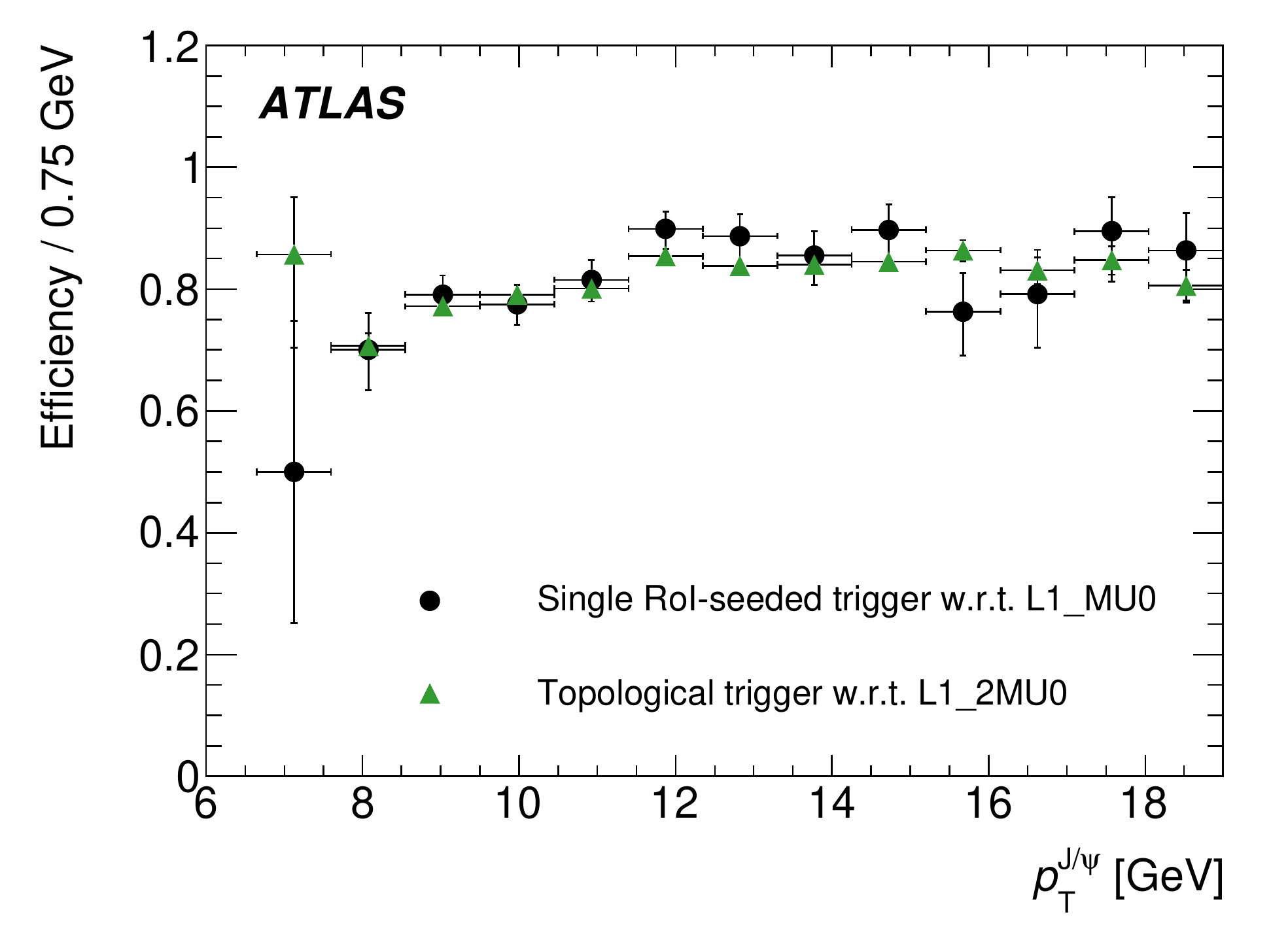}
    \label{fig:JpsiEff_wrt_L1}
  }
  \subfigure[]{
    \includegraphics[width=0.45\textwidth]{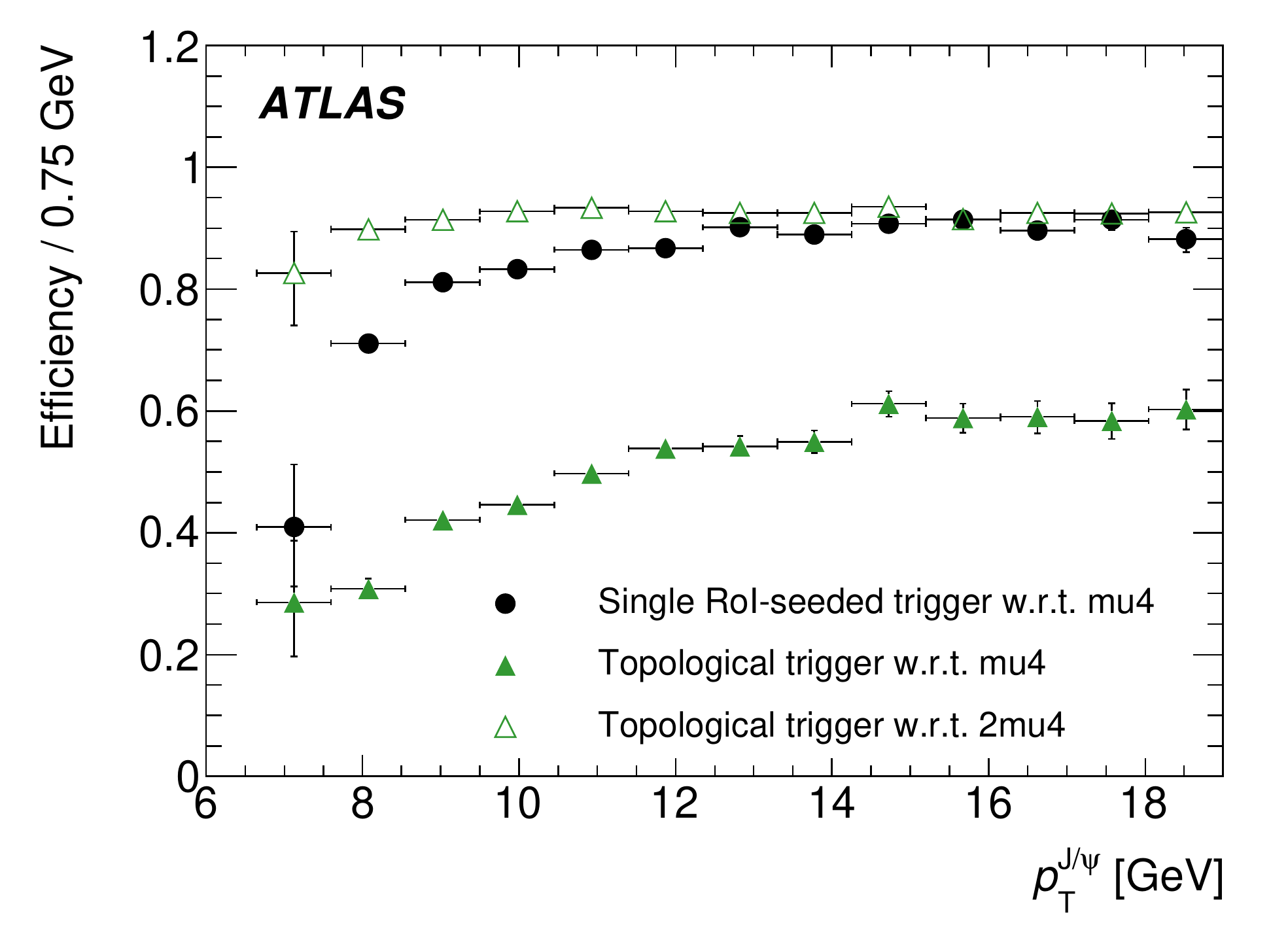}
    \label{fig:JpsiEff_wrt_mu4}
  }
  \caption{Efficiencies for \Jmumu\ events selected offline as a function of the \jpsi\  \pt\  for \subref{fig:JpsiEff_wrt_L1}  the single RoI-seeded mu4\_Jpsimumu trigger with respect to L1\_MU0 and the topological 2mu4\_Jpsimumu trigger with respect to L1\_2MU0 and 
\subref{fig:JpsiEff_wrt_mu4} the mu4\_Jpsimumu trigger with respect to the mu4 trigger and the 
2mu4\_Jpsimumu with
respect to the mu4 and 2mu4 triggers}
  \label{fig:Jpsifig}
\end{figure}

The efficiencies of the $B$-physics triggers have been measured from data using triggers in 
monitoring mode (Section~\ref{sec:commissioning}). The efficiencies of the
mu4\_Jpsimumu trigger 
with respect to L1\_MU0 and the 2mu4\_Jpsimumu trigger with respect to L1\_2MU0 
are shown in Fig.~\ref{fig:JpsiEff_wrt_L1} for events containing a \Jmumu\ decay 
reconstructed offline with 
both muon's $\pt > 4 \GeV$.   The efficiencies shown in Fig.~\ref{fig:JpsiEff_wrt_L1} include the HLT muon trigger efficiencies and
the efficiency of the subsequent \Jmumu\ selection cuts. 
The efficiencies have been determined within a systematic uncertainty of less than 1\%; 
statistical uncertainties are presented in the figures.

\begin{sloppypar}
In order to show the efficiency of
the \Jmumu\ selection itself, independent of the muon trigger, 
Fig.~\ref{fig:JpsiEff_wrt_mu4} shows the efficiency of: 
the single RoI-seeded mu4\_Jpsimumu trigger with respect to the mu4 trigger;
the topological \linebreak 2mu4\_Jpsimumu trigger with respect to the 2mu4 trigger; 
and 
the topological 2mu4\_Jpsimumu trigger with respect to the mu4 trigger.
The mu4\_Jpsimumu trigger has an efficiency of 85\% with respect to mu4 including 
the efficiency to reconstruct the second muon at the HLT,
which causes a reduction of efficiency for low \pt\ \jpsi . 
The benefit of using single RoI triggers is shown by comparing the 
mu4\_Jpsimumu trigger efficiency 
with the lower efficiency of 50\% for the 2mu4\_Jpsimumu trigger with 
respect to the mu4 trigger.
The lower efficiency of the
topological trigger results mainly from 
the requirement for a second L1 muon;
the efficiency of the 2mu4\_Jpsimumu trigger is 92\% for events with a 2mu4 trigger.
\end{sloppypar}

\section{Overall Trigger Performance}\label{sec:global}
\def \figurepath{.}
In this section the overall performance of the ATLAS trigger is presented.  Overall trigger 
performance parameters include the total rates at each trigger level, the CPU processing time per 
event, and the load on CPU resources available at L2 and EF.  To demonstrate these performance 
parameters, a run from period I was selected which took place during the last $pp$ fill of 2010 
and had instantaneous luminosities ranging from $0.85\times\Lumi{32}$ to 
$1.8\times\Lumi{32}$. This run
was 15 hours long and had an integrated 
luminosity of 6.4~\ipb.

\subsection{Total Trigger Rates}

\begin{figure}[!ht]
  \centering
  \subfigure[]{
    \includegraphics[width=0.45\textwidth]{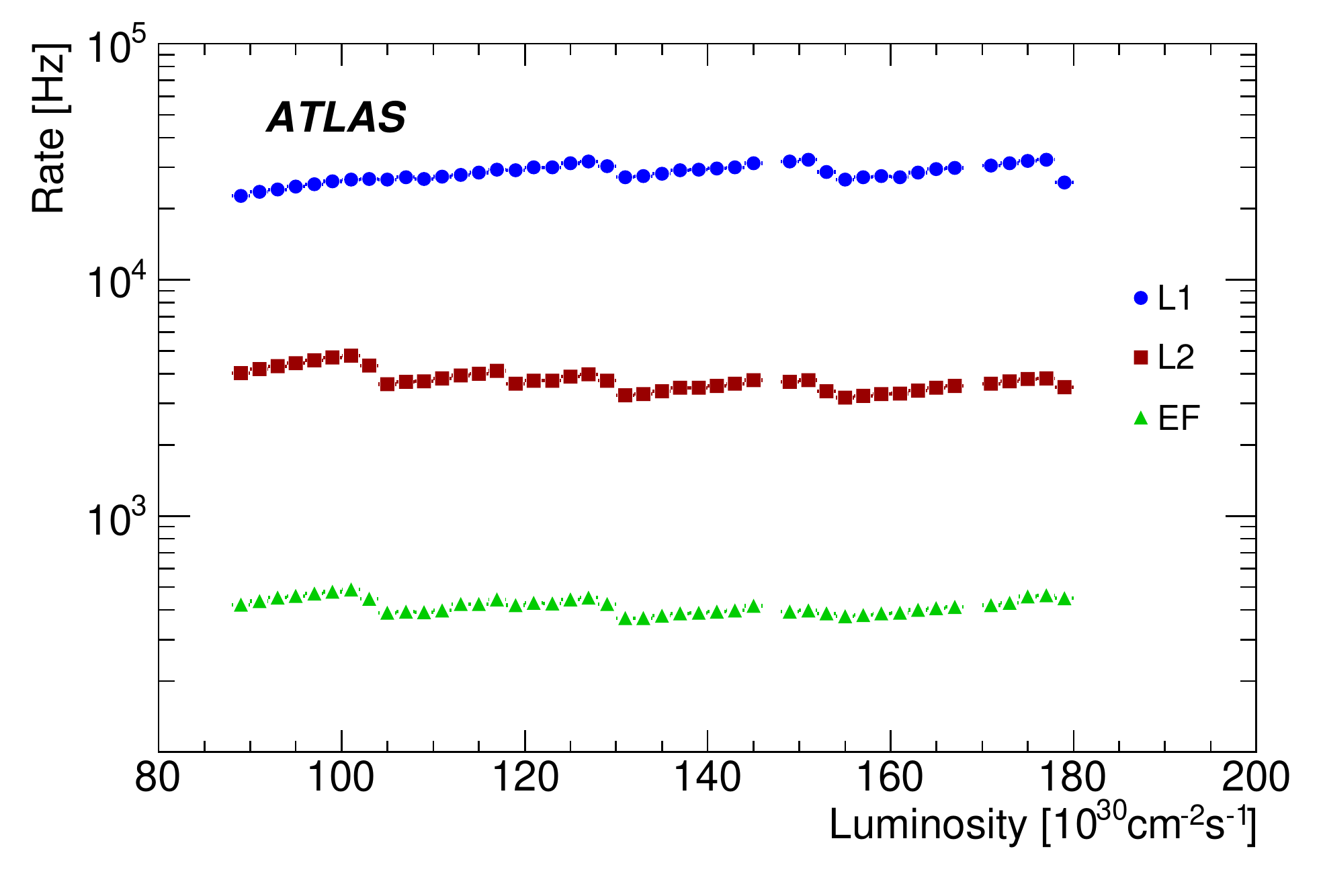}
    \label{fig:ratesTotal}
  }
  \subfigure[]{
    \includegraphics[width= 0.45\textwidth]{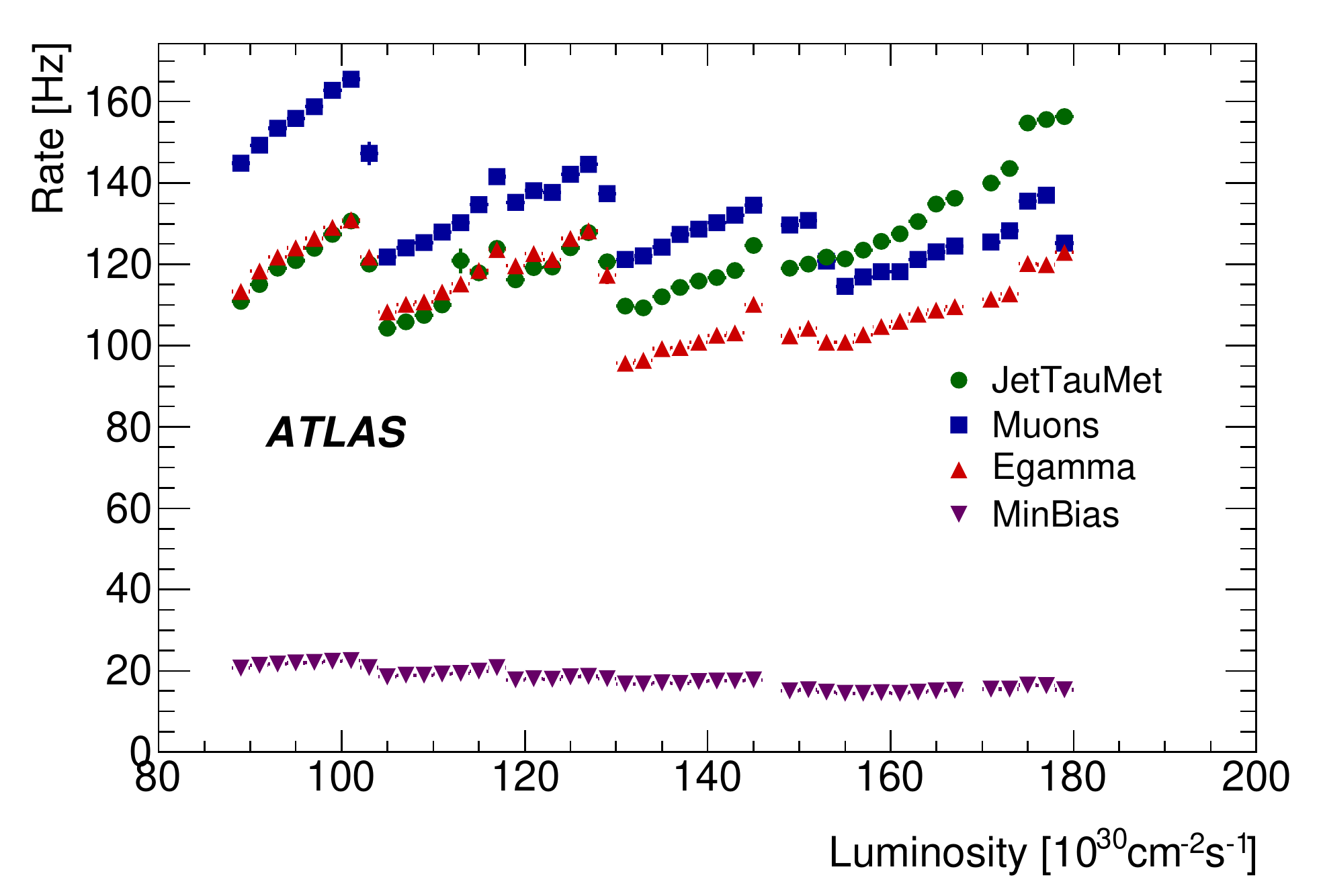}
    \label{fig:ratesStreams}
  }
  \caption{Total output trigger rates as a function of instantaneous luminosity in a sample run 
from period I for \subref{fig:ratesTotal} each trigger level and \subref{fig:ratesStreams} 
each stream. B-jet triggers are included in the JetTauEtmiss stream and B-physics triggers 
are included in the muon stream}
  \label{fig:totalRates}
\end{figure}

The total L1, L2, and EF output rates are given in Fig.~\ref{fig:ratesTotal} as a function of 
instantaneous luminosity for the sample run from period I. By changing prescale factors as 
the luminosity fell, the trigger rates were kept stable throughout the run at 
$\sim$30 kHz (L1), $\sim$4 kHz (L2), and $\sim$450 Hz (EF).  
The prescale factor changes can be seen in the figures as discontinuities in the 
rate as a function of luminosity. Prescale factors at L2 and EF are changed at the same time, 
while L1 prescale factors are set independently.
The output rates for each stream in the same run are given in Fig.~\ref{fig:ratesStreams}.  
The relative fractions of each stream are tuned as a function of instantaneous 
luminosity in order to optimize the total rate and physics yield.   

ATLAS utilizes an \emph{inclusive streaming} scheme, meaning that an event that fires a 
trigger in two different streams will be written twice, once in each stream, creating 
some overlap between different streams.  The only pairs of streams that show a significant overlap 
($>$1\%) at \Lum=\Lumi{32} are: \emph{Egamma-JetTauEtmiss} 14\%, \emph{Egamma-Muons} 2\%, 
and \emph{Muons-JetTauEtmiss} 4\%.  At higher instantaneous luminosity, when the lower \pt\ 
threshold items will have higher \linebreak prescales, the \emph{Egamma-JetTauEtmiss} overlap will 
decrease.  The goal is to keep the total overlap between streams below 10\%.

\subsection{Timing}

\begin{figure}[!ht]
  \centering
  \subfigure[]{
    \includegraphics[width=0.45\textwidth]{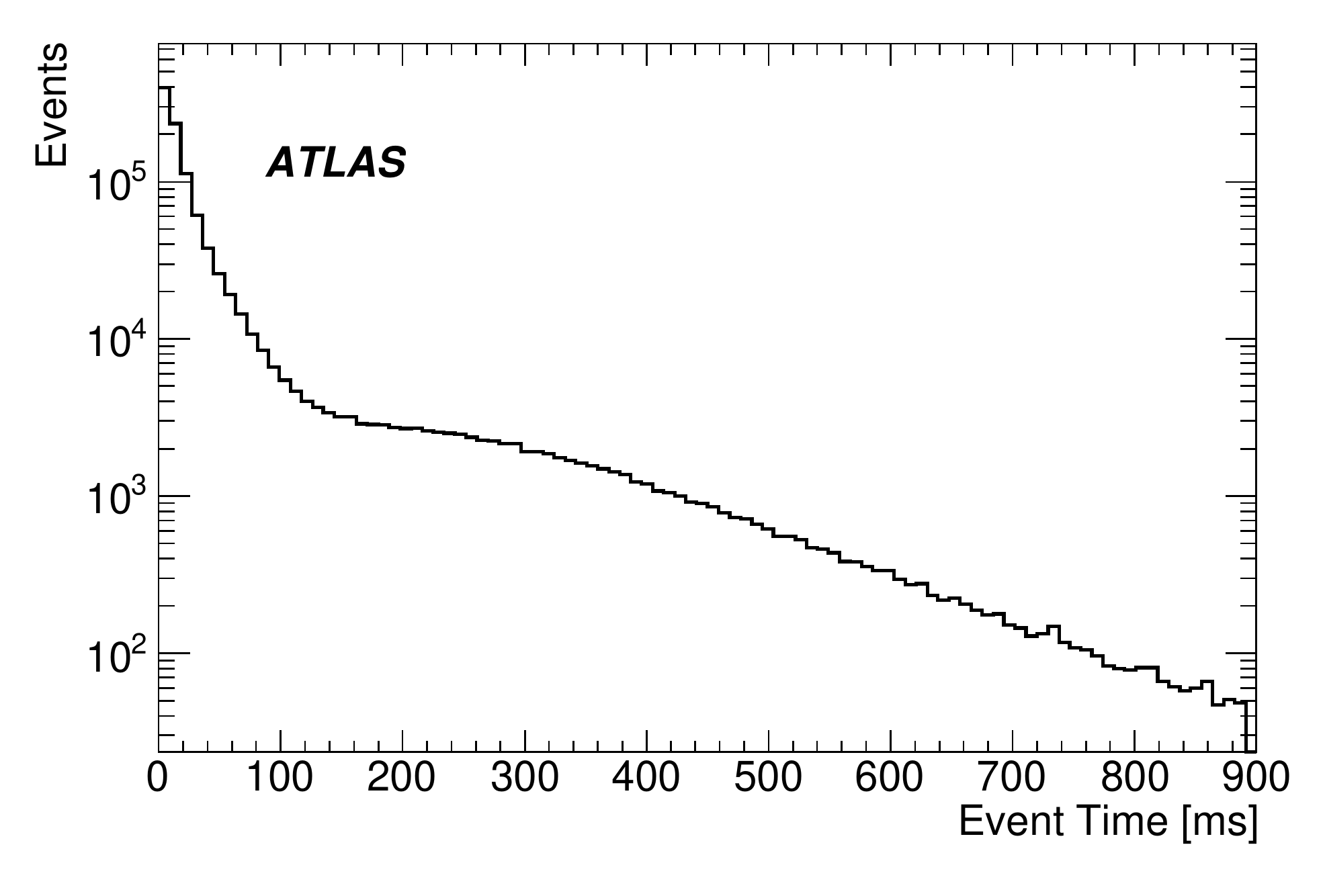}
    \label{fig:timing_L2}
  }
  \subfigure[]{
    \includegraphics[width= 0.45\textwidth]{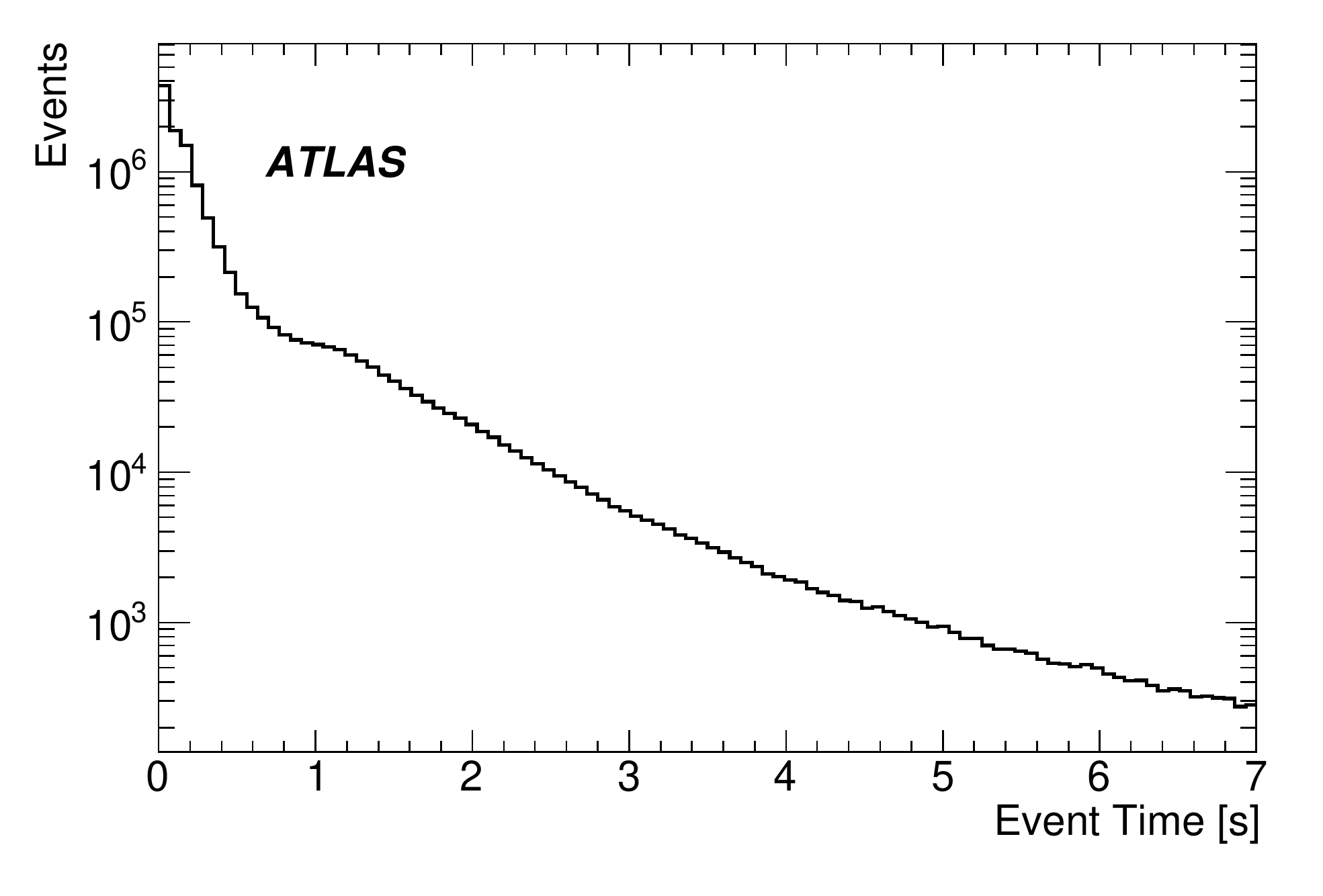}
    \label{fig:timing_EF}
  }
  \caption{ Processing time per event for \subref{fig:timing_L2} L2 and \subref{fig:timing_EF} 
EF in the sample run}
  \label{fig:timing}
\end{figure}

The timing performance of the individual algorithms has been discussed throughout the paper.   
Figure~\ref{fig:timing} shows the total processing time per event in the sample run for L2 and EF.  

\begin{figure}[!ht]
  \centering
  \subfigure[]{
    \includegraphics[width=0.45\textwidth]{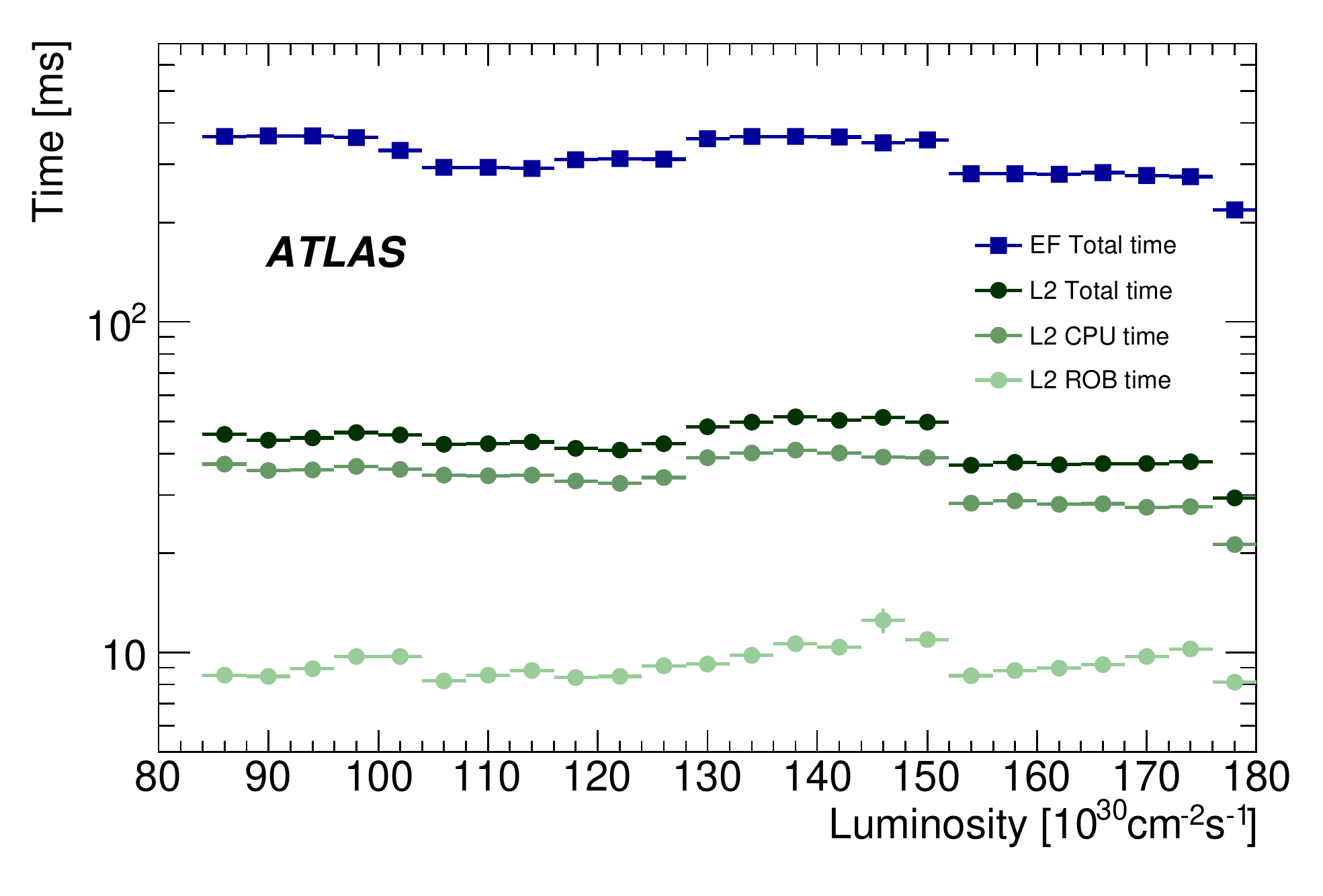}
    \label{fig:timing_a}
  }
  \subfigure[]{
    \includegraphics[width= 0.45\textwidth]{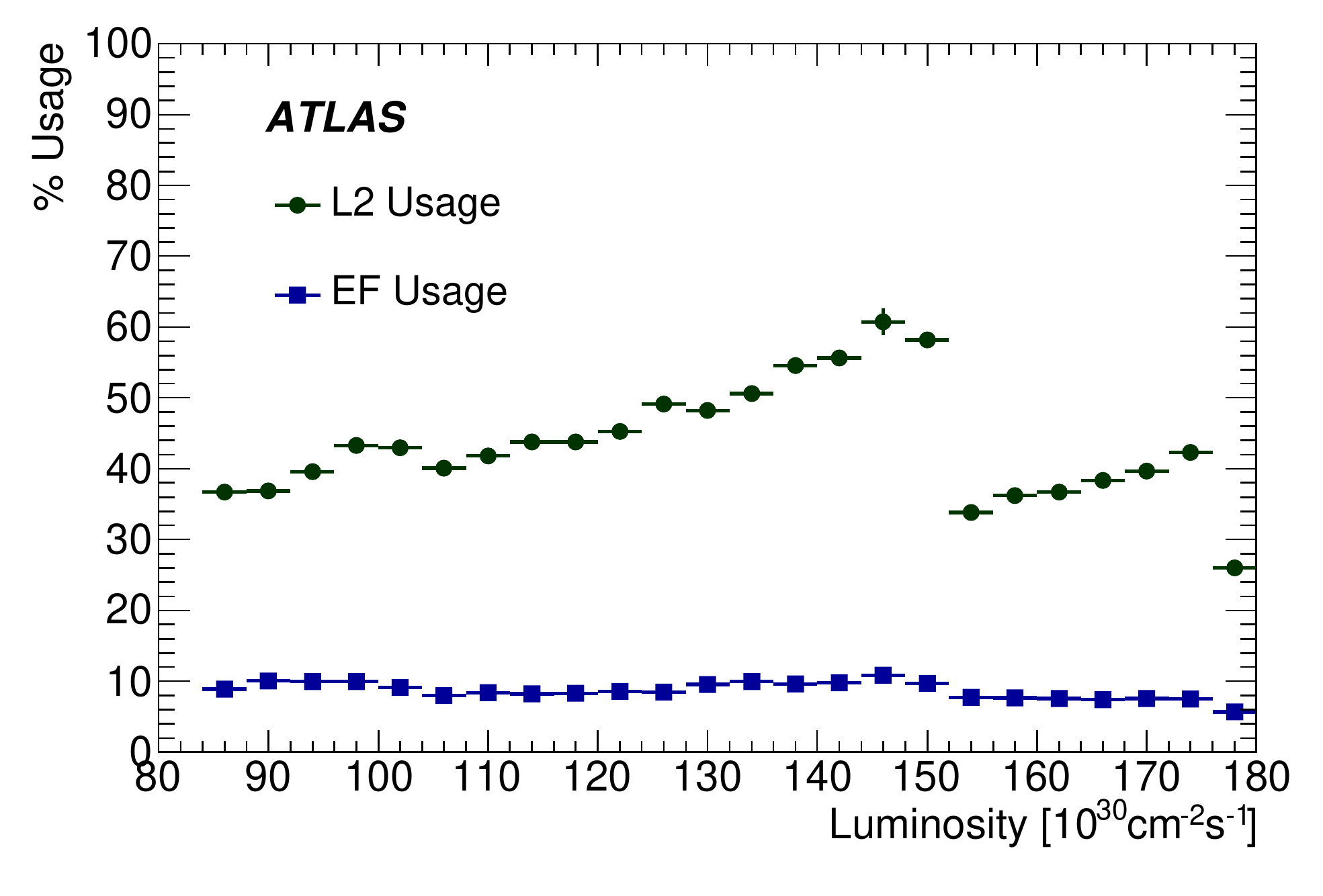}
    \label{fig:timing_b}
  }
  \caption{\subref{fig:timing_a} Mean time per event and \subref{fig:timing_b} fraction of trigger system CPU usage for L2 and EF as function of luminosity in the sample run}
  \label{fig:timing_lumi}
\end{figure}

Figure~\ref{fig:timing_a} presents the mean processing time per event at L2 and EF as a function 
of instantaneous luminosity; L2 is further subdivided into the mean time to retrieve data over 
the network from the Read out Buffers (ROB time) and the computational time taken by the algorithms 
(CPU time).  The figure shows that L2 was running close to the design limit of  $\sim40$ms and EF 
was running at $\sim400$ms, well below the design limit of $\sim4$s.   Figure~\ref{fig:timing_b}, 
reporting the fraction of CPU used in the HLT farm, shows that the HLT farm was well within its 
CPU capacity.  As was the the case for the trigger rates, discontinuities in the CPU usage with luminosity
are due to deliberate changes of prescale sets to control the trigger rate.

\begin{table*}[!ht]
  \begin{center}
  \caption{The bandwidth allocation guidelines per trigger group for 2011 for a total rate of $\sim$200Hz. 
For primary physics triggers, 
the L1 and HLT thresholds and predicted trigger rates are given for a luminosity of \Lumi{33}}
    \begin{tabular}{lcccc}
      \hline \hline
           & \multicolumn{2}{c}{$p_T$ threshold [\GeV]} & \\
      Category      & L1 & HLT & Rate [Hz] & Bandwidth [Hz]\\ 
      \hline
      {\bf minbias, zerobias} & & & &  {\bf 10} \\
      \hline
      {\bf muon} & & & &  {\bf 45} \\
      \hspace{5mm}single muon & 10 & 20 & 25 & \\ 
      \hspace{5mm}di-muon & 4 & 10 &  3 & \\
      \hline
      {\bf $B$-physics} & & & &  {\bf 15} \\
      \hspace{5mm}Jpsimumu &  &  & 8 & \\
      \hspace{5mm}Upsimumu &  &  & 4 & \\
      \hspace{5mm}Rare B decays &  &  & 3 & \\
      \hline
      {\bf \egamma} & & & &  {\bf 55} \\
      \hspace{5mm}single electron & 14 & 20 & 20 & \\
      \hspace{5mm}di-electron  & 7 & 12 & 0.5 & \\
      \hspace{5mm}electron~+~muon & 5(e),4($\mu$) & 10(e),6($\mu$) & 3.5 & \\
      \hspace{5mm}single photon & 14 & 80 & 1 & \\
      \hspace{5mm}di-photon  & 14 & 20 & 1.5 & \\
      \hline
      {\bf tau, tau+{\MET}} & & & &  {\bf 30} \\
      \hspace{5mm}single tau  & 30 & 100 & 4 & \\
      \hspace{5mm}di-tau & 11 & 29 & 4 & \\
      \hspace{5mm}tau+electron & 6($\tau$),10(e) & 16($\tau$),15($\mu$) & 3 &\\
      \hspace{5mm}tau+muon & 6($\tau$),4($\mu$) & 16($\tau$),15($\mu$) & 5 &\\
      \hline
      {\bf jet, $b$-jet, jet+{\MET}} & & & &  {\bf 55} \\
      \hspace{5mm}jet+{\MET} & 50(jet),20({\MET}) & 75(jet),45({\MET}) & 15 & \\ 
      \hspace{5mm}single jet & 75 & 250 & 3 & \\
      \hspace{5mm}single forward jet & 50 & 100 & 3 & \\
      \hspace{5mm}multi-jets (with $b$-tagging) & 10 & & 8 & \\
      \hline
      {\bf {\MET}, {\SumET}, {exotics}} & & & &  {\bf 10} \\
      \hline
      {\bf calibration \& commissioning} & & & &  {\bf 10} \\
      \hline \hline
    \end{tabular}

  \label{tab:principaltriggers2011}
  \end{center}
\end{table*}

\section{Outlook}\label{sec:outlook}
\def \figurepath{.}
The trigger menus for 2011 and 2012 running
 will cover instantaneous luminosities from \mbox{$\sim\Lumi{32}$} to \linebreak \mbox{$\sim5\cdot\Lumi{33}$} at \sTev{7} with around 10-23 \pp\ interactions per bunch crossing and a 50~ns bunch spacing.   
At these instantaneous luminosities 
the main triggers will select electrons and muons with \pt\ above about $20\GeV$, jets with \pt\ above about 
$200\GeV$,
\MET\ above about $50\GeV$, as well as \MET\ in combination with a tau or jet.  The primary triggers are shown in 
Table~\ref{tab:principaltriggers2011} together with the
L1 and HLT thresholds and predicted trigger rates for a luminosity of \Lumi{33}.

The table also shows the bandwidth allocation guidelines
for each group of triggers.
The primary triggers make up
about two thirds of the output bandwidth. The remainder of the bandwidth is filled with
supporting, commissioning, calibration, and monitoring triggers.
Supporting triggers populate the largest part of the remaining
bandwidth.  For example, prescaled jet and photon supporting triggers 
provide an approximately flat event yield as a function of \pt\ to be used for 
measurements limited by systematic uncertainties. 
  In addition, a smaller fraction of bandwidth is allocated to
commissioning triggers specifically intended for the further development of the trigger menu.
 The total number of triggers is
reduced compared to 2010 menus, as many items necessary for commissioning or
lower luminosities are removed.

In contrast to the rapid evolution in 2010, the 2011/12 LHC
conditions will be increasingly stable, and changes in the trigger menu
will be less frequent than in 2010.  Daily changes will be
limited to adjustments of \linebreak prescales, mainly for monitoring and commissioning triggers.
  To improve the stability of the data recorded for physics analysis, changes to 
primary triggers and re-tuning of
the menu is limited to monthly updates.  The trigger
will, however, continue to evolve to match LHC luminosity and beam conditions.


\section{Conclusion}\label{sec:conclusion}

The ATLAS trigger system has been commissioned and has successfully delivered
data for ATLAS physics analysis. 
Efficiencies, which meet the original design criteria, have been determined 
from data. These include overall trigger efficiencies of: greater than $99\%$ 
for electrons and photons with \linebreak
$\ET>25~GeV$; $94-96$\% for muons with $\pT>13~GeV$, in the regions of full acceptance;
greater than $90$\% for tau leptons with $\pT>30~GeV$;
greater than $99$\% for jets with 
$\ET>60~GeV$.
The missing \ET\ trigger was fully efficient above $100~\GeV$ throughout the 2010
data-taking period. 
Quantities calculated online, using fast trigger algorithms, 
show excellent agreement with those reconstructed offline. Data and simulation agree
well for these quantities and for measured 
trigger efficiencies. 

\begin{sloppypar}
The trigger system has been demonstrated to 
function well, satisfying operational requirements and
evolving to meet the demands of rapidly increasing LHC luminosity. 
Trigger menus will
continue to evolve to fulfil future demands
via progressive increase of prescales, tightening of selection cuts,
application of isolation requirements,  and increased use of
 multi-object and combined triggers. The excellent performance
of the trigger system in 2010 and the results of studies 
confirming the scaling to higher luminosities 
give confidence that the ATLAS trigger system will continue to meet 
the challenges of running in 2011 and beyond. 
\end{sloppypar}

\section*{\bf Acknowledgements}

We thank CERN for the very successful operation of the LHC, as well as the
support staff from our institutions without whom ATLAS could not be
operated efficiently.

\begin{sloppypar}
We acknowledge the support of ANPCyT, Argentina; YerPhI, Armenia; ARC,
Australia; BMWF, Austria; ANAS, Azerbaijan; SSTC, Belarus; CNPq and FAPESP,
Brazil; \linebreak NSERC, NRC and CFI, Canada; CERN; CONICYT, Chile; CAS, MOST and
NSFC, China; COLCIENCIAS, Colombia; MSMT CR, MPO CR and VSC CR, Czech
Republic; DNRF, DNSRC and Lundbeck Foundation, Denmark; ARTEMIS, European
Union; IN2P3-CNRS, \linebreak CEA-DSM/IRFU, France; GNAS, Georgia; BMBF, DFG, HGF, MPG
and AvH Foundation, Germany; GSRT, Greece; ISF, MINERVA, GIF, DIP and
Benoziyo Center, Israel; \linebreak INFN, Italy; MEXT and JSPS, Japan; CNRST, Morocco;
FOM and NWO, Netherlands; RCN, Norway; MNiSW, \linebreak Poland; GRICES and FCT,
Portugal; MERYS (MECTS), Romania; MES of Russia and ROSATOM, Russian
Federation; JINR; MSTD, Serbia; MSSR, Slovakia; ARRS and MVZT, Slovenia;
DST/NRF, South Africa; MICINN, Spain; SRC and Wallenberg Foundation,
Sweden; SER, SNSF and Cantons of Bern and Geneva, Switzerland; NSC, Taiwan;
TAEK, Turkey; STFC, the Royal Society and Leverhulme Trust, United Kingdom;
DOE and NSF, United States of America.
\end{sloppypar}

\begin{sloppypar}
The crucial computing support from all WLCG partners is acknowledged
gratefully, in particular from CERN and the ATLAS Tier-1 facilities at
TRIUMF (Canada), NDGF (Denmark, Norway, Sweden), CC-IN2P3 (France),  \linebreak
KIT/GridKA (Germany), INFN-CNAF (Italy), \linebreak NL-T1 (Netherlands), PIC (Spain),
ASGC (Taiwan), RAL (UK) and BNL (USA) and in the Tier-2 facilities
worldwide.
\end{sloppypar}

\begin{sloppypar}
\printnomenclature
\end{sloppypar}

\bibliographystyle{atlasnote}
\begin{sloppypar}
\bibliography{bibliography}
\end{sloppypar}

\onecolumn

\begin{flushleft}
{\Large The ATLAS Collaboration}

\bigskip

G.~Aad$^{\rm 48}$,
B.~Abbott$^{\rm 111}$,
J.~Abdallah$^{\rm 11}$,
A.A.~Abdelalim$^{\rm 49}$,
A.~Abdesselam$^{\rm 118}$,
O.~Abdinov$^{\rm 10}$,
B.~Abi$^{\rm 112}$,
M.~Abolins$^{\rm 88}$,
H.~Abramowicz$^{\rm 153}$,
H.~Abreu$^{\rm 115}$
% ...
\bigskip

% ...
$^{10}$ Institute of Physics, Azerbaijan Academy of Sciences, Baku, Azerbaijan\\
$^{11}$ Institut de F\'isica d'Altes Energies and Departament de F\'isica de la Universitat Aut\`onoma  de Barcelona and ICREA, Barcelona, Spain\\
% ...
$^{48}$ Fakult\"{a}t f\"{u}r Mathematik und Physik, Albert-Ludwigs-Universit\"{a}t, Freiburg i.Br., Germany\\
$^{49}$ Section de Physique, Universit\'e de Gen\`eve, Geneva, Switzerland\\
% ...
$^{88}$ Department of Physics and Astronomy, Michigan State University, East Lansing MI, United States of America\\
% ...
$^{111}$ Homer L. Dodge Department of Physics and Astronomy, University of Oklahoma, Norman OK, United States of America\\
$^{112}$ Department of Physics, Oklahoma State University, Stillwater OK, United States of America\\
% ...
$^{115}$ LAL, Univ. Paris-Sud and CNRS/IN2P3, Orsay, France\\
% ...
$^{118}$ Department of Physics, Oxford University, Oxford, United Kingdom\\
% ...
$^{153}$ Raymond and Beverly Sackler School of Physics and Astronomy, Tel Aviv University, Tel Aviv, Israel\\
% ...
\end{flushleft}

\end{document}